\RequirePackage{ifpdf}
\ifpdf 
\documentclass[pdftex]{sigma1}
\else
\documentclass{sigma1}
\fi

 \numberwithin{equation}{section}
 \numberwithin{theorem}{section}
 \numberwithin{proposition}{section}
 \numberwithin{lemma}{section}
 \numberwithin{corollary}{section}
 \numberwithin{definition}{section}
 \numberwithin{example}{section}
 \numberwithin{remark}{section}
 \numberwithin{note}{section}

 1 

\usepackage{amssymb}
\usepackage[mathscr]{euscript}
\usepackage{graphicx}
\usepackage[all,knot]{xy}
\usepackage{multicol} 
\usepackage{cmap-cyr-vf}
\usepackage{skt}

 \usepackage{yfonts}

\newtheorem{prop}{Proposition}[section]

\newtheorem{defin}{Definition}[subsection]
\newtheorem{lemm}{Lemma}[section]

\newcommand\dttF{\hbox{\sffamily\bfseries\slshape F}}

\newcommand\dttB{\hbox{\sffamily\bfseries\slshape B}}
\newcommand\dttP{\hbox{\sffamily\bfseries\slshape P}}
\newcommand\dttQ{\hbox{\sffamily\bfseries\slshape Q}}
\newcommand\dttJ{\hbox{\sffamily\bfseries\slshape J}}

\newcommand{\Ni}{\hbox{ {\vrule height .22cm}{\leaders\hrule\hskip.2cm} }}
\newcommand{\iN}{\hbox{ {\leaders\hrule\hskip.2cm}{\vrule height .22cm} }}

\newcommand{\BL}{{\Big{[}}}
\newcommand{\BR}{{\Big{]}}}
\newcommand{\bl}{{\big{[}}}
\newcommand{\br}{{\big{]}}}
\newcommand{\bee}{\begin{equation}}
\newcommand{\eee}{\end{equation}}
\newcommand{\dd}{{\hbox{d}}}
\newcommand{\vol}{\hbox{d}{\mathfrak{y}}}

\def\<{\langle} \def\>{\rangle}

\begin{document}

\allowdisplaybreaks

\renewcommand{\PaperNumber}{}


\ShortArticleName{{\sffamily\bfseries\slshape $n$-plectic Maxwell Theory}}

\ArticleName{\hbox{\sffamily\bfseries\slshape{$n$-plectic Maxwell Theory}}}

\Author{\sffamily\bfseries\slshape D. Vey\footnote{\footnotesize{{\sffamily\bfseries\slshape e-mail} :  dim.vey@gmail.com} -  {\sffamily\bfseries\slshape Last update}: v3 o9.o5.2o14} }

\AuthorNameForHeading{\sffamily\bfseries\slshape D. Vey}

{\sffamily\slshape SPHERE Laboratory - UMR 7219 
}

{\sffamily\slshape
Denis Diderot University - Paris, France}

\

\Abstract{{\footnotesize{{{\em This paper  provides a detailed treatment of the $n$-plectic Maxwell theory using the general setting developed in  the work of F. H\'elein and J. Kouneiher.  In particular we explore the DeDonder-Weyl theory, the question of algebraic and dynamical observable forms, the copolarization setting related to the good search of canonical forms.  Finally, we give  some indications for the construction of the higher Lepage-Dedecker correspondence and we emphasize some aspect of the   underlying Grassmannian viewpoint in the {\sffamily 2D}  case.}
}}
}}

\Address{
}






 {\footnotesize{
 
 \tableofcontents

}}

 \thispagestyle{empty}

\


  This paper is dedicated to the application of Multisymplectic Geometry ${\hbox{\sffamily{(MG)}} }$, also termed $n$-plectic Geometry, to field theory, in particular to the Maxwell theory \cite{max}. We apply the formalism developed by F. H\'elein and J. Kouneiher   \cite{FH-01}  \cite{H-02}   \cite{HK-01} \cite{HKHK01} \cite{HK-02}  \cite{HK-03} to the Maxwell     variational problem.
First, as a quick introduction into the subject, we draw the outline of the paper, then we make some    remarks     about ${\hbox{\sffamily{(MG)}} }$  and finally we offer some comments about  the derivation of the  Euler-Lagrange equations or Maxwell's equations.

{\em Outline}. We are interested in the   DeDonder-Weyl theory for the Maxwell equations, see section \eqref{section8}. In particular, we focus on the Maxwell multisymplectic manifold and  the related Dirac primary constraint set  \eqref{oo98}. We obtain the multisymplectic Hamilton equations \eqref{oo981}. Finally, we present the Maxwell theory in the setting of an $n$-phase space \eqref{oo982}.    In section \eqref{section9}, we investigate  observables and canonical variables for {\sffamily 4D} Maxwell theory in the ${\hbox{\sffamily{(DDW)}} }$ setting. First, we give some simple examples of algebraic $3$-forms \eqref{ooobb01},  both position and momenta that are defined on ${\pmb{\cal M}}_{\tiny{\hbox{\sffamily Maxwell}}} \subset {\pmb{\cal M}}_{\tiny{\hbox{\sffamily DDW}}}$, with their related Poisson bracket \eqref{bliocppwwEE}. The following section \eqref{glopsss} is dedicated to the  expression of all algebraic $(n-1)$-forms and their related infinitesimal (multi)symplectomorphisms, also termed $n$-plectomorphisms. This result of this search is given by   proposition \eqref{algebraicallsymplec0}. Then,   we describe dynamical observables \eqref{gliopopo} and algebraic observables in the pre-multisymplectic setting \eqref{aaaaaaap1}. We describe observable functionals in \eqref{Observvvvvv}: their kinematical  and dynamical aspects. Finally, we obtain the Poisson bracket structure. The latter is a result already written by  J. Kijowski and W. Szczyrba  \cite{KS0}  \cite{KS1}. In particular, we describe generalized positions and momenta observable $3$-forms in the pre-multisymplectic case, where the pre-multisymplectic manifold is ${\pmb{\cal M}}^{\hbox{\skt 0}}_{\tiny{\hbox{\sffamily Maxwell}}} \subset {\pmb{\cal M}}_{\tiny{\hbox{\sffamily Maxwell}}}$. In this case, the point of interest is that any algebraic $3$-form is a dynamical observable form. Finally, we offer in section \eqref{stress001} some remarks about the stress-energy tensor.   The following section  \eqref{section999} is dedicated to dynamical equations    and canonical variables. In particular, we make a   brief review of the use of   graded structures and Grassman variables in \eqref{qmapd99}. This is related to various works, for example the ones of  I.V. Kanatchikov  \cite{Kana-OZO}  \cite{Kana-01}  \cite{Kana-014}, M. Forger, C. Paufler and H. R\"{o}mer 
\cite{Forger009} \cite{Forger010}     \cite{Forger011},  F. H\'elein and J. Kouneiher \cite{HK-01} or  S. Hrabak  \cite{Hrabbak01} \cite{Hrabbak02}. In section \eqref{duysid}, we show the interplay between superforms, Grassman variables and dynamical equations, and the relationship with the {\em external bracket} that appear in F. H\'elein and J. Kouneiher \cite{HK-01}.  The section \eqref{qsqsqsq999} is dedicated to the copolarization setting developed by F. H\'elein and J. Kouneiher  in the serie  of papers  \cite{HKHK01} \cite{HK-02}  \cite{HK-03}. Finally we recall a possible polarization for the Maxwell multisymplectic manifold - already found in \cite{HKHK01} \cite{HK-02}  \cite{HK-03}. We give in section \eqref{gliopowww22}   the detailed calculation about the impossibility to   include  
$ \displaystyle
\textswab{C}^{1}_{\tiny{\hbox{\sffamily 2D}}}T^\star{\pmb{\cal M}}^{\tiny{\hbox{\sffamily Maxwell}}}   =         \dd A_{1}  \oplus \dd A_2
$
 in $ { {\dttP}}^1_{\tiny{\hbox{\sffamily 2D}}} T^\star{\pmb{\cal M}}^{\tiny{\hbox{\sffamily Maxwell}}}$, in the {\sffamily 2D}-case. This mathematical result is strongly connected to the  consideration that a given component of the  gauge potential is not observable by itself. Then, the components $A_{\mu}$ of the $1$-forms $A = A_{\mu} \dd x^{\mu}$ are not observables from which we can describe a good copolarization. However, thinking in terms of differential forms directly, we can build a canonical pair of forms $(A,{\pmb{\pi}})$ and a well-defined Poisson bracket between observable functionals related to these canonical forms, see the formula \eqref{abffebe16}.      In section \eqref{section10} we describe the   Lepage-Dedecker correspondence for Maxwell theory in the two dimensional case. We describe in details the Lepage-Dedecker correspondence as well as  the calculation of the Hamiltonian respectively in \eqref{section51} and \eqref{section52}. Then we derive the generalized Hamilton-Maxwell equations  in \eqref{section53}. Notice that in such a context, the Legendre correspondence is non-degenerate. We recall the definition of the {\em enlarged pseudofiber} and the {\em pseudofiber} - see \cite{HKHK01} \cite{HK-02}  \cite{HK-03} - in the last section \eqref{section54} and we illustrate  this notion in the context of the Maxwell {\sffamily 2D} theory.

  {{\em{Multisymplectic Geometry}}}.    Within the context of covariant canonical quantization Multisymplectic Geometry ${\hbox{\sffamily{(MG)}} }$ is a generalization of  symplectic  geometry  for field theory. It allows us to construct a general framework for the calculus of variations with several variables. Historically ${\hbox{\sffamily{(MG)}} }$ was developed in three distinct steps. Its origins are connected with the names of C. Carath\'eodory \cite{cara02}   and  H. Weyl \cite{Weyl}  on one hand and T. De Donder \cite{Donder01} \cite{Donder}  on the other. We make this distinction    since the motivations involved were different:   Carath\'eodory and later Weyl, were involved with the
generalization of the Hamilton-Jacobi equation to several variables and the line of development
stemming from their work is concerned with the solutions of variational problems in the setting ofÊ
the action functional. 
On the other hand,  Cartan \cite{cartan}  recognized   the crucial importance of developing an  invariant language for differential geometry not dependent on local coordinates. T. DeDonder carried this development further. The two approaches merged in the so-called DeDonder-Weyl ${\hbox{\sffamily{(DDW)}} }$ theory based on the  multisymplectic manifold ${\pmb{\cal M}}_{\tiny{\hbox{\sffamily DDW}}}$. The second step arose with the work of T. Lepage and P. Dedecker. As was first noticed by T. Lepage \cite{Lepage}  \cite{Lepage02}, the DeDonder-Weyl  setting  is a special case of  the more general multisymplectic theory that we refer as the Lepage-Dedecker  {\sffamily  (LD)} theory. The geometrical tools permitting a fully general treatment  were provided by  P. Dedecker \cite{KS12}  \cite{KS199}. 
 The next step was taken by the Warsaw school in the seventies which further developed the geometric setting. W. Tulczyjew \cite{Tutu} \cite{Tul}, J. Kijowski \cite{JK-01} \cite{KS0}, K. Gawedski \cite{GAGA} and W. Szczyrba \cite{ KS1} \cite{JKWMT} all formulated important steps. We find already in their work the notion of {\em algebraic form}, and in the work  of J. Kijowski \cite{JK-01} a corresponding formulation of the notion of   {\em dynamical observable} emerges. We emphasize, for the full geometrical multisymplectic approach, two fundamental points: the generalized Legendre correspondence - introduced by T. Lepage and P. Dedecker - and the  issue of observable and Poisson structure, two  cornerstones within the {\em universal} Hamiltonian formalism developed by F. H\'elein,  \cite{FH-01}  \cite{H-02}  and F. H\'elein and J. Kouneiher  \cite{HK-01} \cite{HKHK01} \cite{HK-02}  \cite{HK-03}\footnote{For an introductive   synthesis of the principal motives and results presented in their work, see D. Vey \cite{Vey01} \cite{Vey02}. }.  Hence, for field theory, we are led to think of the solutions of variational problems as  $n$-dimensional submanifolds ${\pmb{\Sigma}} $ embedded in a
$(n+k)$-dimensional manifold $\pmb{\mathfrak{Z}} $. One observes the key role of the Grassmannian bundle as the analogue of the tangent bundle for variational problems for field theory.

  {{\em{Maxwell Theory}}}.   Let us recall the expression of the Euler-Lagrange equations for the Maxwell theory. We are first interested with the vacuum Maxwell  action, given by:
\begin{equation}\label{YM01}
(\mathfrak{i}) \quad 
{\pmb{\cal L}}^{\circ}_{\tiny{\hbox{\sffamily Maxwell}}} = {1\over 2} \int_{{\cal X}}     \dttF \wedge \star \dttF, 
\quad     
\quad  (\mathfrak{ii}) \quad 
{\pmb{\cal L}}^{\circ}_{\tiny{\hbox{\sffamily Maxwell}}}[x,A,\hbox{d}A] = - {1\over 4} \int_{{\cal X}}     \dttF_{\mu\nu} \dttF^{\mu\nu}  \sqrt{-g} \hbox{d}  \mathfrak{y}.
\end{equation}
The Lagrangian density is $\displaystyle {{L}}(A) =  - (1/4) \dttF_{\mu\nu} \dttF^{\mu\nu}  \sqrt{-g} $. We denote $\displaystyle \hbox{vol}_{\cal X}(g)$ a Riemannian volume form such that $\displaystyle \hbox{vol}_{\cal X}(g) = \sqrt{-g} \hbox{d}^{4} x$.
In the case where ${\cal X}$ is the Minkowski space-time we obtain then $\displaystyle \sqrt{-g} = 1$ and then $\displaystyle   {{L}} (A) =  -  (1/4) \dttF_{\mu\nu} \dttF^{\mu\nu} $.  We have  Maxwell vacuum equations:\footnote{With a matter  current   $ \dttJ^{\mu}(x) $ over ${\cal X}$  the Lagrangian is written: 
$ \displaystyle
{\cal L}_{\tiny{\hbox{\sffamily Maxwell}}}[x,A,\hbox{d}A] = {\cal L}^{\circ}_{\tiny{\hbox{\sffamily Maxwell}}}[x,A,\hbox{d}A] - \dttJ^{\mu}(x) A_{\mu},
$, and the Euler-Lagrange equations are written: 
$ \displaystyle
\hbox{d} \dttF  = 0$ and $\displaystyle   \hbox{d} \star \dttF  = \star \dttJ $.}
\begin{equation}
\hbox{d} \dttF = 0 \quad   \quad \quad \hbox{and}   \quad \quad \quad \hbox{d} \star \dttF  = 0.
\end{equation}
The curvature form is written: 
$ \displaystyle
\dttF  = \frac{1}{2} \dttF_{\mu\nu}  \dd x^{\mu} \wedge \dd x^{\nu}$, with $ \dttF_{\mu\nu} = \partial_{\mu} {A}_{\nu}   - \partial_{\nu} {A}_{\mu}$. Thus, we write the Hodge star $\star \dttF$ in components  $\displaystyle {\big{(}} \star \dttF  {\big{)}}_{\rho\sigma} = \frac{1}{2!}  {{\pmb{\varepsilon}}^{\mu\nu}}_{\rho\sigma} \dttF_{\mu\nu} $:
\[
\left.
\begin{array}{rcl}
\displaystyle    \star \dttF     & = &  \displaystyle  \star {\big{(}}  \frac{1}{2!} \dttF_{\mu\nu} \dd x^{\mu} \wedge \dd x^{\nu} {\big{)}}
=    \frac{1}{2!}  {\big{(}} \star \dttF {\big{)}}_{\rho\sigma} \dd x^{\rho} \wedge \dd x^{\sigma} = \frac{1}{2!}  \frac{1}{2!}  { {\pmb{\varepsilon} }^{\mu\nu}}_{\rho\sigma} \dttF_{\mu\nu} \dd x^{\rho} \wedge \dd x^{\sigma}
\\
 \displaystyle     & = &  \displaystyle   
\frac{ \sqrt{-g} }{4}
 {{ {\varepsilon}}^{\mu\nu}}_{\rho\sigma} \dttF_{\mu\nu} \dd x^{\rho} \wedge \dd x^{\sigma}
=
 \frac{ \sqrt{-g} }{4} g_{\alpha\mu} g_{\beta\nu}
 {{ {\varepsilon}}^{\mu\nu}}_{\rho\sigma} \dttF^{\alpha\beta} \dd x^{\rho} \wedge \dd x^{\sigma}.
\end{array}
\right.
\]
 We see the equivalence between   \eqref{YM01}$(\mathfrak{i})$ and \eqref{YM01}$(\mathfrak{ii})$, with\footnote{ $ ^{\lceil}     $  
 {\em Proof} We make a straightforward computation which involves the  Hodge duality.
\[
\left.
\begin{array}{rcl}
\displaystyle   \dttF  \wedge \star \dttF     & = &  \displaystyle
  \BL \frac{1}{2} \dttF_{\lambda\varsigma} \dd x^{\lambda} \wedge \dd x^{\varsigma} \BR \wedge 
 \BL   \frac{ \sqrt{-g} }{4} g_{\alpha\mu} g_{\beta\nu}
 {{ {\varepsilon}}^{\mu\nu}}_{\rho\sigma} \dttF^{\alpha\beta} \dd x^{\rho} \wedge \dd x^{\sigma}  \BR
\\
 \displaystyle     & = &  \displaystyle \frac{1}{8}
 \BL  \dttF_{\lambda\varsigma}  \sqrt{g} g_{\alpha\mu} g_{\beta\nu}
 {{ {\varepsilon}}^{\mu\nu}}_{\rho\sigma} \dttF^{\alpha\beta}  \BR 
\dd  x^{\lambda} \wedge \dd x^{\varsigma} \wedge \dd x^{\rho} \wedge \dd x^{\sigma}.
\end{array}
\right.
\]
 Since $\displaystyle  \hbox{vol}_{\cal X}(g) = \sqrt{g} \hbox{d}\mathfrak{y} = \frac{1}{4!} {\pmb{\varepsilon}}_{\lambda\varsigma\rho\sigma} \dd  x^{\lambda} \wedge
\dd  x^{\varsigma} \wedge
\dd x^{\rho} \wedge
\dd x^{\sigma} 
$, we obtain:
\[
\left.
\begin{array}{rcl}
\displaystyle    \dttF \wedge \star \dttF     & = &  \displaystyle  \frac{1}{8} \BL \dttF_{\lambda\varsigma}  \sqrt{g} g_{\alpha\mu} g_{\beta\nu}
 {{ {\varepsilon}}^{\mu\nu}}_{\rho\sigma} \dttF^{\alpha\beta}  \BR 
\dd  x^{\lambda} \wedge \dd x^{\varsigma} \wedge \dd x^{\rho} \wedge \dd x^{\sigma}
=   \frac{1}{8} \dttF_{\lambda\varsigma}    \dttF^{\alpha\beta}  
 { {\pmb{\varepsilon}} }_{\alpha\beta\rho\sigma}   { \pmb{\varepsilon} }^{\lambda\varsigma\rho\sigma}  { \sqrt{g} }   \hbox{d}\mathfrak{y}   
 \\
\displaystyle     & = &  \displaystyle  -  \frac{1}{2} \delta_{\alpha}^{[\lambda} \delta_{\beta}^{\varsigma]}  \dttF_{\lambda\varsigma} \dttF^{\alpha\beta}    { \sqrt{g} }   \hbox{d}\mathfrak{y}   = -  \frac{1}{2} \frac{1}{2}  \BL \dttF_{\alpha\beta} \dttF^{\alpha\beta}
- \dttF_{\beta\alpha} \dttF^{\alpha\beta} \BR   { \sqrt{g} }   \hbox{d}\mathfrak{y}  = - \frac{1}{2}  \dttF_{\alpha\beta} \dttF^{\alpha\beta}
  { \sqrt{g} }   \hbox{d}\mathfrak{y}  
\end{array}
\right.
\]
where we have used the identity  $ { {\pmb{\varepsilon}} }_{\alpha\beta\rho\sigma}  { \pmb{\varepsilon} }^{\lambda\varsigma\rho\sigma} =   -    2 !  2 ! \delta_{\alpha}^{[\lambda} \delta_{\beta}^{\varsigma]} $ in the last line. $       \rfloor $} $  \displaystyle \dttF \wedge \star \dttF   = -  (1/2)  \dttF_{\alpha\beta} \dttF^{\alpha\beta}
  { \sqrt{g} }   \hbox{d}\mathfrak{y}  $. The Euler-Lagrange equations for the Maxwell theory  are written \eqref{fkqm21}
 \bee\label{fkqm21}
 \underbrace{  \frac{\partial}{\partial A_\nu} {\cal L}_{\tiny{\hbox{\sffamily Maxwell}}} }_{(\mathfrak{i})} =  \underbrace{  \partial_\mu {\Big{(}} \frac{\partial}{ \partial (\partial_\mu A_\nu) } {\cal L}_{\tiny{\hbox{\sffamily Maxwell}}} {\Big{)}} }_{(\mathfrak{ii})} .
 \eee
We recover the Maxwell's equations\footnote{$ ^{\lceil}  $  {\em Proof} We compute $ (\mathfrak{i}) $ and $(\mathfrak{ii}) $.  The first term is  
$ \displaystyle
 (\mathfrak{i}) =  \frac{\partial{\cal L}}{\partial A_\nu} = \dttJ^{\nu} (x)
$.
The second leads to the following calculation:
\[
\left.
\begin{array}{rcl}
\displaystyle         (\mathfrak{ii})  & = &  \displaystyle  \partial_\mu {\Big{(}}   - \frac{1}{4}  \frac{(\dttF_{\alpha\beta} \dttF^{\alpha\beta} )}{\partial (\partial_\mu A_\nu) } {\Big{)}}  = - \frac{1}{4} \partial_\mu  {\Big{(}}   \frac{\partial \dttF_{\alpha\beta} }{\partial (\partial_\mu A_\nu) }   \dttF^{\alpha\beta} + \dttF_{\alpha\beta} \frac{ \partial \dttF^{\alpha\beta} }{ \partial (\partial_\mu A_\nu) } {\Big{)}}   
  \\
 \displaystyle     & = &  \displaystyle  
   - \frac{1}{4} \partial_\mu  {\Big{(}}   \frac{\partial \dttF_{\alpha\beta} }{\partial (\partial_\mu A_\nu) }   \dttF^{\alpha\beta} + \dttF_{\alpha\beta} \frac{  \partial (g^{\alpha\sigma} g^{\beta\rho} \dttF_{\sigma\rho}) }{ \partial (\partial_\mu A_\nu) } {\Big{)}}  
     - \frac{1}{4} \partial_\mu  {\Big{(}}   \frac{\partial \dttF_{\alpha\beta} }{\partial (\partial_\mu A_\nu) }   \dttF^{\alpha\beta} + \dttF^{\sigma\rho} \frac{  \partial (  \dttF_{\sigma\rho}) }{ \partial (\partial_\mu A_\nu) } {\Big{)}}  
   \\
\displaystyle     & = &  \displaystyle
       - \frac{1}{2}  \partial_\mu  {\Big{(}}   \frac{\partial \dttF_{\alpha\beta} }{\partial (\partial_\mu A_\nu) }   \dttF^{\alpha\beta} {\Big{)}}  
       =   - \frac{1}{2} \partial_\mu {\Big{(}} {\big{(}}   \delta^{\mu}_{\alpha}   \delta^{\nu}_{\beta}  -   \delta^{\nu}_{\alpha}   \delta^{\mu}_{\beta}  {\big{)}}  \dttF^{\alpha\beta} {\Big{)}} =  - \frac{1}{2} \partial_\mu { {(}} \dttF^{\mu\nu} - \dttF^{\nu\mu} { {)}}  = - \partial_\mu \dttF^{\mu\nu} .
\end{array}
\right.
 \]
   Then, we obtain Maxwell's equations \eqref{ririri005}.  $  \rfloor$}: 
 \begin{equation}\label{ririri005}
\dttJ^{\nu} (x) = - \frac{\partial}{\partial x^{\mu}} \dttF^{\mu\nu} (x).
\end{equation}

\section{\hbox{\sffamily\bfseries\slshape{Multisymplectic DeDonder-Weyl-Maxwell theory}}}\label{section8}

First, we describe the geometrical setting and precise the notations for the four dimensional case. We consider ${\cal X}$ to be the spacetime manifold with $\hbox{\sffamily dim}({\cal X}) = n = 4$.
Let $ A \in   T^\star {\cal X}$, be the potential $1$-form.  The space of interested is   $\pmb{\mathfrak{Z}} =    T^\star {\cal X}$.  As noticed in \cite{HKHK01} \cite{HK-02}, the more naive approach is to work in a  local trivialization  of a bundle over ${\cal X}$, since  a connection is not a section of a bundle. This is the chosen path here. 
A point $(x,A)$ in $\pmb{\mathfrak{Z}}$  is in the {\em position} configuration space. Any choice $(x,A)$ is equivalent to the data of an $n$-dimensional submanifold in ${\pmb{\mathfrak{Z}}} = T^\star {\cal X}  \overset{\pi}{\longrightarrow}  {\cal X}$ described as a section of the fiber bundle over ${\cal X}$. Let us consider the map     
${\mathfrak{z}}_{A}: {\cal X} \rightarrow {\pmb{\mathfrak Z}}  =  T^\star {\cal X}$ described by \eqref{022h},
\begin{equation}\label{022h}
{\mathfrak{z}}_{A}:
\left\{
\begin{array}{ccl}
\displaystyle  {\cal X} & \rightarrow & \displaystyle {\pmb{\mathfrak Z}}  =    T^\star {\cal X}
\\ 
\displaystyle x & \mapsto & \displaystyle A(x) = A_\mu(x) \dd x^{\mu}.
\end{array}
\right.
\end{equation}
which is simply some section of the related bundle. We associate with $A$, the bundle $\pmb{\cal P}^{A} = A^\star T \pmb{\mathfrak{Z}}  \otimes_{\pmb{\mathfrak{Z}} }   T^\star {\cal X} $.  The useful quantities to describe $\hbox{d}A$   the differential of the map $A$ as sections of the bundle ${\pmb{\cal P}}^{A}$ over ${\cal X}$ are now introduced.  We denote the exterior covariant derivative on the $1$-form $A$ by $ \hbox{d}^{\bf D}A$:
\begin{equation}
\left.
\begin{array}{ccc}
\displaystyle  \hbox{d}^{\bf D}A  & = &   \big{[}   {\hbox{d}}^{\bf D}  A \big{]}_{\mu\nu}  \dd x^{\mu} \wedge \dd x^{\nu},  
\end{array}
\right.
\quad \quad 
\hbox{with}
\quad \quad 
\big{[}   {\hbox{d}}^{\bf D}  {A} \big{]}_{\mu\nu}  = \partial_{[\mu} {A}_{\nu]}.
\end{equation}
We denote $  v_{\mu \nu}   =  \partial_{\mu} {A}_{\nu}  $ so that $  {\hbox{d}}^{\bf D}  A =   { v}_{[\mu \nu]}$. The space of interest, the analogue for tangent space  is   $ \Lambda^n T_{(x,e,\omega)}  {\pmb{\mathfrak Z}}  $, the fiber bundle of $n$-vector fields of $ {\pmb{\mathfrak Z}} $ over ${\cal X}$. For any $(x^\mu,A_\nu) \in   \pmb{{\mathfrak Z}}$, the fiber $ \Lambda^n T_{(x,A)}(   T^\star{\cal X}) = \Lambda^n T_{(x,A)} \pmb{\mathfrak Z} $ can be identified with $ {\pmb{\cal P}} =  A^\star T \pmb{\mathfrak{Z}} \otimes_{\pmb{\mathfrak{Z}} }   T^\star {\cal X} $ via the diffeomorphism:
 \begin{equation}\label{diff}
\left.
\begin{array}{ccc}
\displaystyle {\pmb{\cal P}} \cong  A^\star T \pmb{\mathfrak{Z}} \otimes_{\pmb{\mathfrak{Z}}}   T^\star {\cal X} & \rightarrow & \displaystyle \Lambda^n T_{(x,A)}(   T^\star {\cal X}) 
\\ 
\displaystyle  \sum_{\mu, \nu}  \big{[}   {\hbox{d}}^{\bf D}  A \big{]}_{\mu\nu}  \dd x^{\mu} \otimes \dd x^{\nu}   & \mapsto & \displaystyle  z = z_1 \wedge ... \wedge z_n,
\end{array}
\right.
\end{equation}
where $\displaystyle \forall 1 \leq \eta \leq n $,  $\displaystyle  z_\eta =   \frac{\partial}{\partial x^\alpha} + \sum_{1 \leq \beta \leq n}  {v}_{\alpha\beta}  \frac{\partial}{\partial A_\beta}  
$.
 In order to fix ideas we stress that  we have local coordinates $(x^\mu,A_\mu,  )$ for the configuration  bundle $\pmb{\mathfrak{Z}}$. The data of the local coordinates $(x^\mu,A_\mu  , {v}_{\mu \nu}  )$ - or equivalentely $(x^\mu,A_\mu, z_{\mu \nu}  )$- can be thought as coordinates on $\pmb{\cal P} $ or $ \Lambda^n T_{(x,e,\omega)}(\pmb{\mathfrak{Z}}) $. We identify  $\pmb{\cal P}  \cong \Lambda^n T_{(x,e,\omega)}(\pmb{\mathfrak{Z}}) $.

In this section we develop three points. First, we describe the setting of the ${\hbox{\sffamily{(DDW)}} }$-Maxwell theory,  in section \eqref{oo98}. In particular we consider the Dirac primary constraint set and the related Maxwell multisymplectic manifold ${\pmb{\cal M}}_{\tiny{\hbox{\sffamily Maxwell}}}$,  see \eqref{oo98}. Then we derive the generalized Hamilton equations respectively in the multisymplectic \eqref{oo981} and in the pre-multisymplectic  \eqref{oo982}  settings. In the latter, we observe the connection with the covariant phase space.

\subsection{\hbox{\sffamily\bfseries\slshape{Multisymplectic DeDonder-Weyl-Maxwell theory}}}\label{oo98}

The generalized  Legendre correspondence   is constructed on ${\pmb{{\cal M}}} = \Lambda^n T^\star  {\pmb{\mathfrak{Z}}}$.  For all  $  (q,p) \in {\pmb{\cal M}}$ we introduce the local coordinates on the bundle ${\pmb{{\cal M}}}$.    Let us denote  $\displaystyle (q^{\pmb{\mu}})_{1 \leq {\pmb{\mu}} \leq 2n}$ the local coordinates on  $ {\pmb{\mathfrak{Z}}} =  T^{\star}{\cal{X}}$  and $p_{{\pmb{\mu}}_1 ... {\pmb{\mu}}_n} $ the local coordinates on $\Lambda^{n} T^{\star}_{q}  {\pmb{\mathfrak{Z}}}$ in the basis $(\dd q^{{\pmb{\mu}}_1} \wedge \cdots \wedge \dd q^{{\pmb{\mu}}_n} )_{1 < {{\pmb{\mu}}_1}  < \cdots < {{\pmb{\mu}}_n} < 2n} $, completely antisymmetric in $({\pmb{\mu}}_1 \cdots {\pmb{\mu}}_n)$. The canonical Poincar\'e-Cartan $n$-form is written in local coordinates (here $n=k=4$):
\bee
{\pmb{\theta}} = \sum_{1 < {\pmb{\mu}}_1 < \cdots < {\pmb{\mu}}_n < 2n} p_{{\pmb{\mu}}_1 ... {\pmb{\mu}}_n} \dd  q^{{\pmb{\mu}}_1} \wedge ... \wedge \dd q^{{\pmb{\mu}}_n}.
\eee
We consider the  $\hbox{\sffamily{(DDW)}}$  submanifold  ${\pmb{{\cal M}}}_{\tiny{\hbox{\sffamily DDW}}} \subset {\pmb{{\cal M}}}$, described by:  
\[
{\pmb{{\cal M}}}_{\tiny{\hbox{\sffamily DDW}}} =  {\Big{\{}} (x,{A},p) / x \in {\cal X},  {A} \in  T^\star {\cal X}  ,p \in \Lambda^n T^\star ( T^\star{\cal X})  \quad \hbox{such that} \quad \partial_{{A}_\mu } \wedge \partial_{{A}_\nu } \iN p = 0   {\Big{\}}}.
\]
We   restrict and adapt our notations to the case   ${\pmb{{\cal M}}}_{\tiny{\hbox{\sffamily DDW}}} \subset {\pmb{{\cal M}}}$. All the components $p_{{\pmb{\mu}}_1 \cdots {\pmb{\mu}}_n}$ are taken equal to zero, excepted for 
$
p_{1 ... n } = \mathfrak{e} $ and for the multimomenta $  p_{1 ... (\nu-1) ({A}_\mu ) (\nu+1) ... n }$ denoted ${\pmb{\pi}}^{{A}_{\mu}\nu}  $.  We  define a Legendre correspondence:
\bee
\Lambda^n T( T^\star{\cal X})  \times \Bbb{R}= \Lambda^n T  {\pmb{\mathfrak{Z}}}  \times \Bbb{R} \  {\pmb{\leftrightarrow}}   \ \Lambda^n T^\star (  T^\star{\cal X}) = \Lambda^n T^\star   {\pmb{\mathfrak{Z}}}   : 
(q,v,w)   {\pmb{\leftrightarrow}}   (q,p),
\eee
which is generated by the function
$
{\cal W }:  \Lambda^nT   {\pmb{\mathfrak{Z}}}  \times \Lambda^nT^\star   {\pmb{\mathfrak{Z}}}   \rightarrow \Bbb{R}   
(q,v,p)  \mapsto \langle p , v \rangle - { L} (q,v) $.

 $\hbox{\em{Maxwell $n$-plectic manifold}}$ ${\pmb{\cal M}}_{\tiny{\hbox{\sffamily Maxwell}}}$.    Let us describe  the general construction for the De Donder-Weyl multisymplectic manifold. We consider  ${\pmb{\theta}}^{\tiny{\hbox{\sffamily DDW}}}_{(q,p)}$,  the  Poincar\'e-Cartan $n$-form: 
\begin{equation}\label{thetaDDWMmax}
{\pmb{\theta}}^{\tiny{\hbox{\sffamily DDW}}}_{(q,p)}:= \mathfrak{e}  \hbox{d} \mathfrak{y}  + {\pmb{\pi}}^{{A}_{\mu}\nu}
  \hbox{d} {A}_{\mu} \wedge   \hbox{d} \mathfrak{y}_{\nu}. 
\end{equation}
where $  \hbox{d} \mathfrak{y}= \dd x^1  \wedge ... \wedge \dd  x^n$ is a volume $n$-form defined on ${\cal X}$ and we also denote $\hbox{d} \mathfrak{y}_{\beta}:= \partial_{\beta} \iN   \hbox{d} \mathfrak{y}$. Due to the Legendre correspondence  construction,  the equivalence relation between $(q,v)$ and $(q,p)$ is written: 
\begin{equation}\label{ririri004}
(q,v)   {\pmb{\leftrightarrow}}  (q,p) \quad \quad \Longleftrightarrow \quad \quad \frac{\partial \langle p , v  \rangle  }{\partial v} =  \frac{\partial L(q,v)  }{\partial v}.
\end{equation}
The term $\langle p , v  \rangle$ is understood as the following expression  $\langle p , v  \rangle = {\pmb{\theta}}^{\tiny{\hbox{\sffamily DDW}}}_{p} (\cal Z) $. With ${\cal Z} = {\cal Z}_1 \wedge {\cal Z}_2 \wedge {\cal Z}_3 \wedge {\cal Z}_4 $ and where $\forall \alpha$ $ \displaystyle
{\cal Z}_\alpha =  \frac{\partial}{\partial x^\alpha} +{\cal Z}_{\alpha \mu} \frac{\partial}{\partial {A}_\mu} $. We gives  the straightforward calculation with
${\cal Z}_{\alpha\mu}  =  \partial_{\alpha} {A}_{\mu} $:
\[
\langle p , v  \rangle = {\pmb{\theta}}^{\tiny{\hbox{\sffamily DDW}}}_p ({\cal Z}) = \mathfrak{e} \hbox{d} \mathfrak{y} ({\cal Z})   + {\pmb{\pi}}^{{A}_{\mu}\nu}
\hbox{d}{A}_{\mu} \wedge \hbox{d} \mathfrak{y}_{\nu}  ({\cal Z}).
\]
The expression $\langle p , v  \rangle =  {\pmb{\theta}}^{\tiny{\hbox{\sffamily DDW}}}_p ({\cal Z}) $ is  given, see appendix \eqref{appendixA} by:
 \begin{equation}\label{angleMax0}
\langle p , v  \rangle =   \mathfrak{e}   + {\pmb{\pi}}^{{A}_{\mu}\nu} 
{\cal Z}_{\nu\mu}   = \mathfrak{e}     + {\pmb{\pi}}^{{A}_{\mu}\nu} 
\partial_{\nu} A_\mu . 
 \end{equation}
We obtain: 
 \[
 \frac{\partial \langle p , v  \rangle }{\partial (\partial_\nu A_\mu) } =  \frac{\partial }{\partial (\partial_\nu A_\mu) } {\big{(}}  \mathfrak{e} \hbox{d} \mathfrak{y}   + {\pmb{\pi}}^{{A}_{\mu}\nu} 
\partial_{\nu} A_\mu    {\big{)}} = {\pmb{\pi}}^{{A}_{\mu}\nu}.
 \]
On the other side, since ${\dttF}_{\mu\nu} =  {\partial_\mu {A}_\nu } -  {\partial_\nu {A}_\mu }$: 
\[
 \frac{\partial{L} (q,v)}{\partial (\partial_\nu {A}_{\mu} )}    = 
-
\frac{1}{4}
 \frac{\partial}{\partial (\partial_\nu {A}_{\mu}  ) }  
 {\big{(}}
\eta^{\mu\lambda} \eta^{\nu\sigma}   {\dttF}_{\mu\nu} \dttF_{\lambda\sigma}
 {\big{)}}
 =    - \frac{1}{4}
\eta^{\mu\lambda} \eta^{\nu\sigma}    {\dttF}_{\lambda\sigma}    
\frac{\partial}{\partial (\partial_\nu {A}_{\mu}  ) }  
 {\big{(}}
 {\partial_\mu { A}_\nu }-  {\partial_\nu {A}_\mu }
  {\big{)}}.
\]
The expression of the multimomenta is given by \eqref{ririri002}.
 \begin{equation}\label{ririri002}
 {\pmb{\pi}}^{{  A}_\mu \nu}  =   \eta^{\mu\lambda} \eta^{\nu\sigma}  \dttF_{\lambda\sigma}    =  \dttF^{\mu\nu}
 \end{equation}
The equivalence \eqref{ririri004} is now written \eqref{ririri003}:   
 \begin{equation}\label{ririri003}
 (q,v)   {\pmb{\leftrightarrow}}   (q,p) \quad \quad     \Longleftrightarrow     \quad \quad    {\pmb{\pi}}^{{A}_\mu \nu}  =   \eta^{\mu\lambda} \eta^{\nu\sigma}  \dttF_{\lambda\sigma}  .
 \end{equation}
 Notice that the Legendre transformation is degenerated. We cannot find  a unique correspondence between the multivelocities $v$ and the multimomenta $p$. Given a $v \in T\Bbb{R} \otimes_{   {\pmb{\mathfrak{Z}}}} T^{\star} (T^{\star} {\cal X})$ the equation \eqref{ririri004} has a solution $p \in {\pmb{\cal M}}_{\tiny{\hbox{\sffamily DDW}}}$ if and only if $p \in {\pmb{\cal M}}_{\tiny{\hbox{\sffamily Maxwell}}}$ with:
\begin{align}
{\pmb{\cal M}}_{\tiny{\hbox{\sffamily Maxwell}}} = &
  {\Big{\{}}  (x,{ A}, \mathfrak{e}   \vol +  \eta^{\mu\lambda} \eta^{\nu\sigma}  \dttF_{\lambda\sigma}   \hbox{d} {A}_{\mu} \wedge   \hbox{d} \mathfrak{y}_{\nu}     \ \  /  \ \   (x,A) \in T^{\star}{\cal X} , \mathfrak{e} \in \Bbb{R}   {\Big{\}}}  \subset {\pmb{\cal M}}_{\tiny{\hbox{\sffamily DDW}}}. 
\end{align}
Notice that ${\pmb{\cal M}}_{\tiny{\hbox{\sffamily Maxwell}}} \subset {\pmb{\cal M}}_{\tiny{\hbox{\sffamily DDW}}}  $ is a vector sub-bundle of ${\pmb{\cal M}}_{\tiny{\hbox{\sffamily DDW}}} $. The degenerate feature is related to the contraint  $ {\pmb{\pi}}^{{A}_\nu \mu}  =    \dttF^{\nu\mu} = - \dttF^{\mu\nu}$.
The Legendre transform is recovered  if we impose the compatibility
conditions: $ {\pmb{\pi}}^{{ A}_\nu \mu}  +  {\pmb{\pi}}^{{ A}_\mu \nu}  = 0$. It is an example of a {\em Dirac primary constraint set} \cite{Dirac}. Therefore, we
 restrict to  the submanifold:
\begin{align}\label{Maxmultisympl}
{\pmb{\cal M}}_{\tiny{\hbox{\sffamily Maxwell}}} = &
  {\Big{\{}}  (x,{ A},p) \in {\pmb{\cal M}}_{\tiny{\hbox{\sffamily DDW}}} \ \  /  \ \  {\pmb{\pi}}^{{ A}_\nu \mu}  +  {\pmb{\pi}}^{{ A}_\mu \nu}  = 0   \ \hbox{with} \ {\pmb{\pi}}^{{A}_\nu \mu}  =    \dttF^{\nu\mu} {\Big{\}}}  \subset {\pmb{\cal M}}_{\tiny{\hbox{\sffamily DDW}}} .
\end{align}
In the Maxwell case, the Dirac set are compatibility conditions that allows us to recover a Legendre transform. The   $\hbox{\sffamily (DDW)}$ theory setting is concerned rather with the Legendre correspondence.  We introduce two  different spaces. The first  is the $\hbox{\sffamily (DDW)}$  submanifold ${\pmb{\cal M}}_{\tiny{\hbox{\sffamily DDW}}}$ on which we consider the canonical Cartan-Poincar\'e 4-form: 
\bee
{\pmb{\theta}}^{\tiny{\hbox{\sffamily DDW}}}_{(q,p)}:= \mathfrak{e}  \hbox{d} \mathfrak{y}  + \sum_{\mu} \sum_{\nu} {\pmb{\pi}}^{{A}_{\mu}\nu}
  \hbox{d} {A}_{\mu} \wedge   \hbox{d} \mathfrak{y}_{\nu}, 
  \quad \quad   
 {\pmb{\Omega}}^{\tiny{\hbox{\sffamily DDW}}} = \hbox{d} \mathfrak{e} \wedge  \hbox{d} \mathfrak{y}  +       \hbox{d}  {\pmb{\pi}}^{{A}_{\mu}\nu}  \wedge
 \hbox{d}{A}_{\mu}  \wedge   \hbox{d} \mathfrak{y}_{\nu}.
\eee
The second is ${\pmb{\cal M}}_{\tiny{\hbox{\sffamily Maxwell}}}$, defined by \eqref{Maxmultisympl} (with the imposed constraints ${\pmb{\pi}}^{{ A}_\nu \mu}  +  {\pmb{\pi}}^{{ A}_\mu \nu}  = 0$)\footnote{
 The important point concerns the restriction related to those constraints on the allowed vector fields on the multisymplectic space. In such a context,  the vector fields on ${\pmb{\cal M}}_{\tiny{\hbox{\sffamily Maxwell}}}$ must be written  with the term $ \displaystyle {\Upsilon}_{\alpha}^{A_\mu \nu} {\big{(}} \frac{\partial }{ \partial {\pmb{\pi}}^{A_\mu \nu}}  - \frac{\partial }{ \partial {\pmb{\pi}}^{A_\nu \mu}}  {\big{)}}$ rather than with the term $ \displaystyle {\Upsilon}_{\alpha}^{A_\mu \nu}   \frac{\partial }{ \partial {\pmb{\pi}}^{A_\mu \nu}} $ in the expression \eqref{0008FDGGE} of $X \in \Lambda^{4} T{\pmb{\cal M}}_{\tiny{\hbox{\sffamily Maxwell}}}$.}.

\subsection{\hbox{\sffamily\bfseries\slshape{Hamilton-Maxwell equation in the DeDonder-Weyl framework}}}\label{oo981}

We compute  the Hamiltonian function of the Maxwell theory in the ${\hbox{\sffamily {(DDW)}} }$ case:
\[
{\cal H}^{{\tiny\hbox{\sffamily Maxwell}}} (q,p) =  \langle p , v  \rangle - L(q,v) =  \langle p , v  \rangle + \frac{1}{4}  {\big{(}} \eta^{\mu\lambda} \eta^{\nu\sigma}   {\dttF}_{\mu\nu} {\dttF}_{\lambda\sigma}  {\big{)}}.
\]
Making  use of relation \eqref{angleMax0} we find: 
\[
\begin{array}{lll}
 \displaystyle  {\cal H}^{{\tiny\hbox{\sffamily Maxwell}}}  (q,p)  & = & \displaystyle  \mathfrak{e} + {\pmb{\pi}}^{{A}_{\mu}\nu} 
\partial_{\nu} A_\mu  + \frac{1}{4}    {\pmb{\pi}}^{{A}_\mu \nu}   {\dttF}_{\mu\nu} =  \mathfrak{e}  - \frac{1}{4}    {\pmb{\pi}}^{{A}_\mu \nu}   {\dttF}_{\mu\nu}
\end{array}
\]
Then, the Hamiltonian function \eqref{Max001} is given by:
\begin{equation}\label{Max001}
 {\cal H}^{{\tiny\hbox{\sffamily Maxwell}}}  (q,p)  = \mathfrak{e} -   \frac{1}{4} \eta_{\mu\rho} \eta_{\nu\sigma}   {\pmb{\pi}}^{{A}_\mu \nu}   {\pmb{\pi}}^{{A}_\rho \sigma}  
 \quad \quad  
\hbox{with}   \quad \quad
 {\pmb{\pi}}^{{A}_\mu \nu}  =   \eta^{\mu\lambda} \eta^{\nu\sigma}  \dttF_{\lambda\sigma}    = \dttF^{\mu\nu}
 \end{equation}
 In order to obtain the generalized Hamilton equations $ X \iN  {\pmb{\Omega}}^{{\tiny\hbox{\sffamily DDW}}}    = (-1)^n \hbox{d}  {\cal H}^{{\tiny\hbox{\sffamily Maxwell}}} $, we need to compute $\hbox{d}  {\cal H}^{{\tiny\hbox{\sffamily Maxwell}}}  (q,p)$, the differential of the Hamiltonian function.   Since we work with a degenerate Legendre transform a naive use of the general method leads to incorrect equations of motion. We have: 
 \[
 \hbox{d}  {\cal H}^{{\tiny\hbox{\sffamily Maxwell}}}  (q,p) = \hbox{d}   \mathfrak{e} -   \frac{1}{4} \eta_{\mu\rho} \eta_{\nu\sigma}   \hbox{d}   ( {\pmb{\pi}}^{{A}_\mu \nu}   {\pmb{\pi}}^{{A}_\rho \sigma}  )  =  \hbox{d}   \mathfrak{e} -   \frac{1}{2} \eta_{\mu\rho} \eta_{\nu\sigma}  {\pmb{\pi}}^{{A}_\rho \sigma}    \hbox{d}     {\pmb{\pi}}^{{A}_\mu \nu},\]
  which describes  the right  hand side of the Hamilton equations \eqref{Max002}. We denote from now ${\cal H} (q,p):=  {\cal H}^{{\tiny\hbox{\sffamily Maxwell}}}   (q,p) $.
 \begin{equation}\label{Max002}
 X \iN  {\pmb{\Omega}}^{{\tiny\hbox{\sffamily DDW}}}    = (-1)^n \hbox{d} {\cal H} 
 \end{equation}
 Let us denote $\forall 1 \leq \alpha \leq 4$: 
 \bee\label{0008FDGGE}
X_\alpha = \frac{\partial}{\partial x^\alpha}   + \Theta_{\alpha {\mu}}   \frac{\partial}{\partial A_{\mu}}  + {\Upsilon}_{\alpha} \frac{\partial }{ \partial \mathfrak{e}}
+ {\Upsilon}_{\alpha}^{A_\mu \nu} \frac{\partial }{ \partial {\pmb{\pi}}^{A_\mu \nu}}.
 \eee
 Then we consider a $n$-vector field $X= X_1 \wedge X_{2} \wedge X_{3} \wedge X_{4} \in \Lambda^{4} T^{\star} {\pmb{\cal M}}_{{\tiny\hbox{\sffamily DDW}}}$. 
 \begin{lemm}\label{curva2}
 {\em 
Let $X$ be a $(n-1)$-vector field and let ${\big{\{}} \dd {\pmb{\rho}}_{i} {\big{\}}}_{1 \leq i \leq n}$ be a set of $n$ $1$-forms.  We have:
\[
X \iN  (\bigwedge_{1 \leq i \leq n} \dd {\pmb{\rho}}_{i}) = X \iN \dd {\pmb{\rho}}_{1} \wedge \cdots \wedge \dd {\pmb{\rho}}_{n} = \sum_{j} (-1)^{j+1} {\big{(}}
\dd {\pmb{\rho}}_{1} \wedge \cdots \wedge \dd {\pmb{\rho}}_{j-1} \wedge \dd {\pmb{\rho}}_{j+1} \wedge \cdots \wedge \dd {\pmb{\rho}}_{n} {\big{)}} (X) \dd {\pmb{\rho}}_{j}
\]
}
\end{lemm}
 Thanks to lemma \eqref{curva2}, the left side  of the Hamilton equations \eqref{Max002} is written: 
\[
\begin{array}{lll}
\displaystyle   X \iN  {\pmb{\Omega}}^{{\tiny\hbox{\sffamily DDW}}}    & = &  \displaystyle X \iN  {\big{(}}  \hbox{d} \mathfrak{e} \wedge \hbox{d} \mathfrak{y}  +  \hbox{d} {\pmb{\pi}}^{{A}_{\mu}\nu} \wedge
  \hbox{d} {A}_{\mu} \wedge   \hbox{d} \mathfrak{y}_{\nu}  {\big{)}}
\displaystyle 
 \\
  \displaystyle   & =  & \displaystyle    \hbox{d} \mathfrak{y} (X) \hbox{d}\mathfrak{e} - (\hbox{d} \mathfrak{e} \wedge  \hbox{d} \mathfrak{y}_\rho) (X) \dd x^{\rho}   +    (\hbox{d}{A}_{\mu}  \wedge  \hbox{d} \mathfrak{y}_{\nu} ) (X)   \hbox{d} {\pmb{\pi}}^{{A}_{\mu}\nu}   
 \\
  \displaystyle &  & \displaystyle -  (  \hbox{d} {\pmb{\pi}}^{{A}_{\mu}\nu} \wedge
 \hbox{d} \mathfrak{y}_{\nu}   ) (X)\hbox{d}{A}_{\mu}   + (   \hbox{d} {\pmb{\pi}}^{{A}_{\mu}\nu} \wedge
\dd{A}_{\mu} \wedge  \hbox{d} \mathfrak{y}_{\rho\nu}  ) (X) \dd x^{\rho}.
 \end{array}
 \]
 So that we obtain:
 \bee\label{mmmmmmmmmmqq}
\begin{array}{lll}
\displaystyle   X \iN  {\pmb{\Omega}}^{{\tiny\hbox{\sffamily DDW}}}   & =   & \displaystyle
 \hbox{d} \mathfrak{e}  - \Upsilon_\rho \dd x^{\rho} +         \Theta_{\nu\mu}  \hbox{d} {\pmb{\pi}}^{{A}_{\mu}\nu}  -   {\Upsilon}_{\nu}^{A_\mu \nu} \dd{A}_{\mu} +{\big{(}}  {\Upsilon}_{\rho}^{A_\mu \nu}       \Theta_{\nu\mu}  - {\Upsilon}_{\nu}^{A_\mu \nu}  \Theta_{\rho\mu} {\big{)}} \dd x^\rho.
 \end{array}
\eee
    The decomposition  on the different forms $\dd {\pmb{\pi}}^{A_{\mu} \nu} $, $\dd \mathfrak{e}$, $\dd A_{\mu}$ and $\dd x^{\rho}$ gives:
\begin{equation} 
\left|
\begin{array}{ccl}
\displaystyle  -    \Theta_{\nu\mu}  &  =  & \displaystyle   -   \frac{1}{2} \eta_{\mu\rho} \eta_{\nu\sigma}  {\pmb{\pi}}^{{A}_\rho \sigma} 
\\
 \displaystyle -   {\Upsilon}_{\nu}^{A_\mu \nu} &  =  & \displaystyle 0
\end{array}
\right.
\quad   \hbox{and}   \quad 
\left.
\begin{array}{ccl}
\displaystyle  - \Upsilon_\rho + {\big{(}}  {\Upsilon}_{\rho}^{A_\mu \nu}       \Theta_{\nu\mu}  - {\Upsilon}_{\nu}^{A_\mu \nu}  \Theta_{\rho\mu} {\big{)}}  &  =  & \displaystyle 0.
\end{array}
\right.
\nonumber
\end{equation}
that is equivalent to:
\begin{equation} 
\left|
\begin{array}{ccl}
\displaystyle   \partial_\nu {  A_{\nu}}   &  =  & \displaystyle     \frac{1}{2} \eta_{\mu\rho} \eta_{\nu\sigma}  {\pmb{\pi}}^{{A}_\rho \sigma}  
\\
 \displaystyle   \partial_\nu {\pmb{\pi}}^{A_\mu \nu} &  =  & \displaystyle 0
\end{array}
\right.
\ \hbox{and} \ 
\left.
\begin{array}{ccl}
\displaystyle   - \partial_\rho \mathfrak{e} + {\big{(}} ( \partial_{\rho} {{\pmb{\pi}}^{ {A_\mu} \nu}} )      ( \partial_\nu A_{\mu})  - ({\partial_{\nu}} {\pmb{\pi}}^{{A_\mu} \nu} )( \partial_{\rho} A_{\mu} ){\big{)}}  &  =  & \displaystyle 0.
\end{array}
\right.
\nonumber
\end{equation}
  The second line of the previous system   gives the half of the Maxwell equations. Notice that the Legendre degenerate transform implies ${\pmb{\pi}}^{A_{\mu} \nu} = \dttF^{\mu\nu}$ so that $
\partial_\nu {\pmb{\pi}}^{A_\mu \nu} = \partial_{\nu} \dttF^{\mu\nu} = 0$. However we can not recover the full set of   Maxwell's equations,  since 
$ \displaystyle
   \frac{1}{2} \eta_{\mu\rho} \eta_{\nu\sigma}  {\pmb{\pi}}^{{A}_\rho \sigma}  = \frac{1}{2} \dttF_{\mu\nu} \neq  {\partial_\mu A_\nu}. $  We are {\em not} recovering the usual Euler-Lagrange equations precisely because we work on the {\em degenerate} space.  Now let us consider rather the space ${\pmb{\cal M}}_{\tiny{\hbox{\sffamily Maxwell}}}$. The constraint ${\pmb{\pi}}^{A_{\mu} \nu} + {\pmb{\pi}}^{A_{\nu} \mu} = 0$ selects the authorized directions for the vector fields and the ones we are not allowed to described. In this context, the vector fields involved  in the contraction with the multisymplectic form are   given by \eqref{dmqpapspp001}. We denote   $\forall 1 \leq \alpha \leq 4$:
 \bee\label{dmqpapspp001}
X_\alpha = \frac{\partial}{\partial x^\alpha}   + \Theta_{\alpha {\mu}}   \frac{\partial}{\partial A_{\mu}}  + {\Upsilon}_{\alpha} \frac{\partial }{ \partial \mathfrak{e}}
+ {\Upsilon}_{\alpha}^{A_\mu \nu} {\Big{(}} \frac{\partial }{ \partial {\pmb{\pi}}^{A_\mu \nu}}  - \frac{\partial }{ \partial {\pmb{\pi}}^{A_\nu \mu}}  {\Big{)}}.
 \eee
The Hamilton equations \eqref{Max002} becomes:
$
 X \iN  {\pmb{\Omega}}^{{\tiny\hbox{\sffamily DDW}}}    = (-1)^n \hbox{d} {\cal H} 
$.
\[
\begin{array}{lll}
\displaystyle   X \iN   {\pmb{\Omega}}^{\tiny{\hbox{\sffamily DDW}}} & = &
\displaystyle
X \iN {\big{(}}   \hbox{d} \mathfrak{e} \wedge  \hbox{d} \mathfrak{y} + \dd  {\pmb{\pi}}^{A_{\mu} \nu } 
 \wedge \dd  {A}_{\mu} \wedge   \hbox{d} \mathfrak{y}_{\nu} {\big{)}}
 \\
  \displaystyle    & =  & \displaystyle  \hbox{d} \mathfrak{y} (X) \hbox{d}\mathfrak{e} - (\hbox{d} \mathfrak{e} \wedge  \hbox{d} \mathfrak{y}_\rho) (X) \dd x^{\rho} 
+ (\hbox{d}{A}_{\mu}  \wedge  \hbox{d} \mathfrak{y}_{\nu} ) (X)   \hbox{d} {\pmb{\pi}}^{{A}_{\mu}\nu}   
  \\
  \displaystyle &  & \displaystyle 
 -  (  \hbox{d} {\pmb{\pi}}^{{A}_{\mu}\nu} \wedge
 \hbox{d} \mathfrak{y}_{\nu}   ) (X)\hbox{d}{A}_{\mu} + (   \hbox{d} {\pmb{\pi}}^{{A}_{\mu}\nu} \wedge
\dd{A}_{\mu} \wedge  \hbox{d} \mathfrak{y}_{\rho\nu}  ) (X) \dd x^{\rho}.
\end{array}
\]
Then, we obtain: 
\[
\begin{array}{lll}
 \displaystyle  X \iN   {\pmb{\Omega}}^{\tiny{\hbox{\sffamily DDW}}} & =  & \displaystyle  
 \hbox{d} \mathfrak{e}  - \Upsilon_\rho \dd x^{\rho} +    (  \Theta_{\nu\mu} -  \Theta_{\mu\nu} ) \hbox{d} {\pmb{\pi}}^{{A}_{\mu}\nu}  -   (   {\Upsilon}_{\nu}^{A_\mu \nu}  -  {\Upsilon}_{\nu}^{A_\nu \mu}  ) \dd A_{\mu}   
 \\
  \displaystyle &  & \displaystyle  + {\Big{(}}  {\big{(}}  {\Upsilon}_{\rho}^{A_\mu \nu}       \Theta_{\nu\mu}  - {\Upsilon}_{\nu}^{A_\mu \nu}  \Theta_{\rho\mu} {\big{)}} -   {\big{(}}  {\Upsilon}_{\rho}^{A_\nu \mu}       \Theta_{\mu\nu}  - {\Upsilon}_{\mu}^{A_\nu \mu}  \Theta_{\rho\nu} {\big{)}}  {\Big{)}}  \dd x^{\rho}.
\end{array}
\] 
 The decompositions along $\hbox{d} {\pmb{\pi}}^{{A}_{\mu}\nu}$ and $ \dd A_{\mu}$,  gives:
\begin{equation}\label{zhfoazeob}
\left|
\begin{array}{rcl}
\displaystyle     (  \Theta_{\nu\mu} -  \Theta_{\mu\nu} )   &  =  & \displaystyle   -   \eta_{\mu\rho} \eta_{\nu\sigma}  {\pmb{\pi}}^{{A}_\rho \sigma} 
\\
 \displaystyle -     (   {\Upsilon}_{\nu}^{A_\mu \nu}  -  {\Upsilon}_{\nu}^{A_\nu \mu}  )  &  =  & \displaystyle 0
\end{array}
\right.
\quad \Longrightarrow \quad  
\left|
\begin{array}{rcl}
\displaystyle   \partial_\mu {  A_{\nu}}  -   \partial_\nu {  A_{\mu}} &  =  & \displaystyle    \dttF_{\mu\nu} 
\\
 \displaystyle  \partial_\nu {\big{(}}  {\pmb{\pi}}^{A_\mu \nu} - {\pmb{\pi}}^{A_\nu \mu} {\big{)}} &  =  & \displaystyle 0.
\end{array}
\right.
\end{equation}
Hence, the second line of equation \eqref{zhfoazeob} gives Maxwell's equations:
\bee
 \frac{1}{2} \partial_\nu {\big{(}}  {\pmb{\pi}}^{A_\mu \nu} - {\pmb{\pi}}^{A_\mu \nu} {\big{)}}  = 
 \partial_\nu  {\pmb{\pi}}^{A_\mu \nu}   =  \partial_\nu \dttF^{\mu\nu} = 0.
\eee

 $\hbox{\em{Remark}}.$ We give a detailed calculation  in the next section \eqref{gliopowww22}, - with the example of the {\sffamily 2D}-case - for the expression $X \iN   {\pmb{\Omega}}^{\tiny{\hbox{\sffamily DDW}}} $, where we will consider respectively  the case $X \in \forall X , \overline{X} \in \Lambda^{n} T {\pmb{\cal M}}_{\tiny{\hbox{\sffamily Maxwell}}}$ and $X \in \forall X , \overline{X} \in \Lambda^{n} T {\pmb{\cal M}}_{\tiny{\hbox{\sffamily DDW}}}$.

\subsection{\hbox{\sffamily\bfseries\slshape{Maxwell theory as an $n$-phase space}}}\label{oo982}

We refer to the work of J. Kijowski \cite{JK-01} for the treatment of Maxwell's theory in the setting of a $n$-phase space. Due to the  abelian feature of the Maxwell gauge theory, this treatment is essentially the same that the one exposed in the previous section.
 the notion of a $n$-phase space, inspired by J. Kijowski and W. Szczyrba   \cite{KS1}, and developed further by F. H\'elein \cite{H-03} \cite{H-02}.

\begin{defin}\label{jfzajoezjvzo013}
A $n$-phase space is a triple  $({\pmb{\cal M}}, {\pmb{\Omega}}, {\pmb{\beta}})$  where ${\pmb{\cal M}}$ is a smooth manifold, ${\pmb{\Omega}}$  is a closed (n + 1)-form and ${\pmb{\beta}}$ is an everywhere non-vanishing n-form.
\end{defin}
For a $n$-phase space  $({\pmb{\cal M}}, {\pmb{\Omega}}, {\pmb{\beta}})$, a Hamiltonian $n$-curve is pictured as an oriented $n$-submanifold which satisfies:
\[
\forall m \in {\pmb{\Gamma}}, \forall  X \in \Lambda^{n}T_{m} {\pmb{\Gamma}} \quad X \iN {\pmb{\Omega}}_{m} = 0
\quad \quad \hbox{and} \quad \quad
\forall m \in {\pmb{\Gamma}}, \exists  X \in \Lambda^{n}T_{m} {\pmb{\Gamma}} \quad X \iN {\pmb{\beta}}_{m} \neq 0.
\]
The last condition is an independence condition. We can canonically construct  $n$-phase space data  by   means of the hypersurface of a multisymplectic manifold. We recall that a premultisymplectic $n$-form is closed but may be degenerate. In the general picture of a $n$-phase space we express {\em dynamics} on a {\em level set} of ${\cal H}$.\footnote{We can  construct canonically a $n$-premultisymplectic manifold $({\pmb{\cal{M}}}^{\hbox{\skt 0}},{\pmb{\Omega}}|_{{\pmb{\cal{M}}}^{\hbox{\skt 0}}},{\pmb{\beta}} = {\pmb{\eta}} \iN {\pmb{\Omega}} |_{{\pmb{\cal{M}}}^{\hbox{\skt 0}}} )$. Here the $
 {\pmb{\Omega}}|_{{\pmb{\cal{M}}}^{\hbox{\skt 0}}} = \mathcal{H}^{-1}(0):= \{ (q,p) \in {\pmb{\Omega}}|_{{\pmb{\cal{M}}} } |\ \mathcal{H}(q,p) =0\}
$ and ${\pmb{\eta}}$ is a vector field such that $\hbox{d}\mathcal{H}({\pmb{\eta}}) = 1$.
 In this case we observe the connection between  relativistic dynamical systems and the treatment of the  {\em Hamiltonian constraint}. } We recover the dynamical equations  in the pre-multisymplectic case  \eqref{ZZZZoo}  -  see F. H\'elein  \cite{FH-01}   \cite{H-03} \cite{H-02}.
\begin{equation}\label{ZZZZoo} 
{\forall \Xi \in C^\infty (\pmb{\cal M}, T_m{\pmb{{\cal M}}})  ,\quad  \quad (\Xi \iN {\pmb{\Omega}}) |_{\pmb{\Gamma}} = 0 \quad \quad \quad \quad \hbox{and}  \quad \quad \quad \quad {\pmb{\beta}}|_\Gamma \neq 0}.
\end{equation}
 The canonical pre-multisymplectic form is given by:
\begin{equation}\label{thetapremulti}
{\pmb{\theta}}^{{\tiny{\hbox{\sffamily pre-multi. Maxwell}}}}_{(q,p)}:=  {\pmb{\theta}}^{{\tiny{\hbox{\sffamily Maxwell}}}}_{(q,p)} \big |_{{\cal H} = 0} = \mathfrak{e}  \hbox{d} \mathfrak{y}  + {\pmb{\pi}}^{{A}_{\mu}\nu}
  \hbox{d} {A}_{\mu} \wedge   \hbox{d} \mathfrak{y}_{\nu} \big |_{{\cal H} = 0}.
\end{equation}
We have,
$ \displaystyle
 {\cal H}   (q,p)  = \mathfrak{e} -  {1}/{4} \eta_{\mu\rho} \eta_{\nu\sigma}   {\pmb{\pi}}^{{A}_\mu \nu}   {\pmb{\pi}}^{{A}_\rho \sigma}$ with ${\pmb{\pi}}^{{A}_\mu \nu}  =   \eta^{\mu\lambda} \eta^{\nu\sigma}   \dttF_{\lambda\sigma}    = \dttF^{\mu\nu}$. The imposition of  the Hamiltonian constraint ${\cal H} = 0$ leads us to consider $\mathfrak{e} =  {1}/{4} \eta_{\mu\rho} \eta_{\nu\sigma}   {\pmb{\pi}}^{{A}_\mu \nu}   {\pmb{\pi}}^{{A}_\rho \sigma} = - {H} (x^{\mu}, A_\nu ,  {\pmb{\pi}}^{A_\mu \nu} )$. Hence, the pre-multisymplectic canonical forms ${\pmb{\theta}}^{{\tiny{\hbox{\sffamily  pre-multi}}}}_{(q,p)}$ and ${\pmb{\Omega}}^{{\tiny{\hbox{\sffamily  pre-multi}}}}_{(q,p)}$ are respectively written:
\[
\begin{array}{lll}
 \displaystyle {\pmb{\theta}}^{\tiny{\hbox{\sffamily pre-multi}}}_{(q,p)}    & =  & \displaystyle  
(1/4)  \eta_{\mu\rho} \eta_{\nu\sigma}   {\pmb{\pi}}^{{A}_\mu \nu}   {\pmb{\pi}}^{{A}_\rho \sigma}  \hbox{d} \mathfrak{y}  + {\pmb{\pi}}^{{A}_{\mu}\nu}
  \hbox{d} {A}_{\mu} \wedge   \hbox{d} \mathfrak{y}_{\nu} 
=
(1/4)   {\pmb{\pi}}^{{A}_\mu \nu}   {\pmb{\pi}}_{{A}_\mu \nu}  \hbox{d} \mathfrak{y}  + {\pmb{\pi}}^{{A}_{\mu}\nu}
  \hbox{d} {A}_{\mu} \wedge   \hbox{d} \mathfrak{y}_{\nu}, 
\end{array}
\] 
and:
\[
\begin{array}{lll}
 \displaystyle {\pmb{\Omega}}^{\tiny{\hbox{\sffamily pre-multi}}}_{(q,p)}   & =  & \displaystyle   \dd {\pmb{\theta}}^{\tiny{\hbox{\sffamily pre-multi}}}_{(q,p)} 
=
   \frac{1}{2} \eta_{\mu\rho} \eta_{\nu\sigma}  {\pmb{\pi}}^{{A}_\rho \sigma}    \hbox{d}     {\pmb{\pi}}^{{A}_\mu \nu} \wedge \hbox{d} \mathfrak{y}  +  \hbox{d} {\pmb{\pi}}^{{A}_{\mu}\nu} \wedge
  \hbox{d} {A}_{\mu} \wedge   \hbox{d} \mathfrak{y}_{\nu}.
\end{array}
\] 

We denote, in order to simplify the notations: 
$\displaystyle {\pmb{\theta}}^{\tiny{\hbox{\sffamily pre-multi}}}_{(q,p)} = {\pmb{\theta}}^{{\hbox{\skt 0}}}_{(q,p)}  $ and $\displaystyle {\pmb{\Omega}}^{\tiny{\hbox{\sffamily pre-multi}}}_{(q,p)}  = {\pmb{\Omega}}^{{\hbox{\skt 0}}}_{(q,p)} $.
Therefore, we consider the theory on the {\em pre-multisymplectic} Maxwell space $\displaystyle { {\pmb{\cal M}} }_{\tiny{\hbox{\sffamily Maxwell}}}^{{\hbox{\skt 0}}} $ \eqref{Maxmultisymplpre}:
\begin{align}\label{Maxmultisymplpre}
{ {\pmb{\cal M}} }_{\tiny{\hbox{\sffamily Maxwell}}}^{{\hbox{\skt 0}}} = &
  {\Big{\{}}  (x,{ A},p) \in {\pmb{\cal M}}_{\tiny{\hbox{\sffamily DDW}}} \ \  /  \ \  {\pmb{\pi}}^{{ A}_\nu \mu}  +  {\pmb{\pi}}^{{ A}_\mu \nu}  = 0 \ \hbox{and} \  \mathfrak{e} = \frac{1}{4} \eta_{\mu\rho} \eta_{\nu\sigma}   {\pmb{\pi}}^{{A}_\mu \nu}   {\pmb{\pi}}^{{A}_\rho \sigma} {\Big{\}}}.
\end{align}
We observe the following inclusion of spaces: $\displaystyle { {\pmb{\cal M}} }_{\tiny{\hbox{\sffamily Maxwell}}}^{{\hbox{\skt 0}}} \subset  {\pmb{\cal M}}_{{\tiny{\hbox{Maxwell}}}} \subset {\pmb{\cal M}}_{\tiny{\hbox{\sffamily DDW}}}$. The generalized Hamilton equations are given with the calculation of $\displaystyle X \iN  {\pmb{\Omega}}^{{\hbox{\skt 0}}}$:
  \[
\begin{array}{lll}
\displaystyle    X \iN  {\pmb{\Omega}}^{{\hbox{\skt 0}}}  & = & \displaystyle  X \iN (   {1}/{2} \eta_{\mu\rho} \eta_{\nu\sigma}  {\pmb{\pi}}^{{A}_\rho \sigma}    \hbox{d}     {\pmb{\pi}}^{{A}_\mu \nu}\wedge  \hbox{d} \mathfrak{y} ) +        X \iN (  \hbox{d}  {\pmb{\pi}}^{{A}_{\mu}\nu}  \wedge
 \hbox{d}{A}_{\mu}  \wedge   \hbox{d} \mathfrak{y}_{\nu} )
 \\
  \displaystyle     & =  & \displaystyle   
 {1}/{2} \eta_{\mu\rho} \eta_{\nu\sigma}  {\pmb{\pi}}^{{A}_\rho \sigma} \hbox{d} \mathfrak{y} (X) \dd     {\pmb{\pi}}^{{A}_\mu \nu} - (   {1}/{2} \eta_{\mu\rho} \eta_{\nu\sigma}  {\pmb{\pi}}^{{A}_\rho \sigma}    \hbox{d}     {\pmb{\pi}}^{{A}_\mu \nu} \wedge  \hbox{d} \mathfrak{y}_\rho) (X) \dd x^\rho
   \\
  \displaystyle &  & \displaystyle +  (\hbox{d}{A}_{\mu}  \wedge  \hbox{d} \mathfrak{y}_{\nu} ) (X)   \hbox{d} {\pmb{\pi}}^{{A}_{\mu}\nu} -  (  \hbox{d} {\pmb{\pi}}^{{A}_{\mu}\nu} \wedge
 \hbox{d} \mathfrak{y}_{\nu}   ) (X)\hbox{d}{A}_{\mu}   + (   \hbox{d} {\pmb{\pi}}^{{A}_{\mu}\nu} \wedge
\dd{A}_{\mu} \wedge  \hbox{d} \mathfrak{y}_{\rho\nu}  ) (X) \dd x^\rho.   \\
\end{array}
\]
So that:
  \[
\begin{array}{lll}
\displaystyle    X \iN   {\pmb{\Omega}}^{{\hbox{\skt 0}}} & = & \displaystyle 
\hbox{d} \mathfrak{e}  - \Upsilon_\rho \dd x^{\rho} +   (  \Theta_{\nu\mu} -  \Theta_{\mu\nu} + \eta_{\mu\rho} \eta_{\nu\sigma}  {\pmb{\pi}}^{{A}_\rho \sigma} ) \hbox{d} {\pmb{\pi}}^{{A}_{\mu}\nu}  -   (   {\Upsilon}_{\nu}^{A_\mu \nu}  -  {\Upsilon}_{\nu}^{A_\nu \mu}  ) \dd A_{\mu} 
   \\
  \displaystyle &  & \displaystyle   +  {\Big{(}}  {\big{(}}  {\Upsilon}_{\rho}^{A_\mu \nu}       \Theta_{\nu\mu}  - {\Upsilon}_{\nu}^{A_\mu \nu}  \Theta_{\rho\mu} {\big{)}} -   {\big{(}}  {\Upsilon}_{\rho}^{A_\nu \mu}       \Theta_{\mu\nu}  - {\Upsilon}_{\mu}^{A_\nu \mu}  \Theta_{\rho\nu} {\big{)}} - \eta_{\mu\rho} \eta_{\nu\sigma}  {\pmb{\pi}}^{{A}_\rho \sigma} {\Upsilon}_{\rho}^{A_\mu \nu}  {\Big{)}}  \dd x^\rho.
\end{array}
\]
 Once again, the decompositions along $\hbox{d} {\pmb{\pi}}^{{A}_{\mu}\nu}$ and $ \dd A_{\mu}$ gives:  
\begin{equation} 
\left|
\begin{array}{rcc}
\displaystyle    (  \Theta_{\nu\mu} -  \Theta_{\mu\nu} + \eta_{\mu\rho} \eta_{\nu\sigma}  {\pmb{\pi}}^{{A}_\rho \sigma} )  &  =  & \displaystyle  0 
\\
 \displaystyle -    (   {\Upsilon}_{\nu}^{A_\mu \nu}  -  {\Upsilon}_{\nu}^{A_\nu \mu}  )  &  =  & \displaystyle 0.
\end{array}
\right.
\end{equation}
We recover \eqref{zhfoazeob} and then the same conclusions.

 \section{\hbox{\sffamily\bfseries\slshape{Algebraic Observables  and observable functionals}}}\label{section9}

We being this section with the definition \eqref{022jvzo118} of algebraic observable $(n-1)$-forms and the  set $\mathfrak{sp}_{\circ} ( {\pmb{\cal M}} )$ of infinitesimal symplectomorphisms of the  related multisymplectic manifold $( {\pmb{\cal M}} , {\pmb{\Omega}} ) $.
 \begin{defin}\label{022jvzo118}
 {\em
Let $({\pmb{\cal M}} , {\pmb{\Omega}} ) $ be an $n$-multisymplectic manifold. A $(n-1)$-form ${\pmb{\varphi}}$ is called an algebraic observable $(n-1)$-form if and only if there exists $\Xi_{\pmb{\varphi}}$ such that $\Xi_{\pmb{\varphi}} \iN {\pmb{\Omega}} + \dd {\pmb{\varphi}} = 0 $.}
\end{defin}
We denote ${{\textswab{P}}}_{\circ}^{n-1}({\pmb{\cal M}})$ the set of all algebraic observable $(n-1)$-forms.
This   reflects the {\em symmetry} point of view. It is the natural analogue to the  question of the Poisson bracket for classical mechanics. Then, $\forall {\pmb{\varphi}}, {\pmb{\rho}} \in \textswab{P}_{\circ}^{n-1} ({\pmb{\cal M}})$, we define the Poisson bracket \eqref{gigi0}: 
\begin{equation}\label{gigi0}
{\big{\{}} {\pmb{\varphi}} , {\pmb{\rho}} {\big{\}}} = \Xi_{\pmb{\rho}} \wedge \Xi_{\pmb{\varphi}} \iN {\pmb{\Omega}} = -  \Xi_{\pmb{\rho}}   \iN \hbox{d} {\pmb{\varphi}} =  \Xi_{\pmb{\varphi}}   \iN \hbox{d} {\pmb{\rho}}.
\end{equation}
where,  ${\big{\{}} {\pmb{\varphi}} , {\pmb{\varrho}} {\big{\}}} \in  {{\textswab{P}}}_{\circ}^{n-1} ({\pmb{\cal M}}) $ and the bracket \eqref{gigi0} satisfy the antisymmetry property $ \displaystyle
{\big{\{}}  {\pmb{\varphi}}  ,  {\pmb{\rho}} {\big{\}}} + {\big{\{}}  {\pmb{\rho}}  ,  {\pmb{\varphi}} {\big{\}}} = 0,
$ and Jacobi structure modulo an exact term $\forall {\pmb{\varphi}}, {\pmb{\rho}}, {\pmb{\eta}} \in {{\textswab{P}}}_{\circ}^{n-1} ({\pmb{\cal M}})$:
\bee
{\big{\{}}  \{  {\pmb{\varphi}} , {\pmb{\rho}} \} {\pmb{\eta}}   {\big{\}}} + {\big{\{}}  \{ {\pmb{\rho}} , {\pmb{\eta}}  \} {\pmb{\varphi}} {\big{\}}} + {\big{\{}}  \{ {\pmb{\eta}}   , {\pmb{\varphi}} \} {\pmb{\rho}} {\big{\}}} = \hbox{d} (\xi_{{\pmb{\varphi}}} \wedge \xi_{{\pmb{\rho}}} \wedge \xi_{{\pmb{\eta}} } \iN {\pmb{\Omega}}).
\eee
We have   defined an {\em infinitesimal symplectomorphism} of $({\pmb{\cal M}}, {\pmb{\Omega}})$ to be a vector field $\Xi \in \Gamma({\pmb{\cal M}} , T{\pmb{\cal M}})$ such that 
$
{\cal{L}}_{\Xi} {\pmb{\Omega}} = 0  
$, using the Cartan formula, we obtain: 
\[
{\cal{L}}_{\Xi} {\pmb{\Omega}}  = \hbox{d} (\Xi \iN {\pmb{\Omega}}) + \Xi \iN \hbox{d} {\pmb{\Omega}} = 0.
\]
Now, since the multisymplectic $(n+1)$-form is closed $\hbox{d} {\pmb{\Omega}} = 0$, this relation is equivalent to 
$
\hbox{d} (\Xi \iN {\pmb{\Omega}}) = 0
$. We are looking for vector fields $\Xi \in \Gamma({\pmb{\cal M}} , T{\pmb{\cal M}}) $ such that $\hbox{d} (\Xi \iN {\pmb{\Omega}})= 0$. 
We denote by $\mathfrak{sp}_{\circ} ( {\pmb{\cal M}} ) $ the set of infinitesimal symplectomorphisms of the multisymplectic manifold $( {\pmb{\cal M}} , {\pmb{\Omega}} ) $: 
\bee
\mathfrak{sp}_{\circ} ( {\pmb{\cal M}} ) = 
{\Big{\{}} \Xi \in \Gamma({\pmb{\cal M}} , T{\pmb{\cal M}}) \  /   \  \hbox{d} (\Xi \iN {\pmb{\Omega}})  = 0 {\Big{\}}}.
\eee

\subsection{\hbox{\sffamily\bfseries\slshape{Some algebraic observable $(n-1)$-forms}}}\label{ooobb01}

We are interested in the algebraic observable $(n-1)$-forms and their related infinitesimal symplectomorphisms on the multisymplectic manifold $\displaystyle ({\pmb{\cal M}}_{\tiny{\hbox{\sffamily Maxwell}}} , {\pmb{\Omega}}^{\tiny{\hbox{\sffamily DDW}}})$. First we take some simple examples  and we enter in the general setting step by step. We find two types of algebraic observable $(n-1)$-forms: the (generalized) positions $(n-1)$-forms and the (generalized) momenta observable $(n-1)$-forms.  Let us begin with the following   algebraic observable $(n-1)$-forms:  $\dttP^{\mu} = \dd x^{\mu} \wedge {\pmb{\pi}}$, $\dttP^{\mu}_{{\phi}} = {\phi}(x) \dd x^{\mu} \wedge {\pmb{\pi}}$,  $\dttQ_{\mu\nu}  = A \wedge    \vol_{\mu\nu}$ and  $\dttQ_{\mu\nu}^{\psi}  = {\psi}(x) A \wedge    \vol_{\mu\nu}$. We denote the  Faraday $(n-2)$-form   \cite{HK-01}  \cite{Kana-OZO} \cite{Kana-01} by:
\bee
{\pmb{\pi}} = \frac{1}{2} {\pmb{\pi}}^{A_\mu \nu}   \vol_{\mu\nu} = \frac{1}{2}  \sum_{\mu , \nu} {\pmb{\pi}}^{A_\mu \nu} \frac{\partial}{\partial x^{\mu}} \iN \frac{\partial}{\partial x^{\nu}} \iN   \vol. 
\eee
and    the  potential  $1$-form $A = A_\mu \dd x^{\mu}$. The couple of variables $(A,{\pmb{\pi}})$ depicts the canonical variables for the Maxwell theory \cite{HK-01} \cite{HKHK01}  \cite{HK-02}  \cite{HK-03}  \cite{Kana-OZO} \cite{Kana-01}. Notice that  the Faraday $(n-2)$-form is also written:  $\star \dd A = \eta^{\mu\lambda } \eta^{\nu\sigma} (\partial_\mu A_\nu - \partial_\nu A_\mu) \dd \mathfrak{y}_{\lambda\sigma}$. First, let us focus on  $\dttP^{\mu} =  \dd x^{\mu} \wedge {\pmb{\pi}}$. We have:   
\[
\dttP^{\rho} = 
\dd x^{\rho} \wedge {\pmb{\pi}} = \dd x^{\rho} \wedge {\big{(}} \frac{1}{2} {\pmb{\pi}}^{A_\mu \nu}   \vol_{\mu\nu}  {\big{)}}  
= \frac{1}{2} {\pmb{\pi}}^{A_\mu \nu} {\big{(}}  \delta^{\rho}_{\mu}   \vol_{\nu} -  \delta^{\rho}_{\nu}   \vol_{\mu}  {\big{)}}
=  \frac{1}{2} {\big{(}} {\pmb{\pi}}^{A_\rho \nu}       \vol_{\nu} -  {\pmb{\pi}}^{A_\mu \rho}     \vol_{\mu}  {\big{)}}.
\] 
Using the constraint ${\pmb{\pi}}^{ A_{\mu} \nu} = - {\pmb{\pi}}^{ A_{\nu} \mu} $, we obtain: 
$\dttP^{\mu} 
=  {\pmb{\pi}}^{{A}_{\mu}\nu}   \vol_\nu$. 
Now we compute the exterior derivative $\dd \dttP^{\mu} = \dd {\big{(}} \dd x^{\rho} \wedge {\pmb{\pi}}   {\big{)}} =   \dd {\big{(}} {\pmb{\pi}}^{{A}_{\mu}\nu}     \vol_\nu  {\big{)}} =  \dd  {\pmb{\pi}}^{{A}_{\mu}\nu} \wedge   \vol_\nu
$.
If we consider $\displaystyle \Xi ({\dttP}^{\mu} )  = \frac{\partial }{\partial A_\mu }$ we have $ \dd \dttP^{\mu}   = - \Xi ({\dttP}^{\mu} ) \iN {\pmb{\Omega}}^{\tiny{\hbox{\sffamily DDW}}}$ as shown by the following straightforward calculation: 
\[
\Xi ({\dttP}^{\mu} ) \iN {\pmb{\Omega}}^{\tiny{\hbox{\sffamily DDW}}}  = \frac{\partial }{\partial A_\mu } \iN {\big{(}} \hbox{d}  \mathfrak{e} \wedge \hbox{d} \mathfrak{y} +  \hbox{d}  {\pmb{\pi}}^{{A}_{\mu}\nu} \wedge 
\hbox{d} {A}_{\mu} \wedge  \hbox{d} \mathfrak{y}_{\nu} {\big{)}}  = - \dd  {\pmb{\pi}}^{{A}_{\mu}\nu} \wedge   \vol_\nu .
\]
We prefer to consider $\dttP_{\phi}  =\phi_{\mu} (x)   {\pmb{\pi}}^{A_\mu \nu}   \vol_\nu
$. The exterior derivative $\dd \dttP_{{\phi}}$ is given by:
\bee
\left.
\begin{array}{ccl}
\displaystyle  \dd \dttP_{{\phi}}  &  =  & \displaystyle     \dd {\big{(}} \phi_{\mu} (x)    {\pmb{\pi}}^{A_\mu \nu} {\big{)}} \wedge   \vol_\nu 
= 
{\big{(}}
 {\pmb{\pi}}^{A_\mu \nu}
\frac{\partial \phi_{\mu} }{\partial x^{\alpha}} (x)  \dd x^{\alpha} 
+ \phi_{\mu} (x) \dd  {\pmb{\pi}}^{A_\mu \nu}
{\big{)}}
\wedge  \vol_\nu
 \\
\displaystyle    &  =  & \displaystyle   {\pmb{\pi}}^{A_\mu \nu}
\frac{\partial \phi_{\mu} }{\partial x^{\nu}}   \vol
+ \phi_{\mu}(x) \dd  {\pmb{\pi}}^{A_\mu \nu}
\wedge  \vol_\nu .
\end{array}
\right.
\eee
The related  infinitesimal symplectomorphism is denoted by $\Xi (  \dttP_{{\phi}} )$:
 \bee
\Xi (  \dttP_{{\phi}} ) = \phi_{\mu} (x)  \frac{\partial}{\partial {A} _{\mu}} -  {\big{(}}  \frac{\partial  \phi_{\mu}  }{\partial x^{\nu}}  (x) {\pmb{\pi}}^{{A}_\mu \nu}  {\big{)}}  \frac{\partial}{\partial \mathfrak{e}} 
\eee
Let us compute the contraction $ \Xi (  \dttP_{{\phi}} )  \iN {\pmb{\Omega}}^{\tiny{\hbox{\sffamily DDW}}}$:  
\[
\left.
\begin{array}{ccl}
\displaystyle   
 \Xi (  \dttP_{{\phi}} )  \iN {\pmb{\Omega}}^{\tiny{\hbox{\sffamily DDW}}}   &  =  & \displaystyle  {\big{(}}   \phi_{\mu} (x )  \frac{\partial}{\partial {A} _{\mu}}      -  {\big{(}} \frac{\partial  \phi_{\mu}  }{\partial x^{\nu}} (x)  {\pmb{\pi}}^{{A}_\mu \nu}  {\big{)}}     \frac{\partial}{\partial \mathfrak{e}}   {\big{)}}  \iN {\big{(}} \hbox{d}  \mathfrak{e} \wedge \hbox{d} \mathfrak{y} +  \hbox{d}  {\pmb{\pi}}^{{A}_{\mu}\nu} \wedge 
 { A}_{\mu} \wedge  \hbox{d} \mathfrak{y}_{\nu}  {\big{)}} 
 \\
\displaystyle       &  =  & \displaystyle     -  {\big{(}}  \frac{\partial  \phi_{\mu} (x )  }{\partial x^{\nu}}  {\pmb{\pi}}^{{A}_\mu \nu}         {\big{)}}    \vol  -  \phi_{\mu} (x) \dd  {\pmb{\pi}}^{A_\mu \nu}
\wedge    \vol_\nu = - \dd \dttP_{{\phi}} .
\end{array}
\right.
\]
We focus on some algebraic position $(n-1)$-forms: $ \displaystyle \dttQ^{\psi}  =   \frac{1}{2} \psi^{\mu\nu} (x) A \wedge    \vol_{\mu\nu}  $ with $\psi^{\mu\nu} (x)$ a real function which is antisymmetric in the indices $\mu,\nu$. 
\[
\dttQ^{\psi}   =  \frac{1}{2}  \psi^{\mu\nu} (x) A_{\rho} \dd x^{\rho} \wedge    \vol_{\mu\nu} =  \frac{1}{2}  \psi^{\mu\nu} (x) A_{\rho} {\big{(}} \delta^{\rho}_{\mu}   \vol_{\nu} -\delta^{\rho}_{\nu}   \vol_{\mu} {\big{)}} 
 =  \frac{1}{2}  \psi^{\mu\nu} (x) {\big{(}} A_{\mu}     \vol_{\nu} - A_{\nu}    \vol_{\mu} {\big{)}} 
\]
Since $\psi^{\mu\nu} = - \psi^{\nu\mu}$ then  $\dttQ^{\psi}   =  \psi^{\mu\nu} (x) A_{\mu}     \vol_{\nu}$. We compute $\dd \dttQ^{\psi}$:
\[
\left.
\begin{array}{ccl}
\displaystyle   
\dd \dttQ^{\psi}  
    & = &  \displaystyle    \dd {\big{(}} \psi^{\mu\nu}(x) A_{\mu}     \vol_{\nu}  {\big{)}} A_{\mu} \frac{\partial  \psi^{\mu\nu}}{\partial x^{\sigma}} (x) \dd x^{\sigma}  \wedge     \vol_{\nu}
 + \psi^{\mu\nu}(x) \dd  A_{\mu} \wedge     \vol_{\nu}  
 \\
  \displaystyle   
    & = &  \displaystyle
 A_{\mu} \frac{\partial \psi^{\mu\nu}}{\partial x^{\nu}}     (x)   \vol 
 +
\psi^{\mu\nu} (x) \dd  A_{\mu} \wedge     \vol_{\nu} 
 \end{array}
\right.
\]
The related infinitesimal symplectomorphisms  are denoted $\Xi ( \dttQ^{\psi})$ and are given  by:  
\bee
\Xi ( \dttQ^{\psi})  = - {\big{(}} {{(}} A_\mu \frac{\partial \psi^{\mu\nu}}{\partial x^{\nu}}     {{)}}     \frac{\partial}{\partial \mathfrak{e}} 
+ \psi^{\mu\nu} (x) \frac{\partial}{\partial {\pmb{\pi}}^{A_\mu \nu} } {\big{)}}
\eee
Let  compute $\Xi ( \dttQ^{\psi})  \iN {\pmb{\Omega}}^{\tiny{\hbox{\sffamily DDW}}}$: 
\[
\left.
\begin{array}{ccl}
\displaystyle   
\Xi ( \dttQ^{\psi})  \iN {\pmb{\Omega}}^{\tiny{\hbox{\sffamily DDW}}} & = &  \displaystyle  - {\big{(}}  (A_\mu \frac{\partial \psi^{\mu\nu}}{\partial x^{\nu}}  )  \frac{\partial}{\partial \mathfrak{e}} 
+ \psi^{\mu\nu} (x) \frac{\partial}{\partial {\pmb{\pi}}^{A_\mu \nu} }   {\big{)}}  \iN {\big{(}} \hbox{d}  \mathfrak{e} \wedge \hbox{d} \mathfrak{y} +  \hbox{d}  {\pmb{\pi}}^{{A}_{\mu}\nu} \wedge 
\dd {A}_{\mu} \wedge  \hbox{d} \mathfrak{y}_{\nu}  {\big{)}}
\\
 \displaystyle & = & \displaystyle \  - {\big{(}}
 A_{\mu} \frac{\partial \psi^{\mu\nu}}{\partial x^{\nu}}     (x) {\big{)}}  \vol 
 -
\psi^{\mu\nu} (x) \dd  A_{\mu} \wedge     \vol_{\nu}  = - \dd \dttQ^{\psi}  
\end{array}
\right.
\]
We summarize the results relating the algebraic observable $(n-1)$-forms  $\dttP_{{\phi}},  \dttQ^{\psi}  $ and their related infinitesimal symplectomorphisms $\Xi (  \dttP_{{\phi}} )  ,    \Xi ( \dttQ^{\psi}) $: 
\begin{equation}\label{cxcxcxcx00}
\left|
\begin{array}{lll}
\displaystyle  \dttP_{{\phi}}  &  =  & \displaystyle  \phi_{\mu} (x)  {\pmb{\pi}}^{{A}_{\mu}\nu}   \vol_\nu
\\
\displaystyle    \dttQ^{\psi} &  =  & \displaystyle   \frac{1}{2} \psi^{\mu\nu} (x) A \wedge    \vol_{\mu\nu} 
\end{array}
\right.
\quad    
\left|
\begin{array}{lll}
\displaystyle   \Xi (  \dttP_{{\phi}} )    &  =  & \displaystyle    \phi_{\mu} (x )  \frac{\partial}{\partial {A} _{\mu}}      -  {\big{(}}  \frac{\partial  \phi_{\mu} (x )  }{\partial x^{\nu}}  {\pmb{\pi}}^{{A}_\mu \nu}   {\big{)}}    \frac{\partial}{\partial \mathfrak{e}} 
\\
\displaystyle    \Xi ( \dttQ^{\psi})  &  =  & \displaystyle    -   {\big{(}} A_\mu \frac{\partial \psi^{\mu\nu}}{\partial x^{\nu}}     {\big{)}}    \frac{\partial}{\partial \mathfrak{e}} 
+  \psi^{\mu\nu} (x)    \frac{\partial}{\partial {\pmb{\pi}}^{A_\mu \nu} }  
\end{array}
\right.
\end{equation}
 Let notice that if we  work in the pre-multisymplectic case $ ({\pmb{\cal M}}_{{\hbox{\skt 0}}} , {\pmb{\Omega}}^{\hbox{\skt 0}})$ we have: \begin{equation}\label{cxcxcxcx00bb}
\left|
\begin{array}{ccl}
\displaystyle   \dttP^{\hbox{\skt 0}}_{{\phi}}  &  =  & \displaystyle   {\pmb{\pi}}^{{A}_{\mu}\nu}   \vol_\nu
\\
\displaystyle    \dttQ^{\psi}_{\hbox{\skt 0}} &  =  & \displaystyle    \frac{1}{2} \psi^{\mu\nu} (x) A \wedge    \vol_{\mu\nu} 
\end{array}
\right.
\quad \quad 
\left|
\begin{array}{ccl}
\displaystyle   \Xi (  \dttP^{\hbox{\skt 0}}_{{\phi}} )    &  =  & \displaystyle    \phi_{\mu} (x )  \frac{\partial}{\partial {A} _{\mu}}      
\\
\displaystyle    \Xi ( \dttQ^{\psi}_{\hbox{\skt 0}} )  &  =  & \displaystyle     -    \psi^{\mu\nu} (x) \frac{\partial}{\partial {\pmb{\pi}}^{A_\mu \nu} }  
\end{array}
\right.
\end{equation}
We need a more embracing view to describe more precisely the general  conditions on the functions $\phi_{\mu} (x)$ and $\psi^{\mu\nu}(x)$ and also to consider more general choice of functions. In doing so we  provide a deeper  description of the infinitesimal symplectomorphisms $  \Xi ( \dttQ^{\psi}  ) $, $ \Xi (  \dttP_{{\phi}} )$, $  \Xi ( \dttQ^{\psi}_{\hbox{\skt 0}} ) $ and $ \Xi (  \dttP^{\hbox{\skt 0}}_{{\phi}} )$. It is the subject of the following sections \eqref{glopsss} - \eqref{aaaaaaap1}. Before going to that step, we give in the next section \eqref{bliocppwwEE} the Poisson bracket structure in this   simple case. The objects of interest are ${\big{\{}} \dttQ^{{\psi}} ,  \dttQ^{\tilde{\psi}}   {\big{\}}}$,   ${\big{\{}} \dttP_{\phi} ,  \dttP_{\phi^{\bullet}}  {\big{\}}} $ and $ {\big{\{}} \dttQ^{\psi} ,  \dttP_{\phi}  {\big{\}}}$.

\subsection{\hbox{\sffamily\bfseries\slshape{Poisson Bracket for algebraic $(n-1)$-forms}}}\label{bliocppwwEE}

We begin with the following proposition:

\begin{prop}\label{amqwv1} {\em Let $\phi_{\mu}(x) , {\tilde{\phi}}_{\mu}(x)$ and $\psi^{\mu\nu}(x) , {{\tilde{\psi}}}^{\mu\nu} (x)$ smooth functions with $\psi^{\mu\nu}(x)  = - \psi^{\nu\mu}(x)$ and  ${\tilde{\psi}}^{\mu\nu}(x)  = - {\tilde{\psi}}^{\nu\mu}(x)$. 
  for ${ ({\pmb{\cal M}}_{\tiny{\hbox{\sffamily Maxwell}}} , {\pmb{\Omega}}^{\tiny{\hbox{\sffamily DDW}}})}$ the set of canonical Poisson brackets  is given by:}
\[
\left.
\begin{array}{ccl}
\displaystyle   {\big{\{}} \dttQ^{\psi}  , \dttQ^{{\tilde{\psi}}}   {\big{\}}}  &  =  & \displaystyle    0,
 \\
  \displaystyle  {\big{\{}} \dttP_{\phi}  ,  \dttP_{{{\tilde{\phi}}}}  {\big{\}}}  &  =  & \displaystyle 0 
  \\
 \displaystyle  {\big{\{}} \dttQ^{\psi} ,  \dttP_{\phi}  {\big{\}}}  &  =  & \displaystyle 
    -   \psi^{\mu\nu} (x)    \phi_{\mu} (x ) \hbox{\em \dd}\mathfrak{y}_{\nu}.     
\end{array}
\right.
\]
\end{prop}
 This corresponds to the mathematical setting of the traditional Poisson bracket  for algebraic $(n-1)$-forms:
$
{{\textswab{P}}}_{\circ}^{n-1} ({\pmb{\cal M}})  \times  {{\textswab{P}}}_{\circ}^{n-1} ({\pmb{\cal M}})    \rightarrow      {{\textswab{P}}}_{\circ}^{n-1}  ({\pmb{\cal M}})$.  Let us consider two algebraic position observable $(n-1)$-forms given by \eqref{cxcxcxcx00}, namely:  $\displaystyle \dttQ^{\psi} =  \psi^{\mu\nu}  (x) A_{\mu}   (x)  \vol_{\nu}$ and $\displaystyle \dttQ^{\tilde{\psi}} =  {\tilde{\psi}}^{\mu\nu}  (x) A_{\mu}   (x)  \vol_{\nu} $. We compute the  bracket: 
\[
\left.
\begin{array}{ccl}
\displaystyle  {\big{\{}} \dttQ^{\psi} ,  \dttQ^{{\tilde{\psi}}}  {\big{\}}}   &  =  & \displaystyle  \Xi (\dttQ^{{\psi}}) \iN \Xi (\dttQ^{{\tilde{\psi}}}) \iN {\pmb{\Omega}}^{\hbox{\tiny{\sffamily DDW}}} 
\\
 \displaystyle        
&  =  & \displaystyle   - \Xi (\dttQ^{{\psi}}) \iN  {\big{(}}  {{(}} A_\mu \frac{\partial {\tilde{\psi}}^{\mu\nu} }{\partial x^{\nu}}     {{)}}     \frac{\partial}{\partial \mathfrak{e}} 
+ {\tilde{\psi}}^{\mu\nu}  (x) \frac{\partial}{\partial {\pmb{\pi}}^{A_\mu \nu} } {\big{)}}  \iN {\big{(}}  \hbox{d}  \mathfrak{e} \wedge \hbox{d} \mathfrak{y} +  \hbox{d}  {\pmb{\pi}}^{{A}_{\mu}\nu} \wedge 
\dd {A}_{\mu} \wedge  \hbox{d} \mathfrak{y}_{\nu}  {\big{)}} 
\\
 \displaystyle       
&  =  & \displaystyle     {\big{(}}  {{(}} A_\mu \frac{\partial \psi^{\mu\nu}  }{\partial x^{\nu}}     {{)}}     \frac{\partial}{\partial \mathfrak{e}} 
+ \psi^{\mu\nu} (x) \frac{\partial}{\partial {\pmb{\pi}}^{A_\mu \nu} }   {\big{)}}  \iN    {\big{(}}  {\big{(}} A_\mu \frac{\partial {\tilde{\psi}}^{\mu\nu}}{\partial x^{\nu}}     {\big{)}}   \vol
+ {\tilde{\psi}}^{\mu\nu}  (x) \dd {A}_{\mu} \wedge  \hbox{d} \mathfrak{y}_{\nu}  {\big{)}}  = 0.
\end{array}
\right.
\]
Now we compute ${\big{\{}} \dttP_{\phi} ,  \dttP_{\phi^{\bullet}}  {\big{\}}}$ where the algebraic $(n-1)$-forms  $\dttP_{\phi}$ and the related infinitesimal symplectomorphisms  $\Xi ( \dttP_{\phi} )$ are given by \eqref{cxcxcxcx00}. Hence we have  the  internal  bracket: 
\[
\left.
\begin{array}{ccl}
\displaystyle  {\big{\{}} \dttP_{\phi} ,  \dttP_{\tilde{\phi}}  {\big{\}}}  &  =  & \displaystyle   \Xi ( \dttP_{\phi} ) \iN \Xi ( \dttP_{\tilde{\phi}} ) \iN {\pmb{\Omega}}^{\hbox{\tiny{\sffamily DDW}}}  
\\
 \displaystyle     
&  =  & \displaystyle
   -  \Xi ( \dttP_{\phi} ) \iN {\big{(}}   {\tilde{\phi}}_{\mu}  (x )  \frac{\partial}{\partial {A} _{\mu}}      -  {\big{(}}  \frac{\partial  {\tilde{\phi}}_{\mu} (x )  }{\partial x^{\nu}}  {\pmb{\pi}}^{{A}_\mu \nu} {\big{)}}  \frac{\partial}{\partial \mathfrak{e}}   {\big{)}} \iN {\pmb{\Omega}}^{\hbox{\tiny{\sffamily DDW}}}
\\
 \displaystyle     
&  =  & \displaystyle     -  \Xi ( \dttP_{\phi} ) \iN {\big{(}}  -  {\tilde{\phi}}_{\mu}  (x )  \dd {\pmb{\pi}}^{A_\mu \nu} \wedge \vol_{\nu}    -  {\big{(}}  \frac{\partial  {\tilde{\phi}}_{\mu}  (x )  }{\partial x^{\nu}}  {\pmb{\pi}}^{{A}_\mu \nu}         {\big{)}}    \vol {\big{)}}   = 0 .
\end{array}
\right.
\]
Finally, we compute the last bracket ${\big{\{}} \dttQ^{\psi} ,  \dttP_{\phi}  {\big{\}}}$:
\[
\left.
\begin{array}{ccl}
\displaystyle   {\big{\{}} \dttQ^{\psi} ,  \dttP_{\phi}  {\big{\}}}   &  =  & \displaystyle    -  \Xi ( \dttQ^{\psi^{\circ}} ) \iN {\big{(}}   -  \phi_{\mu} (x )  \dd {\pmb{\pi}}^{A_\mu \nu} \wedge \vol_{\nu}    -  {\big{(}}  \frac{\partial  \phi_{\mu} (x )  }{\partial x^{\nu}}  {\pmb{\pi}}^{{A}_\mu \nu}         {\big{)}}    \vol {\big{)}}    
\\
 \displaystyle  
&  =  & \displaystyle  
- {\big{(}}  {\big{(}} A_\mu \frac{\partial \psi^{\mu\nu}}{\partial x^{\nu}}     {\big{)}}     \frac{\partial}{\partial \mathfrak{e}} 
+ \psi^{\mu\nu} (x) \frac{\partial}{\partial {\pmb{\pi}}^{A_\mu \nu} } {\big{)}}   \iN {\big{(}}   -  \phi_{\mu} (x )  \dd {\pmb{\pi}}^{A_\mu \nu} \wedge \vol_{\nu}    -  {\big{(}}  \frac{\partial  \phi_{\mu} (x )  }{\partial x^{\nu}}  {\pmb{\pi}}^{{A}_\mu \nu}         {\big{)}}    \vol {\big{)}}  
\\
 \displaystyle  
&  =  & \displaystyle   -   \psi^{\mu\nu} (x)    \phi_{\mu} (x )   \vol_{\nu}    .
\end{array}
\right.
\]
Finally,
\[
\left.
\begin{array}{ccl}
\displaystyle  {\big{\{}} \dttQ^{\psi} ,  \dttP_{\phi}  {\big{\}}}    &  =  & \displaystyle    \Xi ( \dttQ^{\psi} ) \iN \dd  \dttP_{\phi}  
\\
 \displaystyle  &  =  & \displaystyle 
- {\big{(}}  {\big{(}} A_\mu \frac{\partial \psi^{\mu\nu}}{\partial x^{\nu}}     {\big{)}}     \frac{\partial}{\partial \mathfrak{e}} 
+ \psi^{\mu\nu} (x) \frac{\partial}{\partial {\pmb{\pi}}^{A_\mu \nu} } {\big{)}}  \iN {\big{(}}   {\pmb{\pi}}^{A_\mu \nu}
\frac{\partial \phi_{\mu} }{\partial x^{\nu}}   \vol
+ \phi_{\mu} (x) \dd  {\pmb{\pi}}^{A_\mu \nu}
\wedge  \vol_\nu  {\big{)}} .
\end{array}
\right.
\]
 So that ${\big{\{}} \dttQ^{\psi} ,  \dttP_{\phi}  {\big{\}}}  = -   \psi^{\mu\nu} (x)    \phi_{\mu} (x )   \vol_{\nu} $. We summarize our results and recover the   proposition \eqref{amqwv1}:
\[
\left.
\begin{array}{ccl}
\displaystyle   {\big{\{}} \dttQ^{\psi_{\circ}} ,  \dttQ^{\psi_{\bullet}}  {\big{\}}}  &  =  & \displaystyle  {\big{\{}} \dttP_{\phi} ,  \dttP_{\phi^{\bullet}}  {\big{\}}}  = 0,
\quad \quad \hbox{and} \quad \quad     {\big{\{}} \dttQ^{\psi} ,  \dttP_{\phi}  {\big{\}}}  
=  -   \psi^{\mu\nu} (x)    \phi_{\mu} (x )   \vol_{\nu}  .
\end{array}
\right.
\]

\subsection{\hbox{\sffamily\bfseries\slshape{All algebraic observable $(n-1)$-forms}}}\label{glopsss}

  In this section we describe the set of all algebraic $(n-1)$-forms and their related   infinitesimal symplectomorphisms $\Xi   \in \Gamma( { {\pmb{\cal M}} }_{\tiny{\hbox{\sffamily Maxwell}}}  , T { {\pmb{\cal M}} }_{\tiny{\hbox{\sffamily Maxwell}}}   )$. 
First we introduce the notations. We consider $\zeta \in \pmb{\mathfrak{Z}}$ and we denote:
\begin{equation}\label{vfz1}
\zeta   = X^{\nu} (x,{{A}}) \frac{\partial}{\partial x^{\nu}}  + \Theta_{\mu} (x,{{A}}) \frac{\partial}{\partial {{A}}_{\mu}}      
\end{equation}
with $  {X}^{\nu},  {\Theta}_{\mu}   $ are smooth real-valued  functions  on $ \pmb{\mathfrak{Z}}  $. We denote   $\Xi_{\tiny{\hbox{\sffamily DDW}}}    \in \Gamma( { {\pmb{\cal M}} }_{\tiny{\hbox{\sffamily DDW}}}  , T { {\pmb{\cal M}} }_{\tiny{\hbox{\sffamily DDW}}}   )$:
\begin{equation}\label{vfm001}
\Xi_{\tiny{\hbox{\sffamily DDW}}}  =       {\pmb{X}}^{\nu} (q,p)  \frac{\partial}{\partial x^{\mu}} 
+
  {\pmb{\Theta}}_{\mu}(q,p)   \frac{\partial}{\partial {{A}}_{\mu}}     +   {\pmb{\Upsilon}}  (q,p)    \frac{\partial}{\partial \mathfrak{e}}   +   {\pmb{\Upsilon}}^{{{A}}_{\mu}\nu}   (q,p)  \frac{\partial}{\partial {\pmb{\pi}}^{{{A}}_{\mu}\nu}}.
\end{equation}
We also denote  vector fields  $\Xi  = \Xi_{\tiny{\hbox{\sffamily Maxwell}}}  \in \Gamma( { {\pmb{\cal M}} }_{\tiny{\hbox{\sffamily Maxwell}}}  , T { {\pmb{\cal M}} }_{\tiny{\hbox{\sffamily Maxwell}}}   )$:
\begin{equation}\label{vfm001}
\Xi_{\hbox{\tiny\sffamily DDW}}   =       
{\pmb{X}}^{\nu} (q,p)  \frac{\partial}{\partial x^{\mu}} +
{\pmb{\Theta}}_{\mu}(q,p)   \frac{\partial}{\partial {{A}}_{\mu}}     +   {\pmb{\Upsilon}} (q,p) \frac{\partial}{\partial \mathfrak{e}}  + {\pmb{\Upsilon}}^{{{A}}_{\mu}\nu}   (q,p) {\Big{(}}  \frac{\partial}{\partial {\pmb{\pi}}^{{{A}}_{\mu}\nu}} -  \frac{\partial}{\partial {\pmb{\pi}}^{{{A}}_{\nu}\mu}} {\Big{)}}. \end{equation}
The objects  ${\pmb{X}}^{\nu} (q,p), {\pmb{\Theta}}_{\mu} (q,p) ,  {\pmb{\Upsilon}} (q,p)  $ and ${\pmb{\Upsilon}}^{{{A}}_{\mu}\nu} (q,p)  $ are smooth functions on $   {\pmb{\cal M}}_{\tiny{\hbox{\sffamily Maxwell}}} \subset {\pmb{\cal M}}_{\tiny{\hbox{\sffamily DDW}}}   \subset \Lambda^n T^{\star} (T^{\star}{\cal X})$, with values in $\Bbb{R}$.  Now we evaluate the expression $\Xi \iN  {\pmb{\Omega}}^{{\tiny\hbox{\sffamily DDW}}}$:
\[
\Xi  \iN  {\pmb{\Omega}}^{{\tiny\hbox{\sffamily DDW}}} =     {\pmb{\Upsilon}}     \hbox{d} \mathfrak{y}   -    {\pmb X}^{\nu}  \hbox{d} \mathfrak{e} \wedge \hbox{d}  \mathfrak{y}_{\nu}  +   {\pmb{\Upsilon}}^{{{A}}_{\mu}\nu}   \hbox{d} {{A}}_{\mu}  \wedge  \hbox{d} \mathfrak{y}_{\nu}  -    {\pmb{\Theta}}_{{\mu}} \hbox{d}  {\pmb{\pi}}^{{{A}}_{\mu}\nu}  \wedge  \hbox{d} \mathfrak{y}_{\nu}  +  {\pmb X}^{\rho} \hbox{d}  {\pmb{\pi}}^{{{A}}_{\mu}\nu}  \wedge
 \hbox{d} {{A}}_{\mu} \wedge \hbox{d}  \mathfrak{y}_{\rho\nu}.  \]
  We lift relations  from the   definition of a symplectomorphism $ \hbox{d} (\Xi \iN  {\pmb{\Omega}}^{{\tiny\hbox{\sffamily DDW}}}) = 0$. We make the following calculation:
\[
\left.
\begin{array}{rcl}
\displaystyle   \hbox{d} (\Xi  \iN  {\pmb{\Omega}}^{{\tiny\hbox{\sffamily DDW}}} )   & = &  \displaystyle
\hbox{d} {\pmb{\Upsilon}}     \wedge   \hbox{d} \mathfrak{y} -  \hbox{d} {\pmb{X}}^{\nu} \wedge \hbox{d} \mathfrak{e} \wedge \hbox{d} \mathfrak{y}_{\nu}  \\
 \displaystyle     &  &  \displaystyle  +
\hbox{d}     {\pmb{\Upsilon}}^{{{A}}_{\mu}\nu}    \hbox{d} {A}_{\mu}    \wedge  \hbox{d} \mathfrak{y}_{\nu}    - \hbox{d}  {\pmb{\Theta}}_{{\mu}}  \wedge \hbox{d}  {\pmb{\pi}}^{{A}_{\mu}\nu}    \wedge  \hbox{d} \mathfrak{y}_{\nu}  + \hbox{d} {\pmb X}^{\rho} \wedge \hbox{d}  {\pmb{\pi}}^{{A}_{\mu}\nu}    \wedge
\hbox{d} {A}_{\mu}  \wedge \hbox{d} \mathfrak{y}_{\rho\nu} .
\end{array}
\right.
\]
Then, we write this expression in the form of a sum  
$
 \hbox{d} (\Xi  \iN  {\pmb{\Omega}}^{{\tiny\hbox{\sffamily DDW}}} )   =  \sum_{i}  {\pmb \iota}_{i}  $
with each terms $ {\pmb \iota}_{i} $ given by:
\[
\left|
\begin{array}{ccl}
\displaystyle  {\pmb \iota}_{1}      & =  & \displaystyle     \hbox{d} {\pmb{\Upsilon}}     \wedge \vol     
 \\
\displaystyle  {\pmb \iota}_{2}      & =  & \displaystyle         -  \hbox{d} {\pmb X}^{\nu} \wedge \hbox{d} \mathfrak{e} \wedge \dd \mathfrak{y}_{\nu} 
 \\
\displaystyle  {\pmb \iota}_{3}     & =  & \displaystyle    \hbox{d}    {\pmb{\Upsilon}}^{{ A}_{\mu}\nu}    \hbox{d} {A}_{\mu}    \wedge  \dd \mathfrak{y}_{\nu}     
 \\
\displaystyle  {\pmb \iota}_{4}     & =  & \displaystyle   - \hbox{d}  {\pmb{\Theta}}_{{\mu}}  \wedge \hbox{d}  {\pmb{\pi}}^{{A}_{\mu}\nu}    \wedge  \hbox{d} \mathfrak{y}_{\nu}  
 \\
\displaystyle  {\pmb \iota}_{5}     & =  & \displaystyle        \hbox{d} {\pmb X}^{\rho} \wedge \hbox{d}  {\pmb{\pi}}^{{A}_{\mu}\nu}    \wedge
\hbox{d} {A}_{\mu}   \wedge \hbox{d} \mathfrak{y}_{\rho\nu} .
\end{array}
\right.
\]
 Since  $ \displaystyle   \hbox{d}   {\pmb{\Upsilon}}   =      \frac{\partial {\pmb{\Upsilon}} }{\partial x^{\alpha}} \dd x^{\alpha} + \frac{\partial {\pmb{\Upsilon}} }{\partial  {{A} _{\beta}}} \hbox{d} {{A} _{\beta}}  + \frac{\partial  {\pmb{\Upsilon}} }{\partial  \mathfrak{e}} \hbox{d} \mathfrak{e} + \frac{\partial   {\pmb{\Upsilon}}  }{\partial  {\pmb{\pi}}^{{A}_{\beta}\alpha}  } \hbox{d} {\pmb{\pi}}^{{A}_{\beta}\alpha} 
   $,  the first term $ {\pmb \iota}_{1}    $ is written: 
   \bee\label{1gkdkdfjsj443}
    {\pmb \iota}_{1}    =
   \frac{\partial {\pmb{\Upsilon}} }{\partial  {{A} _{\beta}}} \hbox{d} {{A} _{\beta}}   \wedge \hbox{d} \mathfrak{y}  +    \frac{\partial  {\pmb{\Upsilon}} }{\partial  \mathfrak{e}}    \hbox{d} \mathfrak{e} \wedge \hbox{d} \mathfrak{y}  +      \frac{\partial   {\pmb{\Upsilon}}  }{\partial  {\pmb{\pi}}^{{ A}_{\beta}\alpha}  }     \hbox{d} {\pmb{\pi}}^{{A}_{\beta}\alpha} 
 \wedge   \hbox{d} \mathfrak{y} .
   \eee
Moreover, since $ \displaystyle   \hbox{d}  {\pmb X}^{\rho}   =      \frac{\partial  {\pmb X}^{\rho} }{\partial x^{\alpha}} \dd x^{\alpha} + \frac{\partial  {\pmb X} }{\partial  {{A} _{\beta}}} \hbox{d} {{A} _{\beta}} + \frac{\partial  {\pmb X}^{\rho} }{\partial  \mathfrak{e}} \hbox{d} \mathfrak{e}   + \frac{\partial  {\pmb X}^{\rho} }{\partial  {\pmb{\pi}}^{{A}_{\beta}\alpha} } \hbox{d} {{\pmb{\pi}}^{ {A}_{\beta}  \alpha } }
   $, the term $  {\pmb \iota}_{2}  
$ is written: 
\bee\label{2gkdkdfjsj443}
\left.
\begin{array}{rcl}
\displaystyle   {\pmb \iota}_{2}   & = &  \displaystyle  -  \frac{\partial  {\pmb {X}}^{\nu} }{\partial x^{\alpha}}   \dd x^{\alpha} \wedge \hbox{d} \mathfrak{e} \wedge \hbox{d} \mathfrak{y}_{\nu}  -   
  \frac{\partial  {\pmb {X}}^{\nu} }{\partial  {{A} _{\beta}}}  \hbox{d} {{A} _{\beta}}  \wedge \hbox{d} \mathfrak{e} \wedge \hbox{d} \mathfrak{y}_{\nu}    -   \frac{\partial  {\pmb {X}}^{\nu} }{\partial  {\pmb{\pi}}^{{  A}_{\beta}\alpha} } \hbox{d} {{\pmb{\pi}}^{ {  A}_{\beta}  \alpha } }    \wedge \hbox{d} \mathfrak{e}   \wedge  \hbox{d} \mathfrak{y}_{\nu} .  \\
\end{array}
\right.
\eee
whereas the term  $ {\pmb \iota}_{5}   $
is written: 
\bee\label{5gkdkdfjsj443}
\left.
\begin{array}{rcl}
\displaystyle   {\pmb \iota}_{5}  & = &  \displaystyle       \frac{\partial  {\pmb X}^{\rho} }{\partial x^{\alpha}}  \dd x^{\alpha} \wedge \hbox{d}  {\pmb{\pi}}^{{A}_{\mu}\nu}    \wedge
\hbox{d} {A}_{\mu}   \wedge \hbox{d} \mathfrak{y}_{\rho\nu}      +     \frac{\partial  {\pmb X}^{\rho} }{\partial  {{A} _{\beta}}}   \hbox{d} {{A} _{\beta}}   \wedge \hbox{d}  {\pmb{\pi}}^{{A}_{\mu}\nu}   \wedge \hbox{d} {A}_{\mu}   \wedge \hbox{d} \mathfrak{y}_{\rho\nu}    
\\
\displaystyle     &  &  \displaystyle 
+   \frac{\partial  {\pmb X}^{\rho} }{\partial  \mathfrak{e}}  \hbox{d} \mathfrak{e}     \wedge \hbox{d}  {\pmb{\pi}}^{{A}_{\mu}\nu}    \wedge
\hbox{d} {A}_{\mu}   \wedge \hbox{d} \mathfrak{y}_{\rho\nu} 
 +    \frac{\partial  {\pmb X}^{\rho} }{\partial  {\pmb{\pi}}^{{A}_{\beta}\alpha} }   \hbox{d} {{\pmb{\pi}}^{ {A}_{\beta}  \alpha } }   \wedge \hbox{d}  {\pmb{\pi}}^{{A}_{\mu}\nu}    \wedge
\hbox{d} {A}_{\mu}   \wedge \hbox{d} \mathfrak{y}_{\rho\nu}  .
\end{array}
\right.
\eee
due to  $ \displaystyle  \hbox{d}   {\pmb{\Upsilon}}^{{A}_{\mu}\nu}   =    \frac{\partial  {\pmb{\Upsilon}}^{{A}_{\mu}\nu}     }{\partial x^{\alpha}} \dd x^{\alpha} + \frac{\partial   {\pmb{\Upsilon}}^{{A}_{\mu}\nu}    }{\partial  {{A} _{\beta}}} \hbox{d} {{A} _{\beta}}  + \frac{\partial   {\pmb{\Upsilon}}^{{A}_{\mu}\nu}    }{\partial  \mathfrak{e}} \hbox{d} \mathfrak{e} + \frac{\partial    {\pmb{\Upsilon}}^{{A}_{\mu}\nu}    }{\partial  {{\pmb{\pi}}^{ {A}_{\beta}  \alpha } }} \hbox{d} {\pmb{\pi}}^{{A}_{\beta}\alpha}  
$ we expand the term $  {\pmb \iota}_{3}   $ : 
\bee\label{3gkdkdfjsj443}
\left.
\begin{array}{rcl}
\displaystyle  {\pmb \iota}_{3}    & = &  \displaystyle     - {\big{(}} \frac{\partial  {\pmb{\Upsilon}}^{{A}_{\mu}\nu}     }{\partial x^{\nu}} - \frac{\partial  {\pmb{\Upsilon}}^{{A}_{\nu}\mu}     }{\partial x^{\nu}}   
{ \big{)}} 
   \hbox{d} {A}_{\mu}    \wedge  \dd \mathfrak{y}_{\nu}     
 +  
 { \big{(}}
   \frac{\partial   {\pmb{\Upsilon}}^{{A}_{\mu}\nu}    }{\partial  {{A} _{\beta}}}  -  \frac{\partial   {\pmb{\Upsilon}}^{{A}_{\nu}\mu}    }{\partial  {{A} _{\beta}}}    
   { \big{)}}
      \hbox{d} {{A} _{\beta}}     \wedge  \hbox{d} {A}_{\mu}    \wedge   \hbox{d}  \mathfrak{y}_{\nu}      
      \\
\displaystyle     &  &  \displaystyle   +  
 { \big{(}}
    \frac{\partial   {\pmb{\Upsilon}}^{{A}_{\mu}\nu}    }{\partial  \mathfrak{e}} 
    -
        \frac{\partial   {\pmb{\Upsilon}}^{{A}_{\nu}\mu}    }{\partial  \mathfrak{e}} 
     { \big{)}}
       \hbox{d} \mathfrak{e} \wedge  \hbox{d} {A}_{\mu}    \wedge  \dd \mathfrak{y}_{\nu}   +
     { \big{(}} 
  \frac{\partial    {\pmb{\Upsilon}}^{{A}_{\mu}\nu}    }{\partial  {{\pmb{\pi}}^{ {A}_{\beta}  \alpha } }} 
  - 
  \frac{\partial    {\pmb{\Upsilon}}^{{A}_{\nu}\mu}    }{\partial  {{\pmb{\pi}}^{ {A}_{\beta}  \alpha } }}  
       { \big{)}}
        \hbox{d} {\pmb{\pi}}^{{A}_{\beta}\alpha}   \wedge  \hbox{d} {A}_{\mu}    \wedge  \hbox{d} \mathfrak{y}_{\nu},       
\end{array}
\right.
\eee
and finally $\displaystyle  \hbox{d} {\pmb{\Theta}}_{\mu}  =    \frac{\partial {\pmb{\Theta}}_{\mu}}{\partial x^{\alpha}} \dd x^{\alpha} + \frac{\partial {\pmb{\Theta}}_{\mu}}{\partial  {{A} _{\beta}}} \hbox{d} {{A} _{\beta}}   + \frac{\partial  {\pmb{\Theta}}_{\mu}}{\partial  \mathfrak{e}} \hbox{d} \mathfrak{e} + \frac{\partial  {\pmb{\Theta}}_{\mu} }{\partial  {\pmb{\pi}}^{{A}_{\beta}\alpha}  }  \hbox{d} {\pmb{\pi}}^{{A}_{\beta}\alpha} $ gives   the last   term ${\pmb \iota}_{4}  
$:
\bee\label{4gkdkdfjsj443}
\left.
\begin{array}{rcl}
\displaystyle  {\pmb \iota}_{4}    & = &  \displaystyle   \frac{\partial {\pmb{\Theta}}_{\mu}}{\partial x^{\nu}}   \hbox{d}  {\pmb{\pi}}^{{A}_{\mu}\nu}    \wedge  \hbox{d} \mathfrak{y}
 -    \frac{\partial {\pmb{\Theta}}_{\mu}}{\partial  {{A} _{\beta}}}    \hbox{d} {{A} _{\beta}}   \wedge  \hbox{d}  {\pmb{\pi}}^{{A}_{\mu}\nu}    \wedge  \hbox{d} \mathfrak{y}_{\nu}
      -    \frac{\partial  {\pmb{\Theta}}_{\mu}}{\partial  \mathfrak{e}}    \hbox{d}  \mathfrak{e} \wedge \hbox{d}  {\pmb{\pi}}^{{A}_{\mu}\nu}    \wedge  \hbox{d} \mathfrak{y}_{\nu}  
      \\
      \displaystyle     &  &  \displaystyle  
       -      \frac{\partial  {\pmb{\Theta}}_{\mu} }{\partial  {\pmb{\pi}}^{{A}_{\beta}\alpha}  }    \hbox{d} {\pmb{\pi}}^{{A}_{\beta}\alpha}    \wedge\hbox{d}  {\pmb{\pi}}^{{A}_{\mu}\nu}    \wedge  \hbox{d} \mathfrak{y}_{\nu} . \\
\end{array}
\right.
\eee
 The decomposition of the terms $\eqref{1gkdkdfjsj443}$-$\eqref{4gkdkdfjsj443}$ on the different ($n+1$)-forms involves  $\hbox{d} \mathfrak{e} \wedge \hbox{d} \mathfrak{y}  $,  $    \hbox{d}  {\pmb{\pi}}^{{A}_{\mu}\nu}    \wedge  \hbox{d} \mathfrak{y}   $, $ \hbox{d}  A_{\mu}  \wedge   \hbox{d} \mathfrak{y}   $, $  \hbox{d} \mathfrak{e} \wedge  \hbox{d} {A}_{\mu}    \wedge  \hbox{d} \mathfrak{y}_{\nu}  $, $ \hbox{d} {\pmb{\pi}}^{{A}_{\beta}\alpha}   \wedge  \hbox{d} {A}_{\mu}    \wedge  \hbox{d}\mathfrak{y}_{\nu} $, $ \hbox{d} \mathfrak{e}     \wedge \hbox{d}  {\pmb{\pi}}^{{A}_{\mu}\nu}    \wedge
\hbox{d} {A}_{\mu}   \wedge \hbox{d} \mathfrak{y}_{\rho\nu}   $, $\hbox{d} {\pmb{\pi}}^{{A}_{\beta}\alpha}    \wedge\hbox{d}  {\pmb{\pi}}^{{A}_{\mu}\nu}    \wedge  \hbox{d} \mathfrak{y}_{\nu} $. We now describe more precisely the different terms.   The decomposition involves the following terms ${\bf j}_{1}-{\bf j}_{12}$: 
 \begin{multicols}{2}
  {\em  |}  ${\bf j}_{1}$  is the term related to the decomposition on  $    \displaystyle {\big{[}} \hbox{d} {\pmb{\pi}}^{{A}_{\beta}\alpha}    \wedge\hbox{d}  {\pmb{\pi}}^{{A}_{\mu}\nu}    \wedge  \hbox{d} \mathfrak{y}_{\nu} {\big{]}}   $  so that: 
\[
{\bf j}_{1} =   -       \frac{\partial  {\pmb{\Theta}}_{\mu} }{\partial  {\pmb{\pi}}^{{A}_{\beta}\alpha}  }    \hbox{d} {\pmb{\pi}}^{{A}_{\beta}\alpha}    \wedge\hbox{d}  {\pmb{\pi}}^{{A}_{\mu}\nu}    \wedge  \hbox{d} \mathfrak{y}_{\nu}  
\]

  {\em  |}  ${\bf j}_{2}$  is the term related to the decomposition on $  {\big{[}}  \hbox{d} {{\pmb{\pi}}^{ {A}_{\beta}  \alpha } }   \wedge \hbox{d}  {\pmb{\pi}}^{{A}_{\mu}\nu}    \wedge
\hbox{d} {A}_{\mu}   \wedge \hbox{d} \mathfrak{y}_{\rho\nu}  {\big{]}}   $
 \[
{\bf j}_{2} =    
   \frac{\partial  {\pmb X}^{\rho} }{\partial  {\pmb{\pi}}^{{A}_{\beta}\alpha} }   \hbox{d} {{\pmb{\pi}}^{ {A}_{\beta}  \alpha } }   \wedge \hbox{d}  {\pmb{\pi}}^{{A}_{\mu}\nu}    \wedge
\hbox{d} {A}_{\mu}   \wedge \hbox{d} \mathfrak{y}_{\rho\nu} 
\]

  {\em  |}  ${\bf j}_{3}$  is the term related to the decomposition on $ {\big{[}}  \hbox{d} \mathfrak{e} \wedge \hbox{d} \mathfrak{y} {\big{]}}  $
\[
{\bf j}_{3} =   { \big{(}} \frac{\partial {\pmb{X}}^{\nu}}{\partial x^{\nu}} +  \frac{\partial  {\pmb{\Upsilon}} }{\partial  \mathfrak{e}}  { \big{)}} \hbox{d} \mathfrak{e} \wedge \hbox{d} \mathfrak{y} 
\]

{\em  |}  ${\bf j}_{4}$  is the term related to the decomposition on $   {\big{[}}  \hbox{d}  {\pmb{\pi}}  \wedge   \hbox{d} \mathfrak{y} {\big{]}}     $
\[
{\bf j}_{4} =  \frac{\partial {\pmb{\Theta}}_{\mu}}{\partial x^{\nu}}    \hbox{d}  {\pmb{\pi}}^{{A}_{\mu}\nu}    \wedge  \hbox{d} \mathfrak{y}   +   \frac{\partial   {\pmb{\Upsilon}}  }{\partial  {\pmb{\pi}}^{{A}_{\beta}\alpha}  }    \hbox{d} {\pmb{\pi}}^{{A}_{\beta}\alpha} 
 \wedge   \hbox{d} \mathfrak{y}  
\]

{\em  |}  ${\bf j}_{5}$  is the term related to the decomposition on 
$   {\big{[}}  \hbox{d}  {A} \wedge    \hbox{d}    \mathfrak{y} {\big{]}}   $
\[
{\bf j}_{5} =    \frac{\partial {\pmb{\Upsilon}} }{\partial  {{A} _{\beta}}}      \hbox{d} {{A} _{\beta}}   \wedge \dd \mathfrak{y}   -    { \big{(}} \frac{\partial  {\pmb{\Upsilon}}^{{A}_{\mu}\nu}     }{\partial x^{\nu}} - \frac{\partial  {\pmb{\Upsilon}}^{{A}_{\nu}\mu}     }{\partial x^{\nu}}   
{ \big{)}}    \hbox{d} {A}_{\mu}    \wedge  \hbox{d} \mathfrak{y}  
\]

{\em  |}  ${\bf j}_{6}$  is the term related to the decomposition on $   {\big{[}}   \hbox{d}  \mathfrak{e} \wedge \hbox{d}    {A} \wedge \dd \mathfrak{y}_{\nu}  {\big{]}}   $
\[
\left.
\begin{array}{rcl}
      \displaystyle  {\bf j}_{6}     &  = &  \displaystyle
   { \big{(}}
    \frac{\partial   {\pmb{\Upsilon}}^{{A}_{\mu}\nu}    }{\partial  \mathfrak{e}} 
    -
        \frac{\partial   {\pmb{\Upsilon}}^{{A}_{\nu}\mu}    }{\partial  \mathfrak{e}} 
     { \big{)}}
    \hbox{d} \mathfrak{e} \wedge  \hbox{d} {A}_{\mu}    \wedge  \hbox{d} \mathfrak{y}_{\nu}   \\
      \displaystyle     &  &  \displaystyle     -   
   \frac{\partial  {\pmb {X}}^{\rho} }{\partial  {{A} _{\beta}}}    \hbox{d} {{A} _{\beta}}  \wedge \hbox{d} \mathfrak{e} \wedge \hbox{d} \mathfrak{y}   
    \end{array}
\right.
\]
 
{\em  |}  ${\bf j}_{7}$  is the term related to the decomposition on $       [ \hbox{d} {{A} _{\beta}}     \wedge  \hbox{d} {A}_{\mu}    \wedge   \hbox{d}  \mathfrak{y}_{\nu} ]$
\[
{\bf j}_{7} =   { \big{(}}
   \frac{\partial   {\pmb{\Upsilon}}^{{A}_{\mu}\nu}    }{\partial  {{A} _{\beta}}}  -  \frac{\partial   {\pmb{\Upsilon}}^{{A}_{\nu}\mu}    }{\partial  {{A} _{\beta}}}    
   { \big{)}} 
       \hbox{d} {{A} _{\beta}}     \wedge  \hbox{d} {A}_{\mu}    \wedge   \hbox{d}  \mathfrak{y}_{\nu} 
\]

{\em  |}  ${\bf j}_{8}$  is the term related to the decomposition on $  {\big{[}}  \hbox{d} {A} \wedge \hbox{d} {\pmb{\pi}} \wedge \hbox{d} \mathfrak{y}_{\nu}   {\big{]}}   $
\[
\left.
\begin{array}{rcl}
      \displaystyle  {\bf j}_{8}     &  = &  \displaystyle    \frac{\partial  {\pmb X}^{\rho} }{\partial x^{\alpha}} \dd x^{\alpha} \wedge \hbox{d}  {\pmb{\pi}}^{{A}_{\mu}\nu}    \wedge
\hbox{d} {A}_{\mu}   \wedge \hbox{d} \mathfrak{y}_{\rho\nu} 
\\
      \displaystyle     &  &  \displaystyle
 -     \frac{\partial {\pmb{\Theta}}_{\mu}}{\partial  {{A} _{\beta}}}    \hbox{d} {{A} _{\beta}}   \wedge  \hbox{d}  {\pmb{\pi}}^{{A}_{\mu}\nu}    \wedge  \hbox{d} \mathfrak{y}_{\nu} 
 \\
       \displaystyle     &  &  \displaystyle
  + 
  \frac{\partial    {\pmb{\Upsilon}}^{{A}_{\mu}\nu}    }{\partial  {{\pmb{\pi}}^{ {A}_{\beta}  \alpha } }}  \hbox{d} {\pmb{\pi}}^{{A}_{\beta}\alpha}   \wedge  \hbox{d} {A}_{\mu}    \wedge  \hbox{d}\mathfrak{y}_{\nu}    
    \end{array}
\right.
\]
 
{\em  |}  ${\bf j}_{9}$  is the term related to the decomposition on $ {\big{[}}  \hbox{d}  \mathfrak{e} \wedge \hbox{d}  {\pmb{\pi}}^{{A}_{\mu}\nu}    \wedge  \hbox{d} \mathfrak{y}_{\nu}  {\big{]}}   $ 
\[
{\bf j}_{9} = -     \frac{\partial  {\pmb{\Theta}}_{\mu}}{\partial  \mathfrak{e}}   \hbox{d}  \mathfrak{e} \wedge \hbox{d}  {\pmb{\pi}}^{{A}_{\mu}\nu}    \wedge  \hbox{d} \mathfrak{y}_{\nu}  
\]

{\em  |}  ${\bf j}_{10}$  is the term related to the decomposition on $  {\big{[}}    \hbox{d} {{\pmb{\pi}}^{ {A}_{\beta}  \alpha } }    \wedge \hbox{d} \mathfrak{e}   \wedge  \dd \mathfrak{y}_{\nu}   {\big{]}}   $ 
\[
{\bf j}_{10} =    -   \frac{\partial  {\pmb X}^{\rho} }{\partial  {\pmb{\pi}}^{{A}_{\beta}\alpha} }  \hbox{d} {{\pmb{\pi}}^{ {A}_{\beta}  \alpha } }    \wedge \hbox{d} \mathfrak{e}   \wedge  \dd \mathfrak{y}_{\nu} 
\]

{\em  |}  ${\bf j}_{11}$  is the term related to the decomposition on $ {\big{[}}   \hbox{d} {A}  \wedge  \hbox{d} {\pmb{\pi}}   \wedge \hbox{d} {A}  \wedge \hbox{d} \mathfrak{y}_{\rho\nu}  {\big{]}}  $ 
\[
{\bf j}_{11} =     \frac{\partial  {\pmb X}^{\rho} }{\partial  {{A} _{\beta}}}    \hbox{d} {{A} _{\beta}}   \wedge \hbox{d}  {\pmb{\pi}}^{{A}_{\mu}\nu}   \wedge \hbox{d} {A}_{\mu}   \wedge \hbox{d} \mathfrak{y}_{\rho\nu} 
\]

{\em  |}  ${\bf j}_{12}$  is the term related to the decomposition on $ {\big{[}} \hbox{d} \mathfrak{e}  \wedge  \hbox{d} {\pmb{\pi}}   \wedge \hbox{d} {  A}  \wedge \hbox{d} \mathfrak{y}_{\rho\nu} {\big{]}}$
   \bee\label{masterAOF}
   {\bf j}_{12} =   \frac{\partial  {\pmb {X}}^{\rho} }{\partial  \mathfrak{e}}   \hbox{d} \mathfrak{e}     \wedge \hbox{d}  {\pmb{\pi}}^{{A}_{\mu}\nu}    \wedge
\hbox{d} {A}_{\mu}   \wedge \hbox{d} \mathfrak{y}_{\rho\nu}  
\eee
\end{multicols}
   The decomposition of $ \hbox{d} (\Xi  \iN  {\pmb{\Omega}}^{{\tiny\hbox{\sffamily DDW}}} )  $ gives us information about the dependence of the involved functions. 
  Hence from \eqref{masterAOF}-${\bf j}_{1}$-${\bf j}_{2}$, we conclude that ${\pmb {X}}^{\rho}$ and ${\pmb{\Theta}}_{\mu}$ are independent of  the variables ${\pmb{\pi}}^{{A}_{\beta}\alpha} $.   
 Then, from the terms  \eqref{masterAOF}-${\bf j}_{9}$-${\bf j}_{10}$ and \eqref{masterAOF}-${\bf j}_{2}$  we observe that  ${\pmb{\Theta}}_{\mu}$ are independent of the variable $\mathfrak{e}$.  From  \eqref{masterAOF}-${\bf j}_{11}$ we find that     ${\bf {X}}^{\rho}$ is independent of  the variables $ {{  A} _{\beta}}$.  
From \eqref{masterAOF}-${\bf j}_{12}$ we find again that  $ {\bf {X}}^{\rho}$ is independent of the variable $\mathfrak{e}$. Due to decompositions \eqref{masterAOF}-${\bf j}_{1}$-${\bf j}_{2}$-${\bf j}_{9}$-${\bf j}_{10}$-${\bf j}_{12}$   we find that ${\pmb {X}}^{\rho} = {\pmb {X}}^{\rho} (x, {A})$ and  $ {\pmb{\Theta}}_{{\mu}}   =  {\pmb{\Theta}}_{{\mu}}   (x,{A})$.  From \eqref{masterAOF}-${\bf j}_{11}$, we   observe  ${\pmb X}^{\rho} = {\pmb {X}}^{\rho} (x)$, so that, due to \eqref{masterAOF}-${\bf j}_{6}$-${\bf j}_{7}$, we obtain $  {\pmb{\Upsilon}}^{{A}_{\mu}\nu}      =  {\pmb{\Upsilon}}^{{A}_{\mu}\nu}     (x,{\pmb{\pi}}) $. We don't have any extra information   on  $ {\pmb{\Upsilon}} = {\pmb{\Upsilon}} (x,{A},\mathfrak{e} , {\pmb{\pi}})$.  The functions $   {\pmb {X}}^{\nu} ,   {\pmb {\Theta}} _{\mu} ,   {\pmb{\Upsilon}} , {\pmb{\Upsilon}}^{{A}_{\mu}\nu}    $ are smooth functions on $ {\pmb{\cal M}}_{\tiny{\hbox{\sffamily DDW}}}  \subset \Lambda^n T^{\star} {\pmb{\mathfrak{Z}}}$ and, from the previous analysis, satisfy the following coordinate dependence:  \begin{equation}\label{qmddmzdzdsss}
\left.
\begin{array}{ccccccc}
\displaystyle  {{\pmb{X}}}^{\nu}   =   \displaystyle    {{\pmb{X}}}^{\nu} (x),  \  &  & \     {{\pmb{ \Theta}}}_{{\mu}}       =   \displaystyle   {{\pmb{ \Theta}}}_{{\mu}}    (x,{A}),  \  &   & \ {\pmb{\Upsilon}}  =   {\pmb{\Upsilon}}   (x,{A}, \mathfrak{e}, {\pmb{\pi}}),
 \  &   & \   {\pmb{\Upsilon}}^{{A}_{\mu}\nu}    =   \displaystyle  {\pmb{\Upsilon}}^{{A}_{\mu}\nu}    (x,{\pmb{\pi}}).
\end{array}
\right.
\end{equation}
We consider the further condition  ${\pmb{\Upsilon}}^{{A}_{\mu}\nu}   (q,p) = - {\pmb{\Upsilon}}^{{A}_{\nu}\mu}  (q,p)  $ so that we are left with equations  \eqref{masterAOF}-${\bf j}_{3}$-${\bf j}_{4}$-${\bf j}_{5}$-${\bf j}_{6}$: 
\begin{equation}
\left|
\begin{array}{ccc}
\displaystyle   \frac{\partial {\pmb{X}}^{\nu}}{\partial x^{\nu}} +  \frac{\partial  {\pmb{\Upsilon}} }{\partial  \mathfrak{e}}   &  =  & \displaystyle   0
\\
 \displaystyle         \frac{\partial {\pmb{\Theta}}_{\mu}}{\partial x^{\nu}}   +   \frac{\partial   {\pmb{\Upsilon}}  }{\partial  {\pmb{\pi}}^{{A}_{\mu}\nu}  }     &  =  & \displaystyle   0
\end{array}
\right.
\quad \quad \quad \quad \quad \quad       
\left|
\begin{array}{ccc}
\displaystyle     \frac{\partial {\pmb{\Upsilon}} }{\partial  {{A} _{\mu}}}     -     \frac{\partial  {\pmb{\Upsilon}}^{{A}_{\mu}\nu}     }{\partial x^{\nu}}      &  =  & \displaystyle   0
 \\ \displaystyle  \frac{\partial   {\pmb{\Upsilon}}^{{A}_{\mu}\nu}    }{\partial  \mathfrak{e}}      -   
   \frac{\partial  {\bf {X}}^{\nu} }{\partial  {{A} _{\mu}}}      &  =  & \displaystyle 0
\end{array}
\right.
\end{equation}
together with the set of equations involving more than two terms \eqref{masterAOF}-${\bf j}_{8}$. We have the following proposition: 
 \begin{prop}\label{algebraicallsymplec0}
  Let $\Xi \in \Gamma ({\pmb{\cal M}} , T {\pmb{\cal M}})$ then $\Xi$ satisfies {\em  $\dd (\Xi \iN {\pmb{\Omega}}^{\tiny{\hbox{\sffamily DDW}}}) = 0 $}  if and only if $\Xi$ is written $ \Xi = \overline{ \zeta}  + \chi   $
with
\bee\label{tyuoi12}
\left.
\begin{array}{ccl}
\displaystyle
\overline{ \zeta}  & = &  \displaystyle   X^{\nu}  \frac{\partial}{\partial x^{\nu}}  + \Theta _{\mu} \frac{\partial}{\partial  {A}_{\mu}}      -  {\Big{(}}           \mathfrak{e}  ( \frac{\partial  X^{\nu} }{\partial x^{\nu}} )  + \frac{\partial  \Theta_{{\mu}}  }{\partial x^{\nu}}  {\pmb{\pi}}^{ {A}_\mu \nu}         {\Big{)}}     \frac{\partial}{\partial \mathfrak{e}}
     \\
 \displaystyle
  &  &  \displaystyle
 -   {\Big{(}}   {\pmb{\pi}}^{ {A}_\rho \sigma}  \delta^{\mu}_{\rho}  {\big{(}}     \big{[}  (\frac{\partial  X^{\nu} }{\partial x^{\sigma}} )   -  \delta^{\nu}_{\sigma}     (\frac{\partial  X^{\lambda} }{\partial x^{\lambda}} )   \big{]}   +     ( \frac{\partial  \Theta_{{\sigma}}  }{\partial  {A}_{\nu} } ) {\big{)}}  -  \mathfrak{e}   ( \frac{\partial  X^{\nu} }{\partial { {A} _{\mu}}  }  )  {\Big{)}}     \frac{\partial}{ \partial {\pmb{\pi}}^{ {A}_\mu \nu}  } 
  \end{array}
\right.
\eee
and $\displaystyle \chi = {\Upsilon}  \frac{\partial}{\partial \mathfrak{e}}
 + {\Upsilon}^{ {A}_{\mu}\alpha}   \frac{\partial}{\partial {\pmb{\pi}}^{ {A}_\mu \alpha}  } $  with $\Upsilon :  {\pmb{\mathfrak{Z}}} \rightarrow \Bbb{R} $ and $ {\Upsilon}^{ {A}_{\mu}\alpha} :   {\pmb{\mathfrak{Z}}} \rightarrow  \Bbb{R}$  smooth functions on $ {\pmb{\mathfrak{Z}}}   $ such that: 
$ \displaystyle
\frac{\partial {{\Upsilon}}  }{\partial   {A} _{\mu}}  - \frac{\partial  {{\Upsilon}}^{ {A}_{\mu}\nu} }{\partial x^{\nu}} = 0
$.
\end{prop}
 The proposition \eqref{algebraicallsymplec0} is the application of the result due to  F. H\'elein and J. Kouneiher \cite{HK-02} \cite{HK-03}, which is recalled in  proposition  \eqref{algebraicallsymplec}. The latter describes the more general   search for all algebraic observable $(n-1)$-forms.
Any  infinitesimal symplectomorphism $ \Xi \in \mathfrak{sp}_{\circ} ( {\pmb{\cal M}} )$ can be written in the form  $
\Xi = \chi + {\bar{\zeta}}
$.
\begin{prop}\label{algebraicallsymplec}
If ${\pmb{\cal M}}$ is an open subset of $\Lambda^nT^{\star}\pmb{\mathfrak{Z}}$, then
the set of all infinitesimal symplectomorphisms
$\Xi$ on ${\pmb{\cal M}}$ are of the form $\Xi = \chi + \overline{ \zeta}$, where
\begin{equation}\label{szszszchixi}
\chi:= \sum_{\beta_1<\cdots <\beta_n}\chi_{\beta_1\cdots \beta_n}(q)
{\partial \over \partial p_{\beta_1\cdots \beta_n}} \quad \hbox{ and }\quad
\overline{ \zeta}:= \sum_\alpha  \zeta^\alpha(q){\partial \over \partial q^\alpha}
- \sum_{\alpha,\beta}{\partial  \zeta^\alpha\over \partial q^\beta}(q){\pmb{{ {\Pi}}}}^\beta_\alpha,
\end{equation}
\end{prop}
with, 

$[\mathfrak{1}]$ the coefficients $\chi_{\beta_1\cdots \beta_n}$ are such that 
$ \hbox{d} (\chi\iN {\pmb{\Omega}})= 0$. 

$[\mathfrak{2}]$ $\displaystyle \zeta:= \sum_\alpha  \zeta^\alpha(q){\partial \over \partial q^\alpha}$
is an arbitrary vector field on $\pmb{\mathfrak Z}$.

$[\mathfrak{3}]$
$  \displaystyle
{\pmb{{ {\Pi}}}}^\beta_\alpha:= \sum_{\beta_1<\cdots <\beta_n}\sum_\mu \delta^\beta_{\beta_\mu}
p_{\beta_1\cdots \beta_{\mu-1}\alpha\beta_{\mu+1}\cdots \beta_n}
{\partial \over \partial p_{\beta_1\cdots \beta_n}}$.
 
We  decompose the vector field $\Xi \in \Gamma( {\pmb{\cal M}}  , T{\pmb{\cal M}} )$ with general coordinates:
\[
\Xi = \Xi^{\alpha} (q,p) \frac{\partial}{\partial q^{\alpha}} + \sum_{\alpha_1 < ... < \alpha_{n}} \Xi_{\alpha_1 ... \alpha_n} (q,p) \frac{\partial}{\partial p_{\alpha_1 ... \alpha_n}}.
\]
Now we adapt our notations for the Maxwell theory, in the {\sffamily (DDW)} setting:
\bee
 \Xi^{\alpha} (q,p) = {\Big{\{}} {\bf {X}}^{\nu} (q,p)   ;  {\pmb{\Theta}}_{\mu} (q,p)   {\Big{\}}} , 
 \quad \quad    
 \Xi_{\alpha_1 ... \alpha_n} (q,p) = {\Big{\{}}  {\pmb{\Upsilon}} (q,p)  ; {\pmb{\Upsilon}}^{{{A}}_{\mu}\nu}   (q,p)   {\Big{\}}} . 
 \eee  
 We denote,
\[
 \chi = { \Xi} - \overline{ \zeta} = {\pmb{X}}^{\nu}   \frac{\partial}{\partial x^{\nu}}  +  {{\pmb{\Theta}}}_{\mu}    \frac{\partial}{\partial {A} _{\mu}}     +   {\pmb{\Upsilon}}     \frac{\partial}{\partial \mathfrak{e}} +      {\pmb{\Upsilon}}^{{A}_{\mu}\nu}      \frac{\partial}{\partial {\pmb{\pi}}^{{A}_{\mu}\nu} }
 - \overline{ \zeta} . 
 \]
We consider the expression \eqref{tyuoi12} so that we obtain a expression for $\chi = \Xi - \bar{\zeta} $
\bee\label{tyuoi1442}
\left.
\begin{array}{ccl}
\displaystyle
  \chi    & = &  \displaystyle 
 \underbrace{ { \big{(}} {\pmb{\Upsilon}} -  {\big{(}}     \mathfrak{e}  ( \frac{\partial  X^{\nu} }{\partial x^{\nu}} )  + \frac{\partial  \Theta_{{\mu}}  }{\partial x^{\nu}}  {\pmb{\pi}}^{{A}_\mu \nu}     {\big{)}}  { \big{)}} }_{\chi^{\mathfrak{e}}} \frac{\partial}{\partial \mathfrak{e}}   
 \\
 \displaystyle
  &  &  \displaystyle
  +\underbrace{  { \Big{(}} {\pmb{\Upsilon}}^{{A}_{\mu}\nu}   -  {\pmb{\pi}}^{{A}_\rho \sigma}  \delta^{\mu}_{\rho}   \BL     \big{[}  (\frac{\partial  X^{\nu} }{\partial x^{\sigma}} )   -  \delta^{\nu}_{\sigma}     (\frac{\partial  X^{\lambda} }{\partial x^{\lambda}} )   \big{]}   +     ( \frac{\partial  \Theta_{{\sigma}}  }{\partial {A}_{\nu} } ) \BR  -  \mathfrak{e}   ( \frac{\partial  X^{\nu} }{\partial {{A} _{\mu}}  }  )  { \Big{)}}      }_{\chi^{{\pmb{\pi}} } }     \frac{\partial}{ \partial {\pmb{\pi}}^{{A}_\mu \nu}  }.  \\
  \end{array}
\right.
\eee
  As announced  in the proposition \eqref{algebraicallsymplec}, we have coefficients of $\chi$ such that $\hbox{d} (\chi \iN {\pmb{\Omega}}^{{\tiny\hbox{\sffamily DDW}}}  ) = 0$. Since
  $\displaystyle \chi = \chi^{\mathfrak{e}}  \frac{\partial}{\partial \mathfrak{e}}   +  \chi^{{\pmb{\pi}}}     \frac{\partial}{ \partial {\pmb{\pi}}^{{A}_\mu \nu}  }  $, 
  the interior product of   $\chi  $ with ${\pmb{\Omega}}^{{\tiny\hbox{\sffamily DDW}}}$  is simply written:
\[
\left.
\begin{array}{ccl}
\displaystyle
\chi \iN {\pmb{\Omega}}^{{\tiny\hbox{\sffamily DDW}}}  & = & \displaystyle { \big{(}} \chi^{\mathfrak{e}}  \frac{\partial}{\partial \mathfrak{e}}   +  \chi^{{\pmb{\pi}}}     \frac{\partial}{ \partial {\pmb{\pi}}^{{A}_\mu \nu}  }     { \big{)}}  \iN  { \big{(}} \hbox{d} \mathfrak{e} \wedge \dd \mathfrak{y} +     \hbox{d} {\pmb{\pi}}^{{A}_\mu \nu}      \wedge
\hbox{d}  {{A}}_{\mu}  \wedge \hbox{d}  \mathfrak{y}_{\nu} 
   { \big{)}}
   \\
  \displaystyle  & = & \displaystyle    \chi^{\mathfrak{e}}  \hbox{d} \mathfrak{y}    + { \big{(}}     \hbox{d} {\pmb{\pi}}^{{A}_\mu \nu}    (\chi^{{\pmb{\pi}}}     \frac{\partial}{ \partial {\pmb{\pi}}^{{A}_\mu \nu}  }  )
\hbox{d}  {{A}}_{\mu}  \wedge \hbox{d}  \mathfrak{y}_{\nu} 
  { \big{)}}
  =    \chi^{\mathfrak{e}}  \hbox{d} \mathfrak{y}     +       \chi^{{\pmb{\pi}}}  
\hbox{d}  {{A}}_{\mu}  \wedge \hbox{d}  \mathfrak{y}_{\nu}  .
     \end{array}
\right.
\]
Now we compute  $ \hbox{d} ( \chi \iN {\pmb{\Omega}}^{{\tiny\hbox{\sffamily DDW}}} ) =   \hbox{d} \chi^{\mathfrak{e}} \wedge \hbox{d} \mathfrak{y} +  \hbox{d}  \chi^{{\pmb{\pi}}}  \wedge \hbox{d} {A} _{\mu} \wedge  \hbox{d} \mathfrak{y}_{\nu}$.
\[
\left.
\begin{array}{ccl}
\displaystyle
\dd (
\chi \iN {\pmb{\Omega}}^{{\tiny\hbox{\sffamily DDW}}} )  & = & \displaystyle \hbox{d}  { \big{(}} {\pmb{\Upsilon}} -  (    \mathfrak{e}   \frac{\partial  X^{\nu} }{\partial x^{\nu}}   + \frac{\partial  \Theta_{{\mu}}  }{\partial x^{\nu}}  {\pmb{\pi}}^{{A}_\mu \nu}   )   { \big{)}} \wedge \dd\mathfrak{y} + \hbox{d}    {\pmb{\Upsilon}}^{{A}_{\mu}\nu}     \wedge \hbox{d} {A} _{\mu} \wedge  \hbox{d} \mathfrak{y}_{\nu} 
   \\
  \displaystyle  & & \displaystyle   -
\hbox{d} {\Big{(}}    {\pmb{\pi}}^{{A}_\rho \sigma}  \delta^{\mu}_{\rho}   {\big{(}}        \big{[}  (\frac{\partial  X^{\nu} }{\partial x^{\sigma}} )   -  \delta^{\nu}_{\sigma}     (\frac{\partial  X^{\lambda} }{\partial x^{\lambda}} )   \big{]}   +     ( \frac{\partial  \Theta_{{\sigma}}  }{\partial {A}_{\nu} } ) \BR  -  \mathfrak{e}   ( \frac{\partial  X^{\nu} }{\partial {{A} _{\mu}}  }  )  {\big{)}}       {\Big{)}}    \wedge \hbox{d} {A} _{\mu} \wedge  \hbox{d} \mathfrak{y}_{\nu} ,
     \end{array}
\right.
\]
due to the expression of    the exterior derivatives $ \hbox{d}    {\pmb{\Upsilon}} $ and $ \hbox{d}   {\pmb{\Upsilon}}^{{A}_{\mu}\nu}  $,  we obtain:
\[
\left.
\begin{array}{rcl}
\displaystyle    \dd (
\chi \iN {\pmb{\Omega}}^{{\tiny\hbox{\sffamily DDW}}} )        & = &  \displaystyle   {\big{(}}    \frac{\partial {\pmb{\Upsilon}} }{\partial  {{A} _{\mu}}} -   \frac{\partial  {\pmb{\Upsilon}}^{{A}_{\mu}\nu}     }{\partial x^{\nu}} {\big{)}} \hbox{d} {A} _{\mu} \wedge  \hbox{d} \mathfrak{y}_{\nu}
+ {\big{(}} \frac{\partial  {\pmb{\Upsilon}} }{\partial  \mathfrak{e}} - \frac{\partial  X^{\nu} }{\partial x^{\nu}}  {\big{)}} \hbox{d} \mathfrak{e} \wedge \hbox{d} \mathfrak{y}  
 \\
 \displaystyle     &   &  \displaystyle + {\big{(}} \frac{\partial   {\pmb{\Upsilon}}  }{\partial  {\pmb{\pi}}^{{A}_{\mu}\nu}  }  -         \frac{\partial  \Theta_{{\mu}}  }{\partial x^{\nu}}    {\big{)}}  \hbox{d} {\pmb{\pi}}^{{A}_\mu \nu}         \wedge \hbox{d} \mathfrak{y} + {\big{(}} \frac{\partial   {\pmb{\Upsilon}}^{{A}_{\mu}\nu}    }{\partial  {{A}_{\beta}}}  {\big{)}} \hbox{d} {{A}_{\beta}}  \wedge \hbox{d} {A} _{\mu} \wedge  \hbox{d} \mathfrak{y}_{\nu}  
\\
\displaystyle     &   &  \displaystyle    + {\big{(}} \frac{\partial   {\pmb{\Upsilon}}^{{A}_{\mu}\nu}    }{\partial  \mathfrak{e}}  -  \frac{\partial  X^{\nu} }{\partial {{A} _{\mu}}  }    {\big{)}} \hbox{d} \mathfrak{e}  \wedge \hbox{d} {A} _{\mu} \wedge  \hbox{d} \mathfrak{y}_{\nu}   
\\
\displaystyle     &   &  \displaystyle + {\Big{(}} (  \frac{\partial    {\pmb{\Upsilon}}^{{A}_{\mu}\nu}    }{\partial  {{\pmb{\pi}}^{ {A}_{\rho}  \sigma } }} 
  )
-  \delta^{\mu}_{\rho} {\big{(}}   [  (\frac{\partial  X^{\nu} }{\partial x^{\sigma}} )   -  \delta^{\nu}_{\sigma}     (\frac{\partial  X^{\lambda} }{\partial x^{\lambda}} )  ] -  ( \frac{\partial  \Theta_{{\nu}}  }{\partial {A}_{\sigma} } )    {\big{)}} {\Big{)}} \hbox{d} {\pmb{\pi}}^{{A}_\rho \sigma}   \wedge \hbox{d} {A} _{\mu} \wedge  \hbox{d} \mathfrak{y}_{\nu} .
\end{array}
\right.
\]
 Now we are interested in terms in which ${\pmb{\Upsilon}}$ is involved: the first three terms in the last equation are concerned.  Let notice that, if we denote $q = \{ x, { A}  \}$ then, $ {\pmb{\Upsilon}} = {\pmb{\Upsilon}} (q,\mathfrak{e} , {\pmb{\pi}})$ and the first two terms in the last equation give: 
\begin{equation}
\left.
\begin{array}{rcl}
\displaystyle  \frac{\partial  {\pmb{\Upsilon}} }{\partial  \mathfrak{e}}  (q, \mathfrak{e} , {\pmb{\pi}} ) -   \frac{\partial {{X}}^{\nu}}{\partial x^{\nu}}   (q)  &  =  & \displaystyle   0,
\\
 \displaystyle  \frac{\partial   {\pmb{\Upsilon}}  }{\partial  {\pmb{\pi}}^{{  A}_{\mu}\nu}  }   (q, \mathfrak{e} , {\pmb{\pi}} ) -         \frac{\partial  \Theta_{{\mu}}  }{\partial x^{\nu}}   (q)&  =  & \displaystyle   0.
\end{array}
\right.
\end{equation}
Hence, it exists $  {{\Upsilon}}   (q) =  {{\Upsilon}}  (x,{  A}) $.
So that we have:
\bee  {\pmb{\Upsilon}} (q, \mathfrak{e} , {\pmb{\pi}} ) =
{\Upsilon}   (q) + (\frac{\partial  \Theta_{{\mu}}  }{\partial x^{\nu}} ) {\pmb{\pi}}^{{  A}_\mu \nu}  + \mathfrak{e}    ( \frac{\partial { {X}}^{\nu}}{\partial x^{\nu}}   )  .
\eee
On the other side    $  {\pmb{\Upsilon}}^{{  A}_{\mu}\nu}    (q,p) =   {\pmb{\Upsilon}}^{{  A}_{\mu}\nu}    (x,{\pmb{\pi}})
$, therefore the interesting information is contained in the set of equations: 
\begin{equation}
\left.
\begin{array}{rcl}
\displaystyle     \frac{\partial   {\pmb{\Upsilon}}^{{  A}_{\mu}\nu}    }{\partial  \mathfrak{e}}  (x, {\pmb{\pi}}) - ( \frac{\partial  X^{\nu} }{\partial {{  A} _{\mu}}  }  )  (q)   &  =  & \displaystyle   0
\\
 \displaystyle  \BL \frac{\partial    {\pmb{\Upsilon}}^{{  A}_{\mu}\nu}     }{\partial  {\pmb{\pi}}^{{  A}_{\mu}\sigma}  }   (q, \mathfrak{e} , {\pmb{\pi}} ) -  \delta^{\mu}_{\rho} {\big{(}}     [  (\frac{\partial  X^{\nu} }{\partial x^{\sigma}} )   -  \delta^{\nu}_{\sigma}     (\frac{\partial  X^{\lambda} }{\partial x^{\lambda}} )  ] -  ( \frac{\partial  \Theta_{{\nu}}  }{\partial {  A}_{\sigma} } )    {\big{)}}      \BR  \hbox{d} {\pmb{\pi}}^{{  A}_\mu \sigma}    \wedge \hbox{d} {  A} _{\mu} \wedge  \hbox{d} \mathfrak{y}_{\nu} &  =  & \displaystyle   0
\end{array}
\right.
\end{equation}

Hence, it exists $  {{\Upsilon}}^{{  A}_{\mu}\nu}      (q) =  {{\Upsilon}}^{{  A}_{\mu}\nu}  (x,{  A}) $.
So that:
\bee
{\pmb{\Upsilon}}^{{  A}_{\mu}\nu}    (q,p) =
{{\Upsilon}}^{{  A}_{\mu}\nu}  (x,{  A})  -     {\pmb{\pi}}^{{  A}_\rho \sigma}     \delta^{\mu}_{\rho} {\Big{(}}     {\big{(}}  (\frac{\partial  X^{\nu} }{\partial x^{\sigma}} {\big{)}}   -  \delta^{\nu}_{\sigma}     (\frac{\partial  X^{\lambda} }{\partial x^{\lambda}} )  \BR -  ( \frac{\partial  \Theta_{{\nu}}  }{\partial {  A}_{\sigma} } )    {\big{\}}} +  \mathfrak{e}   ( \frac{\partial  X^{\nu} }{\partial { {A} _{\mu}}  }  ) {\Big{)}} .
\eee
 The set of infinitesimal symplectomorphisms   $\displaystyle \mathfrak{sp}_{\circ} ({\pmb{\cal M}}_{\tiny{\hbox{\sffamily Maxwell}}} )$ of $\displaystyle ( {\pmb{\cal M}}_{\tiny{\hbox{\sffamily Maxwell}}},{\pmb{\Omega}}^{\tiny{\hbox{\sffamily DDW}}} )$ is   described by vector fields $ \Xi = \Xi\big |_{ {\pmb{\cal M}}_{\tiny{\hbox{\sffamily Maxwell}}} }  = \overline{ \zeta}  + \chi $ with $\displaystyle  \overline{ \zeta}$ described by \eqref{tyuoi12} and $\displaystyle \chi = {\Upsilon}  \frac{\partial}{\partial \mathfrak{e}}
 +{\Upsilon}^{A_{\mu}\alpha}  \frac{\partial}{\partial {\pmb{\pi}}^{\omega_\mu \alpha} }$. Here, $X^{\nu}, \Theta_{\mu},  \Upsilon,  {{\Upsilon}^{A_{\mu}\alpha}}$ are defined on $ \pmb{\mathfrak{Z}}$  and not anymore on the full multisymplectic manifold ${\pmb{\cal M}}_{\tiny{\hbox{\sffamily Maxwell}}} $.   
 \begin{prop}\label{algess}
 If we assume that {\em $dx^{\mu} (\Xi)  = 0$ } - we
throw away the  $X^{\mu}$ which correspond to parts of the stress-energy-tensor - the proposition \eqref{algebraicallsymplec0} gives:
\[
\Xi = \big{(}  {\Upsilon}^{{A}_{\mu}\nu}  -  {\pmb{\pi}}^{{A}_\mu \sigma}     ( \frac{\partial  \Theta_{{\sigma}} }{\partial {A}_{\nu}} )  \big{)}\frac{\partial}{ \partial {\pmb{\pi}}^{A_\mu \nu} } 
 +  \big{(}  {\Upsilon}  - \frac{\partial  \Theta_{{\mu}} }{\partial x^{\nu}}  {\pmb{\pi}}^{A_\mu \nu}      \big{)}  \frac{\partial}{\partial \mathfrak{e}} +   \Theta_{\mu} \frac{\partial}{\partial {A}_{\mu}}   ,
\]
\end{prop}
  ${\Upsilon}^{{A}_{\mu}\nu} $, $ {\Upsilon}  $ and $\Theta_{{\mu}} $ are smooth arbitrary functions of $(x,{A})$ with ${\Upsilon}^{{A}_{\mu}\nu} (q) = - {\Upsilon}^{{A}_{\nu}\mu} (q)$, they satisfy the condition:  
$ \displaystyle
\frac{\partial {{\Upsilon}}  }{\partial  {A}_{\mu}}  - \frac{\partial  {{\Upsilon}}^{{A}_{\mu}\nu}}{\partial x^{\nu}} = 0$.

\begin{prop}\label{alDSDc0}  
Let $\pmb{\varphi} \in \Gamma ({\pmb{\cal M}}_{\tiny{\hbox{\sffamily Maxwell}}} , \Lambda^{n-1} T^{\star} {\pmb{\cal M}}_{\tiny{\hbox{\sffamily Maxwell}}}   )$. The $(n-1)$-form $\pmb{\varphi} $ is an algebraic observable if and only if $\pmb{\varphi} $ is written $\pmb{\varphi}  = \pmb{\varphi}_{X} + \pmb{\varphi}_{A} + \pmb{\varphi}_{\chi} $ where
{\em 
\bee\label{tyuoi1qqss2}
\left|
\begin{array}{ccl}
\displaystyle
 \pmb{\varphi}_{X}  & = &  \displaystyle     \mathfrak{e}  {  X}^{\rho}   \hbox{d}  \mathfrak{y}_{\rho}      - {\pmb{\pi}}^{{  A}_\mu \nu}    {  X}^{\rho}       \hbox{d} {  A}_{\mu}   \wedge  \dd \mathfrak{y}_{\rho\nu}     ,
  \\
 \displaystyle \pmb{\varphi}_{A}  
  & = &  \displaystyle    {\pmb{\pi}}^{{  A}_\mu \nu}      {{  \Theta}}_{{\mu}}        \dd \mathfrak{y}_{\nu} .
  \end{array}
\right.
\eee
}
where $X^{\mu} , \Theta_{\mu}: {\pmb{\mathfrak{Z}}}  \rightarrow \Bbb{R}$ are arbitrary smooth functions on ${\pmb{\mathfrak{Z}}}$  and ${\pmb{\varphi}}_{\chi} $ is a $(n-1)$-form such that 
{\em 
\bee
\dd {\pmb{\varphi}}_{\chi} = \Upsilon \hbox{d} \mathfrak{y} + \Upsilon^{  {  A}_{\mu}  \nu  } \hbox{d}   {  A}_{\mu}   \wedge \hbox{d}  \mathfrak{y}_{\nu} ,
\eee
}
with $\Upsilon $ and $\Upsilon^{  {  A}_{\mu}  \nu  }$ such that $\Upsilon^{  {  A}_{\mu}  \nu  } = - \Upsilon^{  {  A}_{\nu} \mu  }$. The functions  $\Upsilon $ and $\Upsilon^{  {  A}_{\mu}  \nu  }$ satisfy 
\bee
\frac{\partial {{\Upsilon}}  }{\partial  {A}_{\mu}}  - \frac{\partial  {{\Upsilon}}^{{  A}_{\mu}\nu}}{\partial x^{\nu}} = 0 .
\eee
\end{prop}
We notice that $ \pmb{\varphi}_{X} + \pmb{\varphi}_{A} $ are the so-called generalized algebraic momenta $(n-1)$-forms. Recall that an arbitrary vector field on $\pmb{\mathfrak{Z}}$ is written \eqref{vfz1}, 
\[
\zeta:= \sum_\alpha  \zeta^\alpha(q){\partial \over \partial q^\alpha} = X^{\nu} (x,{{A}}) \frac{\partial}{\partial x^{\nu}}  + \Theta_{\mu} (x,{{A}}) \frac{\partial}{\partial {{A}}_{\mu}}  .
\]
 Let denote $\dttP_{\zeta} = \zeta \iN {\pmb{\theta}} $. We have: 
\[
\dttP_{\zeta} =  \zeta \iN  {\big{(}}   \mathfrak{e}   \hbox{d} \mathfrak{y} +       {\pmb{\pi}}^{{A}_\mu \nu}      
\hbox{d}  {{A}}_{\mu}  \wedge \hbox{d}  \mathfrak{y}_{\nu} 
   \big{)} =  \mathfrak{e}  \hbox{d} \mathfrak{y} ( \zeta ) +  {\pmb{\pi}}^{{A}_\mu \nu}       {\big{(}}  ( \zeta \iN \hbox{d}  {{ A}}_{\mu}  ) \wedge  \hbox{d}  \mathfrak{y}_{\nu}  -  \hbox{d}  {{ A}}_{\mu}  \wedge (\zeta \iN \hbox{d}  \mathfrak{y}_{\nu})    {\big{)}} .
\]
Since $ \zeta \iN \hbox{d}  {{  A}}_{\mu}  =  {{  \Theta}}_{{\mu}}  $ and $\displaystyle \zeta \iN \hbox{d}  \mathfrak{y}_{\nu} =  ({  X}^{\rho} \frac{\partial}{\partial x^{\rho}} )  \iN \hbox{d}  \mathfrak{y}_{\nu}  =  {  X}^{\rho}\hbox{d}  \mathfrak{y}_{\rho\nu}  $, we   obtain: 
\[
\dttP_{\zeta}  =    \mathfrak{e}  {  X}^{\rho} \hbox{d}  \mathfrak{y}_{\rho}    +  {\pmb{\pi}}^{{  A}_\mu \nu}     {{  \Theta}}_{{\mu}}      \dd \mathfrak{y}_{\nu}   - {\pmb{\pi}}^{{  A}_\mu \nu}    {  X}^{\rho}       \hbox{d} {  A}_{\mu}  \wedge  \dd \mathfrak{y}_{\rho\nu} =   {\pmb{\varphi}}_{X} +   {\pmb{\varphi}}_{A}.
\]
 $\dttP_{\zeta}$ are the {\em generalized momenta} $(n-1)$-form. We have $\hbox{d} \dttP_{\zeta}  = - \overline{\zeta} \iN \pmb{\Omega}^{\tiny{\hbox{\sffamily DDW}}} $. The  canonical symplectomorphism associated to $\dttP_{\zeta}$ is denoted $\Xi (\dttP_{\zeta}) =   \bar{\zeta}$.
 We evaluate the exterior derivative $\hbox{d} \dttP_{\zeta} = \hbox{d} [{\pmb{\varphi}}_{X} +   {\pmb{\varphi}}_{A}  ] $:
 \[
\left.
\begin{array}{rcl}
 \displaystyle  \hbox{d} \dttP_{\zeta}    & = &  \displaystyle  \hbox{d} {\big{(}}     \mathfrak{e}  {\pmb X}^{\nu}    +  {\pmb{\pi}}^{{  A}_\mu \nu}     {{\pmb \Theta}}_{{\mu}}   {\big{)}}   \wedge \hbox{d}  \mathfrak{y}_{\nu} 
-  \hbox{d}  {\big{(}}  {\pmb{\pi}}^{{  A}_\mu \nu}     {\pmb X}^{\rho}    {\big{)}}   \wedge  \hbox{d} {  A}_{\mu}   \wedge  \dd \mathfrak{y}_{\rho\nu}    
 \\
 \displaystyle     & = &  \displaystyle
\underbrace{ {\pmb X}^{\nu} \hbox{d}  \mathfrak{e}   \wedge \hbox{d}  \mathfrak{y}_{\nu} }_{{\pmb \iota}_{1}}
+  \underbrace{ \mathfrak{e}   \hbox{d}   {\pmb X}^{\nu}  \wedge \hbox{d}  \mathfrak{y}_{\nu} 
}_{{\pmb \iota}_{2}}
  \underbrace{
   +      {\pmb{\pi}}^{{  A}_\mu \nu}       \hbox{d} {{\pmb \Theta}}_{{\mu}}    \wedge \hbox{d} \mathfrak{y}_{\nu}     
     }_{{\pmb \iota}_{3}}
    \underbrace{
   +            {{\pmb \Theta}}_{{\mu}} \hbox{d}  {\pmb{\pi}}^{{  A}_\mu \nu}    \wedge \hbox{d} \mathfrak{y}_{\nu}  
      }_{{\pmb \iota}_{4}}
       \\
 \displaystyle     &  &  \displaystyle   -  \underbrace{ 
   {\pmb X}^{\rho}        \hbox{d} {\pmb{\pi}}^{{A}_\mu \nu}   \wedge  \hbox{d} {  A}_{\mu}   \wedge  \dd \mathfrak{y}_{\rho\nu}     
  }_{{\pmb \iota}_{5}}
  -  \underbrace{ 
   {\pmb{\pi}}^{{  A}_\mu \nu}    \hbox{d}    {\pmb X}^{\rho}  \wedge     \hbox{d} {  A}_{\mu}  \wedge  \dd  \mathfrak{y}_{\rho\nu}     
  }_{{\pmb \iota}_{6}}.  \\
\end{array}
\right.
\]
   Now we expand the objects $\hbox{d}   {\pmb X}^{\rho}$, $  \hbox{d} {{\pmb \Theta}}_{{\mu}} $  so that: 
  \[
 {\pmb \iota}_{2} =  \mathfrak{e}   \hbox{d}   {\pmb X}^{\nu}  \wedge \hbox{d}  \mathfrak{y}_{\nu} 
   = \mathfrak{e}   {\big{(}}   \frac{\partial  {{\pmb X}}^{\nu} }{\partial x^{\alpha}} \dd x^{\alpha} + \frac{\partial  {{\pmb X}}^{\nu} }{\partial  {{  A}_{\beta}}} \hbox{d} {{  A}_{\beta}} + \frac{\partial  {{{\pmb X}}}^{\nu} }{\partial  \mathfrak{e}} \hbox{d} \mathfrak{e}   + \frac{\partial  {{\pmb X}}^{\nu} }{\partial  {\pmb{\pi}}^{{  A}_{\beta}\alpha}} \hbox{d} {{\pmb{\pi}}^{ {  A}_{\beta}  \alpha }}
   {\big{)}}   \wedge \hbox{d}  \mathfrak{y}_{\nu} ,
   \]
 \[
 {\pmb \iota}_{3} =  {\pmb{\pi}}^{{  A}_\mu \nu}        \hbox{d} {{\pmb \Theta}}_{{\mu}}   \wedge \hbox{d} \mathfrak{y}_{\nu}     
   =
  {\pmb{\pi}}^{{A}_\mu \nu}       {\big{(}}  \frac{\partial {\pmb{\Theta}}_{\mu}}{\partial x^{\alpha}} \dd x^{\alpha} + \frac{\partial {\pmb{\Theta}}_{\mu}}{\partial  {{  A}_{\beta}}} \hbox{d} {{  A}_{\beta}}   + \frac{\partial  {\pmb{\Theta}}_{\mu}}{\partial  \mathfrak{e}} \hbox{d} \mathfrak{e} + \frac{\partial  {\pmb{\Theta}}_{\mu} }{\partial  {\pmb{\pi}}^{{  A}_{\beta}\alpha}  }  \hbox{d} {\pmb{\pi}}^{{  A}_{\beta}\alpha}  {\big{)}}   \wedge \hbox{d} \mathfrak{y}_{\nu}    . 
 \]
   Then, 
 \[
\left.
\begin{array}{rcl}
\displaystyle    \hbox{d} \dttP_{\zeta}     & = &  \displaystyle  \underbrace{ {\pmb X}^{\nu} \hbox{d}  \mathfrak{e}   \wedge \hbox{d}  \mathfrak{y}_{\nu} }_{{\pmb \iota}_{1}}
+  \underbrace{ \mathfrak{e}   \hbox{d}   {\pmb X}^{\nu}  \wedge \hbox{d}  \mathfrak{y}_{\nu} 
}_{{\pmb \iota}_{2}}
    \underbrace{
   +            {{\pmb \Theta}}_{{\mu}}  \hbox{d}  {\pmb{\pi}}^{{  A}_\mu \nu}     \wedge \hbox{d} \mathfrak{y}_{\nu}  
      }_{{\pmb \iota}_{4}} 
  -  \underbrace{ 
    \hbox{d} {\pmb{\pi}}^{{  A}_\mu \nu}    {\pmb X}^{\rho}       \hbox{d} {  A}_{\mu}  \wedge  \dd \mathfrak{y}_{\rho\nu}     
  }_{{\pmb \iota}_{5}}   
    \\
\displaystyle     &  &  \displaystyle
+  
   \underbrace{
       {\pmb{\pi}}^{{  A}_\mu \nu}  (\frac{\partial {\pmb{\Theta}}_{\mu}}{\partial x^{\alpha}} )\dd x^{\alpha}   \wedge \hbox{d} \mathfrak{y}_{\nu}    
    }_{{\pmb \iota}_{7}}
 \underbrace{
   +    {\pmb{\pi}}^{{  A}_\mu \nu}  ( \frac{\partial {\pmb{\Theta}}_{\mu}}{\partial  {{ A}_{\beta}}} ) \hbox{d} {{  A}_{\beta}}    \wedge \hbox{d} \mathfrak{y}_{\nu}    
    }_{{\pmb \iota}_{8}}
    \underbrace{
   +    {\pmb{\pi}}^{{  A}_\mu \nu}   (\frac{\partial  {\pmb{\Theta}}_{\mu}}{\partial  \mathfrak{e}}) \hbox{d} \mathfrak{e}     \wedge \hbox{d} \mathfrak{y}_{\nu}  
    }_{{\pmb \iota}_{9}}  
          \\
\displaystyle     & &  \displaystyle  + \underbrace{
      {\pmb{\pi}}^{{  A}_\mu \nu}  (\frac{\partial  {\pmb{\Theta}}_{\mu} }{\partial  {\pmb{\pi}}^{{ A}_{\beta}\alpha}  }  ) \hbox{d} {\pmb{\pi}}^{{A}_{\beta}\alpha}   \wedge \hbox{d} \mathfrak{y}_{\nu}     
    }_{{\pmb \iota}_{10}} 
    -  \underbrace{ 
   {\pmb{\pi}}^{{  A}_\mu \nu}    \frac{\partial  {{\pmb X}}^{\rho} }{\partial x^{\alpha}} \dd x^{\alpha}  \wedge       \hbox{d} {  A}_{\mu}  \wedge  \hbox{d} \mathfrak{y}_{\rho\nu}     
  }_{{\pmb \iota}_{11}}
  -  \underbrace{ 
   {\pmb{\pi}}^{{  A}_\mu \nu}    \frac{\partial  {{\pmb X}}^{\rho} }{\partial  {{  A}_{\beta}}} \hbox{d} {{  A}_{\beta}}  \wedge    \hbox{d} {A}_{\mu}  \wedge  \hbox{d} \mathfrak{y}_{\rho\nu}     
  }_{{\pmb \iota}_{12}}
       \\
\displaystyle     &  &  \displaystyle  -  \underbrace{ 
   {\pmb{\pi}}^{{  A}_\mu \nu}    \frac{\partial  {{\pmb X}}^{\rho} }{\partial  \mathfrak{e}} \hbox{d} \mathfrak{e}      \wedge       \hbox{d} {  A}_{\mu}   \wedge  \hbox{d} \mathfrak{y}_{\rho\nu}     
  }_{{\pmb \iota}_{13}}
  -  \underbrace{ 
   {\pmb{\pi}}^{{  A}_\mu \nu}   \frac{\partial  {{\pmb X}}^{\rho} }{\partial  {\pmb{\pi}}^{{  A}_{\beta}\alpha} } \hbox{d} {{\pmb{\pi}}^{ {  A}_{\beta}  \alpha } }
  \wedge   \hbox{d} {  A}_{\mu}  \wedge  \hbox{d} \mathfrak{y}_{\rho\nu}     
  }_{{\pmb \iota}_{14}}.  
\end{array}
\right.
\]
   Since, ${{\pmb X}} = {{\pmb X}}  (x)$ and $ {\pmb{\Theta}}_{\mu} = {\pmb{\Theta}}_{\mu}  (x, {A})$ - see \eqref{qmddmzdzdsss}, we obtain vanishing contributions from the terms ${\pmb \iota}_{9}, {\pmb \iota}_{10}, {\pmb \iota}_{12}, {\pmb \iota}_{13}$ and ${\pmb \iota}_{14}$. Therefore: 
 \[
\left.
\begin{array}{rcl}
\displaystyle    \hbox{d} \dttP_{\zeta}       & = &  \displaystyle
\underbrace{ {\pmb X}^{\nu} \hbox{d}  \mathfrak{e}   \wedge \hbox{d}  \mathfrak{y}_{\nu} }_{{\pmb \iota}_{1}}
+  \underbrace{ \mathfrak{e}   \hbox{d}   {\pmb X}^{\nu}  \wedge \hbox{d}  \mathfrak{y}_{\nu} 
}_{{\pmb \iota}_{2}}
    \underbrace{
   +            {{\pmb \Theta}}_{{\mu}}  \hbox{d}  {\pmb{\pi}}^{{  A}_\mu \nu}     \wedge \hbox{d} \mathfrak{y}_{\nu}  
      }_{{\pmb \iota}_{4}} 
  -  \underbrace{ 
    \hbox{d} {\pmb{\pi}}^{{  A}_\mu \nu}    {\pmb X}^{\rho}       \hbox{d} {A}_{\mu}   \wedge  \dd \mathfrak{y}_{\rho\nu}     
  }_{{\pmb \iota}_{5}}
 \\
 \displaystyle     &  &  \displaystyle +  \underbrace{
       {\pmb{\pi}}^{{  A}_\mu \nu}  (\frac{\partial {\pmb{\Theta}}_{\mu}}{\partial x^{\alpha}} )\dd x^{\alpha}   \wedge \hbox{d} \mathfrak{y}_{\nu}    
    }_{{\pmb \iota}_{7}}
 \underbrace{
   +    {\pmb{\pi}}^{{A}_\mu \nu}  ( \frac{\partial {\pmb{\Theta}}_{\mu}}{\partial  {{  A}_{\beta}}} ) \hbox{d} {{  A}_{\beta}}    \wedge \hbox{d} \mathfrak{y}_{\nu}    
    }_{{\pmb \iota}_{8}}
 -  \underbrace{ 
   {\pmb{\pi}}^{{  A}_\mu \nu}   ( \frac{\partial  {{\pmb X}}^{\rho} }{\partial x^{\alpha}} )  \dd x^{\alpha}  \wedge       \hbox{d} {  A}_{\mu}  \wedge  \dd \mathfrak{y}_{\rho\nu}     
  }_{{\pmb \iota}_{11}}  .
 \\
\end{array}
\right.
\]
 On the other hand, the general expression for a canonical symplectomorphism is: 
 \[
\left.
\begin{array}{rcl}
\displaystyle   \bar{\zeta} \iN  {\pmb{\Omega}}^{{\tiny\hbox{\sffamily DDW}}}       & = &  \displaystyle \bar{\zeta}  \iN  \hbox{d} \mathfrak{e} \wedge \dd \mathfrak{y} +  \bar{\zeta} \iN {\big{[}} \hbox{d} {\pmb{\pi}}^{{  A}_{\mu} \nu }    \wedge \hbox{d}    {{  A}}_{\mu}  \wedge \dd \mathfrak{y}_{\nu} 
  {\big{]}}  
  \\
\displaystyle     & = &  \displaystyle - {\big{(}}    \mathfrak{e}  ( \frac{\partial  X^{\nu} }{\partial x^{\nu}} )  + \frac{\partial  \Theta_{{\mu}}  }{\partial x^{\nu}}  {\pmb{\pi}}^{{  A}_\mu \nu}    {\big{)}}     \dd \mathfrak{y}  \underbrace{ - {{\pmb X}}^{\nu}  \hbox{d} \mathfrak{e} \wedge  \dd \mathfrak{y}_{\nu} }_{-{\pmb \iota}_{1}}  \\
\displaystyle     &   &  \displaystyle  + \hbox{d} {\pmb{\pi}}^{{A}_{\mu} \nu }   (\bar{\zeta}) {\hbox{d} {  A}_{\mu} }  \wedge \dd \mathfrak{y}_{\nu} - {\hbox{d} {  A}_{\mu}  } (\bar{\zeta})    \hbox{d} {\pmb{\pi}}^{{  A}_{\mu} \nu }    \wedge \dd \mathfrak{y}_{\nu} +   \hbox{d} x^{\rho}  (\bar{\zeta}) \hbox{d} {\pmb{\pi}}^{{  A}_{\mu} \nu }  \wedge
{\hbox{d} {  A}_{\mu}  } \wedge \dd \mathfrak{y}_{\rho\nu}  
\\
\displaystyle     & = &  \displaystyle 
 \underbrace{  -    \mathfrak{e}  ( \frac{\partial  {\pmb X}^{\nu} }{\partial x^{\nu}} )     \hbox{d} \mathfrak{y} }_{-  {{\pmb \iota}_{2}}  }   \underbrace{  - \frac{\partial  \Theta_{{\mu}}}{\partial x^{\nu}}  {\pmb{\pi}}^{{  A}_\mu \nu}  \hbox{d} \mathfrak{y} }_{ -{{\pmb \iota}_{7}}  }    \underbrace{ - {{\pmb X}}^{\nu}  \hbox{d} \mathfrak{e} \wedge \dd \mathfrak{y}_{\nu} }_{- {{\pmb \iota}_{1}}    } + \hbox{d} {\pmb{\pi}}^{{  A}_{\mu} \nu }    (\bar{\zeta}) {\hbox{d} {  A}_{\mu}  }  \wedge \dd \mathfrak{y}_{\nu}  
 \\
 \displaystyle     &   &  \displaystyle 
  - {\hbox{d} {  A}_{\mu} } (\bar{\zeta})    \hbox{d} {\pmb{\pi}}^{{  A}_{\mu} \nu }  \wedge \dd \mathfrak{y}_{\nu} 
 \underbrace{  + {\pmb X}^{\rho}  \hbox{d} {\pmb{\pi}}^{{  A}_{\mu} \nu }   \wedge
{\hbox{d} {  A}_{\mu}  } \wedge \dd \mathfrak{y}_{\rho\nu} 
  }_{- {{\pmb \iota}_{5}}   } .
  \\
\end{array}
\right.
\]
 Since $\hbox{d} {  A}_{\mu} (\bar{\zeta})   = {\pmb \Theta}_{\mu}$,   we   observe: 
\[
 - {\hbox{d} {  A}_{\mu} } (\bar{\zeta})    \hbox{d} {\pmb{\pi}}^{{  A}_{\mu} \nu } \wedge \dd \mathfrak{y}_{\nu}  = - {\pmb \Theta}_{\mu}   \hbox{d} {\pmb{\pi}}^{{  A}_{\mu} \nu } \wedge \hbox{d} \mathfrak{y}_{\nu} =  -     {\pmb \iota}_{4}   .
\]
Let   denote  $ (\mathfrak{i}) = {{\pmb \iota}_{1}}  + {{\pmb \iota}_{2}} + {{\pmb \iota}_{4}}  + {{\pmb \iota}_{5}}   + {{\pmb \iota}_{7}}   $. We notice that  
$
\bar{\zeta} \iN {\pmb{\Omega}}^{{\tiny\hbox{\sffamily DDW}}}       =  - (\mathfrak{i})  +  \hbox{d} {\pmb{\pi}}^{{  A}_{\mu} \nu }   (\bar{\zeta}) {\hbox{d} {  A}_{\mu} }  \wedge \dd \mathfrak{y}_{\nu}  $.
Let also notice that $ \hbox{d} {\pmb{\pi}}^{{  A}_{\mu} \nu }    (\bar{\zeta}) = \bar{\zeta}^{{\pmb{\pi}}}$, with 
\bee
\overline{ \zeta} = {\pmb X}^{\nu}  \frac{\partial}{\partial x^{\nu}}  +  { {\pmb \Theta}_{\rho} }  \frac{\partial}{\partial {{  A}_{\rho}  } }  +  {\bar{\zeta}}^{\mathfrak{e}}    \frac{\partial}{\partial \mathfrak{e}}
  +   {\bar{\zeta}}^{{\pmb{\pi}}}   {\partial \over \partial  {\pmb{\pi}}^{{  A}_\mu \sigma}   },
\eee
we are left with the term $ {\bar{\zeta}}^{{\pmb{\pi}}}  $: 
\bee
 {\bar{\zeta}}^{{\pmb{\pi}}} =  {\pmb{\pi}}^{{A}_\mu \sigma}   \BL   \big{[}  (\frac{\partial  X^{\nu} }{\partial x^{\sigma}} )   -  \delta^{\nu}_{\sigma}     (\frac{\partial  X^{\lambda} }{\partial x^{\lambda}} )   \big{]}   +     ( \frac{\partial  \Theta_{{\sigma}} }{\partial {  A}_{\nu} } ) \BR  -  \mathfrak{e}   ( \frac{\partial  X^{\nu} }{\partial {{  A}_{\mu}}  }  )    ,
\eee
so that: 
$
\bar{\zeta} \iN {\pmb{\Omega}}^{{\tiny\hbox{\sffamily DDW}}}    =  - (\mathfrak{i}) +   {\big{(}} {\bar{\zeta}}^{{\pmb{\pi}}}  {\big{)}}^{\mu\nu}  {\hbox{d} {  A}_{\mu} }  \wedge \hbox{d} \mathfrak{y}_{\nu}  
$. Finally   we   denote the last remaining terms $\displaystyle (\mathfrak{ii})  =    {\pmb \iota}_{8} +      {\pmb \iota}_{11}  $ so that  the equality   $\hbox{d} [{\pmb{\varphi}}_{X} +   {\pmb{\varphi}}_{A}  ]  = (\mathfrak{i}) + (\mathfrak{ii}) $ holds. Therefore, in order to prove the equality $\displaystyle \bar{\zeta} \iN  {\pmb{\Omega}}^{{\tiny\hbox{\sffamily DDW}}}   = - \hbox{d}  [{\pmb{\varphi}}_{X} +   {\pmb{\varphi}}_{A}  ]  $, we only need to prove that $ \displaystyle  \hbox{d} {\pmb{\pi}}^{{A}_{\mu} \nu }   (\bar{\zeta}) {\hbox{d} {  A}_{\mu} }  \wedge \hbox{d} \mathfrak{y}_{\nu} 
 = - (\mathfrak{ii})$.  Since $\dd x^{\alpha}   \wedge   \hbox{d} \mathfrak{y}_{\rho\nu}   = \delta^{\alpha}_{\rho}  \hbox{d} \mathfrak{y}_{ \nu}  - \delta^{\alpha}_{\nu}  \hbox{d} \mathfrak{y}_{\rho} $,
 \[
\left.
\begin{array}{rcl}
 \displaystyle    {\pmb \iota}_{11}   & = &  \displaystyle  - {\pmb{\pi}}^{{  A}_\mu \nu}    \frac{\partial  {{\bf X}}^{\rho} }{\partial x^{\alpha}} \dd x^{\alpha}  \wedge       \hbox{d} {  A}_{\mu}   \wedge  \dd \mathfrak{y}_{\rho\nu}   = 
     {\pmb{\pi}}^{{  A}_\mu \nu}     \frac{\partial  {{\bf X}}^{\rho} }{\partial x^{\alpha}}  
     \hbox{d} {  A}_{\mu}   \wedge 
     \BL  \delta^{\alpha}_{\rho}  \hbox{d} \mathfrak{y}_{ \nu}  - \delta^{\alpha}_{\nu}  \hbox{d} \mathfrak{y}_{\rho}  \BR   
     \\
\displaystyle     &  = &  \displaystyle  {\pmb{\pi}}^{{A}_\mu \nu}    \frac{\partial  {{\bf X}}^{\rho} }{\partial x^{\rho}}  
     \hbox{d} {A}_{\mu}   \wedge 
       \hbox{d} \mathfrak{y}_{ \nu} 
-  {\pmb{\pi}}^{{A}_\mu \nu}   \frac{\partial  {{X}}^{\rho} }{\partial x^{\nu}}  
     \hbox{d} {A}_{\mu}   \wedge 
     \hbox{d} \mathfrak{y}_{\rho} =   {\pmb{\pi}}^{{A}_\mu \sigma}  \BL  (\frac{\partial  X^{\nu} }{\partial x^{\sigma}} )   -  \delta^{\nu}_{\sigma}     (\frac{\partial  X^{\lambda} }{\partial x^{\lambda}} )  \BR .
\end{array}
\right.
\]

 and also, on the same vein, 
  \[
    {\pmb \iota}_{8}  =
 {\pmb{\pi}}^{{ A}_\mu \sigma} ( \frac{\partial  \Theta_{{\nu}} }{\partial {  A}_{\sigma} } )   -  \mathfrak{e}   ( \frac{\partial  X^{\nu} }{\partial {{  A}_{\mu}}  }  ) .
 \]
 so that we found the wanted result.

 \subsection{\hbox{\sffamily\bfseries\slshape{Dynamical observable $(n-1)$-forms}}}\label{gliopopo}

We continue the investigation with the following two propositions \eqref{sfsf4321} and \eqref{algebraFF44}. 
 
 \begin{prop}\label{sfsf4321}  Let ${\pmb{\Xi}} \in \Gamma ({\pmb{\cal M}} , T {\pmb{\cal M}})$ then ${\pmb{\Xi}} $ satisfies {\em $\dd ({\pmb{\Xi}}  \iN {\pmb{\Omega}}^{\tiny{\hbox{\sffamily DDW}}})$} and {\em $\dd {\cal H} ({\pmb{\Xi}} ) = 0 $} if and only if ${\pmb{\Xi}} $ is written 
 ${\pmb{\Xi}}  = {\pmb{\Xi}}_{X} + {\pmb{\Xi}}_{A}$
 with 
\bee\label{tyuoi1qqss2}
\left.
\begin{array}{ccl}
\displaystyle
  {\pmb{\Xi}}_{X}   & = &  \displaystyle X^{\mu} \partial_{\mu} \iN   {\pmb{\theta}} , 
     \end{array}
\right.
\eee
  a vector field on the Minkowski space-time ${\cal X}$ -  a generator of the
action of the Poincar\'e group - and
\bee\label{tyuoi1qqs3s2}
\left.
\begin{array}{ccl}
\displaystyle
  {\pmb{\Xi}}_{A}   & = &  \displaystyle  
  \Theta _{\mu} \frac{\partial}{\partial  {A}_{\mu}}      -  {\Big{(}}           \frac{\partial  \Theta_{{\mu}}  }{\partial x^{\nu}}  {\pmb{\pi}}^{ {A}_\mu \nu}         {\Big{)}}     \frac{\partial}{\partial \mathfrak{e}}
   +   \Upsilon^{  {  A}_{\mu}  \nu  }    \frac{\partial}{ \partial {\pmb{\pi}}^{ {A}_\mu \nu}  } .
   \end{array}
\right.
\eee
 \end{prop}

\begin{prop}\label{algebraFF44}
Let $\pmb{\rho} \in \Gamma ({\pmb{\cal M}}_{\tiny{\hbox{\sffamily Maxwell}}} , \Lambda^{n-1} T^{\star} {\pmb{\cal M}}_{\tiny{\hbox{\sffamily Maxwell}}}   )$. The $(n-1)$-form $\pmb{\rho} $ is a  dynamical observable if and only if $\pmb{\rho} $ is written $\pmb{\rho}  = \pmb{\rho}_{X} + \pmb{\rho}_{A}   $ where,
{\em 
\bee\label{tyuoi1qqss2}
\left|
\begin{array}{ccl}
\displaystyle
 \pmb{\rho}_{X}  & = &  \displaystyle     \mathfrak{e}  {  X}^{\rho}   \hbox{d}  \mathfrak{y}_{\rho}      - {\pmb{\pi}}^{{  A}_\mu \nu}    {  X}^{\rho}       \hbox{d} {  A}_{\mu}   \wedge  \dd \mathfrak{y}_{\rho\nu}     
  \\
 \displaystyle \pmb{\rho}_{A}  
  & = &  \displaystyle    {\pmb{\pi}}^{{  A}_\mu \nu}      {{  \Theta}}_{{\mu}}        \dd \mathfrak{y}_{\nu}  +    \Upsilon^{  {  A}_{\mu}  \nu  }       A_{\mu}       \hbox{d}  \mathfrak{y}_{\nu} , \\
    \end{array}
\right.
\eee
}
  $X^{\mu} , \Theta_{\mu} , \Upsilon^{  {  A}_{\mu}  \nu }: {\pmb{\mathfrak{Z}}}  \rightarrow \Bbb{R}$ are arbitrary smooth functions     such that 
  \[
  \Upsilon^{  {  A}_{\mu}  \nu  } = - \Upsilon^{  {  A}_{\nu} \mu  } \quad \hbox{with}  \quad \frac{\partial  {{\Upsilon}}^{{  A}_{\mu}\nu}}{\partial x^{\nu}} = 0.
\]
\end{prop}
  Since  $\displaystyle \hbox{d}  {\cal H}^{{\tiny\hbox{Maxwell}}}  (q,p) = \hbox{d}  {\cal H}   (q,p)   = \hbox{d}   \mathfrak{e} -   \frac{1}{2} \eta_{\mu\rho} \eta_{\nu\sigma}  {\pmb{\pi}}^{{A}_\rho \sigma}    \hbox{d}     {\pmb{\pi}}^{{A}_\mu \nu}$, we consider $\hbox{d}  {\cal H}  (\Xi)$ as a polynomial expression depending on the variables $( \mathfrak{e},{\pmb{\pi}}^{A_{\mu} \nu} ) $. We have:
\[ 
\left.
\begin{array}{ccl}
\displaystyle
  \hbox{d}  {\cal H}  (\Xi)  & = &  \displaystyle  
    \hbox{d}  {\cal H}   (\chi + \overline{\zeta}) ,  
   \\
    \displaystyle & = &  \displaystyle  
      \hbox{d}   \mathfrak{e} {\big{(}} {\Upsilon}  \frac{\partial}{\partial \mathfrak{e}}
  {\big{)}}
 -   \frac{1}{2} \eta_{\mu\rho} \eta_{\nu\sigma}  {\pmb{\pi}}^{{A}_\rho \sigma}    \hbox{d}     {\pmb{\pi}}^{{A}_\mu \nu}
 {\big{(}}
   {\Upsilon}^{ {A}_{\alpha} \beta}   \frac{\partial}{\partial {\pmb{\pi}}^{ {A}_\alpha \beta}  }
   {\big{)}}
 - \dd \mathfrak{e}     {\Big{(}}          {\big{(}}        \mathfrak{e}  ( \frac{\partial  X^{\nu} }{\partial x^{\nu}} )  + \frac{\partial  \Theta_{{\mu}}  }{\partial x^{\nu}}  {\pmb{\pi}}^{ {A}_\mu \nu}         {\big{)}}     \frac{\partial}{\partial \mathfrak{e}}   {\Big{)}}    
             \\
        \displaystyle &  &  \displaystyle  -   \frac{1}{2} \eta_{\mu\rho} \eta_{\nu\sigma}  {\pmb{\pi}}^{{A}_\rho \sigma}    \hbox{d}     {\pmb{\pi}}^{{A}_\mu \nu}    {\Big{(}}   {\Big{(}}   {\pmb{\pi}}^{ {A}_\rho \sigma}  \delta^{\mu}_{\rho}  {\big{(}}     \big{[}  (\frac{\partial  X^{\nu} }{\partial x^{\sigma}} )   -  \delta^{\nu}_{\sigma}     (\frac{\partial  X^{\lambda} }{\partial x^{\lambda}} )   \big{]}   +     ( \frac{\partial  \Theta_{{\sigma}}  }{\partial  {A}_{\nu} } ) {\big{)}}  -  \mathfrak{e}   ( \frac{\partial  X^{\nu} }{\partial { {A} _{\mu}}  }  )  {\Big{)}}     \frac{\partial}{ \partial {\pmb{\pi}}^{ {A}_\mu \nu}  } 
 {\Big{)}} .
 \\
\end{array}
\right.
\]
So that we obtain: 
 \bee\label{tyussss}
\left.
\begin{array}{ccl}
\displaystyle
  \hbox{d}  {\cal H}   (\Xi)  & = &  \displaystyle  \Upsilon  +   {\big{[}}  \mathfrak{e}  {\big{]}}    {\big{[}}   - ( \frac{\partial  X^{\nu} }{\partial x^{\nu}} )   {\big{]}}  
   +
     {\big{[}}   {\pmb{\pi}}^{ {A}_\mu \nu}     {\big{]}}       {\big{[}}   -   \frac{1}{2} \eta_{\mu \rho} \eta_{\nu \sigma}      {\Upsilon}^{ {A}_{\rho} \sigma}       -     \frac{\partial  \Theta_{{\mu}}  }{\partial x^{\nu}}     {\big{]}}      
   + {\big{[}}    \mathfrak{e}    {\pmb{\pi}}^{{A}_\lambda \kappa}      {\big{]}}   {\big{[}}    \frac{1}{2} \eta_{\mu\lambda} \eta_{\nu\kappa}   ( \frac{\partial  X^{\nu} }{\partial { {A} _{\mu}}  }  )    {\big{]}}     
        \\
                 \displaystyle &  &  \displaystyle +
            {\big{[}}        {\pmb{\pi}}^{{A}_\lambda \kappa}         {\pmb{\pi}}^{ {A}_\rho \sigma}  {\big{]}}  
            {\big{[}} 
                 -   \frac{1}{2} \eta_{\mu\lambda} \eta_{\nu\kappa} \delta^{\mu}_{\rho}     {\big{(}}   [  (\frac{\partial  X^{\nu} }{\partial x^{\sigma}} )   -  \delta^{\nu}_{\sigma}     (\frac{\partial  X^{\lambda} }{\partial x^{\lambda}} )  ]     +     ( \frac{\partial  \Theta_{{\sigma}}  }{\partial  {A}_{\nu} } ) {\big{)}}    {\big{]}}   . 
\end{array}
\right.
\eee
  Thanks to \eqref{tyussss} we have $\Upsilon  = 0 $ and   the following relations \eqref{flso9448} as coefficient expression respectively of $\mathfrak{e}$, $ {\pmb{\pi}}^{ {A}_\mu \nu} $, $    \mathfrak{e}    {\pmb{\pi}}^{{A}_\lambda \kappa}  $ and $       {\pmb{\pi}}^{{A}_\lambda \kappa}         {\pmb{\pi}}^{ {A}_\rho \sigma}$: 
    \bee\label{flso9448}
\left|
\begin{array}{lll}
\displaystyle        0 & = &  \displaystyle     {\big{[}}   - ( \frac{\partial  X^{\nu} }{\partial x^{\nu}} )   {\big{]}}       \\
\displaystyle     0 & = &  \displaystyle      {\big{[}}   -   \frac{1}{2} \eta_{\mu \rho} \eta_{\nu \sigma}      {\Upsilon}^{ {A}_{\rho} \sigma}       -     \frac{\partial  \Theta_{{\mu}}  }{\partial x^{\nu}}     {\big{]}}     \\
\displaystyle      0 & = &  \displaystyle      {\big{[}}    \frac{1}{2} \eta_{\mu\lambda} \eta_{\nu\kappa}   ( \frac{\partial  X^{\nu} }{\partial { {A} _{\mu}}  }  )    {\big{]}}     \\
\displaystyle         0 & = &  \displaystyle       {\big{[}}      -   \frac{1}{2} \eta_{\mu\lambda} \eta_{\nu\kappa} \delta^{\mu}_{\rho}     {\big{(}}   [  (\frac{\partial  X^{\nu} }{\partial x^{\sigma}} )   -  \delta^{\nu}_{\sigma}     (\frac{\partial  X^{\lambda} }{\partial x^{\lambda}} )  ]     +     ( \frac{\partial  \Theta_{{\sigma}}  }{\partial  {A}_{\nu} } ) {\big{)}}  {\big{]}}  .  \\
\end{array}
\right.
\eee

The relation \eqref{flso9448} - $ {\big{[}}    \mathfrak{e}    {\pmb{\pi}}^{{A}_\lambda \kappa}      {\big{]}}  $  leads to $X^{\nu} (x,A) = X^{\nu} (x)$. From    \eqref{flso9448} - $   {\big{[}}   {\pmb{\pi}}^{ {A}_\mu \nu}     {\big{]}} $   we obtain the following relation: 
\bee\label{7755667766}
\left|
\begin{array}{ccl}
 ({1}/{2}) \eta_{\rho\mu} \eta_{\sigma\nu}      {\Upsilon}^{ {A}_{\rho} \sigma}    & = &  \displaystyle   -     {\partial_{\nu}  \Theta_{{\mu}}  }     
     \\ 
 -     ({1}/{2})  \eta_{\sigma\nu} \eta_{\rho\mu}      {\Upsilon}^{ {A}_{\sigma} \rho}    & = &  \displaystyle            {\partial_{\mu}  \Theta_{{\nu}}  }.      
     \\
  \end{array}
\right.
\eee 
If we sum the last two equations \eqref{7755667766} we obtain: 
\[
\eta_{\rho\mu} \eta_{\sigma\nu}   ({1}/{2})  (  {\Upsilon}^{ {A}_{\rho} \sigma} -   {\Upsilon}^{ {A}_{\sigma} \rho}  ) =    \eta_{\rho\mu} \eta_{\sigma\nu} {\Upsilon}^{ {A}_{\rho} \sigma} = \partial_{\mu} \Theta_{\nu} - \partial_{\nu} \Theta_{\mu}  .
\]
Finally, due to relations  \eqref{flso9448}, we obtain:
\[
\left.
\begin{array}{ccl}
\displaystyle
\overline{ \zeta}  & = &  \displaystyle   X^{\nu}  \frac{\partial}{\partial x^{\nu}}  + \Theta _{\mu} \frac{\partial}{\partial  {A}_{\mu}}      -  {\Big{(}}           \frac{\partial  \Theta_{{\mu}}  }{\partial x^{\nu}}  {\pmb{\pi}}^{ {A}_\mu \nu}         {\Big{)}}     \frac{\partial}{\partial \mathfrak{e}},
  \end{array}
\right.
\]
and $ \displaystyle \chi =    {\Upsilon}^{ {A}_{\mu}\alpha}   \frac{\partial}{\partial {\pmb{\pi}}^{ {A}_\mu \alpha}  } $  with  $ \displaystyle \Upsilon :  {\pmb{\mathfrak{Z}}} \rightarrow \Bbb{R} $ and $ {\Upsilon}^{ {A}_{\mu}\alpha} = \partial_{\mu} \Theta_{\nu} - \partial_{\nu} \Theta_{\mu} :   {\pmb{\mathfrak{Z}}} \rightarrow  \Bbb{R}$  smooth functions on $ {\pmb{\mathfrak{Z}}}   $ such that: 
\begin{equation}
 \frac{\partial  {{\Upsilon}}^{ {A}_{\mu}\nu} }{\partial x^{\nu}} = 0 .
\end{equation}
Hence ${\pmb{\Xi}} \in \Gamma ({\pmb{\cal M}} , T {\pmb{\cal M}})$ such that  $\dd ({\pmb{\Xi}}  \iN {\pmb{\Omega}}^{\tiny{\hbox{\sffamily DDW}}}) = \dd {\cal H} ({\pmb{\Xi}} ) = 0 $ if and only if ${\pmb{\Xi}} $ is written as announced in proposition \eqref{sfsf4321}: 
\[
{\pmb{\Xi}} = {\pmb{\Xi}}_{X} + {\pmb{\Xi}}_{A} = X^{\mu} \frac{\partial }{\partial x^{\mu}}   + \Theta _{\mu} \frac{\partial}{\partial  {A}_{\mu}}      -  {\Big{(}}           \frac{\partial  \Theta_{{\mu}}  }{\partial x^{\nu}}  {\pmb{\pi}}^{ {A}_\mu \nu}         {\Big{)}}     \frac{\partial}{\partial \mathfrak{e}}  +  [ \partial_{\mu} \Theta_{\nu} - \partial_{\nu} \Theta_{\mu} ]    \frac{\partial}{\partial {\pmb{\pi}}^{ {A}_\mu \nu}  }.
\]

\subsection{\hbox{\sffamily\bfseries\slshape{Algebraic observable $(n-1)$-forms in the pre-multisymplectic case}}}\label{aaaaaaap1}

We   enter into some details, considering the pre-multisymplectic case. For the ${\hbox{\sffamily{(DDW)}}}$ theory   and without taking into account the decomposition on the space-time variables - so that we forget the stress-energy tensor part. We focus on the following  infinitesimal symplectomorphisms, $\Xi^{{\hbox{\skt 0}}}   \in \Gamma( { {\pmb{\cal M}} }_{\tiny{\hbox{\sffamily DDW}}}^{{\hbox{\skt 0}}}  , T { {\pmb{\cal M}} }_{\tiny{\hbox{\sffamily DDW}}}^{{\hbox{\skt 0}}}  )$, written in the form:
\begin{equation}\label{vfm001}
\Xi^{{\hbox{\skt 0}}}_{\hbox{\tiny\sffamily DDW}}   =       
{\pmb{\Theta}}_{\mu}(q,p)   \frac{\partial}{\partial {{A}}_{\mu}}     +   {\pmb{\Upsilon}}^{{{A}}_{\mu}\nu}   (q,p)  \frac{\partial}{\partial {\pmb{\pi}}^{{{A}}_{\mu}\nu}}.
\end{equation}
Notice that due to   the Dirac primary constraint set, we must consider the following object $\Xi^{{\hbox{\skt 0}}}   \in \Gamma( { {\pmb{\cal M}} }_{\tiny{\hbox{\sffamily Maxwell}}}^{{\hbox{\skt 0}}}  , T { {\pmb{\cal M}} }_{\tiny{\hbox{\sffamily Maxwell}}}^{{\hbox{\skt 0}}}  )$ which is given by the interplay of some forbidden directions:
\begin{equation}\label{vfm001efee}
\Xi^{{\hbox{\skt 0}}}   =       {\pmb{\Theta}}_{\mu}(q,p)   \frac{\partial}{\partial {{A}}_{\mu}}      +   {\pmb{\Upsilon}}^{{{A}}_{\mu}\nu}   (q,p) {\Big{(}}  \frac{\partial}{\partial {\pmb{\pi}}^{{{A}}_{\mu}\nu}} -  \frac{\partial}{\partial {\pmb{\pi}}^{{{A}}_{\nu}\mu}} {\Big{)}}.
\end{equation}
 $ {\pmb{\Theta}}_{\mu} (q,p)$ and ${\pmb{\Upsilon}}^{{{A}}_{\mu}\nu} (q,p)  $ are smooth functions on $ { {\pmb{\cal M}} }_{\tiny{\hbox{\sffamily Maxwell}}}^{{\hbox{\skt 0}}}  \subset {\pmb{\cal M}}_{\tiny{\hbox{\sffamily Maxwell}}} \subset {\pmb{\cal M}}_{\tiny{\hbox{\sffamily DDW}}}   \subset \Lambda^n T^{\star} (T^{\star}{\cal X})$, with values in $\Bbb{R}$. We evaluate the expression $\Xi^{{\hbox{\skt 0}}} \iN  {\pmb{\Omega}}^{{\hbox{\skt 0}}}$: 
 \[
 \left.
\begin{array}{rcl}
\displaystyle  \Xi^{{\hbox{\skt 0}}} \iN  {\pmb{\Omega}}^{{\hbox{\skt 0}}}        & = &  \displaystyle  \Xi^{{\hbox{\skt 0}}} \iN (  \frac{1}{2} \eta_{\mu\rho} \eta_{\nu\sigma}  {\pmb{\pi}}^{{A}_\rho \sigma}    \hbox{d}     {\pmb{\pi}}^{{A}_\mu \nu}\wedge  \hbox{d} \mathfrak{y} ) + \Xi^{{\hbox{\skt 0}}} \iN (  \hbox{d}  {\pmb{\pi}}^{{A}_{\mu}\nu}  \wedge
 \hbox{d}{A}_{\mu}  \wedge   \hbox{d} \mathfrak{y}_{\nu} ) 
 \\
\displaystyle     & = &  \displaystyle  {\big{(}} {\pmb{\Upsilon}}^{{{A}}_{\mu}\nu} - {\pmb{\Upsilon}}^{{{A}}_{\nu}\mu}  {\big{)}}  \hbox{d} {{A}}_{\mu}  \wedge  \hbox{d} \mathfrak{y}_{\nu}  -    {\pmb{\Theta}}_{{\mu}} \hbox{d}  {\pmb{\pi}}^{{{A}}_{\mu}\nu}  \wedge  \hbox{d} \mathfrak{y}_{\nu}  +  \frac{1}{2} \eta_{\mu\rho} \eta_{\nu\sigma}  {\pmb{\pi}}^{{A}_\rho \sigma}  {\big{(}} {\pmb{\Upsilon}}^{{{A}}_{\mu}\nu} - {\pmb{\Upsilon}}^{{{A}}_{\nu}\mu}  {\big{)}} \vol  .
\end{array}
\right.
\]
 Now due to  the definition of the symplectomorphism via the formula $ \hbox{d} (\Xi^{{\hbox{\skt 0}}} \iN  {\pmb{\Omega}}^{{\hbox{\skt 0}}}  ) = 0$, we make the following calculation:
 \[
\left.
\begin{array}{ccl}
\displaystyle 
\hbox{d} (\Xi^{{\hbox{\skt 0}}} \iN  {\pmb{\Omega}}^{{\hbox{\skt 0}}}  ) 
    &  =  & \displaystyle      
      \hbox{d}   {\big{(}} {\pmb{\Upsilon}}^{{{A}}_{\mu}\nu} - {\pmb{\Upsilon}}^{{{A}}_{\nu}\mu}  {\big{)}} \wedge \hbox{d} {A}_{\mu}    \wedge  \hbox{d} \mathfrak{y}_{\nu}    - \hbox{d}  {\pmb{\Theta}}_{{\mu}}  \wedge \hbox{d}  {\pmb{\pi}}^{{A}_{\mu}\nu}    \wedge  \hbox{d} \mathfrak{y}_{\nu}      
\\
 \displaystyle   
&    & \displaystyle    +  \frac{1}{2} \eta_{\mu\rho} \eta_{\nu\sigma}  {\pmb{\pi}}^{{A}_\rho \sigma} \dd  {\big{(}} {\pmb{\Upsilon}}^{{{A}}_{\mu}\nu} - {\pmb{\Upsilon}}^{{{A}}_{\nu}\mu}  {\big{)}}  \wedge \vol
\\
 \displaystyle   
&    & \displaystyle  
 +   \frac{1}{2} \eta_{\mu\rho} \eta_{\nu\sigma}  {\big{(}} {\pmb{\Upsilon}}^{{{A}}_{\mu}\nu} - {\pmb{\Upsilon}}^{{{A}}_{\nu}\mu}  {\big{)}}   \dd  {\pmb{\pi}}^{{A}_\rho \sigma}  \wedge \vol .
\end{array}
\right.
\]
Using the decomposition of $\hbox{d} {\pmb{\Theta}}_{\mu}$ and $\hbox{d}   {\pmb{\Upsilon}}^{{A}_{\mu}\nu} $, we obtain:
\[
  \hbox{d} {\pmb{\Theta}}_{\mu}  =    \frac{\partial {\pmb{\Theta}}_{\mu}}{\partial x^{\alpha}} \dd x^{\alpha} + \frac{\partial {\pmb{\Theta}}_{\mu}}{\partial  {{A} _{\beta}}} \hbox{d} {{A} _{\beta}}   +  \frac{\partial  {\pmb{\Theta}}_{\mu} }{\partial  {\pmb{\pi}}^{{A}_{\beta}\alpha}  }  \hbox{d} {\pmb{\pi}}^{{A}_{\beta}\alpha} \ ,
\quad   
 \hbox{d}   {\pmb{\Upsilon}}^{{A}_{\mu}\nu}   =  \frac{\partial  {\pmb{\Upsilon}}^{{A}_{\mu}\nu}     }{\partial x^{\alpha}} \dd x^{\alpha} + \frac{\partial   {\pmb{\Upsilon}}^{{A}_{\mu}\nu}    }{\partial  {{A} _{\beta}}} \hbox{d} {{A} _{\beta}}   + \frac{\partial    {\pmb{\Upsilon}}^{{A}_{\mu}\nu}    }{\partial  {{\pmb{\pi}}^{ {A}_{\beta}  \alpha } }} \hbox{d} {\pmb{\pi}}^{{A}_{\beta}\alpha}  
\]
The different decompositions  of the involved ($n+1$)-forms are written:

{\em  |}  ${\bf j}_{1}$  is the term related to the decomposition on  $    \displaystyle {\big{[}}   \hbox{d} {\pmb{\pi}}^{{A}_{\beta}\alpha}    \wedge\hbox{d}  {\pmb{\pi}}^{{A}_{\mu}\nu}    \wedge  \hbox{d} \mathfrak{y}_{\nu} {\big{]}}   $  so that: 
\[
{\bf j}_{1} =     -       \frac{\partial  {\pmb{\Theta}}_{\mu} }{\partial  {\pmb{\pi}}^{{A}_{\beta}\alpha}  }    \hbox{d} {\pmb{\pi}}^{{A}_{\beta}\alpha}    \wedge\hbox{d}  {\pmb{\pi}}^{{A}_{\mu}\nu}    \wedge  \hbox{d} \mathfrak{y}_{\nu}  
\]

{\em  |}  ${\bf j}_{2}$  is the term related to the decomposition on $  {\big{[}}    \hbox{d}   {A}  \wedge \hbox{d} {A}  \wedge \hbox{d} \mathfrak{y}_{\nu} {\big{]}}   $
  \[
{\bf j}_{2} =    
     {\big{(}}   \frac{\partial   {\pmb{\Upsilon}}^{{A}_{\mu}\nu}    }{\partial  {{A} _{\beta}}}  - \frac{\partial   {\pmb{\Upsilon}}^{{A}_{\nu}\mu}    }{\partial  {{A} _{\beta}}}     {\big{)}}      \hbox{d} {{A} _{\beta}}     \wedge  \hbox{d} {A}_{\mu}    \wedge   \hbox{d}  \mathfrak{y}_{\nu} 
\]

{\em  |}  ${\bf j}_{3}$  is the term related to the decomposition on $ {\big{[}}  \hbox{d}  {A} \wedge    \hbox{d}    \mathfrak{y}  {\big{]}}  $
\[
{\bf j}_{3} =       \frac{1}{2} \eta_{\mu\rho} \eta_{\nu\sigma}  {\pmb{\pi}}^{{A}_\rho \sigma}  {\big{(}}    \frac{\partial   {\pmb{\Upsilon}}^{{A}_{\mu}\nu}    }{\partial  {{A} _{\beta}}} -  \frac{\partial   {\pmb{\Upsilon}}^{{A}_{\nu}\mu}    }{\partial  {{A} _{\beta}}} {\big{)}} \dd A_{\beta}  \wedge \vol 
+ {\big{(}}   \frac{\partial  {\pmb{\Upsilon}}^{{A}_{\nu}\mu}     }{\partial x^{\nu}}    - \frac{\partial  {\pmb{\Upsilon}}^{{A}_{\mu}\nu}     }{\partial x^{\nu}}   {\big{)}}     \hbox{d} {A}_{\mu}    \wedge  \dd \mathfrak{y} 
\]

{\em  |}  ${\bf j}_{4}$  is the term related to the decomposition on $   {\big{[}}  \hbox{d}  {\pmb{\pi}}  \wedge   \hbox{d} \mathfrak{y}   {\big{]}}     $
 \[
\left.
\begin{array}{ccl}
\displaystyle 
{\bf j}_{4}
    &  =  & \displaystyle 
  \frac{1}{2} \eta_{\mu\rho} \eta_{\nu\sigma}  {\pmb{\pi}}^{{A}_\rho \sigma}   {\big{(}}  \frac{\partial    {\pmb{\Upsilon}}^{{A}_{\mu}\nu}    }{\partial  {{\pmb{\pi}}^{ {A}_{\beta}  \alpha } }}  -      \frac{\partial    {\pmb{\Upsilon}}^{{A}_{\nu}\mu}    }{\partial  {{\pmb{\pi}}^{ {A}_{\beta}  \alpha } }}  {\big{)}} \hbox{d} {\pmb{\pi}}^{{A}_{\beta}\alpha}  \wedge \vol 
+      \frac{\partial {\pmb{\Theta}}_{\mu}}{\partial x^{\nu}}    \hbox{d}  {\pmb{\pi}}^{{A}_{\mu}\nu}    \wedge  \hbox{d} \mathfrak{y}
\\
 \displaystyle   
&    & \displaystyle 
 + \frac{1}{2} \eta_{\mu\rho} \eta_{\nu\sigma}  {\big{(}} {\pmb{\Upsilon}}^{{{A}}_{\mu}\nu} - {\pmb{\Upsilon}}^{{{A}}_{\nu}\mu}  {\big{)}}    \dd  {\pmb{\pi}}^{{A}_\rho \sigma}  \wedge \vol  
\end{array}
\right.
\]
  
{\em  |}  ${\bf j}_{5}$  is the term related to the decomposition on 
$   \hbox{d} {A} \wedge \hbox{d} {\pmb{\pi}} \wedge \hbox{d} \mathfrak{y}_{\nu}   $
\bee\label{masterAOFII}
{\bf j}_{5} =   -     \frac{\partial {\pmb{\Theta}}_{\mu}}{\partial  {{A} _{\beta}}}    \hbox{d} {{A} _{\beta}}   \wedge  \hbox{d}  {\pmb{\pi}}^{{A}_{\mu}\nu}    \wedge  \hbox{d} \mathfrak{y}_{\nu} +   {\big{(}}
  \frac{\partial    {\pmb{\Upsilon}}^{{A}_{\mu}\nu}    }{\partial  {{\pmb{\pi}}^{ {A}_{\beta}  \alpha } }} -  \frac{\partial    {\pmb{\Upsilon}}^{{A}_{\nu}\mu}    }{\partial  {{\pmb{\pi}}^{ {A}_{\beta}  \alpha } }} {\big{)}}  \hbox{d} {\pmb{\pi}}^{{A}_{\beta}\alpha}   \wedge  \hbox{d} {A}_{\mu}    \wedge  \hbox{d}\mathfrak{y}_{\nu}     
\eee
   The mathematical requirement on  the infinitesimal symplectomorphism  $\hbox{d} (\Xi^{{\hbox{\skt 0}}} \iN  {\pmb{\Omega}}^{{\hbox{\skt 0}}}  ) = 0 $ allows us to precise the conditions on the functions ${\pmb{\Theta}}_{\mu}$ and $ {\pmb{\Upsilon}}^{{A}_{\mu}\nu}$.  The equation \eqref{masterAOFII}-${\bf j}_{1}$ gives   that ${\pmb{\Theta}}_{\mu} $ is independent of momenta,   ${\pmb{\Theta}}_{\mu} = {\pmb{\Theta}}_{\mu} (x ,A ) $. The equation \eqref{masterAOFII}-${\bf j}_{2}$ gives  ${\pmb{\Upsilon}}^{{A}_{\mu}\nu}   =    {\pmb{\Upsilon}}^{{A}_{\mu}\nu}  (x,{\pmb{\pi}})$. Since we have equation \eqref{masterAOFII}-${\bf j}_{3}$ we obtain the following condition: 
\bee
\partial_{\nu}
{\big{(}}     {\pmb{\Upsilon}}^{{A}_{\nu}\mu}     -  {\pmb{\Upsilon}}^{{A}_{\mu}\nu}    {\big{)}} = 0 .
\eee
We recover from relations that emerge from \eqref{masterAOFII}-${\bf j}_{1}$-${\bf j}_{5}$ the results of  J. Kijowski \cite{KS0} and J. Kijowski and W. Szczyrba \cite{KS1}.

\subsection{\hbox{\sffamily\bfseries\slshape{Observable functionals}}}\label{Observvvvvv}

First, we recall the general setting for describing the kinematical and dynamical observable functionals. Then, we construct the dynamical observable functional for Maxwell theory.

${\hbox{\em{ Kinematical observable functionals}}}$.  The  important objects for the needs of physics are observable functionals.  This provides  a bridge   with the classical or quantum observables of field theory. We describe a multisymplectic manifold $({\pmb{\cal M}} , {\pmb{\Omega}})$ together with an Hamiltonian ${\cal H}$. We   denote by ${\cal E}^{\cal H}$ the set of Hamiltonian $n$-curves. This picture is the generalization of an Hamiltonian system\footnote{An Hamiltonian system  $({\pmb{\cal M}}, {\pmb{\Omega}}  , {\cal H})$   is the data of a symplectic manifold $({\cal M}, {{\pmb{\Omega}}} )$, together with a smooth Hamiltonian function ${\cal H}$.} $({\pmb{\cal M}}, {{\pmb{\Omega}}} , {\cal H})$ to the $n$-dimensional case where the dynamical data are  $({\pmb{\cal M}} , {\pmb{\Omega}} , {\cal H}) $. Before giving the definition of an observable functional, we introduce  the notion of {\em slice}.  The quantities of  physical interest are functionals on the set of Hamiltonian $n$-curves ${\cal E}^{\cal H}$. We  construct such observable functionals by integration of an algebraic observable $(n-1)$-form over a submanifold ${\pmb{\Sigma}} \subset {\pmb{\Gamma}}$ of codimension $1$ of a Hamiltonian $n$-curve ${\pmb{\Gamma}}$. Here we recover the picture of  observable functionals, in   classical (or quantum) field theory, as smeared integrals over a spacelike hypersurface.
\begin{defin}\label{jfzajoezjvzo004}
A slice of codimension $1$ is a submanifold ${\pmb{\Sigma}} \subset {\pmb{\cal M}}$ such that $T_m{\pmb{\cal M}} / T_{m}{\pmb{\Sigma}}$ is smoothly oriented with respect to m and, such that for any ${\pmb{\Gamma}} \in {\cal E}^{\cal H}$, ${\pmb{\Sigma}}$ is transverse to ${\pmb{\Gamma}}$. 
\end{defin}
  This definition allows us to give an orientation on ${\pmb{\Sigma}} \cap {\pmb{\Gamma}}$. 
 If ${\pmb{\Sigma}}$ is a slice of codimension $1$ and ${\pmb{\rho}}$ is a $(n-1)$-form on ${\pmb{\cal M}}$, {\em e.g} ${\pmb{\rho}} \in \Gamma({\cal M} , \Lambda^{n-1}T^{\star} {\pmb{\cal M}})$, we   define the concept of functional $\displaystyle \hbox{\sffamily\bfseries\slshape F} _{\pmb{\rho}}:= \int_{\pmb{\Sigma}} {\pmb{\rho}}$. This object is described as $\displaystyle \int_{\pmb{\Sigma}} {\pmb{\rho}}: {\cal E}^{\cal H} \longrightarrow \Bbb{R}$ on the set of Hamiltonian $n$-curves by   means of: 
\bee
 \hbox{\sffamily\bfseries\slshape F}_{\pmb{\rho}}:= \int_{\pmb{\Sigma}} {\pmb{\rho}}: {\pmb{\Gamma}} \mapsto \int_{{\pmb{\Sigma}}\cap {\pmb{\Gamma}} } {\pmb{\rho}}.
\eee
We can integrate the $(n-1)$-form ${\pmb{\rho}}$ on ${\pmb{\Sigma}} \cap {\pmb{\Gamma}}$. To reach the object of interest, we pass from those functionals to observable functionals whose form ${\pmb{\rho}}$ is an algebraic observable.
  \begin{defin}\label{jfzajoezjvzo005}
Let ${\pmb{\Sigma}}$ be a slice of codimension $1$ and let be ${\pmb{\varphi}}$ an algebraic observable $(n-1)$-form.
An observable
functional $\displaystyle \hbox{\sffamily\bfseries\slshape F}_{\pmb{\varphi}} = \int_{\pmb{\Sigma}} {\pmb{\varphi}}$ defined on the set of $n$-dimensional submanifolds ${\cal E}^{\cal H}$ is given  by the map: 
\bee\label{qnvnaeva014} 
 \hbox{\sffamily\bfseries\slshape F}_{\pmb{\varphi}} = \int_{\pmb{\Sigma}} {\pmb{\varphi}}: 
\left\{
\begin{array}{ccl}
\displaystyle     {\cal E}^{\cal H} &  \longrightarrow  &    \displaystyle  \Bbb{R}
 \\ 
\displaystyle     {\pmb{\Gamma}}  & \mapsto  &      \displaystyle  \hbox{\sffamily\bfseries\slshape F}   ({\pmb{\Gamma}}) = {\int_{{\pmb{\Sigma}}\cap {\pmb{\Gamma}}} {\pmb{\varphi}}}
\end{array}
\right.
\eee  
\end{defin}
 Then for any $ {\pmb{\varphi}}, {\pmb{\eta}} \in {{\textswab{P}}}_{\circ}^{n-1} ({\pmb{\cal M}}) $ the Poisson bracket - which coincides with the standard bracket for field theory - between two observable functionals  $\displaystyle  \int_\Sigma {\pmb{\varphi}}$ and $\displaystyle \int_\Sigma  {\pmb{\eta}}$ is defined such that $\forall \ {\pmb{\Gamma}} \in {\cal E}^{\cal H}$ we have \eqref{PPFKS}.
\begin{equation}\label{PPFKS}
 \left\{\int_{\pmb{\Sigma}}  {\pmb{\varphi}} ,\int_{\pmb{\Sigma}} {\pmb{\eta}}  \right\} ({\pmb{\Gamma}}):= \int_{ {\pmb{\Sigma}}  \cap {\pmb{\Gamma}} } \big{\{} {\pmb{\varphi}} ,{\pmb{\eta}}  \big{\}} 
\end{equation}
This Poisson bracket satisfies the Jacobi identity.  Let us consider $ {\pmb{\varphi}}, {\pmb{\rho}}, {\pmb{\eta}} \in {{\textswab{P}}}_{\circ}^{n-1} ({\pmb{\cal M}})$. From previous considerations, we know that 
\bee
{\big{\{}} {\pmb{\varphi}}  ,  \{ {\pmb{\rho}} , {\pmb{\eta}}  \} {\big{\}}} +
{\big{\{}} {\pmb{\eta}} ,  \{  {\pmb{\varphi}} , {\pmb{\rho}} \}    {\big{\}}} +  {\big{\{}}  {\pmb{\rho}}  ,  \{ {\pmb{\eta}}   , {\pmb{\varphi}} \}  {\big{\}}} = - \hbox{d} (\xi_{{\pmb{\varphi}}} \wedge \xi_{{\pmb{\rho}}} \wedge \xi_{{\pmb{\eta}} } \iN {\pmb{\Omega}}).
\eee
Therefore, restricting ourselves to the study of functional observables along Hamiltonian $n$-curves ${\pmb{\Gamma}}$ such that ${\partial}{\pmb{\Gamma}} = \emptyset $ we have the  Jacobi identity: 
\bee
 \left\{ \int_{\pmb{\Sigma}}  {\pmb{\varphi}} , \left\{\int_{\pmb{\Sigma}}  {\pmb{\varrho}} ,\int_{\pmb{\Sigma}} {\pmb{\eta}} \right\}  \right\}  +  \left\{ \int_{\pmb{\Sigma}}  {\pmb{\eta}} , \left\{\int_{\pmb{\Sigma}}  {\pmb{\varphi}} ,\int_{\pmb{\Sigma}} {\pmb{\varrho}} \right\}  \right\} +  \left\{ \int_{\pmb{\Sigma}}  {\pmb{\varrho}} , \left\{\int_{\pmb{\Sigma}}  {\pmb{\eta}} ,\int_{\pmb{\Sigma}} {\pmb{\varphi}} \right\}  \right\}  = 0. 
\eee

${\hbox{\em{Dynamical observable functionals}}}$.  The question of dynamical observable functionals hold the key to a {\em fully covariant theory}. In the perspective of a  covariant theory, we would like to define a bracket  over two different slices ${\pmb{\Sigma}}_\circ $ and $ {{\pmb{\Sigma}}_{\bullet}}$. The bracket defined previously in \eqref{PPFKS} depends on the choice of the given slice ${\pmb{\Sigma}}$. Given ${\pmb{\varphi}} ,  {\pmb{\eta}}  \in {{\textswab{P}}}_{\circ}^{n-1} ({\pmb{\cal M}}) $ we define the following bracket: 
\begin{equation}\label{PdddP}
 \left\{\int_{{\pmb{\Sigma}}_\circ} {\pmb{\varphi}}  ,\int_{{\pmb{\Sigma}}_\bullet}  {\pmb{\eta}} \right\} ({\pmb{\Gamma}}):= \int_{{\pmb{\Sigma}}_{\circ} \cap {\pmb{\Gamma}}} \big{\{} {\pmb{\varphi}} ,{\pmb{\eta}}  \big{\}}  .
\end{equation}
 Therefore, we are interested in dynamical observable functionals.  It is precisely for  {\em dynamical observables}  that we can construct a fully covariant bracket   \eqref{PdddP}.  We consider   an algebraic observable $(n-1)$-form ${\pmb{\varphi}} \in {{\textswab{P}}}_{\circ}^{n-1}({\pmb{\cal M}}) $ via  its related infinitesimal symplectomorphism  ${\Xi_{\pmb{\varphi}} } \in C^{\infty}( {\pmb{\cal M}} , T{\pmb{\cal M}})$ which is the unique vector field such that  ${\Xi_{\pmb{\varphi}} }  \iN {\pmb{\Omega}} = - \dd {\pmb{\varphi}}$. The algebraic observable $(n-1)$-form  becomes a  dynamical observable if we have  the additional condition:   
\bee
{\Xi_{\pmb{\varphi}} } \iN \dd {\cal H} = \hbox{d} {\cal H}( {\Xi_{\pmb{\varphi}} } ) = 0.
\eee
 This condition  reflects    a homological feature:  if ${\pmb{\Gamma}}$ is a Hamiltonian $n$-curve, then this functional $ \hbox{\sffamily\bfseries\slshape F}   ({\pmb{\Gamma}}) $ depends only on the homology class of ${\pmb{\Sigma}}$ \cite{HK-02} \cite{HK-03}. More precisely, following F. H\'elein \cite{FH-01} \cite{H-03} we show that this result follows from: 
\begin{prop} Let ${\pmb{\rho}} \in \Gamma ({\pmb{\cal M}} , \Lambda^{n-1} T^{\star} {\cal M})$ be a dynamical $(n-1)$-form. Let ${\pmb{\Sigma}}_{\circ}$ and ${\pmb{\Sigma}}_{\bullet}$ be two slices such that there exists an open subset ${\cal D} \subset {\pmb{\cal M}}$ which verifies ${\partial {\cal D}} = {\pmb{\Sigma}}_{\circ} - {\pmb{\Sigma}}_{\bullet} $. We have the following equality: 
\[
\int_{{\pmb{\Sigma}}_{\circ}}{\pmb{\rho}} = \int_{ {\pmb{\Sigma}}_{\bullet}} {\pmb{\rho}}.
\]
\end{prop}

${\hbox{\em{Time slice and   Minkowski space}}}$.  A slice of codimension $1$ is thought to be as a slice of time: an hypersurface of type ${\pmb{\Sigma}}_{\circ} = {\big{\{}} t = \tau_{\circ}  {\big{\}}}$ where $\tau_{\circ}$ is a constant. In such a context, we consider the spacetime manifold ${\cal X}_{\hbox{\tiny{Mink}}} = {\cal X}$ to be the flat Minkowski space. In this case we denote ${\cal X} = \Bbb{R}^{1,3}$ endowed with a constant metric $  {\eta}_{\mu\nu}  $, given in the canonical basis by the matrix $\hbox{diag} (1,-1,-1,-1)$. We denote the coordinates on ${\pmb{\cal M}}_{\tiny{\hbox{\sffamily DDW}}}$ by 
$
(x^{\mu}, A_{\mu} , \mathfrak{e} , {\pmb{\pi}}^{A_{\mu} \nu} ) =  (t , x^{1} , x^{2} , x^{3} , A_{\mu} , \mathfrak{e} , {\pmb{\pi}}^{A_{\mu} \nu} )
$.  For any $\tau_{\circ} \in \Bbb{R}$,  
\bee\label{minkowlso}
{\pmb{\Sigma}}_{\circ}:= {\big{\{}}   (x^{\mu},A_{\mu}, \mathfrak{e} , {\pmb{\pi}}^{A_{\mu} \nu} )  
\in  {\pmb{\cal M}}_{\tiny{\hbox{\sffamily DDW}}}  \ / \ t = \tau_{\circ} 
{\big{\}}} ,
\eee  are slices of codimension $1$.  The slices ${\pmb{\Sigma}}_{\circ} $ are oriented by the following condition: 
\bee
\partial_{\circ} \iN \vol  { {\big{|}} }_{{\pmb{\Sigma}}_{\circ}} > 0
\eee
Notice that in such a setting we also denote $x = (t , {\pmb x} ) $ where 
${\pmb x} = (x^{\mu})_{1 \leq \mu \leq n-1 }$ and $t = x^{\circ} $.

{\em Dynamical observable functionals}.  We follow the method developed by D.R. Harrivel   \cite{FFFDCX}. Let $ A: T^{\star} {\cal X} \rightarrow \Bbb{R}$ a smooth application and let $ \hbox{\sffamily\bfseries\slshape h} \in \Bbb{R}$. We associate to $(A,\hbox{\sffamily\bfseries\slshape h})$ the application $\pmb{\varsigma}_{A,\hbox{\sffamily\bfseries\slshape h}}: {\cal X} \rightarrow {\pmb{\cal M}}$ defined $\forall x \in {\cal X}$ by:
\[
\pmb{\varsigma}_{A,   \hbox{\sffamily\bfseries\slshape h}} (x):=  {\big{(}} x , A(x) ,{\cal E}_{A,\hbox{\sffamily\bfseries\slshape h}} (x) \vol + {\pmb{\pi}}   (x , A(x) , \dd A (x) )  {\big{)}} \cong  {\big{(}} x , A(x) ,{\cal E}_{A,\hbox{\sffamily\bfseries\slshape h}} (x)  ,  {\pmb{\pi}}^{A_{\mu} \nu}     (x)   {\big{)}}    \in {\pmb{\cal M}},
 \]
 where ${\pmb{\pi}}   (x , A(x) , \dd A (x) )  =  {\pmb{\pi}}^{A_{\mu} \nu} \dd A_{\mu} \wedge \vol_{\nu} (x , A(x) , \dd A (x) )$
and the function ${\cal E}_{A,\hbox{\sffamily\bfseries\slshape h}}: {\cal X} \rightarrow \Bbb{R} $ is defined by \eqref{functione}. 
\bee\label{functione}
{\cal E}_{A,\hbox{\sffamily\bfseries\slshape h}} (x):   \hbox{\sffamily\bfseries\slshape h} - {\cal H} (x,A(x),0 , {\pmb{\pi}} (x,A(x), \dd A(x) )   )  =  \hbox{\sffamily\bfseries\slshape h}  +     \frac{1}{4} \eta_{\mu\rho} \eta_{\nu\sigma}   {\pmb{\pi}}^{{A}_\mu \nu}   {\pmb{\pi}}^{{A}_\rho \sigma} 
\eee
  The graph of the application $\pmb{\varsigma}_{A,   \hbox{\sffamily\bfseries\slshape h}} (x)$ is written:
\bee
{\bf G} [ \pmb{\varsigma}_{A,   \hbox{\sffamily\bfseries\slshape h}}  ] =  {\big{(}} x , A(x) ,{\cal E}_{A,\hbox{\sffamily\bfseries\slshape h}} (x) \vol + {\pmb{\pi}}   (x , A(x) , \dd A (x) )  {\big{)}}   
\eee
 Then we consider ${\pmb{\Gamma}}_{A,  \hbox{\sffamily\bfseries\slshape h} } \subset {\pmb{\cal M}} $ the image of the application $\pmb{\varsigma}_{A,   \hbox{\sffamily\bfseries\slshape h}} $. Notice that since the expression for $\displaystyle \dd {\cal H}  |_{{\pmb{\Gamma}}_{A,  \hbox{\sffamily\bfseries\slshape h} }  }$ is written \eqref{glidh4},
\bee\label{glidh4}
\dd {\cal H}  |_{{\pmb{\Gamma}}_{A,  \hbox{\sffamily\bfseries\slshape h} }  }  = {\big{(}} \frac{\partial {\cal H} }{\partial x^{\nu}} \dd x^{\nu} + \frac{\partial {\cal H} }{\partial A_{\mu} } \dd A_{\mu} 
+  \frac{\partial {\cal H} }{\partial \mathfrak{e} } \dd   \mathfrak{e}   +  \frac{\partial {\cal H} }{\partial {\pmb{\pi}}^{A_{\mu}  \nu}} \dd {\pmb{\pi}}^{A_{\mu} \nu}  {\big{)}}  |_{{\pmb{\Gamma}}_{A,  \hbox{\sffamily\bfseries\slshape h} }  },  
\eee
and thanks  to the Hamilton equations, we observe that ${\cal H}  |_{{\pmb{\Gamma}}_{A,  \hbox{\sffamily\bfseries\slshape h}}}$ is constant and equal to $ \hbox{\sffamily\bfseries\slshape h } $ on the graph of an Hamiltonian function.  Notice that on the graph of an Hamiltonian function we have\footnote{For any $A $ and a map $x \mapsto p (x)$ such that  $(x, A(x) , \dd A (x) )  {\pmb{\leftrightarrow}}   (x, A(x) , p(x) ) $.} the following relations: 
\bee\label{77766}
\left|
\begin{array}{ccl}
    \displaystyle    {\pmb{\pi}}^{A_{\mu} \nu} (x)  & = &  \displaystyle   \frac{\partial L}{\partial ( \partial_{\nu} A_{\mu} )  } (x, A(x) , \dd A (x) ) ,
     \\ 
\displaystyle \mathfrak{e} (x) & = &  \displaystyle   \hbox{\sffamily\bfseries\slshape h }  + L (x, A(x) , \dd A (x) )  - \sum_{\mu = 0}^{n-1}  \sum_{\nu = 0}^{n-1}  {\pmb{\pi}}^{A_{\mu} \nu} \frac{\partial A_{\mu} }{ \partial  x^{\nu} } (x) ,
  \end{array}
\right.
\eee 
  equivalently: 
\bee\label{7755667766}
\left|
\begin{array}{ccl}
    \displaystyle    {\pmb{\pi}}^{A_{\mu} \nu} (x)  & = &  \displaystyle    \dttF^{\mu\nu} (x),
     \\ 
\mathfrak{e} (x)  & = &  \displaystyle     \hbox{\sffamily\bfseries\slshape h }  +       \frac{1}{4} \eta_{\mu\rho} \eta_{\nu\sigma}    \dttF^{\mu\nu} (x)   \dttF^{ \rho \sigma} (x) .
     \\
  \end{array}
\right.
\eee 
Now we consider a slice ${\pmb{\Sigma}}$ of codimension $1$. We describe, following  \cite{FFFDCX},  {\em slices}  of  type ${\pmb{\Sigma}} = {\big{(}}{\pmb{\varkappa}} \circ \pi^{{\cal X}} {\big{)}}^{-1} (0) $, where $\pi^{{\cal X}}$ is the natural projection $\pi^{{\cal X}}: {\pmb{\cal M}}_{\tiny{\hbox{\sffamily Maxwell}}} \rightarrow  {\cal X} $ and where $ {\pmb{\varkappa}}: {\cal X} \rightarrow \Bbb{R} $
 is a smooth function without any critical point, such that ${\pmb{\varkappa}}^{-1} (0) \neq 0 $.  Thanks to proposition \eqref{algebraFF44},  the dynamical observable $(n-1)$-forms are described by  $\pmb{\rho}  = \pmb{\rho}_{X} + \pmb{\rho}_{A}   $ where $ \pmb{\rho}_{X} $ and $ \pmb{\rho}_{A} $ are described by relations \eqref{tyuoi1qqss2}. Since  $
\pmb{\rho}_{X}   =      \mathfrak{e}  {  X}^{\rho}   \hbox{d}  \mathfrak{y}_{\rho}      - {\pmb{\pi}}^{{  A}_\mu \nu}    {  X}^{\rho}       \hbox{d} {  A}_{\mu}   \wedge  \dd \mathfrak{y}_{\rho\nu}, 
$
\bee\label{888777}
\int_{ {\pmb{\Sigma}} \cap {\pmb{\Gamma}}_{A, h}   }  \pmb{\rho}_{X}   =  \int_{\pi^{\cal X} ({\pmb{\Sigma}})}     {\big{(}}      \hbox{\sffamily\bfseries\slshape h }  +    \frac{1}{4} \eta_{\mu\rho} \eta_{\nu\sigma}    \dttF^{\mu\nu} (x)   \dttF^{ \rho \sigma} (x)   {\big{)}}  {  X}^{\lambda}  \hbox{d}  \mathfrak{y}_{\lambda}      - \underbrace{  \dttF^{\mu\nu}    {  X}^{\rho}       \frac{\partial A_{\mu} }{\partial x^{\alpha} } \dd x^{\alpha}   \wedge  \dd \mathfrak{y}_{\rho\nu}      }_{(\mathfrak{i}) } .
\eee
Since the term $(\mathfrak{i})$ is written, 
\bee
\left.
\begin{array}{rcl}
\displaystyle     (\mathfrak{i}) & = &  \displaystyle  \dttF^{\mu\nu}    {  X}^{\rho}       \frac{\partial A_{\mu} }{\partial x^{\alpha} }( \delta^{\alpha}_{\rho} \dd \mathfrak{y}_{\nu}  
-  \delta^{\alpha}_{\nu} \dd \mathfrak{y}_{\rho}   )    =     
  \dttF^{\mu\nu}    {  X}^{\rho}       \frac{\partial A_{\mu} }{\partial x^{\rho} } \dd \mathfrak{y}_{\nu}    
  -  \dttF^{\mu\nu}    {  X}^{\rho}       \frac{\partial A_{\mu} }{\partial x^{\nu} } \dd \mathfrak{y}_{\rho} 
 \\
 \displaystyle     & = &  \displaystyle \dttF^{\mu\nu}    {  X}^{\rho}       \frac{\partial A_{\mu} }{\partial x^{\rho} } \dd \mathfrak{y}_{\nu}    
+
  \dttF^{\mu\rho}    {  X}^{\nu}       \frac{\partial A_{\mu} }{\partial x^{\rho} } \dd \mathfrak{y}_{\nu} 
  =
{\big{(}}  \dttF^{\mu\nu}    {  X}^{\rho}     + \dttF^{\mu\rho}    {  X}^{\nu}    {\big{)}}    \frac{\partial A_{\mu} }{\partial x^{\rho} } \dd \mathfrak{y}_{\nu}  , 
 \\
\end{array}
\right.
\eee
we obtain for \eqref{888777}: 
\bee\label{glso99}
\int_{ {\pmb{\Sigma}} \cap {\pmb{\Gamma}}_{A, h}   }  \pmb{\rho}_{X}   =  \int_{\pi^{\cal X} ({\pmb{\Sigma}})}   {\Big{(}}       \hbox{\sffamily\bfseries\slshape h }     +  {\big{(}}    \frac{1}{4} \eta_{\mu\alpha} \eta_{\nu\beta}    \dttF^{\mu\nu} (x)   \dttF^{ \alpha\beta} (x)      
+
   \dttF^{\mu\rho}     {\big{)}}    {  X}^{ \lambda}       \frac{\partial A_{\mu} }{\partial x^{\rho} }    
-
 \dttF^{\mu \lambda}    {  X}^{\rho}       \frac{\partial A_{\mu} }{\partial x^{\rho} } {\Big{)}}  \hbox{d}  \mathfrak{y}_{\lambda} .
\eee
On the other side $
   \pmb{\rho}_{A}  =    {\big{(}}   {\pmb{\pi}}^{{A}_\mu \nu}      {{  \Theta}}_{{\mu}}        
   + \Upsilon^{  {  A}_{\mu}  \nu  }   {  A}_{\mu}   {\big{)}}    \hbox{d}  \mathfrak{y}_{\nu}
$, so that: 
\bee\label{biodjfh55}
\int_{ {\pmb{\Sigma}} \cap {\pmb{\Gamma}}_{A, h}   }  \pmb{\rho}_{A}    =  \int_{\pi^{\cal X} ({\pmb{\Sigma}})}    {\big{(}}   {\pmb{\pi}}^{{A}_\mu \nu}      {{  \Theta}}_{{\mu}}        
   + \Upsilon^{  {  A}_{\mu}  \nu  }   {  A}_{\mu}   {\big{)}}    \hbox{d}  \mathfrak{y}_{\nu}   
    =  \int_{\pi^{\cal X} ({\pmb{\Sigma}})}   {\big{(}}   \dttF^{\mu\nu}      {{  \Theta}}_{{\mu}}        
   + (\partial_{\mu} \Theta_{\nu}-\partial_{\nu} \Theta_{\mu})  {  A}_{\mu}   {\big{)}}     \hbox{d}  \mathfrak{y}_{\nu}  .   
\eee
  Now we focus on the second observable functional \eqref{biodjfh55}. We   integrate on a time slice ${\pmb{\Sigma}}_{\circ}$, with $\hbox{\sffamily\bfseries\slshape h} = 0 $ which in turn is equivalent to  consider the functional $\displaystyle \int_{ {\pmb{\Sigma}}_{\circ} \cap {\pmb{\Gamma}}_{A, 0}   }  \pmb{\varphi}_{A}$: 
 \bee\label{dmqlso0987}
\int_{ {\pmb{\Sigma}} \cap {\pmb{\Gamma}}_{A, 0}   }  \pmb{\varphi}_{A}       
 =
     \int_{ {\pmb{\Sigma}}_{t = \tau_\circ}}    {\big{(}}   \dttF^{\mu \circ}      {{  \Theta}}_{{\mu}}    + \Upsilon^{  {  A}_{\mu} \circ}   {  A}_{\mu}   {\big{)}}    \hbox{d}  \mathfrak{y}_{\circ} .  
\eee
As noticed by J. Kijowski \cite{KS0} and J. Kijowski and W. Szczyrba \cite{KS1} if we consider    
  $\Theta_{i} \big |_{{\pmb{\Sigma}}_{\circ}} = - \delta^{k}_{i} \cdot {\pmb{\varsigma}}$ and $ \Upsilon^{  {  A}_{0}  i  } \big |_{{\pmb{\Sigma}}_{\circ}} =0$ - with these conditions,  the observable
  functional  \eqref{dmqlso0987} is now denoted ${\bf D}^{k}    ( {\pmb{\varsigma}} ) $ and is  the $k$-th component of the electric field $\hbox{\sffamily\bfseries\slshape E}^{k}$ smeared with the test function ${\pmb{\varsigma}}$ - we obtain:
  \bee\label{one00}
\left.
\begin{array}{rcl}
\displaystyle   {\bf D}^{k}    ( {\pmb{\varsigma}} )        & = &  \displaystyle 
 \int_{ {\pmb{\Sigma}}_{\circ}}    {\big{(}}   \dttF^{i \circ}  (- \delta^{k}_{i} \cdot {\pmb{\varsigma}} )    -  \Upsilon^{  {  A}_{\circ} i }   {  A}_{i}   {\big{)}}    \hbox{d}  \mathfrak{y}_{\circ}  
     =
       \int_{ {\pmb{\Sigma}}_{\circ} } \dttF^{{{k}  \circ  } } \cdot  {\pmb{\varsigma}} \ \vol_{ \circ }   
=  \int_{ {\pmb{\Sigma}}_{\circ} } \hbox{\sffamily\bfseries\slshape E}^{k}  \cdot  {\pmb{\varsigma}} \ \vol_{ \circ }   .
\end{array}
\right.
\eee
On the other side, if we consider the conditions $\Theta_{i} \big |_{{\pmb{\Sigma}}_{\circ}} = 0$ and $ \Upsilon^{  {  A}_{0}  i  } \big |_{{\pmb{\Sigma}}_{\circ}} = - \varepsilon^{ijk} \partial_{j} {\pmb{\varsigma}}$, the observable
  functional  \eqref{dmqlso0987} is now denoted $\dttB^{k}    ( {\pmb{\varsigma}} ) $ and is  the $k$-th component of the magnetic field $\hbox{\sffamily\bfseries\slshape B}^{k}$ smeared with the test function ${\pmb{\varsigma}}$.
 \bee\label{one001}
\dttB^{k}    ( {\pmb{\varsigma}} )     
    =
     \int_{ {\pmb{\Sigma}}_{\circ}}    {\big{(}}   \dttF^{i \circ}  (- \delta^{k}_{i} \cdot {\pmb{\varsigma}} )    -  \Upsilon^{  {  A}_{\circ} i }   {  A}_{i}   {\big{)}}    \hbox{d}  \mathfrak{y}_{\circ}  
       =
- \frac{1}{2} \varepsilon^{kij } \int_{ {\pmb{\Sigma}}_{\circ} } \dttF_{ij } \cdot  {\pmb{\varsigma}} \ \vol_{ \circ } .
\eee
The objects $  {\bf E}^{k}    ( {\pmb{\varsigma}} )  $ and $ \dttB^{k}    ( {\pmb{\varsigma}} )  $ are observable-valued distributions ${\pmb{\varsigma}} \in {C}^{\infty}_{0} ({\pmb{\Sigma}}_{\circ}) \rightarrow  {\bf E}^{k}    ( {\pmb{\varsigma}} ), \dttB^{k}    ( {\pmb{\varsigma}} )   $ \cite{KS0} \cite{KS1}. 
From \eqref{one00} and \eqref{one001} we obtain the following Poisson brackets: 
\bee
\left.
\begin{array}{rcl}
\displaystyle    {\big{\{}}   \hbox{\sffamily\bfseries\slshape E}^{k} ({\pmb{\varsigma}}_1)  ,  \hbox{\sffamily\bfseries\slshape E}^{i} ({\pmb{\varsigma}}_2)    {\big{\}}}    & = &  
\displaystyle    {\big{\{}}  \hbox{\sffamily\bfseries\slshape B}^{k} ({\pmb{\varsigma}}_1) ,  \hbox{\sffamily\bfseries\slshape B}^{i} ({\pmb{\varsigma}}_2)      {\big{\}}}   =  0,  
\\
\displaystyle    {\big{\{}} \hbox{\sffamily\bfseries\slshape B}^{k} ({\pmb{\varsigma}}_1) , \hbox{\sffamily\bfseries\slshape E}^{i} ({\pmb{\varsigma}}_2)    {\big{\}}}  & = &  \displaystyle  - \varepsilon^{kij}      \int_{ {\pmb{\Sigma}}_{\circ}}    ( \partial_{j} {\pmb{\varsigma}}_1 ) {\pmb{\varsigma}}_2 \vol_{\circ}.
\end{array}
\right.
\eee
 We obtain the set of equal-time Poisson bracket,\footnote{Notice that in such a setting we   denote $x = (t , {\pmb x} ) $ where 
${\pmb x} = (x^{\mu})_{1 \leq \mu \leq n-1 }$ and $t = x^{\circ} $.
}  using observable valued distribution $\hbox{\sffamily\bfseries\slshape E}^{k} ({\pmb{x}}_{\mathfrak{1}})$  and $\hbox{\sffamily\bfseries\slshape B}^{k} ({\pmb{x}}_{\mathfrak{1}})$, see J. Kijowski and W. Szczyrba \cite{KS0} \cite{KS1}: 
\bee
\left.
\begin{array}{rcl}
\displaystyle    {\big{\{}}   \hbox{\sffamily\bfseries\slshape E}^{k} ({\pmb{x}}_{\mathfrak{1}})  ,  \hbox{\sffamily\bfseries\slshape E}^{i} ({\pmb{x}}_{\mathfrak{2}})    {\big{\}}}    & = &  \displaystyle    {\big{\{}}  \hbox{\sffamily\bfseries\slshape B}^{k} ({\pmb{x}}_{\mathfrak{1}}) ,  \hbox{\sffamily\bfseries\slshape B}^{i} ({\pmb{x}}_{\mathfrak{2}})      {\big{\}}}   =  0, \\  
\displaystyle    {\big{\{}} \hbox{\sffamily\bfseries\slshape B}^{k} ({\pmb{x}}_{\mathfrak{1}}) , \hbox{\sffamily\bfseries\slshape E}^{i} ({\pmb{x}}_{\mathfrak{2}})    {\big{\}}}  & = &  \displaystyle  - \varepsilon^{kij} \partial_{j} \delta({\pmb{x}}_{\mathfrak{1}} - {\pmb{x}}_{\mathfrak{2}}).  
\end{array}
\right.
\eee

 \subsection{\hbox{\sffamily\bfseries\slshape{Stress-energy tensor}}}\label{stress001} 
 
 We next examine the relation with the stress energy tensor $  {{{\mathfrak{S}}}}_{\mu\nu} $, see relations \eqref{dddlsqdd44}. First, we focus on some preparatory work. Let   denote:  
\bee
\left.
\begin{array}{rcl}
\displaystyle {\dttP}_{\partial_{\circ}}       & = &  \displaystyle  \partial_{\circ} \iN {\pmb{\theta}}^{\tiny{\hbox{\sffamily DDW}}}  =  \partial_{t} \iN {\pmb{\theta}}^{\tiny{\hbox{\sffamily DDW}}}, \quad \quad \hbox{and} \quad \quad -  \partial_{\circ} \iN {\big{(}} {\pmb{\theta}}^{\tiny{\hbox{\sffamily DDW}}} - {\cal H} (q,p) \vol {\big{ )}}  .
\end{array}
\right.
\eee

 \begin{lemm}\label{curva2dfrfderf}
 {\em We denote $\vol = \dd x^{\circ} \wedge \cdots \wedge \dd x^{n-1}$, $\vol_{\mu} = \partial_{\mu} \iN \vol$ and ${\pmb{\pi}} =  {\pmb{\pi}}^{{A}_\mu \nu}      
\hbox{d}  {{A}}_{\mu}  \wedge \hbox{d}  \mathfrak{y}_{\nu} $. We have the following relation: 
\[
\left.
\begin{array}{rcl}
\displaystyle   \partial_{\circ} \iN {\pmb{\pi}}      & = &  \displaystyle     \sum_{\mu = 0}^{n-1}     \sum_{\nu = 1}^{n-1}  {\pmb{\pi}}^{{A}_\mu \nu} 
   \dd A_{\mu}  \wedge  
   {\big{(}} \partial_{\nu} \iN \vol_{\circ} 
  {\big{ )}}
  \\
  \displaystyle      & = &  \displaystyle       \sum_{\mu = 0}^{n-1}     \sum_{\nu = 1}^{n-1}  {\pmb{\pi}}^{{A}_\mu \nu}      
  \dd x^{1}
  \wedge \dots \wedge
  \dd x^{\nu-1}  \wedge \dd A_{\mu} \wedge \dd x^{\nu+1} \cdots \wedge  \dd x^{n-1}.
\end{array}
\right.
\]

}
\end{lemm}
 
  $ ^{\lceil}     $  {\em Proof}. Since, $ {\partial_{\circ}}  \iN \hbox{d}  {{  A}}_{\mu}  =  0  $ and ${\partial_{\circ}}  \iN \hbox{d}  \mathfrak{y}_{\nu}  = {\partial_{\circ}}  \iN \partial_{\nu} \iN \vol  =  (  {\partial_{\circ}}  \wedge \partial_{\nu} ) \iN \vol  =   - \partial_{\nu} \iN \vol_{\circ}$.
We have: 
\[
\left.
\begin{array}{rcl}
\displaystyle      \partial_{\circ} \iN {\pmb{\pi}}      & = &  \displaystyle   {\partial_{\circ}}  \iN  
       {\pmb{\pi}}^{{A}_\mu \nu}      
\hbox{d}  {{A}}_{\mu}  \wedge \hbox{d}  \mathfrak{y}_{\nu} 
=
   {\pmb{\pi}}^{{A}_\mu \nu}       {\big{(}}  ( {\partial_{\circ}}  \iN \hbox{d}  {{ A}}_{\mu}  ) \wedge  \hbox{d}  \mathfrak{y}_{\nu}  -  \hbox{d}  {{ A}}_{\mu}  \wedge ({\partial_{\circ}}  \iN \hbox{d}  \mathfrak{y}_{\nu})    {\big{)}}   
\\
\displaystyle     & = &  \displaystyle -  {\pmb{\pi}}^{{A}_\mu \nu}         \hbox{d}  {{ A}}_{\mu}  \wedge ({\partial_{\circ}}  \iN \hbox{d}  \mathfrak{y}_{\nu}) 
=
  {\pmb{\pi}}^{{A}_\mu \nu}         \hbox{d}  {{ A}}_{\mu}  \wedge ({\partial_{\nu}}  \iN \hbox{d}  \mathfrak{y}_{\circ})  .
\end{array}
\right.
\]
Now, since $\vol_{\circ} = \dd x^{1} \wedge \cdots \wedge \dd x^{n-1}$ we obtain: 
\[
\left.
\begin{array}{rcl}
\displaystyle      \partial_{\circ} \iN {\pmb{\pi}}      & = &     \displaystyle  {\pmb{\pi}}^{{A}_\mu \nu}         \hbox{d}  {{ A}}_{\mu}  \wedge ({\partial_{\nu}}  \iN ( \dd x^{1} \wedge \cdots \wedge \dd x^{n-1} ))  
\\
 \displaystyle     & = &  \displaystyle
  {\pmb{\pi}}^{{A}_\mu \nu}         \hbox{d}  {{ A}}_{\mu}  \wedge (-1)^{\nu-1}   ( \dd x^{1} \wedge \cdots \wedge \dd x^{\nu-1} \wedge \dd x^{\nu+1} \wedge \cdots \wedge \dd x^{n-1} )  
     \\
    \displaystyle     & = &  \displaystyle
\sum_{\mu = 0}^{n-1}     \sum_{\nu = 1}^{n-1}  {\pmb{\pi}}^{{A}_\mu \nu}   \dd x^{1} \wedge \cdots \wedge \dd x^{\nu-1} \wedge  \hbox{d}  {{ A}}_{\mu}  \wedge   \dd x^{\nu+1} \wedge \cdots \wedge \dd x^{n-1}  .        \rfloor 
\end{array}
\right.  
\]
 From lemma \eqref{curva2dfrfderf}, we have the expression of ${\dttP}_{\partial_{\circ}} $ and ${\pmb{\eta}}_{\circ}$:
\[
\left.
\begin{array}{rcl}
\displaystyle    {\dttP}_{\partial_{\circ}}    & = &  \displaystyle   {\partial_{\circ}}  \iN  
{\big{(}}   \mathfrak{e}   \hbox{d} \mathfrak{y} +       {\pmb{\pi}}^{{A}_\mu \nu}      
\hbox{d}  {{A}}_{\mu}  \wedge \hbox{d}  \mathfrak{y}_{\nu} 
   \big{)} =  {\partial_{\circ}}  \iN  
{\big{(}}   \mathfrak{e}   \hbox{d} \mathfrak{y} +       {\pmb{\pi}} 
   \big{)}  
   \\
\displaystyle     & = &  \displaystyle    \mathfrak{e}   \hbox{d} \mathfrak{y}_{\circ} +    \sum_{\mu = 0}^{n-1}     \sum_{\nu = 1}^{n-1}  {\pmb{\pi}}^{{A}_\mu \nu}      
  \dd x^{1}
  \wedge \dots \wedge
  \dd x^{\nu-1}  \wedge \dd A_{\mu} \wedge \dd x^{\nu+1} \cdots \wedge  \dd x^{n-1} ,
\end{array}
\right.
\]
 and also: 
\[
\left.
\begin{array}{rcl}
 \displaystyle    { \pmb{\eta} }_{\circ}  & = &  \displaystyle {\cal H} (q,p) \vol_{\circ}  -  \partial_{\circ} \iN {\pmb{\theta}}^{\tiny{\hbox{\sffamily DDW}}}
=  {\cal H} (q,p) \vol_{\circ}  -  \dttP_{\partial_{\circ}}  
\\
\displaystyle     & = &  \displaystyle  {\cal H} (q,p) \vol_{\circ}  -  
 {\Big{(}}
  \mathfrak{e}   \hbox{d} \mathfrak{y}_{\circ} +    \sum_{\mu = 0}^{n-1}     \sum_{\nu = 1}^{n-1}  {\pmb{\pi}}^{{A}_\mu \nu}      
  \dd x^{1}
  \wedge \dots \wedge
  \dd x^{\nu-1}  \wedge \dd A_{\mu} \wedge \dd x^{\nu+1} \cdots \wedge  \dd x^{n-1}
  {\Big{ )}}.
  \end{array}
\right.
\]
We consider the Hamiltonian curve ${\pmb{\Gamma}}:= {\big{\{}}  (x^{\nu},A_{\mu} (x), \mathfrak{e}(x) ,{\pmb{\pi}}^{A_{\mu} \nu} (x) ) {\big{\}}} \subset {\pmb{\cal M}}_{\tiny{\hbox{\sffamily Maxwell}}}$ and the instantaneous slices
${\pmb{\Sigma}}_{\circ}  = {\pmb{\Gamma}}  \ \cap \ {\big{\{}} x^{\circ} = t {\big{\}}}$, so that: 
\[
\left.
\begin{array}{rcl}
\displaystyle  \int_{{\pmb{\Sigma}}_{\circ}}    { \pmb{\eta} }_{\circ}        & = &  \displaystyle \int_{{\pmb{\Sigma}}_{\circ}}     \displaystyle  {\cal H} (q,p) \vol_{\circ} 
  \\
 \displaystyle     &  &  \displaystyle
 -  \int_{{\pmb{\Sigma}}_{\circ}}
{\Big{(}}
  \mathfrak{e}   \hbox{d} \mathfrak{y}_{\circ} +    \sum_{\mu = 0}^{n-1}     \sum_{\nu = 1}^{n-1}  {\pmb{\pi}}^{{A}_\mu \nu}      
  \dd x^{1}
  \wedge \dots \wedge
  \dd x^{\nu-1}  \wedge \dd A_{\mu} \wedge \dd x^{\nu+1} \cdots \wedge  \dd x^{n-1}
  {\Big{ )}}
    \\
\displaystyle     & = &  \displaystyle  \int_{{\pmb{\Sigma}}_{\circ}}     \displaystyle  {\cal H} (q,p) \vol_{\circ}  -  \int_{{\pmb{\Sigma}}_{\circ}}
  \mathfrak{e}   \hbox{d} \mathfrak{y}_{\circ} +    \sum_{\mu = 0}^{n-1}     \sum_{\nu = 1}^{n-1}  {\pmb{\pi}}^{{A}_\mu \nu} 
   \dd A_{\mu}  \wedge  
   {\big{(}} \partial_{\nu} \iN \vol_{\circ} 
  {\big{ )}}.
\end{array}
\right.
\]
 
 \

$\hbox{\em{Stress-energy tensor}}$.  The canonical stress energy tensor ${ {{\mathfrak{S}}}}_{\mu\nu} $ and the symmetric stress-energy tensor $ {\overline{{\mathfrak{S}}}}^{\mu\nu} $ for the electromagnetic field are described by: 
\bee\label{dddlsqdd44}
\left|
\begin{array}{rcl}
\displaystyle    {{{\mathfrak{S}}}}_{\mu\nu}     & = &  \displaystyle  - {\big{(}}   \dttF_{\mu\lambda}  {\partial_{\nu} A^{\lambda} }
-
 \frac{1}{4} \eta_{\mu\nu} \dttF^{\alpha\beta} \dttF_{\alpha\beta}
{\big{)}},
  \\
 \displaystyle {\overline{{\mathfrak{S}}}}^{\mu\nu}    & = &  \displaystyle
{\overline{{\mathfrak{S}}}}^{\mu\nu}  + \dttF^{\mu\lambda} \partial_{\lambda} A^{\nu} 
=
-
{\big{(}}
 \dttF^{\mu\lambda}  {\dttF^{\nu}}_{\lambda}  
-
 \frac{1}{4} \eta^{\mu\nu} \dttF^{\alpha\beta} \dttF_{\alpha\beta} {\big{)}}.
\end{array}
\right.
\eee
   The symmetric stress-energy tensor $  {{{\mathfrak{S}}}}_{\mu\nu} $ is  obtained by  adding a term  $\partial_{\lambda} {\pmb{\kappa}}^{\mu\nu\lambda}$    with $  {\pmb{\kappa}}^{\mu\nu\lambda} = -  {\pmb{\kappa}}^{\nu \mu\lambda}$. 
 \bee\label{klioi99}
  {\overline{{\mathfrak{S}}}}_{\mu\nu}   =  {{{\mathfrak{S}}}}_{\mu\nu}   + 
\partial_{\lambda} {\pmb{\kappa}}^{\mu\nu\lambda} .
 \eee
 The relation \eqref{klioi99} is known as the Belinfante-Rosenfeld formula \cite{Bel} \cite{Ros}.
The canonical stress energy tensor  $ { {{\mathfrak{S}}}}_{\mu\nu}   $ associated to $A: T^{\star} {\cal X} \rightarrow \Bbb{R} $ is written: 
\bee
{ {{\mathfrak{S}}}}_{\mu\nu}    = \delta^{\alpha}_{\beta} L (x,A(x) , \dd A (x) ) - \frac{\partial L}{\partial v_{\lambda\alpha} } (x,A(x),\dd A (x) ) \frac{\partial A_{\lambda} }{\partial x^{\beta}} (x).
\eee
Hence,
\[
{ {{\mathfrak{S}}}}_{\mu\nu}   = \delta^{\alpha}_{\beta} (  - 1/4 \dttF_{\mu\nu} \dttF^{\mu\nu}  )- \eta^{\lambda\rho} \eta^{\alpha\sigma}  \dttF_{\rho\sigma}    \frac{\partial A_{\lambda} }{\partial x^{\beta}}  =     
 \delta^{\alpha}_{\beta} (  - 1/4 \dttF_{\mu\nu} \dttF^{\mu\nu}  ) - {\big{(}}
\dttF^{\lambda\alpha}    \frac{\partial A_{\lambda} }{\partial x^{\beta}}  
{\big{)}}
\]
so that we find back \eqref{dddlsqdd44}.
For example  the term  ${\mathfrak{S} }^{\circ}_{\circ}$  is given by $\displaystyle {\mathfrak{S} }^{\circ}_{\circ} =       
 (  - 1/4 \dttF_{\mu\nu} \dttF^{\mu\nu}  ) - {\big{(}}
\dttF^{\lambda\circ}    \frac{\partial A_{\lambda} }{\partial x^{\circ}}  
{\big{)}} $. We describe, see \cite{HK-01} the Hamiltonian counterpart of the stress-energy tensor described as the Hamiltonian tensor:  
\bee\label{hamtensor01}
{{\mathfrak{H}}} (q,p) = \sum_{\alpha,\beta} {{\mathfrak{H}}}^{\alpha}_{\beta} (q,p)  \partial_{\alpha} \otimes \dd x^{\beta}  
\quad \hbox{with} \quad
{{\mathfrak{H}}}^{\alpha}_{\beta} (q,p)  = \frac{\partial L}{\partial v_{\mu\alpha} } (q,{\cal V} (q,p)) {\cal V}_{\mu\beta} - \delta^{\alpha}_{\beta} L (q,{\cal V}(q,p))
\eee
Notice that if $ ( x,A(x) , \dd A(x) ,  \hbox{\sffamily\bfseries\slshape h} )  {\pmb{\leftrightarrow}}  (q,p) $ then ${{\mathfrak{H}}}^{\alpha}_{\beta} (q,p) = - {{\mathfrak{S}}}^{\alpha}_{\beta} (x) $, so that: 
\[
\left.
\begin{array}{rcl}
\displaystyle    {{\mathfrak{H}}}^{\mu}_{\nu} (q,p)     & = &  \displaystyle    \delta^{\mu}_{\nu} {\cal H}(q,p)
-
 \Big{\langle}  p
 ,     {\cal Z}_{1} (q,p) \wedge \cdots \wedge
  {\cal Z}_{\mu-1} (q,p) \wedge
\frac{\partial}{\partial x^{\nu} }
\wedge  {\cal Z}_{\mu+1} (q,p) \wedge \cdots \wedge
 {\cal Z}_{n} (q,p)    \Big{\rangle} 
 \\
 \displaystyle     & = &  \displaystyle \delta^{\mu}_{\nu} {\cal H}(q,p) - \frac{\partial \langle p,z \rangle}{\partial z^{\nu}_{\mu}} \big |_{z={\cal Z}(q,p)}.  \\
\end{array}
\right.
\]
First, we are interested in the component $ {{\mathfrak{H}}}^{\circ}_{\circ} (x,A)   $: 
\bee\label{gloip0090}
 {{\mathfrak{H}}}^{\circ}_{\circ} (x,A)   = {\cal H} (q,p) -   \langle p , \frac{\partial }{\partial x^{\circ}}  \wedge z_{1} \wedge \cdots \wedge z_{n-1} \rangle.
 \eee
 Let us evaluate:  ${\pmb{\theta}}^{\tiny{\hbox{\sffamily DDW}}}_p (\underset{\circ}{\cal Z} )  := \langle p , \frac{\partial }{\partial x^{\circ}}  \wedge z_{1} \wedge \cdots \wedge z_{n-1} \rangle =  \langle p ,  \underset{\circ}{\cal Z} \rangle $ with $ \underset{\circ}{\cal Z}  =  {\partial_{\circ} }   \wedge z_{1} \wedge \cdots \wedge z_{n-1} $
 \[
 \left.
\begin{array}{rcl}
\displaystyle       \langle p , \underset{\circ}{\cal Z}  \rangle & = &  \displaystyle  
 \mathfrak{e} \hbox{d} \mathfrak{y} (\underset{\circ}{\cal Z} )   + {\pmb{\pi}}^{{A}_{\mu}\nu}
\hbox{d}{A}_{\mu} \wedge \hbox{d} \mathfrak{y}_{\nu}  (\underset{\circ}{\cal Z})
=
\mathfrak{e} \hbox{d} \mathfrak{y}  (\underset{\circ}{\cal Z})     + {\pmb{\pi}}^{{A}_{\mu}\circ}
\hbox{d}{A}_{\mu} \wedge \hbox{d} \mathfrak{y}_{\circ} (\underset{\circ}{\cal Z})
 + {\pmb{\pi}}^{{A}_{\mu}1}
\hbox{d}{A}_{\mu} \wedge \hbox{d} \mathfrak{y}_{1}  (\underset{\circ}{\cal Z})
\\
 \displaystyle     &  &  \displaystyle
 + {\pmb{\pi}}^{{A}_{\mu}2}
\hbox{d}{A}_{\mu} \wedge \hbox{d} \mathfrak{y}_{2} (\underset{\circ}{\cal Z})
 + {\pmb{\pi}}^{{A}_{\mu}3}
\hbox{d}{A}_{\mu} \wedge \hbox{d} \mathfrak{y}_{3}  (\underset{\circ}{\cal Z})
 \\
 \displaystyle     & = &  \displaystyle   \mathfrak{e} \hbox{d} \mathfrak{y} (\partial_\circ \wedge \partial_2 \wedge \partial_3 \wedge \partial_4 )  
 + {\pmb{\pi}}^{{A}_{\mu}1}
\hbox{d}{A}_{\mu} \wedge \hbox{d} \mathfrak{y}_{1}  (  {\cal Z}_{1\mu}\partial_1 \wedge \partial_3 \wedge \partial_4 \wedge \frac{\partial}{\partial  A_{\mu}}  ) 
 \\
  \displaystyle    &  & \displaystyle     + {\pmb{\pi}}^{{A}_{\mu}2}
\hbox{d}{A}_{\mu} \wedge \hbox{d} \mathfrak{y}_{2}  (- {\cal Z}_{2\mu}\partial_1 \wedge \partial_2 \wedge \partial_4 \wedge \frac{\partial}{\partial A_{\mu} }  ) 
  \\
  \displaystyle    &  & \displaystyle  + {\pmb{\pi}}^{{A}_{\mu}3}
\hbox{d}{A}_{\mu} \wedge \hbox{d} \mathfrak{y}_{3}  (  {\cal Z}_{3\mu }  \partial_1 \wedge \partial_2 \wedge \partial_3 \wedge \frac{\partial}{\partial  A_{\mu} }  )   
\end{array}
\right.
 \]
 we finally find: $ \displaystyle
\begin{array}{lll}
 \displaystyle  \langle p , \underset{\circ}{\cal Z}   \rangle  &  = & \displaystyle  \mathfrak{e}   
 + {\pmb{\pi}}^{{A}_{\mu}1}
 {\cal Z}_{2\mu} 
 + {\pmb{\pi}}^{{A}_{\mu}2}
  {\cal Z}_{3\mu} 
 + {\pmb{\pi}}^{{A}_{\mu}3}
   {\cal Z}_{4\mu } 
  = \mathfrak{e}  +    \sum_{\mu = 0}^{n-1} \sum_{\alpha = 1}^{n-1}   {\pmb{\pi}}^{A_{\mu} \alpha} \partial_{\alpha} A_{\mu} 
\end{array}
$. Then, the expression $ {{\mathfrak{H}}}^{\circ}_{\circ} (x,A)$  given by \eqref{gloip0090} is written: 
\[
 {{\mathfrak{H}}}^{\circ}_{\circ} (x,A)   = {\cal H} (q,p) -  
 {\big{(}} \mathfrak{e}  + \sum_{\mu = 0}^{n-1} \sum_{\alpha = 1}^{n-1} {\pmb{\pi}}^{A_{\mu} \alpha} \partial_{\alpha} A_{\mu} 
 {\big{)}}.
  \]
  and we finally notice that if we integrate the corresponding $(n-1)$-forms on a slice of time defined by \eqref{minkowlso}: 
  \bee
  \int_{{\pmb{\Sigma}}_{\circ}}  {{\mathfrak{H}}}^{\circ}_{\circ}  \vol_{\circ} = 
    \int_{{\pmb{\Sigma}}_{\circ}}  {\pmb{\eta}}_{\circ}
  \eee
  Notice that we also recover the other component of the stress energy tensor \eqref{hamtensor01} via the study of the following terms: 
\[
\left.
\begin{array}{rcl}
 \displaystyle     {{\mathfrak{H}}}^{i}_{\circ} (x,A)     & = &  \displaystyle -   \langle p , z_{\circ} \wedge z_{1} \wedge \cdots 
  \wedge z_{i-1}
  \wedge  \partial_{\circ }  \wedge z_{i+1}  \wedge \cdots \wedge z_{n-1} \rangle \\
\displaystyle    {{\mathfrak{H}}}^{\circ}_{i} (x,A)      & = &  \displaystyle  -   \langle p ,  \partial_{i}  \wedge z_{1} \wedge \cdots 
  \wedge z_{n-1} \rangle \\
  \displaystyle    {{\mathfrak{H}}}^{i}_{j} (x,A)      & = &  \displaystyle \delta^{i}_{j} {\cal H}(q,p)    -   \langle p , z_{\circ} \wedge z_{1} \wedge \cdots 
  \wedge z_{i-1}
  \wedge  \partial_{j }  \wedge z_{i+1}  \wedge \cdots \wedge z_{n-1} \rangle \end{array}.
\right.
 \]
 Notice that some related work on the Noether theorem for covariant field theory is found in \cite{Forger}  \cite{GOtaystress}.

 \section{\hbox{\sffamily\bfseries\slshape{Dynamical equations    and canonical variables}}}\label{section999}

\subsection{\hbox{\sffamily\bfseries\slshape{Graded structures and Grassman variables}}}\label{qmapd99}

 In this section, we  heuristically  illustrate the tension between the {\em graded structure} and the {\em copolarization} process. One of the main interests in field theory is the search for the good Poisson structure. The copolarization process and the modern classification  concerning the distinction between {\em algebraic observable forms} $({\hbox{\sffamily AOF}})$ and {\em observable forms} $({\hbox{\sffamily OF}})$   appear in the work of F. H\'elein and J. Kouneiher \cite{HK-02} \cite{HK-03}  emerged from the question raised by the setting  of graded structure and the related non-uniqueness construction for     $(p-1)$-forms of various degree.  First, we offer some   remarks on  the Graded structure, the Grassman variables and the notion of {\em superform} found in \cite{HK-01}. Then,   we give some aspects of the copolarization process in \eqref{gliopowww22}.

 ${\hbox{\em{Graded structures.}}}$    They appear in the traditional  ${\hbox{\sffamily{(DDW)}} }$ setting as the algebraic structures related to algebraic forms of arbitrary degree.    We emphasize two main group of references.
   The first is found in  the  work of  I.V. Kanatchikov  \cite{Kana-OZO}  \cite{Kana-01}  \cite{Kana-014} where    interesting ideas on the graded setting are   developed, in particular in connection with the dynamical evolution for forms of lower degrees. The second  concerns the closely related work of M. Forger, C. Paufler and H. R\"{o}mer 
\cite{Forger009} \cite{Forger010}     \cite{Forger011}.   The equation under consideration  is $\overset{r}{X}_{\pmb{\varphi}} \iN {\pmb{\Omega}} =  \dd \overset{n-r}{\pmb{\varphi}} $ so  we say that the  Hamiltonian multivector field  $\overset{r}{X}_{\pmb{\varphi}}$ is  associated with the Hamiltonian form $ \overset{n-r}{\pmb{\varphi}}$. Neither $\overset{r}{X}_{\pmb{\varphi}}$ nor $\overset{n-r}{\pmb{\varphi}}$ are uniquely defined. Equivalently, the kernel of ${\pmb{\Omega}}$ on multivector fields is non-trivial. This simple fact reflects  a non unique correspondence between Hamiltonian multivector fields and Hamiltonian forms \cite{Forger009} \cite{Forger010}.  We concentrate  on  the example of  Kanatchikov's bracket with main focus on graded antisymmetric bracket:\footnote{ We denote $r = n-p$ and $s=n-q$. }
  \bee
   {\big{\{}}  \overset{r}{ {\pmb{\varphi}}  }  ,  \overset{s}{ {\pmb{\varrho}}  }  {\big{\}}}  =  - (-1)^{ (n-r-1)(n-s-1) }
{\big{\{}}  \overset{s}{ {\pmb{\varrho}}  }  ,  \overset{r}{ {\pmb{\varphi}}  }   {\big{\}}},
\eee
 where\footnote{Notice that we term algebraic observable $(n-1)$-forms and denote the set of {\sffamily (AOF)} by ${{\textswab{P}}}^{n-1}_{\circ} ({\pmb{\cal M}}) $ respectively what is more traditionally called (in the related work of I.V. Kanatchikov, M. Forger {\em et al.}...) Hamiltonian $(n-1)$-forms  and usually denoted $ {\Omega}_{\tiny{\hbox{\sffamily ham}}}^{\tiny{\hbox{n-1}}} ({\pmb{\cal M}}) $ },
\bee
 \begin{array}{ccl}
\displaystyle  {{\textswab{P}}}^{r}_{\circ} ({\pmb{\cal M}})     \times   {{\textswab{P}}}^{s}_{\circ} ({\pmb{\cal M}})  &  \rightarrow  &    {{\textswab{P}}}^{r+s-n+1}_{\circ} ({\pmb{\cal M}})      
\\
\displaystyle  (\overset{r}{ {\pmb{\varphi}}} , \overset{s}{ {\pmb{\varrho}}})    & \mapsto   &  \displaystyle  {\big{\{}}  \overset{r}{ {\pmb{\varphi}}  }  ,  \overset{s}{ {\pmb{\varrho}}  }  {\big{\}}}  = 
(-1)^{n-r} \overset{p}{ X_{\pmb{\varphi}} } \iN  \overset{q}{ X_{\pmb{\varrho}} }  \iN {\pmb{\Omega}}  .
\end{array}
\eee
Notice that, as told before, we have   a fundamental ambiguity in the search for Poisson structure for  forms of arbitrary degree $p$ and $q$. To be more precise, the  ambiguity lays   in the choice of the {\em objects} themselves, namely {\em in} the choice of the Hamiltonian multivectors fields $ \overset{p}{ X_{\pmb{\varphi}} }$ and $\overset{q}{ X_{\pmb{\varrho}} } $. This ambiguity takes its origin in the pairing of forms and vector fields via the study of the equation: $\dd {\pmb{\varphi}} = - \Xi_{{\pmb{\varphi}}} \iN {\pmb{\Omega}}$, and beyond by the study of the map:
$ \displaystyle
\left|
\begin{array}{ccl}
\displaystyle 
 \Lambda^{p}  T_{m} {\pmb{\cal M}}  & \rightarrow  & \displaystyle  \Lambda^{n+1-p}T^{\star}_{m} {\pmb{\cal M}}  
  \\
 \displaystyle   \Xi  & \mapsto & \displaystyle   
   \Xi \iN {\pmb{\Omega}}
 \end{array}
\right.
$.

 ${\hbox{\em{Grassman variables.}}}$   They appear in the {\sffamily{BRST}} and  {{\sffamily BV}} formalisms with the concept of ghosts and anti-ghosts.\footnote{We find the connection to the {conceptual} setting of the  supersymmetric landscape \cite{deligne} - where additional virtual {matter} degree of freedom is related to the notion of ghost.   From the mathematical perspective, the graded scenario and  the Gerstenhaber algebra \cite{Gersten} are important geometric structure}   In the multisymplectic landscape, we   find  Grassman-odd variables in the work of  F. H\'elein and J. Kouneiher \cite{HK-01}  and of  S. Hrabak  \cite{Hrabbak01} \cite{Hrabbak02}.

The distinction found in    F. H\'elein and J. Kouneiher  \cite{HK-01},   between the  {\em internal}, the {\em external} and  the $^{\mathfrak{s}} \mathfrak{p}$-{\em bracket}, is related to   considerations for the good expression of the dynamics.  
We delimitate two directions in connection with this. The first is the relation between the dynamical equations   and the    {external} bracket $\{   {\cal H}   \vol  ,  {\pmb{\lambda}} \} $ \cite{HK-01}\footnote{ see also the dynamical evolution equations given by I.V. Kanatchikov \cite{Kana-OZO}  \cite{Kana-01}.}. The second is  the introduction of Grassman additional  variables in    \cite{HK-01} - which  makes   connection with the work of  S. Hrabak  \cite{Hrabbak01} \cite{Hrabbak02}.    We   write dynamical equations in the form:\footnote{where ${\bf{d}}$ is the differential along a graph $\Gamma$ of a solution of the Hamilton equations.} 
\begin{equation}\label{abebe02}
{\hbox{\bf d}} {A} = \{  {\cal H} \hbox{d}\mathfrak{y} , {A}  \}
\quad \quad  \quad    \hbox{and} \quad   \quad \quad 
{\hbox{\bf d}} {{\pmb{\pi}}} = \{  {\cal H} \hbox{d}\mathfrak{y} , {\pmb{\pi}} \} , 
\end{equation}
where we need  the definition of   a Poisson bracket between ${\cal H}  \vol \in \Gamma({\pmb{\cal M}} , \Lambda^{n} T^{\star}{\pmb{\cal M}})$ and $(p-1)$-forms, with $1 \leq p \leq n-1$. We adopt here the terminology developed in \cite{HK-01} where we find the following different brackets: the {external} $\mathfrak{p}$-brackets, the {internal} $\mathfrak{p}$-brackets and the $^{\mathfrak{s}} \mathfrak{p}$-bracket.

$(\mathfrak{i})$ ${\hbox{\em{Internal $\mathfrak{p}$-bracket}}}$. If ${\pmb{\lambda}}, {\pmb{\kappa}}  \in {{\textswab{P}}}_{\circ}^{n-1}({\pmb{\cal M}})$ with  ${{\textswab{P}}}_{\circ}^{n-1}({\pmb{\cal M}})$ the set of all algebraic observable $(n-1)$-forms, we   define the {\em internal} $\mathfrak{p}$-bracket on  ${{\textswab{P}}}_{\circ}^{n-1}({\pmb{\cal M}})$:
\begin{equation}
\{  {\pmb{\lambda}} ,  {\pmb{\kappa}} \} = \Xi ( {\pmb{\kappa}}) \iN \Xi ( {\pmb{\lambda}}) \iN {\pmb{\Omega}}.
\end{equation}
The internal bracket is basically defined on algebraic $(n-1)$-forms.

$(\mathfrak{ii})$ ${\hbox{\em{External $\mathfrak{p}$-bracket}}}$. Now we   extend the previous definition to the case where $\pmb{\varphi} \in \Gamma ({\pmb{\cal M}} , \Lambda^{p} T^{\star} {\pmb{\cal M}})$, with $1 \leq p \leq n $
and ${\pmb{\lambda}} \in {{\textswab{P}}}_{\circ}^{n-1} ({\pmb{\cal M}})$ we obtain the {\em external} $\mathfrak{p}$-bracket:  
\[
\left|
 \begin{array}{ccl}
\displaystyle   \Gamma ({\pmb{\cal M}} , \Lambda^{p} T^{\star} {\pmb{\cal M}})    \times  {{\textswab{P}}}_{\circ}^{n-1} ({\pmb{\cal M}})  &  \rightarrow  &      \Gamma ({\pmb{\cal M}} , \Lambda^{p} T^{\star} {\pmb{\cal M}}) 
\\
 \displaystyle  ( \varphi ,  {\pmb{\lambda}})    & \mapsto   &  \displaystyle  {\big{\{}}   {\pmb{\lambda}} ,    \varphi  {\big{\}}}  = 
- \Xi ( {\pmb{\lambda}}) \iN \dd \varphi  
\end{array}
\right.
  \]
  Notice that $\{   \varphi ,  {\pmb{\lambda}} \} =  - \{   {\pmb{\lambda}} ,    \varphi   \}  = \Xi ( {\pmb{\lambda}}) \iN \dd \varphi $. The interesting case for dynamical evolution is when, $\varphi = {\cal H}   \vol $. 
Then we notice that for any ${\pmb{\lambda}}   \in {{\textswab{P}}}_{\circ}^{n-1}  ({\pmb{\cal M}})$ we have the following relation $\{   {\cal H}   \vol  ,  {\pmb{\lambda}} \}  = - \Xi ( {\pmb{\lambda}}) \iN \dd {\cal H} \wedge   \vol$.

\

 $(\mathfrak{iii})$ ${\hbox{\em{$^{\mathfrak{s}} \mathfrak{p}$-bracket}}}$. The method developed in \cite{HK-01} for  the  construction of a bracket between $(p-1)$ forms for $p$ of arbitrary degree ($1 \leq p \leq n $) is now briefly described. We introduce anticommuting Grassman variables ${\pmb{\tau}}_1 \cdots {\pmb{\tau}}_n$ that behave under change of coordinates like $\partial_1 \cdots \partial_n $. In this case, an arbitrary  form $\pmb{\varphi} \in \Gamma ({\pmb{\cal M}} , \Lambda^{p} T^{\star} {\pmb{\cal M}})$ depends on the set of variables\footnote{For a more detailed view of the  Grassmannian variables ${\pmb{\tau}}_{\alpha} $  and their intrinsic geometrical meaning  see \cite{HK-01}.} $({\pmb{\tau}}_{\alpha}, x^{\alpha}, A_{\mu} , \mathfrak{e} , {\pmb{\pi}}^{A_\mu \alpha} )$.   Grassman variables  ${\pmb{\tau}}_{\alpha}$ are related to  the notion of {\em superform} in \cite{HK-01}. For any  ${\pmb{\lambda}} \in {{\textswab{P}}}^{p-1}_{\circ} ({\pmb{\cal M}})$ such that for all $1 \leq \alpha_1 \leq \cdots \leq \alpha_{n-p} \leq n $ we have:
$
\dd x^{\alpha_1} \wedge \cdots \wedge \dd x^{\alpha_{n-p}} \wedge {\pmb{\lambda}} \in {{\textswab{P}}}^{n-1}_{\circ}   ({\pmb{\cal M}})
$.
We define   the superform:
\bee
^{\mathfrak{s}} {\pmb{\lambda}} =  \sum_{\alpha_1 < \cdots < \alpha_{n-p}} {\pmb{\tau}}_{\alpha_1} \cdots {\pmb{\tau}}_{\alpha_{n-p}} \dd x^{\alpha_1} \wedge \cdots \wedge \dd  x^{\alpha_{n-p}} \wedge {\pmb{\lambda}}.
\eee
We define also a $^{\mathfrak{s}} \mathfrak{p}$-bracket for ${\pmb{\varphi}} \in \Gamma({\cal M}, \Lambda^{n} T^{\star} {\cal M})$,
\[
 {\big{\{}} {\pmb{\varphi}} , ^{\mathfrak{s}}{\pmb{\lambda}}  {\big{\}}}_{\mathfrak{s}}
 = - \Xi (^{\mathfrak{s}}{\pmb{\lambda}}) \iN \dd {\pmb{\varphi}}
   =  - \sum_{\alpha_1 < \cdots < \alpha_{n-p}} {\pmb{\tau}}_{\alpha_1} \cdots {\pmb{\tau}}_{\alpha_{n-p}} \Xi ( \dd  x^{\alpha_1} \wedge \cdots \wedge \dd  x^{\alpha_{n-p}} \wedge {\pmb{\lambda}} ) \iN \dd {\pmb{\varphi}}.
\] 
Let ${\pmb{\lambda}}$ be an admissible form  \cite{HK-01}, and let ${\pmb{\Gamma}}$ be a $n$-dimensional submanifold of ${\pmb{\cal M}}$ which is a graph over ${\cal X}$, then for any oriented ${\pmb{\Sigma}}^{p} \subset {\pmb{\Gamma}}$ with $\hbox{dim} ({\pmb{\Sigma}}^{p} ) = p$, we have:
\begin{equation}
\int_{{\pmb{\Sigma}}^{p} }  {\big{\{}} {\cal H}   \vol   ,  \  ^{\mathfrak{s}}{\pmb{\lambda}}  {\big{\}}}_{\mathfrak{s}} =
\int_{{\pmb{\Sigma}}^{p} }  {\big{\{}} {\cal H}   \vol   ,  {\pmb{\lambda}}  {\big{\}}} .
\end{equation}
However, we do not insist  on this   notions of  the $^{\mathfrak{s}} \mathfrak{p}$-bracket   since for  good treatment of the dynamics, we will choose for adequate bracket a slightly different object.\footnote{The construction   based on  {\em copolarization} of the multisymplectic manifold allows us to define observable forms of any degree collectively. Then in the next section we find good bracket described  by F. H\'elein and J. Kouneiher  without this {\em superform artifact}.} The philosophy which underlines the $^{\mathfrak{s}} \mathfrak{p}$-bracket   is strongly connected to the one found in the work of   S. Hrabak  \cite{Hrabbak01} \cite{Hrabbak02} (and concerns  the multisymplectic  formulation of the classical {\sffamily{BRST}} symmetry for first order field theories.\footnote{Here lay the connection with the conceptual setting for the  {\em ghosts} and the {\em anti-ghosts}   in the  {\sffamily{BRST}}  formalism developed by C. Becchi, A. Rouet, R. Stora, I.V. Tyutin  \cite{Becchi} \cite{Tyutin} and the related  {\sffamily{BV}}  setting of I.A. Batalin and   G.A. Vilkovisky \cite{Batalin}.}) 

\subsection{\hbox{\sffamily\bfseries\slshape{Dynamical equations}}}\label{duysid}

 In this subsection we recover the dynamical equations via two methods. First, we illustrate the dynamical equation with    the     superforms' tool and the $^{\mathfrak{s}} \mathfrak{p}$-bracket. Then, we use  the external brackets ${\big{\{}} {\cal H} \vol ,  \dttP_{\phi}  {\big{\}}} $ and ${\big{\{}} {\cal H} \vol  , \dttQ^{\psi}    {\big{\}}}$.

 ${\hbox{\em{Superforms, Grassman variables and dynamical equations}}}$.   In the context of Maxwell theory - we refer to \cite{HK-01} for detailed calculation - the $1$-form $A$ and  the Faraday $(n-2)$-form  ${\pmb{\pi}}$,  lead - via the use of the {\em superform} $^{\mathfrak{s}}A$ and $^\mathfrak{s} {\pmb{\pi}}$ - to the following dynamical equations: 
  \[
 (\mathfrak{i}) \quad {\pmb{\dd}} A =   {\big{\{}} {\cal H}   \vol , A {\big{\}}}   \quad \quad   \quad      \quad \quad \quad \quad    \quad \quad \quad \quad   (\mathfrak{ii}) \quad {\pmb{\dd}} {\pmb{\pi}} =  {\big{\{}} {\cal H}   \vol , {\pmb{\pi}} {\big{\}}}  
\]
\[
 (\mathfrak{i}) \quad {\pmb{\dd}} A  = \sum_{\alpha < \beta} \eta_{\alpha\mu} \eta_{\beta\nu} {\pmb{\pi}}^{A_\mu \nu}  \dd x^{\alpha} \wedge \dd x^{\beta} \quad \quad \quad \quad  (\mathfrak{ii}) \quad {\pmb{\dd}} \pi  = \dttJ^{\alpha} \vol_\alpha
 \]
   Canonical bracket is described via  the    $^{\mathfrak{s}} \mathfrak{p}$-bracket $ {\big{\{}} {^{\mathfrak{s}}{\pmb{\pi}}} , ^{\mathfrak{s}}A  {\big{\}}}_{\mathfrak{s}}$.   
 
 ${\hbox{\em{External bracket and dynamical equations}}}$.   We also  recover the dynamical equations using the following external brackets ${\big{\{}} {\cal H} \vol ,  \dttP_{\phi}  {\big{\}}} $ and ${\big{\{}} {\cal H} \vol  , \dttQ^{\psi}    {\big{\}}}$.  Let us compute the bracket ${\big{\{}} {\cal H} \vol ,  \dttP_{\phi}  {\big{\}}}$. By definition, 
\[
\left.
\begin{array}{rcl}
\displaystyle {\big{\{}} {\cal H} \vol ,  \dttP_{\phi}  {\big{\}}}   & = &  \displaystyle 
 - \Xi (\dttP_{\phi} ) \iN \dd  {\cal H} \wedge \vol   
=  
  - \bl   \phi_{\mu} (x )  \frac{\partial}{\partial {A}_{\mu}}      -  {\big{(}}  \frac{\partial  \phi_{\mu}   }{\partial x^{\nu}} (x)  {\pmb{\pi}}^{{A}_\mu \nu}         {\big{)}}     \frac{\partial}{\partial \mathfrak{e}} \br \iN \dd  {\cal H} \wedge \vol  
\\
 \displaystyle     & = &  \displaystyle   \bl -      \phi_{\mu} (x )  \frac{\partial {\cal H}}{\partial {  A} _{\mu}} +     \frac{\partial  \phi_{\mu} (x )  }{\partial x^{\nu}}  {\pmb{\pi}}^{{  A}_\mu \nu}  \br   \vol .
 \end{array}
 \right.
\]
Along the graph of a solution, we have
$ \displaystyle
{\big{\{}} {\cal H} \vol ,  \dttP_{\phi}  {\big{\}}}  \big |_{\pmb{\Gamma}}  = \bl -      \phi_{\mu}  \dttJ^{\mu} + \partial_\nu  \phi_{\mu}    \dttF^{\mu\nu} \br   \vol.
$
On the other side, since
$ \displaystyle {\pmb{\dd}} ( \dttP_{\phi}  ) =  \bl  {\pmb{\pi}}^{A_\mu \nu}
\frac{\partial \phi_{\mu} }{\partial x^{\nu}}   \vol
+ \phi_{\mu} (x) \dd  {\pmb{\pi}}^{A_\mu \nu}
\wedge  \vol_\nu  \br, $ we obtain: 
\bee
{\pmb{\dd}} ( \dttP_{\phi}  )  \big |_{\pmb{\Gamma}}  =   \dttF^{\mu\nu}
\frac{\partial \phi_{\mu} }{\partial x^{\nu}}   \vol
+ \phi_{\mu} (x) \frac{\partial \dttF^{\mu\nu}}{ \dd x^{\nu}}  (x) \vol = \bl  \dttF^{\mu\nu}
 \partial_\nu \phi_\mu 
+ \phi_{\mu}   \partial_\nu \dttF^{\mu\nu} \br  \vol.
 \eee
 So that we finally observe the first set of dynamical evolution equations, along a graph of generalized Hamilton equations: 
  \bee\label{dlqmsdlqlm00}
 {\pmb{\dd}} ( \dttP_{\phi}  )  \big |_{\pmb{\Gamma}}  = {\big{\{}} {\cal H} \vol ,  \dttP_{\phi}  {\big{\}}}  \big |_{\pmb{\Gamma}} 
 \quad   \Longleftrightarrow   \quad   \partial_\nu \dttF^{\mu\nu}  = \dttJ^{\mu}.
 \eee
Now we are intersted in the second bracket: 
\[
\left.
\begin{array}{rcl}
\displaystyle  {\big{\{}} {\cal H} \vol  , \dttQ^{\psi}    {\big{\}}}      & = &  \displaystyle 
  - \Xi ( {\dttQ}^{\psi} ) \iN \dd  {\cal H} \wedge \vol  
=
        - \bl {\big{(}} A_\mu \frac{\partial \psi^{\mu\nu}}{\partial x^{\nu}}  (x)   {\big{)}}     \frac{\partial}{\partial \mathfrak{e}} 
+ \psi^{\mu\nu} (x) \frac{\partial}{\partial {\pmb{\pi}}^{A_\mu \nu} }   \br \iN \dd  {\cal H} \wedge \vol  
\\
 \displaystyle    & = &  \displaystyle 
  \bl {\big{(}} A_\mu \frac{\partial \psi^{\mu\nu}}{\partial x^{\nu}}     {\big{)}}     
+ \psi^{\mu\nu} (x) \frac{\partial {\cal H}}{\partial {\pmb{\pi}}^{A_\mu \nu} }   \br  \vol.
\end{array}
\right.
\] 
So that along a graph $\pmb{\Gamma}$ of a solution of the Hamilton equations we find the following relation:
$ \displaystyle
{\big{\{}} {\cal H} \vol  , \dttQ^{\psi}    {\big{\}}}  \big |_{\pmb{\Gamma}}  =     \bl {\big{(}} A_\mu \frac{\partial \psi^{\mu\nu}}{\partial x^{\nu}}     {\big{)}}     
+ \psi^{\mu\nu} (x) \frac{\partial {\cal H}}{\partial {\pmb{\pi}}^{A_\mu \nu} }   \br  \vol
$.
Whereas the expression of ${\pmb{\dd}} ( \dttQ^{\psi}  ) $ is written:
\bee
\left.
\begin{array}{rcl}
\displaystyle {\pmb{\dd}} ( \dttQ^{\psi}  )   & = &  \displaystyle   
\bl
 A_{\mu} \frac{\partial \psi^{\mu\nu}}{\partial x^{\nu}}     (x) \br  \vol 
 +
\psi^{\mu\nu} (x) \dd  A_{\mu} \wedge     \vol_{\nu} 
=
\bl
 A_{\mu} \frac{\partial \psi^{\mu\nu}}{\partial x^{\nu}}     (x)  
 +
\psi^{\mu\nu} (x) \partial_\nu  A_{\mu} \br  \vol 
\end{array}
\right.
\eee
So that we finally observe: 
 \bee\label{dlqmsdlqlm01}
 {\pmb{\dd}} (  \dttQ^{\psi}   )  \big |_{\pmb{\Gamma}}  = {\big{\{}} {\cal H} \vol ,   \dttQ^{\psi}    {\big{\}}}  \big |_{\pmb{\Gamma}} 
 \quad \Longleftrightarrow \quad  \dttF_{\mu\nu}  = \partial_{\mu} A_{\nu} - \partial_{\nu} A_{\mu}
 \eee
 The dynamical evolution is encapsulated by the relations \eqref{dlqmsdlqlm00} and \eqref{dlqmsdlqlm01}:
\bee
\left|
\begin{array}{lll}
\displaystyle     {\pmb{\dd}} (  \dttQ^{\psi}   )  \big |_{\pmb{\Gamma}}   
&  =  & \displaystyle {\big{\{}} {\cal H} \vol ,   \dttQ^{\psi}    {\big{\}}}  \big |_{\pmb{\Gamma}}  
\\
 \displaystyle    {\pmb{\dd}} ( \dttP_{\phi}  )  \big |_{\pmb{\Gamma}}     &  =  & \displaystyle  {\big{\{}} {\cal H} \vol ,  \dttP_{\phi}  {\big{\}}}  \big |_{\pmb{\Gamma}}
\end{array}
\right.
\eee

 \subsection{\hbox{\sffamily\bfseries\slshape{Copolarization and observables $(p-1)$-forms}}}\label{qsqsqsq999}

${\hbox{\em{Copolarization}}}$.  The copolarization   corresponds to the  collective definition of  observable forms and emerges  from the blend of the Relativity Principle  and dynamics.     For details we refer  to F. H\'elein and J. Kouneiher   \cite{HKHK01} \cite{HK-02}  \cite{HK-03}. Here we set out  some key features of  the notion of {\em copolarization}. In particular we give the general definition:  
\begin{defin}\label{jddzo0045}
Let $( {\pmb{\cal M}} , {\pmb{\Omega}} ) $ be a multisymplectic manifold. 
A copolarization on $( {\pmb{\cal M}} , {\pmb{\Omega}} ) $ is a smooth vector sub-bundle $\dttP^{\ast}_{\bullet}T^{\star} {\pmb{{\cal M}}} \subset \Lambda^{\ast} T^{\star} {\pmb{\cal M}}$ which satisfies: 

\

   $[\mathfrak{1}]$ $\dttP^{\ast}_{\bullet}  T^{\star}{\pmb{{\cal M}}}  = \bigoplus_{1 \leq i \leq n} \dttP^{i}_{\bullet} T^{\star} {\pmb{\cal M}}$

 $[\mathfrak{2}]$ Locally, for any $m \in {\pmb{{\cal M}}} $, ${\big{(}} \dttP^{\ast}_{\bullet} T^{\star}_{m} {\pmb{{\cal M}}}  , + , \wedge {\big{)}}$ is a subalgebra of ${\big{(}} \Lambda^{n} T^{\star}_{m} {\pmb{{\cal M}}}  , + , \wedge {\big{)}}$
 
 $[\mathfrak{3}]$ $\forall m \in {\pmb{\cal M}}, \forall \phi \in \Lambda^{n} T^{\star}_{m} {\cal M}, \phi \in \dttP^{n}_{\bullet} T^{\star}_{m} {\pmb{\cal M}} $ if and only if $\forall X, \tilde{X} \in {\cal O}_{m}, \ X \iN {\pmb{\Omega}} = \tilde{X}  \iN {\pmb{\Omega}} \Rightarrow \phi(X) = \phi(\tilde{X})$.
\end{defin}
  We say that a multisymplectic manifold $({\pmb{\cal M}} , {\pmb{\Omega}})$ is equipped with the copolarization $\dttP^{\ast}_{\bullet} T^{\star} {\pmb{\cal M}}$. The notion of copolarization intrinsically defines for any $1 \leq p \leq n$ the set ${{\textswab{P}}}^{p-1}_{\bullet} (\pmb{\cal M})$, namely the set of observable $(p-1)$-forms ${\pmb{\varphi}}$ by $\forall m \in {\pmb{\cal M}} , \dd {\pmb{\varphi}}_{m} \in \dttP^{p}_{\bullet} T^{\star}_{m} {\pmb{\cal M}}$. We refer to \cite{HK-01} \cite{HKHK01} \cite{HK-02}  \cite{HK-03} for the construction of the {\em standard copolarization}. The copolarization     is the natural geometrical setting to describe the canonical forms for field theory based on canonical variables such as a potential $1$-form.

 In the case of Maxwell theory, the two canonical forms are the potential $1$-form $A = A_{\mu} dx^{\mu} $ and  the so-called Faraday $2$-form (in the $4$ dimensional case):
 \bee
  {\pmb{\pi}} =  {1}/{2} {\pmb{\pi}}^{A_\mu \nu}   \vol_{\mu\nu} =  {1}/{2}  \sum_{\mu , \nu} {\pmb{\pi}}^{A_\mu \nu}  {\partial_{\mu}} \iN  {\partial_{\nu}} \iN   \vol .
  \eee
In a more general perspective - for gravity\footnote{see the companion paper on $n$-plectic Vielbein Gravity \cite{VVV}. } or non-abelian Yang-Mills theories   -  canonical forms are described by a couple   $(\omega,{\pmb{\varpi}})$. The general setting allows us to construct a well-defined Poisson bracket between observable functionals related to the canonical forms  $(\omega,{\pmb{\varpi}})$, see \cite{HK-02} \cite{HK-03} \cite{Vey01}:
\begin{equation}\label{abffebe16}
{\Big{\{}} \int_{\Sigma \cap {\pmb{\gamma}}_{\varkappa} } {\pmb{\varpi}}  , \int_{\Sigma \cap {\pmb{\gamma}}_{\varsigma} } \omega 
{\Big{\}}} (\Gamma) = \sum_{ m \in \Sigma \cap {\pmb{\gamma}}_{\varkappa}  \cap {\pmb{\gamma}}_{\varsigma}  \cap \Gamma } 
\mathfrak{S} (m).
\end{equation}
We refer to \cite{HK-02} \cite{HK-03} \cite{Vey01} for more details later  about the bracket~\eqref{abffebe16} and the related geometrical objects $ \Sigma \cap {\pmb{\gamma}}_{\varkappa}  $, $ \Sigma \cap {\pmb{\gamma}}_{\varsigma} $  and $ \Sigma \cap {\pmb{\gamma}}_{\varkappa}  \cap {\pmb{\gamma}}_{\varsigma}  \cap \Gamma$, as well as the counting object $\mathfrak{S} (m)$. Notice that the study of $(p-1)$-forms involves analogous definitions for  {\em slices} in this case. We have the following definition \eqref{jfzajoezjvzo0045}:
\begin{defin}\label{jfzajoezjvzo0045}
A slice of codimension $(n-p+1)$ is a submanifold ${\pmb{\Sigma}} \subset {\pmb{\cal M}}$  of codimension $(n-p+1)$ such that $T_m{\pmb{\cal M}} / T_{m}{\pmb{\Sigma}}$ is smoothly oriented with regard to $m$ and, such that for any ${\pmb{\Gamma}} \in {\cal E}^{\cal H}$, ${\pmb{\Sigma}}$ is transverse to ${\pmb{\Gamma}}$. 
\end{defin}
 We refer to \cite{HK-02} \cite{HK-03} for the question of the orientation of the intersection ${\pmb{\Sigma}} \cap {\pmb{\Gamma}}$. The straightforward analogue of definition \eqref{jfzajoezjvzo005} for the case of arbitrary $(p-1)$-forms is:
 \begin{defin}\label{jfzajoezjvzo005}
Let ${\pmb{\Sigma}}$ be a slice of codimension   $(n-p+1)$  and let ${\pmb{\varphi}}$ be an algebraic observable $(p-1)$-form.
An observable
functional $  \hbox{\sffamily\bfseries\slshape F}  = \int_{\pmb{\Sigma}} {\pmb{\varphi}}$ defined on the set of $n$-dimensional submanifolds ${\cal E}^{\cal H}$   given  by the map: 
\bee\label{qnvnaeva0145} 
 \hbox{\sffamily\bfseries\slshape F}_{\pmb{\varphi}} = \int_{\pmb{\Sigma}} {\pmb{\varphi}}: 
\left\{
\begin{array}{ccl}
\displaystyle     {\cal E}^{\cal H} &  \longrightarrow  &    \displaystyle  \Bbb{R}
 \\ 
 \displaystyle     {\pmb{\Gamma}}  & \mapsto  &      \displaystyle   \hbox{\sffamily\bfseries\slshape F} ({\pmb{\Gamma}}) = {\int_{\Sigma\cap \Gamma} {\pmb{\varphi}}}
\end{array}
\right.
\eee  
\end{defin}
 The notion of  {\em copolarization} definitively emerges from the philosophy of ({\sffamily GR}). This highlights the    fact that we can not evaluate $\dd {\pmb{\varphi}}$ along a Hamiltonian $n$-vector $X$.   If $1 \leq p < n$ then an arbitrary $(p-1)$-form is necessarily  of maximum degree $(n-2)$. Indeed, the interesting question resides on the interpretation of the object  $\dd {\pmb{\varphi}} {\big{|}}_{X}$. This lack is supply  precisely  through   the notion of copolarization.  We construct  a set of $n$ $0$-forms ${\{} \pmb{\rho}_{i} {\}}_{1 \leq i \leq n}$. These $n$ $0$-forms are found in the copolarization of the multisymplectic manifold  $({\pmb{\cal M}},{\pmb{\Omega}})$. These are  observables $0$-forms: $\forall  1 \leq i \leq n, \pmb{\rho}_{i} \in {{{{\textswab{P}}}}}^{0}_{\bullet} ({\pmb{\cal M}})$. Locally we write for $m \in {\pmb{\cal M}}$ 
$\forall  1 \leq i \leq n, \dd \pmb{\rho}_{i} \in \dttP^{1}_{\bullet m} T^{\star}_{m}({\pmb{\cal M}})$. Hence, we reach the full dynamical duality: the evaluation of $ \displaystyle\bigwedge_{1 \leq i \leq n} \dd {\pmb{\rho}}_{i}$  on a Hamiltonian vector field $X$. The fact that $ \displaystyle \bigwedge_{1 \leq i \leq n} \dd {\pmb{\rho}}_{i} (X)$   only depends on $\dd{\cal H}_{m}$   means that $ \displaystyle \bigwedge_{1 \leq i \leq n} \dd {\pmb{\rho}}_{i} = \dd \pmb{\rho}_1 \wedge \cdots \wedge \pmb{\rho}_{n} (X)$ is a copolar form.

In the philosophy of ({\sffamily GR}) this is fully acceptable since we never measure an observable {\em per se} but  we  only compare observable quantities  between each others.  Following \cite{HK-01} \cite{HKHK01} \cite{HK-02}  \cite{HK-03}, the   $(p-1)$-bracket is related to  an equivalence class\footnote{If $X \sim \tilde{X}$, we have for any $1\leq p \leq n$ and $\phi \in \dttP^{p}_{\bullet}T^{\star}_{m}{\pmb{\cal M}}$, $X \Ni \phi \sim \tilde{X} \Ni \phi$ so that we define the equivalence class $[X] \Ni \phi = [X \Ni \phi] \in \dttP^{n-p}_{\bullet}T{\pmb{\cal M}}$, see  \cite{HK-02} \cite{HK-03} for details.} of (decomposable) Hamiltonian vector fields  $[X]^{\cal H}$.  We also have the notion of {\em algebraic copolarization}. This involves the same set of rules but with the replacement of $\dttP^{n}_{\bullet} T^{\star}_{m} {\pmb{\cal M}}$ by $\dttP^{n}_{\circ} T^{\star}_{m} {\pmb{\cal M}} \subset \dttP^{n}_{\bullet} T^{\star}_{m} {\pmb{\cal M}}$ and where the point  $[\mathfrak{3}]$ in \eqref{jddzo0045} is replaced by  $[\mathfrak{3}]_{\circ}$:

 $[\mathfrak{3}]_{\circ}$ $\forall m \in {\pmb{\cal M}}, \forall \phi \in \Lambda^{n} T^{\star}_{m} {\cal M}, \phi \in \dttP^{n}_{\circ} T^{\star}_{m} {\pmb{\cal M}} $ if and only if $ \forall X, \tilde{X} \in \Lambda^{n} T_{m} {\pmb{\cal M}}, \ X \iN {\pmb{\Omega}} = \tilde{X}  \iN {\pmb{\Omega}} \Rightarrow \phi(X) = \phi(\tilde{X})$.

\subsection{\hbox{\sffamily\bfseries\slshape{Copolarization and canonical variables}}}\label{gliopowww22}

We recall the result obtained by F. H\'elein and J. Kouneiher  \cite{HK-01}, they give a possible copolarization of $({\pmb{\cal M}}_{\tiny{\hbox{\sffamily Maxwell}}} , {\pmb{\Omega}}^{\tiny{\hbox{\sffamily DDW}}}  )$ for the Maxwell theory - with ${\pmb{\Omega}}^{\tiny{\hbox{\sffamily DDW}}}   = \dd \mathfrak{e} \wedge \vol + \dd {\pmb{\pi}} \wedge \dd A  $:
 \bee\label{bliopopopo}
\left.
\begin{array}{rcl}
 \displaystyle  \dttP^1 T^\star{\pmb{\cal M}}^{\tiny{\hbox{\sffamily Maxwell}}} & = &  \displaystyle  \bigoplus_{0\leq \mu\leq 3}   \dd x^{\mu} 
 \\
  \displaystyle  \dttP^2 T^\star{\pmb{\cal M}}^{\tiny{\hbox{\sffamily Maxwell}}} & = &  \displaystyle   \bigoplus_{0\leq \mu_1<\mu_2\leq 3}
\dd x^{\mu_1} \wedge \dd x^{\mu_2}  
\oplus   \hbox{d}{A}   
\\
   \displaystyle  \dttP^3 T^\star{\pmb{\cal M}}^{\tiny{\hbox{\sffamily Maxwell}}} & = &  \displaystyle  \bigoplus_{0\leq \mu_1<\mu_2<\mu_3\leq 3}
\dd x^{\mu_1}\wedge \dd x^{\mu_2}\wedge \dd x^{\mu_3}  \oplus \bigoplus_{0\leq \mu\leq 3}  \dd x^{\mu}\wedge \hbox{d}{A}
\oplus   \hbox{d}{{\pmb{\pi}}}  
  \\
    \displaystyle  \dttP^4 T^\star{\pmb{\cal M}}^{\tiny{\hbox{\sffamily Maxwell}}} & = &  \displaystyle \hbox{d} 
\mathfrak{y}  \oplus
\bigoplus_{0\leq \mu\leq 3}  {\partial \over \partial x^\mu}\iN {\pmb{\theta}}^{\hbox{\tiny{DDW}}}
\oplus \bigoplus_{0\leq \mu_1<\mu_2\leq 3}
\dd x^{\mu_1}\wedge \dd x^{\mu_2}\wedge \hbox{d}{A} 
\oplus \bigoplus_{0\leq \mu\leq 3}  \dd x^{\mu}\wedge \hbox{d}{{\pmb{\pi}}} .
\end{array}
\right.
\eee
  The notion of copolarization  describe the good data for forms of various degrees.   We can find several copolarizations for a given theory. The construction of the previous copolarization \eqref{bliopopopo}  allows us to construct a well-defined Poisson bracket between observable functionals related to the canonical forms  $(A,{\pmb{\pi}})$:
\begin{equation}\label{abffebe16}
{\Big{\{}} \int_{ {\pmb{\Sigma}} \cap  {\pmb{\gamma}}_{\varkappa} }  {\pmb{\pi} }     , \int_{{\pmb{\Sigma}} \cap {\pmb{\gamma}}_{\varsigma} } A 
{\Big{\}}} ({\pmb{\Gamma}}) = \sum_{ m \in {\pmb{\Sigma}} \cap {\pmb{\gamma}}_{\varkappa}  \cap {\pmb{\gamma}}_{\varsigma}  \cap {\pmb{\Gamma}} } 
\mathfrak{S} (m).
\end{equation}
 In \cite{HK-02} \cite{HK-03},  it is emphasized  that there is strong reasons to {\em not} include  
$ \displaystyle
\textswab{C}^{1}T^\star{\pmb{\cal M}}^{\tiny{\hbox{\sffamily Maxwell}}}   =       \bigoplus_{0\leq \mu\leq 3}   \dd A_{\mu} ,
$
 in $ { {\dttP}}^1 T^\star{\pmb{\cal M}}^{\tiny{\hbox{\sffamily Maxwell}}}  $. However we  consider this set  of $1$-forms $\textswab{C}^{1}T^\star{\pmb{\cal M}}^{\tiny{\hbox{\sffamily Maxwell}}}  $ and construct the set  
$ \displaystyle
  {\overline{\dttP}}^1 T^\star{\pmb{\cal M}}^{\tiny{\hbox{\sffamily Maxwell}}}   =  \dttP^1 T^\star{\pmb{\cal M}}^{\tiny{\hbox{\sffamily Maxwell}}}  
\oplus  \textswab{C}^{1}T^\star{\pmb{\cal M}}^{\tiny{\hbox{\sffamily Maxwell}}} 
$. Then, we consider the  smooth vector sub-bundle:
$ \displaystyle
{\overline\dttP}^{\ast}   T^{\star}{\pmb{{\cal M}}}^{\tiny{\hbox{\sffamily Maxwell}}}  = \bigoplus_{1 \leq i \leq 4} {\overline\dttP}^{i}  T^{\star} {\pmb{\cal M}}  \subset \Lambda^{\ast} T^{\star} {\pmb{\cal M}}^{\tiny{\hbox{\sffamily Maxwell}}},
$
as a  copolarization candidate  of 
$({\pmb{\cal M}}_{\tiny{\hbox{\sffamily Maxwell}}} , {\pmb{\Omega}}^{\tiny{\hbox{\sffamily DDW}}}  )$. Here, 
$ \displaystyle  \left.
\begin{array}{rcl}
 \displaystyle   {\overline{\dttP}}^i T^\star{\pmb{\cal M}}^{\tiny{\hbox{\sffamily Maxwell}}}  & = &  \displaystyle \dttP^i T^\star{\pmb{\cal M}}^{\tiny{\hbox{\sffamily Maxwell}}}  
\oplus  \textswab{C}^{i}T^\star{\pmb{\cal M}}^{\tiny{\hbox{\sffamily Maxwell}}}  
\end{array}
\right.
$ with   $  \textswab{C}^{i}T^\star{\pmb{\cal M}}^{\tiny{\hbox{\sffamily Maxwell}}} $   respectively written for $i=2,3,4$,
 \[
\left|
\begin{array}{rcl}
  \displaystyle    \textswab{C}^{2}T^\star{\pmb{\cal M}}^{\tiny{\hbox{\sffamily Maxwell}}}   & = &  \displaystyle
  \bigoplus_{0\leq \mu_1<\mu_2\leq 3}
\dd x^{\mu_1} \wedge \dd A_{\mu_2}  
\oplus \bigoplus_{0\leq \mu_1<\mu_2\leq 3}
\dd A_{\mu_1} \wedge \dd A_{\mu_2}   
\end{array}
\right.
 \]
  \[
\left|
\begin{array}{rcl}
\displaystyle    \textswab{C}^{3}T^\star{\pmb{\cal M}}^{\tiny{\hbox{\sffamily Maxwell}}}   & = &  \displaystyle  \bigoplus_{0\leq \mu_1<\mu_2<\mu_3\leq 3}
\dd x^{\mu_1}\wedge \dd x^{\mu_2}\wedge \dd A_{\mu_3}
\oplus
\bigoplus_{0\leq \mu_1<\mu_2<\mu_3\leq 3}
\dd x^{\mu_1}\wedge \dd A_{\mu_2}\wedge \dd A_{\mu_3} 
\\
\\
\displaystyle    &  &  \displaystyle
\oplus
\bigoplus_{0\leq \mu_1<\mu_2<\mu_3\leq 3}
\dd A_{\mu_1}\wedge \dd A_{\mu_2}\wedge \dd A_{\mu_3}
\end{array}
\right.
 \]
 and, 
  \[
\left|
\begin{array}{rcl}
\displaystyle 
   \textswab{C}^{4}T^\star{\pmb{\cal M}}^{\tiny{\hbox{\sffamily Maxwell}}}  & = & \displaystyle    \bigoplus_{0\leq \mu_1<\mu_2  < \mu_3<\mu_4 \leq 4}
\dd x^{\mu_1}\wedge \dd x^{\mu_2} \wedge \dd x^{\mu_3} \wedge \hbox{d}{A} _{\mu_4}  
\\
\\
\displaystyle       &  & \displaystyle 
 \oplus \bigoplus_{0\leq \mu_1<\mu_2  < \mu_3<\mu_4 \leq 4}
\dd x^{\mu_1}\wedge \dd x^{\mu_2} \wedge \hbox{d}{A} _{\mu_3}   \wedge \hbox{d}{A} _{\mu_4}  
\\
\\
\displaystyle       &  & \displaystyle 
\oplus
\bigoplus_{0\leq \mu_1<\mu_2<\mu_3\leq 3}
\dd x^{\mu_1} \wedge
\dd A_{\mu_1}\wedge \dd A_{\mu_2}\wedge \dd A_{\mu_3} 
  \oplus \bigoplus_{0\leq \mu\leq 4}  \dd A_{\mu}\wedge    \hbox{d}{{\pmb{\pi}}} 
\end{array}
\right.
 \]
 In fact, there are several obstructions for   
$ \displaystyle
  {\overline\dttP}^{\ast}   T^{\star}{\pmb{{\cal M}}}^{\tiny{\hbox{\sffamily Maxwell}}}      
$  to describe a  good copolarization. We focus, for a given $1 \leq \mu \leq n$, on the following form   ${\pmb{\rho}}_{\mu} = \dd A_{\mu} \wedge \dd {\pmb{\pi}}  $.  The exterior derivative $\dd {\pmb{\rho}}_{\mu}$ is written: 
\[
\dd {\pmb{\rho}}_{\mu} = \dd ( \dd A_{\mu} \wedge \dd {\pmb{\pi}} )  =  \dd^{2} A_{\mu} \wedge \dd {\pmb{\pi}} - \dd A_{\mu} \wedge \dd {\pmb{\pi}} = - \dd A_{\mu} \wedge \dd {\big{(}}  \frac{1}{2}  {\pmb{\pi}}^{A_\rho \nu}       \vol_{\rho\nu}      {\big{)}}
= -  \frac{1}{2}   \dd A_{\mu} \wedge \dd {\pmb{\pi}}^{A_\rho \nu}  \wedge     \vol_{\rho\nu}  .
\]
The form $\pmb{\rho}_{\mu}$  is not an $\hbox{\sffamily{(OF)}}$ $(n-1)$-form: $  \pmb{\rho}_{\mu}  \notin {{\textswab{P}}}^{n-1}_{\bullet} ({\pmb{\cal M}}^{\hbox{\tiny{\sffamily Maxwell}}}) 
$ so that $\rho_{\mu}$ is not an algebraic observable $(n-1)$-form $\hbox{\sffamily{(AOF)}}$,   $  \pmb{\rho}_{\mu}  \notin {{\textswab{P}}}^{n-1}_{\circ} ({\pmb{\cal M}}^{\hbox{\tiny{\sffamily Maxwell}}}) 
$.  Let us consider two decomposable vector fields $  X , \overline{X} \in \Lambda^{n} T {\pmb{\cal M}}^{\hbox{\tiny{\sffamily Maxwell}}}$.  The $n$-vector $X\in {\bf D}^{n}_{m} {\pmb{\cal M}} \subset \Lambda^nT_{(q,p)}{\pmb{\cal M}}$ is written $X = X_1 \wedge ... \wedge X_n $ and $\forall \nu = 1 ... n $,   
\bee
 X_\alpha =  {\partial (q(x),p(x))\over \partial x^{\alpha} }     = \frac{\partial}{\partial x^\alpha}   + \Theta_{\alpha {\mu}}   \frac{\partial}{\partial A_{\mu}}  + {\Upsilon}_{\alpha} \frac{\partial }{ \partial \mathfrak{e}}
+ {\Upsilon}_{\alpha}^{A_\mu \nu} \frac{\partial }{ \partial {\pmb{\pi}}^{A_\mu \nu}},
\eee
where $\forall 1 \leq \alpha  \leq n$ and ${\Upsilon}_{\alpha}^{A_\mu \nu}  = - {\Upsilon}_{\alpha}^{A_\nu \mu} $.
From \eqref{mmmmmmmmmmqq}, we have the expression of $X \iN {\pmb{\Omega}}^{{\tiny\hbox{\sffamily DDW}}} $
and $ {\overline{X}} \iN  {\pmb{\Omega}}^{{\tiny\hbox{\sffamily DDW}}}  $: \[
\begin{array}{lll}
\displaystyle   X \iN  {\pmb{\Omega}}^{{\tiny\hbox{\sffamily DDW}}}    & = &  \displaystyle   
 \hbox{d} \mathfrak{e}  - \Upsilon_\rho \dd x^{\rho} +         \Theta_{\nu\mu}  \hbox{d} {\pmb{\pi}}^{{A}_{\mu}\nu}  -   {\Upsilon}_{\nu}^{A_\mu \nu} \dd{A}_{\mu} +{\big{(}}  {\Upsilon}_{\rho}^{A_\mu \nu}       \Theta_{\nu\mu}  - {\Upsilon}_{\nu}^{A_\mu \nu}  \Theta_{\rho\mu} {\big{)}} \dd x^\rho ,
  \\
 \displaystyle   {\overline{X}} \iN  {\pmb{\Omega}}^{{\tiny\hbox{\sffamily DDW}}}    & = &  \displaystyle   
 \hbox{d} \mathfrak{e}  - {\overline{\Upsilon}}_\rho \dd x^{\rho} +         {\overline{\Theta}}_{\nu\mu}  \hbox{d} {\pmb{\pi}}^{{A}_{\mu}\nu}  -   {\overline{{\Upsilon}}}_{\nu}^{A_\mu \nu} \dd{A}_{\mu} +{\big{(}}  {\overline{{\Upsilon}}}_{\rho}^{A_\mu \nu}       {\overline{\Theta}}_{\nu\mu}  - {\overline{{\Upsilon}}}_{\nu}^{A_\mu \nu}  {\overline{\Theta}}_{\rho\mu} {\big{)}} \dd x^\rho.
 \end{array}
 \]
 so that  $X \iN {\pmb{\Omega}}^{\tiny{\hbox{\sffamily DDW}}}
= \overline{X} \iN {\pmb{\Omega}}^{\tiny{\hbox{\sffamily DDW}}}$ gives us the following relations: 
\bee\label{smqp000}
\left|
\begin{array}{lll}
   \displaystyle   \Theta_{\nu\mu}  & =   &  \displaystyle   {\overline{\Theta}}_{\nu\mu} 
   \\ 
 \displaystyle    {{{\Upsilon}}}_{\nu}^{A_\mu \nu}  & =   &  \displaystyle  {\overline{{\Upsilon}}}_{\nu}^{A_\mu \nu}    
\end{array}
\right.
\quad   \hbox{and}   \quad  - {{\Upsilon}}_{\rho}  +{\big{(}}  { {{\Upsilon}}}_{\rho}^{A_\mu \nu}       { {\Theta}}_{\nu\mu}  - { {{\Upsilon}}}_{\nu}^{A_\mu \nu}  { {\Theta}}_{\rho\mu} {\big{)}}
)    = - {\overline{\Upsilon}}_{\rho}   +{\big{(}}  {\overline{{\Upsilon}}}_{\rho}^{A_\mu \nu}       {\overline{\Theta}}_{\nu\mu}  - {\overline{{\Upsilon}}}_{\nu}^{A_\mu \nu}  {\overline{\Theta}}_{\rho\mu} {\big{)}}.
\eee
Given the relations \eqref{smqp000},  we want to know if we observe $\dd {\pmb{\rho}}_{\mu} (X) = \dd {\pmb{\rho}}_{\mu} (\overline{X}) $ $\forall X , \overline{X} \in \Lambda^{n} T {\pmb{\cal M}}^{\hbox{\tiny{\sffamily Maxwell}}}$. In the following we treat the example of copolarization for Maxwell theory. The four dimensional case is a straightforward application of the calculation exposed below. First, we focus on the {\sffamily 2D}-case without the imposition of the Dirac constraint set (we work on ${\pmb{\cal M}}^{\hbox{\tiny{\sffamily DDW}}} $). In that case with $X , \overline{X} \in  {\bf D}^{n}_{m} {\pmb{\cal M}}^{\hbox{\tiny{\sffamily DDW}}} \subset  \Lambda^{n} T {\pmb{\cal M}}^{\hbox{\tiny{\sffamily DDW}}} $, 
\bee
\left.
\begin{array}{rcl}
\displaystyle  X_1       & = &  \displaystyle   \frac{\partial}{\partial x^1}   + \Theta_{1 {\mu}}   \frac{\partial}{\partial A_{\mu}}  + {\Upsilon}_{1} \frac{\partial }{ \partial \mathfrak{e}}
+ {\Upsilon}_{1}^{A_\mu \nu}  \frac{\partial }{ \partial {\pmb{\pi}}^{A_\mu \nu}}   ,
\\
\displaystyle    X_2 & = &  \displaystyle  \frac{\partial}{\partial x^2}   + \Theta_{2 {\mu}}   \frac{\partial}{\partial A_{\mu}}  + {\Upsilon}_{2} \frac{\partial }{ \partial \mathfrak{e}}
+ {\Upsilon}_{2}^{A_\mu \nu}   \frac{\partial }{ \partial {\pmb{\pi}}^{A_\mu \nu}}   .
\end{array}
\right.
\eee
We denote $\displaystyle \frac{\partial}{\partial x^{\mu} } = \partial_{\mu}$, 
$\displaystyle  \frac{\partial}{\partial A_{\mu}}  = \partial^{A_{\mu}} $,
$\displaystyle  \frac{\partial}{\partial \mathfrak{e} }  = {\pmb{\partial}}_{\mathfrak{e} } $ and finally
$\displaystyle  \frac{\partial}{\partial {\pmb{\pi}}^{A_\mu \nu} }  = {\pmb{\partial}}_{A_{\mu} \nu} $.  Then, the relations \eqref{smqp000}, see  appendix \eqref{appendixB} ({\em First case}) for details, are written: 
\bee\label{gliopopo88}
 \left|
\begin{array}{lll}
   \displaystyle   \Theta_{11}  & =   &  \displaystyle   {\overline{\Theta}}_{11} 
   \\ 
      \displaystyle   \Theta_{12}  & =   &  \displaystyle   {\overline{\Theta}}_{12} 
    \end{array}
\right.
\quad \quad \quad  
  \left|
\begin{array}{lll}
        \displaystyle   \Theta_{21}  & =   &  \displaystyle   {\overline{\Theta}}_{21} 
   \\ 
      \displaystyle   \Theta_{22}  & =   &  \displaystyle   {\overline{\Theta}}_{22} 
  \end{array}
\right.
\quad \quad \quad 
  \left|
\begin{array}{lll}
   \displaystyle  {\Upsilon}_{1}^{A_1 1}  +  {\Upsilon}_{2}^{A_1 2}         & =   &  \displaystyle   {\overline{\Upsilon}}_{1}^{A_1 1}  +  {\overline{\Upsilon}}_{2}^{A_1 2}   
   \\ 
      \displaystyle        {\Upsilon}_{1}^{A_2 1}  +  {\Upsilon}_{2}^{A_2 2}   & =   &  \displaystyle        {\overline{\Upsilon}}_{1}^{A_2 1}  +  {\overline{\Upsilon}}_{2}^{A_2 2} 
\end{array}
\right.
\eee
and, 
\bee\label{gliopopo89}
  X \iN  {\pmb{\Omega}}^{{\tiny\hbox{\sffamily DDW}}} \big{|}_{\hbox{\footnotesize{\dd}} x^{1}} =   {\overline{X}} \iN  {\pmb{\Omega}}^{{\tiny\hbox{\sffamily DDW}}} \big{|}_{\hbox{\footnotesize{\dd}} x^{1}} \quad \quad \hbox{and} 
  \quad \quad 
    X \iN  {\pmb{\Omega}}^{{\tiny\hbox{\sffamily DDW}}} \big{|}_{\hbox{\footnotesize{\dd}} x^{2}} =   {\overline{X}} \iN  {\pmb{\Omega}}^{{\tiny\hbox{\sffamily DDW}}} \big{|}_{\hbox{\footnotesize{\dd}} x^{2}}
\eee
with,
\bee
\left.
\begin{array}{rcl}
\displaystyle    X \iN  {\pmb{\Omega}}^{{\tiny\hbox{\sffamily DDW}}} \big{|}_{\hbox{\footnotesize{\dd}} x^{1}}    
& = &  \displaystyle
  - \Upsilon_1  + {\big{(}}  { {{\Upsilon}}}_{1}^{A_1 2}       { {\Theta}}_{21}  + { {{\Upsilon}}}_{1}^{A_2 2}       { {\Theta}}_{22}    -  \Theta_{1 {1}}   {\Upsilon}_{2}^{A_1 2}  -  \Theta_{1 {2}}   {\Upsilon}_{2}^{A_2 2}      
      {\big{)}}   
        \\
\displaystyle    X \iN  {\pmb{\Omega}}^{{\tiny\hbox{\sffamily DDW}}} \big{|}_{\hbox{\footnotesize{\dd}} x^{2}}
& = &  \displaystyle
   - \Upsilon_2  +  {\big{(}}  { {{\Upsilon}}}_{2}^{A_1 1}       { {\Theta}}_{11} +  { {{\Upsilon}}}_{2}^{A_2 1}       { {\Theta}}_{12}   - { {{\Upsilon}}}_{1}^{A_1 1}  { {\Theta}}_{21}     - { {{\Upsilon}}}_{1}^{A_2 1}  { {\Theta}}_{22}  {\big{)}}   
\end{array}
\right.
\eee
\bee
\left.
\begin{array}{rcl}
\displaystyle   {\overline{X}} \iN  {\pmb{\Omega}}^{{\tiny\hbox{\sffamily DDW}}} \big{|}_{\hbox{\footnotesize{\dd}} x^{1}}  & = &  \displaystyle  - {\overline{\Upsilon}}_{1} + {\big{(}}  { {\overline{\Upsilon}}}_{1}^{A_1 2}       {\overline{\Theta}}_{21}  + { {\overline{\Upsilon}}}_{1}^{A_2 2}       {\overline{\Theta}}_{22}    -  {\overline{\Theta}}_{1 {1}}   {\overline{\Upsilon}}_{2}^{A_1 2}  -  {\overline{\Theta}}_{1 {2}}   {\overline{\Upsilon}}_{2}^{A_2 2}      
      {\big{)}}  \\
\displaystyle     {\overline{X}} \iN  {\pmb{\Omega}}^{{\tiny\hbox{\sffamily DDW}}} \big{|}_{\hbox{\footnotesize{\dd}} x^{2}}  & = &  \displaystyle    - {\overline{\Upsilon}}_2 + 
 {\big{(}}  { {\overline{\Upsilon}}}_{2}^{A_1 1}       { \overline{\Theta}}_{11} +  { {\overline{\Upsilon}}}_{2}^{A_2 1}       {\overline{\Theta}}_{12}   - { {\overline{\Upsilon}}}_{1}^{A_1 1}  {\overline{\Theta}}_{21}     - { {\overline{\Upsilon}}}_{1}^{A_2 1}  {\overline{\Theta}}_{22}  {\big{)}}
\end{array}
\right.
\eee
 We   write the previous conditions in  matrix notations, we denote 
  \bee
 {\pmb{\Theta}}   =  {\pmb{\Theta}}_{\nu \mu}  =  \left(
 \begin{array}{c@{}c}
{{\Theta}}_{11} \ \ & \ \     {{\Theta}}_{21}  \ \\
{{\Theta}}_{12} \ \ & \ \    {{\Theta}}_{22} \ \\
\end{array} \right),  \quad \hbox{and} \quad    {\pmb{\overline{\Theta}}}  =  {\pmb{\overline{\Theta}}}_{\nu \mu}  =  \left(
 \begin{array}{c@{}c}
{\overline{\Theta}}_{11} \ \ & \ \    {\overline{\Theta}}_{21}  \ \\
{\overline{\Theta}}_{12} \ \ & \ \      {\overline{\Theta}}_{22} \ \\
\end{array} \right) 
\eee
Let us denote for $1 \leq \mu \leq 2  $,
  \bee
    \pmb{\Upsilon}^{[A_\mu]}  = {{\pmb{\Upsilon}}}_{\rho}^{A_\mu \nu}  =  \left(
 \begin{array}{c@{}c}
 {{{\Upsilon}}}_{1}^{A_\mu 1} \ \ & \ \   {{{\Upsilon}}}_{2}^{A_\mu 1}  \ \\
 {{{\Upsilon}}}_{1}^{A_\mu 2}  \ \ & \ \   {{{\Upsilon}}}_{2}^{A_\mu 2} \ \\
\end{array} \right),  \quad \hbox{and} \quad   {\overline{\pmb{\Upsilon}}}^{[A_\mu]}  = {\overline{\pmb{\Upsilon}}}_{\rho}^{A_\mu \nu}  =  \left(
 \begin{array}{c@{}c}
 {{\overline{\Upsilon}}}_{1}^{A_\mu 1} \ \ & \ \    {{\overline{\Upsilon}}}_{2}^{A_\mu 1}  \ \\
 {{\overline{\Upsilon}}}_{1}^{A_\mu 1}  \ \ & \ \   {{\overline{\Upsilon}}}_{2}^{A_\mu 2} \ \\
\end{array} \right),
\eee
so that 
$ \displaystyle
{{{\Upsilon}}}_{\nu}^{A_\mu \nu}  =  {\overline{{\Upsilon}}}_{\nu}^{A_\mu \nu}    
$
is written in matrix notation for $1 \leq \mu \leq 2  $: 
\bee
 \displaystyle \hbox{tr} (    {\pmb{\Upsilon}}^{[A_\mu]}  ) =   \sum_{\nu} {{{\Upsilon}}}_{\nu}^{A_\mu \nu}     =    \sum_{\nu} {\overline{\Upsilon}}_{\nu}^{A_\mu \nu}   =   \hbox{tr}  (  {\overline{{\pmb{\Upsilon}}}}^{[A_\mu]}  ) .
 \eee
 Hence relations \eqref{gliopopo88} are written: 
 \bee
  {\pmb{\Theta}} =  \overline{\pmb{\Theta}},
  \quad \quad   \quad \quad \hbox{and} \quad    \quad \quad
  \quad
  \hbox{tr} (    {\pmb{\Upsilon}}^{[A_\mu]}  )  =   \hbox{tr}  (  {\overline{{\pmb{\Upsilon}}}}^{[A_\mu]}  ) .
 \eee
Finally we look at   the last   relation \eqref{gliopopo89}. Let us denote for $1 \leq \mu \leq 2  $
  \bee
    \pmb{\Upsilon}^{[\nu]}  = {{\pmb{\Upsilon}}}_{\rho}^{A_\mu [\nu]}  
    =  \left(
 \begin{array}{c@{}c}
 {{{\Upsilon}}}_{1}^{A_1 \nu} \ \ & \ \   {{{\Upsilon}}}_{2}^{A_1 \nu}  \ \\
 {{{\Upsilon}}}_{1}^{A_2 \nu}  \ \ & \ \   {{{\Upsilon}}}_{2}^{A_2 \nu} \ \\
\end{array} \right) , \quad \hbox{and}
 \quad   {\overline{\pmb{\Upsilon}}}^{[\nu]}  = {\overline{\pmb{\Upsilon}}}_{\rho}^{A_\mu [\nu]}  =  \left(
 \begin{array}{c@{}c}
 {{\overline{\Upsilon}}}_{1}^{A_1 \nu} \ \ & \ \    {{\overline{\Upsilon}}}_{2}^{A_1 \nu}  \ \\
 {{\overline{\Upsilon}}}_{1}^{A_2 \nu}  \ \ & \ \   {{\overline{\Upsilon}}}_{2}^{A_2 \nu} \ \\
\end{array} \right).
\eee
Since, 
\[
   \left(
 \begin{array}{c@{}c}
 {{{\Upsilon}}}_{1}^{A_1 1} \ \ & \ \   {{{\Upsilon}}}_{2}^{A_1 1}  \ \\
 {{{\Upsilon}}}_{1}^{A_2 1}  \ \ & \ \   {{{\Upsilon}}}_{2}^{A_2 1} \ \\
\end{array} \right)      \left(
 \begin{array}{c@{}c}
- \Theta_{21} \  & \  - \Theta_{22}  \ \\
 \Theta_{11} \  & \    \Theta_{12} \ \\
\end{array} \right)
    =
       \left(
 \begin{array}{c@{}c}
-  {{{\Upsilon}}}_{1}^{A_1 1} \Theta_{21} + {{{\Upsilon}}}_{2}^{A_1 1}  \Theta_{11} \ \ & \ \  -  {{{\Upsilon}}}_{1}^{A_1 1} \Theta_{22}  +  {{{\Upsilon}}}_{2}^{A_1 1} \Theta_{12}  \ \\
- {{{\Upsilon}}}_{1}^{A_2 1}    \Theta_{21} +  {{{\Upsilon}}}_{2}^{A_2 1} \Theta_{11}  \ \ & \ \    -{{{\Upsilon}}}_{1}^{A_2 1} \Theta_{22} +  {{{\Upsilon}}}_{2}^{A_2 1} \Theta_{12} \ \\
\end{array} \right) ,
\]
\[
   \left(
 \begin{array}{c@{}c}
 {{{\Upsilon}}}_{1}^{A_1 2} \ \ & \ \   {{{\Upsilon}}}_{2}^{A_1 2}  \ \\
 {{{\Upsilon}}}_{1}^{A_2 2}  \ \ & \ \   {{{\Upsilon}}}_{2}^{A_2 2} \ \\
\end{array} \right)      \left(
 \begin{array}{c@{}c}
- \Theta_{21}  \ & \   - \Theta_{22}  \ \\
 \Theta_{11}  \ & \    \Theta_{12} \ \\
\end{array} \right)
    =
       \left(
 \begin{array}{c@{}c}
-  {{{\Upsilon}}}_{1}^{A_1 2} \Theta_{21} + {{{\Upsilon}}}_{2}^{A_1 2}  \Theta_{11} \ \ & \ \  -  {{{\Upsilon}}}_{1}^{A_1 2} \Theta_{22}  +  {{{\Upsilon}}}_{2}^{A_1 2} \Theta_{12}  \ \\
- {{{\Upsilon}}}_{1}^{A_2 2}    \Theta_{21} +  {{{\Upsilon}}}_{2}^{A_2 2} \Theta_{11}  \ \ & \ \    -{{{\Upsilon}}}_{1}^{A_2 2} \Theta_{22} +  {{{\Upsilon}}}_{2}^{A_2 2} \Theta_{12} \ \\
\end{array} \right),
\]
  the relations \eqref{gliopopo89} are written, with $\displaystyle   {\pmb{\Upsilon}}_{[\rho]}^{\mathfrak{e}}    =     \left(
 \begin{array}{c@{}c}
0  \ &  \  0  \\
0  \ &  \ -  {\Upsilon}_{\rho}    \\
\end{array} \right)       $ and $\displaystyle {\pmb{\Sigma}} =   \left(
 \begin{array}{c@{}c}
- \Theta_{21}  \ &  \  - \Theta_{22}   \\
 \Theta_{11}  \ &  \   \Theta_{12}  \\
\end{array} \right)  $: 
\bee
\left.
\begin{array}{rcl}
\displaystyle     \hbox{tr} {\big{(}}   {\pmb{\Upsilon}}_{[1]}^{\mathfrak{e}}     + \pmb{\Upsilon}^{[1]}   {\pmb{\Sigma}}  {\big{)}}   & = &  \displaystyle 
  \hbox{tr} {\big{(}}  {\overline{{\pmb{\Upsilon}}}}_{[1]}^{\mathfrak{e}}     +  {\overline{\pmb{\Upsilon}}}^{[1]}     \overline{\pmb{\Sigma}}   {\big{)}}  ,
\\
\displaystyle     \hbox{tr} {\big{(}}  {\pmb{\Upsilon}}_{[2]}^{\mathfrak{e}}      + \pmb{\Upsilon}^{[2]}   {\pmb{\Sigma}}  {\big{)}}   & = &  \displaystyle 
  \hbox{tr} {\big{(}}  {\overline{{\pmb{\Upsilon}}}}_{[2]}^{\mathfrak{e}}    +  {\overline{\pmb{\Upsilon}}}^{[2]}     \overline{\pmb{\Sigma}}   {\big{)}}  .
\end{array}
\right.
\eee
Now, we evaluate $\dd {\pmb{\rho}}_{\mu} (X) =   -   {1}/{2}   \dd A_{\mu} \wedge \dd {\pmb{\pi}}^{A_\rho \nu}  \wedge     \vol_{\rho\nu}   (X) =    {1}/{2}     \dd {\pmb{\pi}}^{A_\rho \nu}  \wedge  \dd A_{\mu} \wedge     \vol_{\rho\nu}   (X)$.
We obtain via a straightforward calculation: 
\[
  \dd {\pmb{\rho}}_{\mu}  =    {1}/{2}     \dd {\pmb{\pi}}^{A_1 2}  \wedge  \dd A_{\mu} \wedge     \vol_{12} +    {1}/{2}     \dd {\pmb{\pi}}^{A_2 1}  \wedge  \dd A_{\mu} \wedge     \vol_{21}  -   {1}/{2}     \dd {\pmb{\pi}}^{A_1 2}  \wedge  \dd A_{\mu} +    {1}/{2}     \dd {\pmb{\pi}}^{A_2 1}  \wedge  \dd A_{\mu}
  \]
or equivalently, 
\bee\label{kiolhogofo}
\left.
\begin{array}{rcl}
\displaystyle     \dd {\pmb{\rho}}_{1}   & = &   \displaystyle -   {1}/{2}     \dd {\pmb{\pi}}^{A_1 2}  \wedge  \dd A_{1} +    {1}/{2}     \dd {\pmb{\pi}}^{A_2 1}  \wedge  \dd A_{1} ,  \\
\displaystyle     \dd {\pmb{\rho}}_{2}   & = &   \displaystyle -   {1}/{2}     \dd {\pmb{\pi}}^{A_1 2}  \wedge  \dd A_{2} +    {1}/{2}     \dd {\pmb{\pi}}^{A_2 1}  \wedge  \dd A_{2} .  \\
\end{array}
\right.
\eee
Then,
\bee
\left.
\begin{array}{rcl}
\displaystyle     \dd {\pmb{\rho}}_{1}   (X)& = &   \displaystyle  {\Upsilon}_{2}^{A_1 2} {\Theta}_{11} -   {\Upsilon}_{1}^{A_1 2} \Theta_{2 {1}}  
-
{\Upsilon}_{2}^{A_2 1} {\Theta}_{11}  + {\Upsilon}_{1}^{A_2 1} {\Theta}_{21} ,
 \\
\displaystyle     \dd {\pmb{\rho}}_{2} (X)  & = &   \displaystyle  
{\Upsilon}_{2}^{A_1 2}  \Theta_{12} - {\Upsilon}_{1}^{A_1 2}   \Theta_{22} 
-    {\Upsilon}_{2}^{A_2 1}   \Theta_{1 {2}}  +  {\Upsilon}_{1}^{A_2 1}   \Theta_{22}.
  \\
\end{array}
\right.
\eee 
We clearly   see that the information contained in the comparison of the contraction of two vector fields  $X , \overline{X} \in  {\bf D}^{n}_{m} {\pmb{\cal M}}^{\hbox{\tiny{\sffamily DDW}}} \subset  \Lambda^{n} T {\pmb{\cal M}}^{\hbox{\tiny{\sffamily DDW}}} $ with the multisymplectic form $  {\pmb{\Omega}}^{{\tiny\hbox{\sffamily DDW}}} $ - that is  $X \iN {\pmb{\Omega}}^{{\tiny\hbox{\sffamily DDW}}}  =  {\overline{X}} \iN  {\pmb{\Omega}}^{{\tiny\hbox{\sffamily DDW}}}  $ -  equivalent to: 
\bee
\left|
\begin{array}{rcl}
\displaystyle      {\pmb{\Theta}}    & = &  \displaystyle \overline{\pmb{\Theta}}  \\
\displaystyle      \hbox{tr} (    {\pmb{\Upsilon}}^{[A_\mu]}  )    & = &  \displaystyle \hbox{tr}  (  {\overline{{\pmb{\Upsilon}}}}^{[A_\mu]}  )   \\
\end{array}
\right.
\quad \quad \hbox{and} \quad \quad
\left|
\begin{array}{rcl}
\displaystyle     \hbox{tr} {\big{(}}   {\pmb{\Upsilon}}_{[1]}^{\mathfrak{e}}     + \pmb{\Upsilon}^{[1]}   {\pmb{\Sigma}}  {\big{)}}   & = &  \displaystyle 
  \hbox{tr} {\big{(}}  {\overline{{\pmb{\Upsilon}}}}_{[1]}^{\mathfrak{e}}     +  {\overline{\pmb{\Upsilon}}}^{[1]}     \overline{\pmb{\Sigma}}   {\big{)}}  
\\
\displaystyle     \hbox{tr} {\big{(}}  {\pmb{\Upsilon}}_{[2]}^{\mathfrak{e}}      + \pmb{\Upsilon}^{[2]}   {\pmb{\Sigma}}  {\big{)}}   & = &  \displaystyle 
  \hbox{tr} {\big{(}}  {\overline{{\pmb{\Upsilon}}}}_{[2]}^{\mathfrak{e}}    +  {\overline{\pmb{\Upsilon}}}^{[2]}     \overline{\pmb{\Sigma}}   {\big{)}}    ,
\end{array}
\right.
\eee
is not sufficient to conclude that $ \dd {\pmb{\rho}}_{1}   (X) =   \dd {\pmb{\rho}}_{1}   (\overline{X})$ or $  \dd {\pmb{\rho}}_{2}   (X) =   \dd {\pmb{\rho}}_{2}   (\overline{X})$. In addition, in the case of the Maxwell theory, we   consider the antisymmetry of multimomenta due to the Dirac primary constraint set. In that case we prefer to consider the following vector fields  $X , \overline{X} \in  {\bf D}^{n}_{m} {\pmb{\cal M}}^{\hbox{\tiny{\sffamily Maxwell}}} \subset  \Lambda^{n} T {\pmb{\cal M}}^{\hbox{\tiny{\sffamily Maxwell}}}$. We denote, 
\bee
\left.
\begin{array}{rcl}
\displaystyle  X_1       & = &  \displaystyle   \frac{\partial}{\partial x^1}   + \Theta_{1 {\mu}}   \frac{\partial}{\partial A_{\mu}}  + {\Upsilon}_{1} \frac{\partial }{ \partial \mathfrak{e}}
+ {\Upsilon}_{1}^{A_\mu \nu} {\big{(}} \frac{\partial }{ \partial {\pmb{\pi}}^{A_\mu \nu}}  - \frac{\partial }{ \partial {\pmb{\pi}}^{A_\nu \mu}}  {\big{)}} ,
\\
\displaystyle    X_2 & = &  \displaystyle  \frac{\partial}{\partial x^2}   + \Theta_{2 {\mu}}   \frac{\partial}{\partial A_{\mu}}  + {\Upsilon}_{2} \frac{\partial }{ \partial \mathfrak{e}}
+ {\Upsilon}_{2}^{A_\mu \nu} {\big{(}} \frac{\partial }{ \partial {\pmb{\pi}}^{A_\mu \nu}}  - \frac{\partial }{ \partial {\pmb{\pi}}^{A_\nu \mu}}  {\big{)}}.
\end{array}
\right.
\eee
In such a context, see appendix \eqref{appendixB} ({\em Second case}),  the relations $X \iN {\pmb{\Omega}}^{\tiny{\hbox{\sffamily DDW}}}
= \overline{X} \iN {\pmb{\Omega}}^{\tiny{\hbox{\sffamily DDW}}} $ are written: 
\bee\label{gliopopo8810}
\left|
\begin{array}{lll}
        \displaystyle   \Theta_{12}  & =   &  \displaystyle   {\overline{\Theta}}_{12} 
   \\ 
      \displaystyle   \Theta_{21}  & =   &  \displaystyle   {\overline{\Theta}}_{21} 
  \end{array}
\right.
\quad \quad \quad \quad \quad \quad 
\left|
\begin{array}{lll}
   \displaystyle    {\Upsilon}_{2}^{A_2 1}   -  {\Upsilon}_{2}^{A_1 2}            & =   &  \displaystyle      {\overline{\Upsilon}}_{2}^{A_2 1}   -  {\overline{\Upsilon}}_{2}^{A_1 2}    
   \\ 
      \displaystyle       {\Upsilon}_{1}^{A_1 2}  - {\Upsilon}_{1}^{A_2 1}     & =   &  \displaystyle           {\overline{\Upsilon}}_{1}^{A_1 2}  - {\overline{\Upsilon}}_{1}^{A_2 1}  
\end{array}
\right.
\eee
and 
 \bee\label{gliopopo8910} 
\left.
\begin{array}{rcl}
\displaystyle     {\big{(}} ( {\Upsilon}_{2}^{A_1 2}  \Theta_{1 {2}}  - {\Upsilon}_{2}^{A_2 1}  \Theta_{1 {2}}   ) - (  \Theta_{2 {2}}   {\Upsilon}_{1}^{A_1 2}  -  \Theta_{2 {2}}   {\Upsilon}_{1}^{A_2 1} ) {\big{)}}   & = &  \displaystyle   {\big{(}} ( {\overline{\Upsilon}}_{2}^{A_1 2}  {\overline{\Theta}}_{1 {2}}  - {\overline{\Upsilon}}_{2}^{A_2 1}  {\overline{\Theta}}_{1 {2}}   )  
 \\
 \displaystyle &   &  \displaystyle  - (  {\overline{\Theta}}_{2 {2}}   {\overline{\Upsilon}}_{1}^{A_1 2}  -  {\overline{\Theta}}_{2 {2}}   {{\overline{\Upsilon}}}_{1}^{A_2 1} ) {\big{)}} \\
\displaystyle   {\big{(}} ( {\Upsilon}_{1}^{A_1 2}  \Theta_{2 {1}}  - {\Upsilon}_{1}^{A_2 1}  \Theta_{2 {1}}   ) - (  \Theta_{1 {1}}   {\Upsilon}_{2}^{A_2 1}  -  \Theta_{1 {1}}   {\Upsilon}_{2}^{A_1 2} ) {\big{)}}  & = &  \displaystyle 
 {\big{(}} ( {{{\overline{\Upsilon}}}}_{1}^{A_1 2}  {\overline{\Theta}}_{2 {1}}  - {{{\overline{\Upsilon}}}}_{1}^{A_2 1}  {\overline{\Theta}}_{2 {1}}   ) 
 \\
 \displaystyle &   &  \displaystyle  - (  {\overline{\Theta}}_{1 {1}}   {{{\overline{\Upsilon}}}}_{2}^{A_2 1}  -  {\overline{\Theta}}_{1 {1}}   {{{\overline{\Upsilon}}}}_{2}^{A_1 2} ) {\big{)}}
\end{array}
\right.
\eee
 We set $\displaystyle   {\pmb{\Upsilon}}_{[1]}^{\mathfrak{e}}    =     \left(
 \begin{array}{c@{}c}
0  \ &  \  0  \\
0  \ &  \ -  {\Upsilon}_{1}    \\
\end{array} \right)  =  {\pmb{\Upsilon}}_{[2]}^{\mathfrak{e}}    =     \left(
 \begin{array}{c@{}c}
0  \ &  \  0  \\
0  \ &  \ -  {\Upsilon}_{2}    \\
\end{array} \right)  = 0$ since it do not entails the general result. We also introduce the following matrix notation: $\displaystyle     {\pmb{\Theta}}_{\nu \mu}^{\hbox{\tiny{\sffamily Maxwell}}}   =   \left(
 \begin{array}{c@{}c}
0 \ \ & \ \     {{\Theta}}_{21}   \\
{{\Theta}}_{12} \ \ & \ \   0 \\
\end{array} \right)$,  $   {\pmb{\overline{\Theta}}}_{\nu \mu}^{\hbox{\tiny{\sffamily Maxwell}}}      = \displaystyle       \left(
 \begin{array}{c@{}c}
0  \ \ & \ \    {\overline{\Theta}}_{21}   \\
{\overline{\Theta}}_{12} \ \ & \ \    0 \\
\end{array} \right) 
$, $ \displaystyle  {\big{(}} {\pmb{\Upsilon}}_{[\rho]}^{A_{\mu} \nu} {\big{)}}^{\hbox{\tiny{\sffamily Maxwell}}}      =  \left(
 \begin{array}{c@{}c}
   {\Upsilon}_{\rho}^{A_1 2}    \ \ & \ \   0  \\
0 \ \ & \ \   -   {\Upsilon}_{\rho}^{A_2 1}    \\
\end{array} \right)  $ and $\displaystyle   {\big{(}} {\overline{{\pmb{\Upsilon}}}}_{[\rho]}^{A_{\mu} \nu} {\big{)}}^{\hbox{\tiny{\sffamily Maxwell}}}  =    \left(
 \begin{array}{c@{}c}
   {\overline{{{\Upsilon}}}}_{\rho}^{A_1 2}    \ \ & \ \   0   \\
0 \ \ & \ \    -  {\overline{{{\Upsilon}}}}_{\rho}^{A_2 1}    \\
\end{array} \right)  $ so that the relation \eqref{gliopopo8810} and \eqref{gliopopo8910} are written: 
 \bee
  {\pmb{\Theta}}^{\hbox{\tiny{\sffamily Maxwell}}}  =  \overline{\pmb{\Theta}}^{\hbox{\tiny{\sffamily Maxwell}}} ,
  \quad \quad       \hbox{and}        \quad
  \quad
  \hbox{tr} (    {\big{(}} {\pmb{\Upsilon}}_{[\rho]}^{A_{\mu} \nu} {\big{)}}^{\hbox{\tiny{\sffamily Maxwell}}}    )  =   \hbox{tr}  (  {\big{(}} {\overline{{\pmb{\Upsilon}}}}_{[\rho]}^{A_{\mu} \nu} {\big{)}}^{\hbox{\tiny{\sffamily Maxwell}}}   ) .
 \eee
 Due to \eqref{kiolhogofo}, we have the expression of $ \dd {\pmb{\rho}}_{1} (X)$ and $ \dd {\pmb{\rho}}_{2} (X)$
  \bee
\left.
\begin{array}{rcl}
\displaystyle     \dd {\pmb{\rho}}_{1}   (X)& = &   \displaystyle  {\Upsilon}_{2}^{A_1 2} {\Theta}_{11}  -  {\Upsilon}_{2}^{A_2 1} {\Theta}_{11}  -  {\Upsilon}_{1}^{A_1 2} \Theta_{21} +  {\Upsilon}_{1}^{A_2 1}  \Theta_{21} 
\\
\displaystyle       &  &   \displaystyle 
+
 {\Upsilon}_{2}^{A_1 2}  \Theta_{11}
 - {\Upsilon}_{2}^{A_2 1} \Theta_{11} 
 -  {\Upsilon}_{1}^{A_1 2} \Theta_{21}
 + {\Upsilon}_{1}^{A_2 1}  \Theta_{21} , \\
\displaystyle     \dd {\pmb{\rho}}_{2} (X)  & = &   \displaystyle  
{\Upsilon}_{2}^{A_1 2}  \Theta_{12} - {\Upsilon}_{1}^{A_1 2}   \Theta_{22} 
-    {\Upsilon}_{2}^{A_2 1}   \Theta_{1 {2}}  +  {\Upsilon}_{1}^{A_2 1}   \Theta_{22}
\\
\displaystyle       &  &   \displaystyle 
+{\Upsilon}_{2}^{A_1 2}  \Theta_{12} - {\Upsilon}_{1}^{A_1 2}   \Theta_{22} 
-    {\Upsilon}_{2}^{A_2 1}   \Theta_{1 {2}}  +  {\Upsilon}_{1}^{A_2 1}   \Theta_{22}.
  \\
\end{array}
\right.
\eee
\bee
\left.
\begin{array}{rcl}
\displaystyle     \dd {\pmb{\rho}}_{1}   (X)& = &   \displaystyle 2  ( {\Upsilon}_{2}^{A_1 2}   -  {\Upsilon}_{2}^{A_2 1} ) {\Theta}_{11}    + 2 ( {\Upsilon}_{1}^{A_2 1}  -  {\Upsilon}_{1}^{A_1 2}  )  \Theta_{21} , \\
\displaystyle     \dd {\pmb{\rho}}_{2} (X)  & = &   \displaystyle  
- 2  ( {\Upsilon}_{1}^{A_1 2}   -  {\Upsilon}_{1}^{A_2 1} ) {\Theta}_{22} 
 -  2  ( {\Upsilon}_{2}^{A_2 1}   -  {\Upsilon}_{2}^{A_1 2} ) {\Theta}_{12}  .
  \\
\end{array}
\right.
\eee 
Once again, this is not sufficient to conclude that $ \dd {\pmb{\rho}}_{1}   (X) =   \dd {\pmb{\rho}}_{1}   (\overline{X})$ or $  \dd {\pmb{\rho}}_{2}   (X) =   \dd {\pmb{\rho}}_{2}   (\overline{X})$ since we dot not have necessarily   $\Theta_{\mu\nu} =  {\overline{\Theta}}_{\mu\nu}$ for $(\mu,\nu) = (1,1)$ or  $(\mu,\nu) = (2,2)$.

\section{\hbox{\sffamily\bfseries\slshape{Lepage-Dedecker for two dimensional Maxwell theory}}}\label{section10}

  In the sections \eqref{section51}-\eqref{section53} - as opposed to the ultimate one \eqref{section54} - we   work with indices notation, in particular with the tedious but straightforward computation of the Hamiltonian. It is just to emphasize the huge amount of calculations for $({\hbox{\sffamily LD}})$ theories - even in a simple case $n=2$  - for the   setting of    Maxwell theory. We refer to H.A. Katstrup \cite{Katstrup} and   F. H\'elein and J. Kouneiher \cite{HK-02} \cite{HK-03} for some aspects of the two dimensional Lepage-Dedecker Maxwell theory.

\subsection{\hbox{\sffamily\bfseries\slshape{Lepage-Dedecker correspondence}}}\label{section51}

Now we perform a Lepage-Dedecker correspondence for the Maxwell {\sffamily 2D} theory. First we express the Lagrangian density $\displaystyle L(x,A,\hbox{d}A) =  - (1/4) \eta^{\mu\lambda} \eta^{\nu\sigma}   \dttF_{\mu\nu} \dttF_{\lambda\sigma} $ so that: 
\[
L(A) =
-\frac{1}{4} {\big{(}}
 \eta^{1\lambda} \eta^{2\sigma}   \dttF_{12} \dttF_{\lambda\sigma} + \eta^{2\lambda} \eta^{1\sigma}   \dttF_{21} \dttF_{\lambda\sigma} {\big{)}}
= -\frac{1}{4} {\big{(}}  \eta^{11} \eta^{22} {\big{(}}  \dttF_{12} {\big{)}}^2  + \eta^{22} \eta^{11} {\big{(}}  \dttF_{21} {\big{)}}^2   {\big{)}} = \frac{1}{2} {\big{(}}   \dttF_{12}  {\big{)}}^2,
\]
then, the Lagrangian is written: 
\bee
L(x,A,\dd A) = \frac{1}{2} {\big{(}} \partial_1 A_2 - \partial_2 A_1  {\big{)}} {\big{(}} \partial_1 A_2 - \partial_2 A_1  {\big{)}}
= \frac{1}{2} {\big{(}} {\big{(}} \partial_1 A_2  {\big{)}}^2  + {\big{(}} \partial_2 A_1  {\big{)}}^2 {\big{)}} - {\big{(}} \partial_1 A_2  {\big{)}} {\big{(}}  \partial_2 A_1 {\big{)}}.
\nonumber
\eee
Now we construct a {\em non degenerate} Legendre {\em correspondence} in the {\sffamily 2D}-case via the following Poincar\'e-Cartan form\footnote{We use here the following notation : $ {\pmb{\theta}}^{{[\mathfrak{2}]} \mid  {[\mathfrak{2}]} }_{(q,p)}  $ means we specify the canonical Poincar\'e-Cartan form for Maxwell theory in the 2 dimensional  case and taking into account forms that involves $2$ fields. (namely forms of the type $ { {\varsigma}}  \hbox{d} A_{1} \wedge  \hbox{d} A_{2}$) Following this logic we     write  the previous canonical form as ${\pmb{\theta}}^{\tiny{\hbox{\sffamily DDW}}}_{(q,p)}:= {\pmb{\theta}}^{ {[\mathfrak{1}]}  \mid  {[\mathfrak{4}]} }_{(q,p)}$}~\eqref{Max003} $ {\pmb{{\theta}}}^{\tiny\hbox{{\sffamily Lepage-Dedecker}}}_{(q,p)}   =  {\pmb{\theta}}^{{[\mathfrak{2}]}  \mid  {[\mathfrak{2}]} }_{(q,p)}  =  {\pmb{\theta}}^{{[\mathfrak{2}]}  }_{(q,p)}$:
\begin{equation}\label{Max003}
 {\pmb{\theta}}^{{[\mathfrak{2}]} }_{(q,p)}:= \mathfrak{e}  \hbox{d} \mathfrak{y}  + \pi^{{A}_{\mu}\nu} 
  \hbox{d} {A}_{\mu} \wedge   \hbox{d} \mathfrak{y}_{\nu} + { {\varsigma}}  \hbox{d} A_{1} \wedge  \hbox{d} A_{2}.
\end{equation}
and the related Multisymplectic $3$-form: 
\begin{equation}\label{abibi10}
 {\pmb{\Omega}}^{ {[\mathfrak{2}]} }_{(q,p)} := \dd \mathfrak{e}  \wedge \hbox{d} \mathfrak{y}  +\dd \pi^{{A}_{\mu}\nu}  \wedge
  \hbox{d} {A}_{\mu} \wedge   \hbox{d} \mathfrak{y}_{\nu} +\dd { {\varsigma}}  \wedge \hbox{d} A_{1} \wedge  \hbox{d} A_{2}.
\end{equation}
Then, we concentrate on the expression of $\langle p , v  \rangle $,
\begin{equation}
\langle p , v  \rangle =   {\pmb{\theta}}^{{[\mathfrak{2}]} }_{(q,p)}  ({\cal Z}) = \mathfrak{e}  \hbox{d} \mathfrak{y}  ({\cal Z})  +  \pi^{{A}_{\mu}\nu}  \hbox{d} {A}_{\mu} \wedge   \hbox{d} \mathfrak{y}_{\nu}  ({\cal Z})
+  { {\varsigma}}   \hbox{d} A_{1} \wedge  \hbox{d} A_{2} ({\cal Z}).
\end{equation}
We demonstrate by direct calculation\footnote{ $ ^{\lceil}     $  
 {\em Proof}. Since $ \displaystyle
{\cal Z}_\nu =  \frac{\partial}{\partial x^\nu} +{\cal Z}_{\nu\mu} \frac{\partial}{\partial {A}_\mu}   
$, we have: $\displaystyle {\cal Z}_{1} = \partial_1 + {\cal Z}_{1 \mu_1} \frac{\partial }{\partial A_{\mu_1}} $ and $ \displaystyle
{\cal Z}_{2} = \partial_2 + {\cal Z}_{2 \mu_2} \frac{\partial }{\partial A_{\mu_2}} 
$ so that we compute ${\cal Z} = {\cal Z}_1 \wedge  {\cal Z}_2$:
\[
\left.
\begin{array}{rcl}
\displaystyle   {\cal Z}  & = &   \displaystyle
  \sum_{ {\mu_1}  < \mu_2 } {\cal Z}^{{\mu_1} {\mu_2} }_{1 2}
\frac{\partial}{\partial q^{{\mu_1}}} \wedge \frac{\partial}{\partial q^{{\mu_2}}}
=
 \sum_{ {\mu_1}  < \mu_2 }
\left| \begin{array}{ccc}
{ {\cal Z}^{\mu_1}_{1} }  & { {\cal Z}^{\mu_1}_{2}   }\\
 & \\
{{\cal Z}^{\mu_2}_{1} }  & {{\cal Z}^{\mu_2}_{2} }
\end{array}\right|  
\frac{\partial}{\partial q^{{\mu_1}}} \wedge \frac{\partial}{\partial q^{{\mu_2}}}
\\
 \displaystyle & = &   \displaystyle
 {\cal Z}^{12}_{12} \partial_1 \wedge \partial_2 + {\cal Z}^{1\mu_2}_{12} \partial_1 \wedge \frac{\partial}{\partial A_{\mu_2} }+ {\cal Z}^{2\mu_2}_{12}  \partial_2 \wedge \frac{\partial}{\partial A_{\mu_2} } +  {\cal Z}^{\mu_1\mu_2}_{12}  \frac{\partial}{\partial A_{\mu_1} } \wedge \frac{\partial}{\partial A_{\mu_2} }.
\end{array}
\right.
\] 
 With, the different terms:  $\displaystyle {\cal Z}^{ 12 }_{12}  =  1 $, $\displaystyle {\cal Z}^{ 2 {\mu_2} }_{12}   
=
 \left| \begin{array}{cccc}
{ 0 } & { 1 } \\
{ {\cal Z}^{\mu_2}_{1} } &{ {\cal Z}^{\mu_2}_{2} } \\
\end{array}\right|  = - {\cal Z}^{\mu_2}_{1}   $, $\displaystyle  {\cal Z}^{ 1 {\mu_2} }_{12}   =    \left| \begin{array}{cccc}
{ 1 } & { 0 } \\
{ {\cal Z}^{\mu_2}_{1} } &{ {\cal Z}^{\mu_2}_{2} } \\
\end{array}\right|  = {\cal Z}^{\mu_2}_{2}  $ and finally  $\displaystyle       
 {\cal Z}^{  \mu_1 {\mu_2} }_{12} =
 \left| \begin{array}{cccc}
{ {\cal Z}^{\mu_1}_{1} } &{ {\cal Z}^{\mu_1}_{2} } \\
{ {\cal Z}^{\mu_2}_{1} } &{ {\cal Z}^{\mu_2}_{2} } \\
\end{array}\right|  =
\BL {\cal Z}^{\mu_1}_{1}  {\cal Z}^{\mu_2}_{2}   - {\cal Z}^{\mu_2}_{1} {\cal Z}^{\mu_1}_{2}  \BR$.  We     make the following calculation:
\[
\langle p , v  \rangle =  \mathfrak{e}  + 
\underbrace{  
  \pi^{{A}_{\mu}\nu}  \hbox{d} {A}_{\mu} \wedge   \hbox{d} \mathfrak{y}_{\nu}  ( {\cal Z}^{1\mu_2}_{12} \partial_1 \wedge \frac{\partial}{\partial A_{\mu_2} }+ {\cal Z}^{2\mu_2}_{12}  \partial_2 \wedge \frac{\partial}{\partial A_{\mu_2} })
 }_{ (\mathfrak{i}) } 
+
\underbrace{  
 { {\varsigma}}  \hbox{d} A_{1} \wedge  \hbox{d} A_{2} (   {\cal Z}^{\mu_1\mu_2}_{12}  \frac{\partial}{\partial A_{\mu_1} } \wedge \frac{\partial}{\partial A_{\mu_2} } )
 }_{(\mathfrak{ii})  }.
\]
The first term in the last equation is given by: $ \displaystyle (\mathfrak{i})  =    
     \sum_{\mu , \nu}
\pi^{{A}_{\mu}\nu}  \hbox{d} {A}_{\mu} \wedge   \hbox{d} \mathfrak{y}_{\nu}  ( {\cal Z} ) $
\[ 
\left.
\begin{array}{ccl}
\displaystyle
(\mathfrak{i})     & = &  \displaystyle  
     \pi^{{A}_{1}1}  \hbox{d} {A}_{1} \wedge   \hbox{d} \mathfrak{y}_{1}  ( {\cal Z} ) 
+ \pi^{{A}_{1}2}  \hbox{d} {A}_{1} \wedge   \hbox{d} \mathfrak{y}_{2}  ( {\cal Z} ) 
  + \pi^{{A}_{2}1}  \hbox{d} {A}_{2} \wedge   \hbox{d} \mathfrak{y}_{1}  ( {\cal Z} )  
+ \pi^{{A}_{2}2}  \hbox{d} {A}_{2} \wedge   \hbox{d} \mathfrak{y}_{2}  ( {\cal Z} )  
\\
\displaystyle  & = &  \displaystyle  \pi^{{A}_{1}1}  \hbox{d} {A}_{1} \wedge   \hbox{d} x^{2}  ( {\cal Z}^{2\mu_2}_{12}  \partial_2 \wedge \frac{\partial}{\partial A_{\mu_2} } ) 
- \pi^{{A}_{1}2}  \hbox{d} {A}_{1} \wedge   \hbox{d} x^{1}  ( {\cal Z}^{1\mu_2}_{12} \partial_1 \wedge \frac{\partial}{\partial A_{\mu_2} } ) 
\\
 \displaystyle  &   &  \displaystyle  + \pi^{{A}_{2}1}  \hbox{d} {A}_{2} \wedge   \hbox{d} x^{2}  ( {\cal Z}^{2\mu_2}_{12}  \partial_2 \wedge \frac{\partial}{\partial A_{\mu_2} })   - \pi^{{A}_{2}2}  \hbox{d} {A}_{2} \wedge   \hbox{d} x^{1}  (  {\cal Z}^{1\mu_2}_{12} \partial_1 \wedge \frac{\partial}{\partial A_{\mu_2} } ) 
   \\
\displaystyle  & = &  \displaystyle     
 \pi^{{A}_{1}1}   {\cal Z}_{11}
+ \pi^{{A}_{1}2}  {\cal Z}_{21}  
+ \pi^{{A}_{2}1}  {\cal Z}_{12}
+ \pi^{{A}_{2}2}   {\cal Z}_{22}
=   \pi^{A_\mu \nu}  \partial_\nu A_\mu . \\
\end{array}
\right.
\]
   Whereas the second term is given by:
\[ 
\left.
\begin{array}{ccl}
\displaystyle
(\mathfrak{ii}) & = &  \displaystyle  
  { {\varsigma}} \dd A_{1} \wedge  \hbox{d} A_{2} ({\cal Z}) =  { {\varsigma}}    \hbox{d} A_{1} \wedge  \hbox{d} A_{2}  ( {\cal Z}^{\mu_1\mu_2}_{12}  \frac{\partial}{\partial A_{\mu_1} } \wedge \frac{\partial}{\partial A_{\mu_2} } )  
  \\
 \displaystyle  & = &  \displaystyle    { {\varsigma}}     \hbox{d} A_{1} \wedge  \hbox{d} A_{2}  ( \BL {\cal Z}^{\mu_1}_{1}  {\cal Z}^{\mu_2}_{2}   - {\cal Z}^{\mu_2}_{1} {\cal Z}^{\mu_1}_{2}  \BR  \frac{\partial}{\partial A_{\mu_1} } \wedge \frac{\partial}{\partial A_{\mu_2} } )   
 =
   {{ {\varsigma}}   }  {\big{(}} {\cal Z}_{11}  {\cal Z}_{22}   - {\cal Z}_{12} {\cal Z}_{21}  {\big{)}} .
 \end{array}.
 \right.
\]
Since,
\[
2 {{ {\varsigma}}   }     \varepsilon^{\mu\nu}  {\cal Z}_{1[\mu}  {\cal Z}_{2\nu]}   = {{ {\varsigma}}   }     \varepsilon^{\mu\nu}  {\cal Z}_{1\mu}  {\cal Z}_{2\nu} - {\cal Z}_{1\nu}  {\cal Z}_{2\mu}  =  {{ {\varsigma}}   }     {\big{(}}    {\cal Z}_{11}  {\cal Z}_{22} - {\cal Z}_{12}  {\cal Z}_{21}  {\big{)}}     -   {{ {\varsigma}}   }   {\big{(}}  {\cal Z}_{12}  {\cal Z}_{21} - {\cal Z}_{11}  {\cal Z}_{22}   {\big{)}} = 2 {{ {\varsigma}}   }  {\big{(}} {\cal Z}_{11}  {\cal Z}_{22}   - {\cal Z}_{12} {\cal Z}_{21}  {\big{)}} ,
\]
we write the term $ (\mathfrak{ii})   =  {{\varsigma}}     \varepsilon^{\mu\nu}  {\cal Z}_{1[\mu}  {\cal Z}_{2\nu]}  .         \rfloor$} that: 
\[
\langle p , v  \rangle =   {\pmb{\theta}}^{{[\mathfrak{2}]} }_{(q,p)}   ({\cal Z}) = \pi^{A_\mu \nu}  \partial_\nu A_\mu  
  + 2   { {\varsigma}}  {\big{(}}   {\cal Z}_{11}  {\cal Z}_{22} -     {\cal Z}_{12}  {\cal Z}_{21}  {\big{)}}.
\]
  Then we have the expression of $\langle p , v  \rangle $:
\bee\label{onzzenozv}
\langle p , v  \rangle  = \pi^{{A}_{1}1}   {\cal Z}_{11}
+ \pi^{{A}_{1}2}  {\cal Z}_{21}  
+ \pi^{{A}_{2}1}  {\cal Z}_{12}
+ \pi^{{A}_{2}2}   {\cal Z}_{22} 
 +  \varsigma  {\big{(}} {\cal Z}_{11}  {\cal Z}_{22}   - {\cal Z}_{12} {\cal Z}_{21}  {\big{)}}
  \eee
We can equivalently write in   contracted notation:
  $
  \langle p , v  \rangle =   {\pmb{\theta}}^{{[\mathfrak{2}]} }_{(q,p)}   ({\cal Z})   =  \pi^{A_\mu \nu}  \partial_\nu A_\mu
+  {{ {\varsigma}}   }     \varepsilon^{\mu\nu}  {\cal Z}_{1[\mu}  {\cal Z}_{2\nu]}   
  $. With the notation ${\cal Z}_{\nu\mu} = \partial_\nu A_{\mu}$,
\[
  \langle p , v  \rangle =  {\pmb{\theta}}^{{[\mathfrak{2}]} }_{(q,p)}   ({\cal Z})   =  \pi^{A_\mu \nu}  \partial_\nu A_\mu
+  {{ {\varsigma}}   }     \varepsilon^{\mu\nu}  \partial_{1} A_{[\mu}  {\partial}_{2} A_{\nu]}   .
\]
Let us denote:
\bee
 {\pmb{\kappa}}_{\mu\nu}  =  {\pmb{\kappa}}_{\mu\nu}^{{[\mathfrak{2}]} }   =   \frac{\partial \langle p , v  \rangle  }{\partial (\partial_\mu {A}_{\nu} )} =    \frac{\partial  }{\partial (\partial_\mu {A}_{\nu} )}   {\pmb{\theta}}^{{[\mathfrak{2}]} }_{(q,p)}   ({\cal Z})   , 
   \eee
    we   work in coordinate expression so that we use the expression~\eqref{onzzenozv}:
    \[
 {\pmb{\theta}}^{{[\mathfrak{2}]} }_{(q,p)}   ({\cal Z}) = 
    \pi^{{A}_{1}1}  \partial_1 A_1
+ \pi^{{A}_{1}2}  \partial_2 A_1  
+ \pi^{{A}_{2}1}  \partial_1 A_2  
+ \pi^{{A}_{2}2}  \partial_2 A_2   
 +  {{ {\varsigma}}   }  {\big{(}} \partial_1 A_1\partial_2 A_2 - \partial_1 A_2 \partial_2 A_1  {\big{)}}. 
    \]
Hence, we find the relations~\eqref{abibi04}$(\mathfrak{i})$. 
\begin{equation} \label{abibi04}
(\mathfrak{i}) \quad
\left|
\begin{array}{ccc}
\displaystyle     {\pmb{\kappa}}_{\mu\nu}  { {{|}} }_{ \mu = 1, \nu = 1} &  =  & \displaystyle   \pi^{{A}_{1}1}    +  \varsigma \partial_2 A_2
\\ 
\displaystyle    {\pmb{\kappa}}_{\mu\nu}   { { {|}} }_{ \mu = 1, \nu = 2}   & =  &    \displaystyle  \pi^{{A}_{2}1}    -  \varsigma \partial_2 A_1   
\\ 
\displaystyle    {\pmb{\kappa}}_{\mu\nu}  { { {|}} }_{ \mu = 2, \nu = 1}   & =  &    \displaystyle  \pi^{{A}_{1}2}   -   \varsigma  \partial_1 A_2  
\\
\displaystyle     {\pmb{\kappa}}_{\mu\nu}  { { {|}} }_{ \mu = 2, \nu = 2} 
& =  &    \displaystyle  \pi^{{A}_{2}2}    +  \varsigma  \partial_1 A_1
\end{array}
\right.
 \quad \quad   (\mathfrak{ii}) \quad 
\left|
\begin{array}{ccl}
\displaystyle     {\pmb{\lambda}}_{\mu\nu} { {{|}} }_{ \mu = 1, \nu = 1} &  =  & \displaystyle   0 
\\ 
\displaystyle    {\pmb{\lambda}}_{\mu\nu} { {{|}} }_{ \mu = 1, \nu = 2}   & =  &    \displaystyle   \partial_1 A_2 - \partial_2 A_1  
\\ 
\displaystyle    {\pmb{\lambda}}_{\mu\nu} { {{|}} }_{ \mu = 2, \nu = 1}   & =  &    \displaystyle  \partial_2 A_1 - \partial_1 A_2
 \\ 
\displaystyle     {\pmb{\lambda}}_{\mu\nu} { {{|}} }_{ \mu = 2, \nu = 2} 
& =  &    \displaystyle  0 
\end{array}
\right.
\end{equation}
On the other side, we denote    $ \displaystyle \frac{\partial L}{\partial (\partial_\mu A_\nu) }  = {\pmb{\lambda}}_{\mu\nu}$. We use the coordinate expression of $L(x,A,\dd A)$ and we obtain~\eqref{abibi04}$(\mathfrak{ii})$. 
The condition for the Legendre transform is: 
\[
\frac{\partial L}{\partial (\partial_\mu A_\nu) } =  \frac{\partial \langle p , v  \rangle  }{\partial (\partial_\mu {A}_{\nu} )} .
\]
We   obtain \eqref{abibi07}($\mathfrak{i}$) and,   choosing to work in  the case   $\varsigma = 1$, we then obtain the relations   \eqref{abibi07}($\mathfrak{ii}$).
\begin{equation} \label{abibi07}
({\mathfrak{i}}) \quad 
\left|
\begin{array}{ccl}
\displaystyle  0 &  =  & \displaystyle    \pi^{{A}_{1}1}    +  \varsigma \partial_2 A_2 
\\ 
\displaystyle    \partial_1 A_2 - \partial_2 A_1 & =  &    \displaystyle   \pi^{{A}_{2}1}    -  \varsigma \partial_2 A_1     
\\ 
\displaystyle  \partial_2 A_1 - \partial_1 A_2   & =  &    \displaystyle   \pi^{{A}_{1}2}   -   \varsigma  \partial_1 A_2 
\\
\displaystyle     0
& =  &    \displaystyle   \pi^{{A}_{2}2}   +  \varsigma  \partial_1 A_1 
\end{array}
\right.
\quad \quad \overset{ \varsigma = 1 }{{\Longrightarrow}  } \quad \quad
({\mathfrak{ii}}) \quad 
\left|
\begin{array}{ccl}
\displaystyle  \pi^{{A}_{1}1} &  =  & \displaystyle     -   \partial_2 A_2
\\ 
\displaystyle  \pi^{{A}_{2}1} & =  &    \displaystyle     \partial_1 A_2 
\\ 
\displaystyle  \pi^{{A}_{1}2}   & =  &    \displaystyle     \partial_2 A_1
 \\ 
\displaystyle     \pi^{{A}_{2}2} 
& =  &    \displaystyle   - \partial_1 A_1 
\end{array}
\right.
\end{equation}
The generalized Legendre correspondence is non degenerate. It is always possible to invert the multimomenta from multivelocities.
 Now, we give the expression of the Hamiltonian function.
From  \eqref{abibi07}($\mathfrak{i}$) we obtain,\footnote{ $ ^{\lceil}     $  
 {\em Proof}
We focus on the second line in~\eqref{abibi07dddd33}, from the second line in~\eqref{abibi07}
we find: 
\bee\label{dznxxxnz}
\partial_{1} A_2 =  \pi^{{A}_{2}1} +  {\big{(}}   1     -  \varsigma  {\big{)}}   \partial_2 A_1 .
\eee
The   third line of~\eqref{abibi07} is written: 
\bee\label{gldkskqj}
 \partial_2 A_1 - \partial_1 A_2    =        \pi^{{A}_{1}2}   -   \varsigma  \partial_1 A_2 \quad \Longrightarrow \quad
 \partial_2 A_1      =        \pi^{{A}_{1}2}  +   {\big{(}}   1     -  \varsigma  {\big{)}} \partial_1 A_2 .
\eee
We insert~\eqref{dznxxxnz} in~\eqref{gldkskqj} so that: 
$ \displaystyle
\partial_2 A_1      =        \pi^{{A}_{1}2}  +   {\big{(}}   1     -  \varsigma  {\big{)}} {\big{(}}  \pi^{{A}_{2}1} +  {\big{(}}   1     -  \varsigma  {\big{)}}   \partial_2 A_1  {\big{)}}
$ and then,
\[
\partial_2 A_1   {\big{(}}  1 -  {\big{(}}   1     -  \varsigma  {\big{)}}^2   {\big{)}}   =        \pi^{{A}_{1}2}  +   {\big{(}}   1     -  \varsigma  {\big{)}}   \pi^{{A}_{2}1}   \quad \Longleftrightarrow \quad
 \partial_2 A_1   \varsigma  {\big{(}}      2   -  \varsigma    {\big{)}}   =        \pi^{{A}_{1}2}  +   {\big{(}}   1     -  \varsigma  {\big{)}}   \pi^{{A}_{2}1} . \
  \rfloor 
\] }  
 \begin{equation} \label{abibi07dddd33}
\left|
\begin{array}{ccl}
\displaystyle \partial_2 A_2 &  =  & \displaystyle  -  ( {  \varsigma}^{-1} )   \pi^{{A}_{1}1}      
\\ 
\displaystyle    \partial_2 A_1 & =  &    \displaystyle  {\big{(}} {  \varsigma  (      2   -  \varsigma  )   {\big{)}}   }^{-1}   {\big{(}}      \pi^{{A}_{1}2}  +   {\big{(}}   1     -  \varsigma  {\big{)}}   \pi^{{A}_{2}1}       {\big{)}}
\\ 
\displaystyle  \partial_1 A_2   & =  &     \displaystyle  {\big{(}} {  \varsigma  (      2   -  \varsigma  )   {\big{)}}   }^{-1}    {\big{(}}      \pi^{{A}_{2}1}  +   {\big{(}}   1     -  \varsigma  {\big{)}}   \pi^{{A}_{1}2}       {\big{)}}  
\\
\displaystyle      \partial_1 A_1
& =  &    \displaystyle   -  ( {  \varsigma}^{-1} ) \pi^{{A}_{2}2}       
\end{array}
\right.
\eee

\subsection{\hbox{\sffamily\bfseries\slshape{Calculation of the Hamiltonian}}}\label{section52}

We are interested in the expression of the Hamiltonian:
\bee\label{gopqsz11}
\displaystyle {\cal H} =  {\pmb{\theta}}^{{[\mathfrak{2}]} }_{(q,p)}    ({\cal Z})  - L ,
\eee
where, $ {\pmb{\theta}}^{{[\mathfrak{2}]} }_{(q,p)}    ({\cal Z})   =   {{\bf k}}_{1} + \cdots +  {{\bf k}}_{6}  
$ and $-L =   {{\bf k}}_{7} + {{\bf k}}_{8}  +  {{\bf k}}_{9}  $,
with
\bee
\left|
\begin{array}{rcl}
\displaystyle {{\bf k}}_{1} & = &  \displaystyle      \pi^{{A}_{1}1}  \partial_1 A_1  
\\
\displaystyle  {{\bf k}}_{2}  & = &  \displaystyle    \pi^{{A}_{1}2}  \partial_2 A_1 
\\
\displaystyle {{\bf k}}_{3}  & = &  \displaystyle         \pi^{{A}_{2}1}  \partial_1 A_2 
\end{array}
\right.
\quad \quad \quad  
\left|
\begin{array}{rcl}
\displaystyle  {{\bf k}}_{4}   & = &  \displaystyle     \pi^{{A}_{2}2}  \partial_2 A_2 
\\
\displaystyle  {{\bf k}}_{5}   & = &  \displaystyle  {{ {\varsigma}}   }  \partial_1 A_1\partial_2 A_2 
\\
\displaystyle  {{\bf k}}_{6}   & = &  \displaystyle  - {{ {\varsigma}}   }   \partial_1 A_2 \partial_2 A_1  
\end{array}
\right.
\quad \quad \quad  
\left|
\begin{array}{rcl}
\displaystyle  {{\bf k}}_{7}  & = &  \displaystyle   -  {1}/{2}  {\big{(}} \partial_1 A_2  {\big{)}}^2
\\
\displaystyle {{\bf k}}_{8}  & = &  \displaystyle       -  {1}/{2}  {\big{(}} \partial_2 A_1  {\big{)}}^2 
\\
\displaystyle  {{\bf k}}_{9}   & = &  \displaystyle        {\big{(}} \partial_1 A_2  {\big{)}} {\big{(}}  \partial_2 A_1 {\big{)}} .
\end{array}
\right.
\eee
  We finally obtain the expression of the Hamiltonian:
\bee\label{qgmh003gg444}
{\cal H} = 
-  \frac{1}{ \varsigma} \overline{\pi}^{\circ\bullet}  +
  \frac{1}{2}  \frac{1}{  \varsigma  {\big{(}}      2   -  \varsigma    {\big{)}}   } 
  \BL   \pi^{\circ\circ}      
 +         \pi^{\bullet\bullet}    
 \BR
   + \frac{  {\big{(}}  \varsigma - 3 {\big{)}}  }{    {\big{(}}      2   -  \varsigma    {\big{)}}^2   }
   \pi^{\circ\bullet} 
  +     \frac{1}{  \varsigma   {\big{(}}      2   -  \varsigma    {\big{)}}^2   } 2 \pi^{\circ\bullet} .
\eee
If we use the transform with $ \varsigma = 1$ then~\eqref{qgmh003gg444}($\mathfrak{ii}$) gives the following Hamiltonian: 
\bee\label{qgmh003}
{\cal H} = -
  \overline{\pi}^{\circ\bullet}
+  \frac{1}{2} {\big{(}} \pi^{\circ\circ}       
+       \pi^{\bullet\bullet}     {\big{)}} .
      \eee
   $ ^{\lceil}     $  
 {\em Proof}. We compute in coordinate the straightforward calculation:
\[
\left.
\begin{array}{rcl}
\displaystyle  {\cal H}       & = &  \displaystyle  
 \pi^{A_\mu \nu} \partial_\nu A_\mu  
  +      { {\varsigma}} \bl   {\cal Z}_{11}  {\cal Z}_{22} -     {\cal Z}_{12}  {\cal Z}_{21}  \br -  \frac{1}{2}  {\big{(}} \partial_1 A_2  {\big{)}}^2  
    - \frac{1}{2}  {\big{(}} \partial_2 A_1  {\big{)}}^2 
      + {\big{(}} \partial_1 A_2  {\big{)}} {\big{(}}  \partial_2 A_1 {\big{)}} 
\\
\displaystyle         & = &  \displaystyle  
( \pi^{{A}_{1}2} )^2 +  ( \pi^{{A}_{2}1} )^2
- 2
 \pi^{{A}_{1}1}  \pi^{{A}_{2}2}  
+ \pi^{{A}_{1}1}  \pi^{{A}_{2}2}     -    \pi^{{A}_{1}2}   \pi^{{A}_{2}1}   
-  \frac{1}{2}  {\big{(}} \pi^{{A}_{2}1} {\big{)}}^2  
\\
\displaystyle         &   &  \displaystyle      - \frac{1}{2}  {\big{(}} \pi^{{A}_{1}2}   {\big{)}}^2 
      + \pi^{{A}_{1}2} \pi^{{A}_{2}1} 
\\
\displaystyle         & = &  \displaystyle  
 -
 \pi^{{A}_{1}1}  \pi^{{A}_{2}2} 
+  
  \frac{1}{2}  {\big{(}} \pi^{{A}_{2}1} {\big{)}}^2  
    + \frac{1}{2}  {\big{(}} \pi^{{A}_{1}2}   {\big{)}}^2       \quad      \rfloor
    \end{array}
\right.
\]
     And the Hamiltonian~\eqref{qgmh003} agrees with the general case~\eqref{qgmh003gg444}.

 \subsection{\hbox{\sffamily\bfseries\slshape{Equations of movement}}}\label{section53}

Now let us derive the generalized Hamilton equations. 
      The general form of a vector field is given by:
      \bee
X_\alpha = \frac{\partial}{\partial x^\alpha}   + \Theta_{\alpha {\mu}}   \frac{\partial}{\partial A_{\mu}}  + {\Upsilon}_{\alpha} \frac{\partial }{ \partial \mathfrak{e}}
+ {\Upsilon}_{\alpha}^{A_\mu \nu} \frac{\partial }{ \partial \pi^{A_\mu \nu}}
+ {\Upsilon}_{\alpha}^{A_\mu A_\nu} \frac{\partial }{ \partial  { {\varsigma}}^{A_\mu A_\nu} }
\eee so that $X = X_1 \wedge X_2 $ is written: 
\[ 
\left.
\begin{array}{rcl}
\displaystyle  X  & = &  \displaystyle  
 \partial_1 \wedge \partial_2 +  
\partial_1 \wedge   \Theta_{2 {\mu}}   \frac{\partial}{\partial A_{\mu}}
+\partial_1 \wedge  {\Upsilon}_{2} \frac{\partial }{ \partial \mathfrak{e}}
+\partial_1 \wedge  {\Upsilon}_{2}^{A_\mu \nu} \frac{\partial }{ \partial \pi^{A_\mu \nu}}
\\
 \displaystyle  &  &  \displaystyle   + \Theta_{1 {\mu}}   \frac{\partial}{\partial A_{\mu}}   \wedge \partial_2
 +
  \Theta_{1 {\mu}}   \frac{\partial}{\partial A_{\mu}}   \wedge \Theta_{2 {\mu}}   \frac{\partial}{\partial A_{\mu}}  
  +
   \Theta_{1 {\mu}}   \frac{\partial}{\partial A_{\mu}}   \wedge {\Upsilon}_{2} \frac{\partial }{ \partial \mathfrak{e}}
   +
    \Theta_{1 {\mu}}   \frac{\partial}{\partial A_{\mu}}   \wedge  {\Upsilon}_{2}^{A_\mu \nu} \frac{\partial }{ \partial \pi^{A_\mu \nu}} 
     \\
    \displaystyle  &  &  \displaystyle   +
 {\Upsilon}_{1} \frac{\partial }{ \partial \mathfrak{e}} \wedge \partial_2
 +
  {\Upsilon}_{1} \frac{\partial }{ \partial \mathfrak{e}} \wedge \Theta_{2 {\mu}}   \frac{\partial}{\partial A_{\mu}}  
    +
   {\Upsilon}_{1} \frac{\partial }{ \partial \mathfrak{e}} \wedge {\Upsilon}_{2} \frac{\partial }{ \partial \mathfrak{e}}
   +
    {\Upsilon}_{1} \frac{\partial }{ \partial \mathfrak{e}} \wedge  {\Upsilon}_{2}^{A_\mu \nu} \frac{\partial }{ \partial \pi^{A_\mu \nu}}
 \\
    \displaystyle  &  &  \displaystyle   + {\Upsilon}_{1}^{A_\mu \nu} \frac{\partial }{ \partial \pi^{A_\mu \nu}} \wedge \partial_2
+ {\Upsilon}_{1}^{A_\mu \nu} \frac{\partial }{ \partial \pi^{A_\mu \nu}} \wedge   \Theta_{2 {\mu}}   \frac{\partial}{\partial A_{\mu}}  
+ {\Upsilon}_{1}^{A_\mu \nu} \frac{\partial }{ \partial \pi^{A_\mu \nu}} \wedge {\Upsilon}_{2} \frac{\partial }{ \partial \mathfrak{e}}
\\
     \displaystyle  &  &  \displaystyle   
+ {\Upsilon}_{1}^{A_\mu \nu} \frac{\partial }{ \partial \pi^{A_\mu \nu}} \wedge  {\Upsilon}_{2}^{A_\mu \nu} \frac{\partial }{ \partial \pi^{A_\mu \nu}}  .
\end{array}
\right.
\]
We compute the first part of generalized Hamilton equations, namely $ X \iN  {\pmb{\Omega}}^{ {[\mathfrak{2}]} }$:
 \[ 
\left.
\begin{array}{rcl}
\displaystyle  X \iN  {\pmb{\Omega}}^{ {[\mathfrak{2}]} }    & = &  \displaystyle  
  X \iN {\Big{(}}  \dd \mathfrak{e} \wedge \hbox{d} \mathfrak{y}  +\dd \pi^{{A}_{\mu}\nu} \wedge
  \hbox{d} {A}_{\mu} \wedge   \hbox{d} \mathfrak{y}_{\nu} +\dd \varsigma \wedge \hbox{d} A_{1} \wedge  \hbox{d} A_{2}
 {\Big{)}}
\\
\displaystyle      & = &  \displaystyle 
   \dd \mathfrak{e} -  (\dd \mathfrak{e} \wedge   \vol_\mu )(X) \dd x^{\mu}
+( \hbox{d} {A}_{\mu} \wedge   \hbox{d} \mathfrak{y}_{\nu}  )(X) \dd \pi^{{A}_{\mu}\nu} - (  \dd \pi^{{A}_{\mu}\nu} \wedge   \hbox{d} \mathfrak{y}_{\nu}  )(X)  \hbox{d} {A}_{\mu} 
\\
\displaystyle      &  &  \displaystyle 
+ (\dd \pi^{A_{\mu} \nu} \wedge \dd A_{\mu} \wedge   \vol_{\rho\nu} ) \dd x^{\rho}  + (  \hbox{d} {A}_{1} \wedge   \hbox{d} \mathfrak{y}_{2}  )(X) \dd \varsigma 
\\
\displaystyle      &  &  \displaystyle 
-
( \dd \varsigma  \wedge   \hbox{d}{A}_{2}  )(X)  \hbox{d} {A}_{1} 
+ (\dd \varsigma  \wedge \dd A_{1} ) \dd A_{2}.
\end{array}
\right.
\]
Since $\dd \varsigma = 0$ the multisymplectic form is written $ {\pmb{\Omega}}^{ {[\mathfrak{2}]} }  { {\big{|}} }_{ \varsigma  = 1}  = \dd \mathfrak{e} \wedge \hbox{d} \mathfrak{y}  +\dd \pi^{{A}_{\mu}\nu} \wedge
  \hbox{d} {A}_{\mu} \wedge   \hbox{d} \mathfrak{y}_{\nu}$. So that   $X \iN  {\pmb{\Omega}}^{ {[\mathfrak{2}]} }  { {\big{|}} }_{ \varsigma  = 1} $ is given by: 
 \[ 
\left.
\begin{array}{rcl}
\displaystyle  X \iN  {\pmb{\Omega}}^{ {[\mathfrak{2}]} } { {\big{|}} }_{ \varsigma  = 1}   & = &  \displaystyle  
 \dd \mathfrak{e} -  (\dd \mathfrak{e} \wedge   \vol_\mu )(X) \dd x^{\mu}
+
(  \hbox{d} {A}_{\mu} \wedge   \hbox{d} \mathfrak{y}_{\nu}  )(X) \dd \pi^{{A}_{\mu}\nu} 
-
(  \dd \pi^{{A}_{\mu}\nu} \wedge   \hbox{d} \mathfrak{y}_{\nu}  )(X)  \hbox{d} {A}_{\mu} 
\\
\displaystyle      &  &  \displaystyle 
+ (\dd \pi^{A_{\mu} \nu} \wedge \dd A_{\mu} \wedge   \vol_{\rho\nu} ) \dd x^{\rho}
\\
\displaystyle      & =   &  \displaystyle   \hbox{d} \mathfrak{e}  - \Upsilon_\rho \dd x^{\rho} 
+   \Theta_{\nu\mu}   \hbox{d} \pi^{{A}_{\mu}\nu}    -  {\Upsilon}_{\nu}^{A_\mu \nu}    \dd A_{\mu}  
+    {\big{(}}  {\Upsilon}_{\rho}^{A_\mu \nu}       \Theta_{\nu\mu}  - {\Upsilon}_{\nu}^{A_\mu \nu}  \Theta_{\rho\mu} {\big{)}}   \dd x^{\rho}.
\end{array}
\right.
\]
 we only keep the interesting part on the decompositions along $\hbox{d} \pi^{{A}_{\mu}\nu}$ and $ \dd A_{\mu}$.
\begin{equation}\label{dlvqslvqd43}
\left|
\begin{array}{ccc}
\displaystyle    \Theta_{\nu\mu}   \hbox{d} \pi^{{A}_{\mu}\nu}    &  =  & \displaystyle  \dd{\cal H}  
\\
 \displaystyle -   {\Upsilon}_{\nu}^{A_\mu \nu}    &  =  & \displaystyle 0
\end{array}
\right.
\quad \quad \quad
\left|
\begin{array}{lcc}
\displaystyle  (\mathfrak{i}) \quad  \Theta_{1\mu}   \hbox{d} \pi^{{A}_{\mu}1} +   \Theta_{2\mu}   \hbox{d} \pi^{{A}_{\mu}2}      &  =  & \displaystyle \dd {\cal H}  
\\
 \displaystyle (\mathfrak{ii}) \quad \quad \quad   -   {\Upsilon}_{\nu}^{A_\mu \nu}    &  =  & \displaystyle 0
\end{array}
\right.
\end{equation}
With $\dd {\cal H} { {\big{|}} }_{ \varsigma  = 1}  =  -
  \overline{\pi}^{\circ\bullet}
+ 1/2 {\big{(}} \pi^{\circ\circ}       
+       \pi^{\bullet\bullet}   {\big{)}}
  = -
 \pi^{{A}_{1}1} \dd \pi^{{A}_{2}2} 
 -  \pi^{{A}_{2}2}  \dd  \pi^{{A}_{1}1}
+   \pi^{{A}_{2}1} \dd \pi^{{A}_{2}1} 
   +  \pi^{{A}_{1}2}   \dd \pi^{{A}_{1}2} 
      $. 
Finally we obtain from~\eqref{dlvqslvqd43}$(\mathfrak{i})$ the Legendre transform given by:
\begin{equation}\label{nervnozszd}
 \partial_1 {  A_{1}}   =     - \pi^{A_{2} 2}
\quad \quad \quad 
 \partial_2 {  A_{2}}     =   
 -  \pi^{A_1 1}
 \quad \quad \quad 
 \partial_1 {  A_{2}}     =    
 \pi^{{A}_{2}1}  
 \quad \quad \quad 
 \partial_2 {  A_{1}}     =  
     \pi^{{A}_{1}2} 
\end{equation}
Whereas from~\eqref{dlvqslvqd43}$(\mathfrak{ii})$ we obtain the Maxwell's equations: 
  \begin{equation}
 \partial_\mu    \pi^{A_\nu \mu}    =        0.
\end{equation}

\subsection{\hbox{\sffamily\bfseries\slshape{Grassmannian viewpoint and pseudofibers}}}\label{section54} 
 
{{\em Enlarged pseudofibers - Pseudofibers}}.     We introduce the following fundamental objects: the {\em enlarged pseudofiber} and the {\em pseudofiber}. Following  \cite{HKHK01} \cite{HK-02}  \cite{HK-03} the enlarged pseudofiber  is defined to be:  
\bee
\dttP_{q} (z) = {\Big{\{}} p \in \Lambda^{n} T^{\star}_{q}  {\pmb{\mathfrak{Z}}} \ / \ \frac{\partial {\cal W}}{\partial z} (q,z,p) = 0 {\Big{\}}}
\eee
The enlarged pseudofiber is understood as the space of $n$-forms $\dttP_{q} (z) \subset \Lambda^{n} T^{\star}_{q}  {\pmb{\mathfrak{Z}}} $ such that the generalized Legendre correspondence is satisfied: $(q,z)  {\pmb{\leftrightarrow}}  (q,p)$. We refer to \cite{HK-01} \cite{HKHK01} \cite{HK-02}  for further details. The key point is that $\dttP_{q} (z)$ is an affine subspace of $\Lambda^{n} T^{\star}_{q}  {\pmb{\mathfrak{Z}}} $ with $ \displaystyle \hbox{dim} (\dttP_{q} (z)) = \frac{(n+k)!}{n!k!} - nk$. Finally,  for a given $(q,z) \in {\bf D}^{\tiny{\hbox{\dd}} \mathfrak{y}}  {\pmb{\mathfrak{Z}}}$, we can find {\em at the same time} an element $p \in \dttP_{q} (z)$ and  choose the value of ${\cal H}(q,p)$. Therefore, we find the definition of the {\em pseudofiber} to be the space defined by \eqref{fdkso99}:
 \bee\label{fdkso99}
 \dttP_{q}^{h} (z) = {\Big{\{}} p \in  \dttP_{q} (z) \ / \ {\cal H} (q,p) = h
 {\Big{\}}}.
 \eee
Notice  that $\hbox{dim} ( \dttP_{q}^{h} (z))=  \hbox{dim} ( \dttP_{q}  (z) ) - 1$ and that  $ \dttP_{q}  (z)$ and $ \dttP_{q}^{h} (z)$ are affine subspaces parallel to $[ T_{z}{\bf D}^{\tiny{\hbox{\dd}} \mathfrak{y}}_{q}  {\pmb{\mathfrak{Z}}} ]^{\perp}$ and $[T_{z} {\bf D}^{n}_{q}  {\pmb{\mathfrak{Z}}} ]^{\perp}$ where the spaces $[ T_{z}{\bf D}^{\tiny{\hbox{\dd}} \mathfrak{y}}_{q}  {\pmb{\mathfrak{Z}}} ]^{\perp} , [T_{z} {\bf D}^{n}_{q}  {\pmb{\mathfrak{Z}}} ]^{\perp} \subset \Lambda^{n} T^{\star}_{q}  {\pmb{\mathfrak{Z}}} $ are respectively defined by \eqref{qmsldkfjgh88}:
 \bee\label{qmsldkfjgh88}
 \left.
\begin{array}{lcl}
\displaystyle  [ T_{z}{\bf D}^{\tiny{\hbox{\dd}} \mathfrak{y}}_{q}  {\pmb{\mathfrak{Z}}} ]^{\perp}   & = &  \displaystyle  
 {\Big{\{}} p \in  \Lambda^{n} T^{\star}_{q}  {\pmb{\mathfrak{Z}}}  \ / \ \forall \xi \in  {T_{z}{\bf D}^{\tiny{\hbox{\dd}} \mathfrak{y}}_{q}  {\pmb{\mathfrak{Z}}} } \ , \ p(\xi) = 0 {\Big{\}}}
\\
 \displaystyle
  [ T_{z}{\bf D}^{n}_{q}  {\pmb{\mathfrak{Z}}} ]^{\perp} 
  & = &  \displaystyle  
 {\Big{\{}} p \in  \Lambda^{n} T^{\star}_{q}  {\pmb{\mathfrak{Z}}}  \ / \ \forall \xi \in  {T_{z}{\bf D}^{n}_{q}  {\pmb{\mathfrak{Z}}} } \ , \ p(\xi) = 0 {\Big{\}}}
 \end{array}
 \right.
 \eee

In the general case we have the following dimension for the involved spaces: ${\bf Gr}^{n}  {\pmb{\mathfrak{Z}} } $, $\Lambda^{n} T^{\star} {\pmb{\mathfrak{Z}} } $, $ \Lambda^{n} T^{\star}_{q} {\pmb{\mathfrak{Z}} } $, $ {\bf D}^{\tiny{\hbox{\dd}} \mathfrak{y}}_{q} {\pmb{\mathfrak{Z}} }$ and of the  enlarged pseudofiber  $\dttP_{q} (z)$ and the pseudofiber  $\dttP_{q}^{h} (z)$. 
\[
\left|
\begin{array}{l}
\displaystyle       \hbox{dim} [{\bf Gr}^{n}  {\pmb{\mathfrak{Z}} }  ] = n+k+nk  \\
\displaystyle     \hbox{dim} [ \Lambda^{n} T^{\star} {\pmb{\mathfrak{Z}} } ] = n+k+ \frac{(n+k)!}{n!k!}    
\end{array}
\right.
\quad \quad \quad \quad
\left|
\begin{array}{l}
\displaystyle    \hbox{dim} [ {\bf D}^{\tiny{\hbox{\dd}} \mathfrak{y}}_{q} {\pmb{\mathfrak{Z}} }] = nk    
\\
\displaystyle      \hbox{dim} [ \Lambda^{n} T^{\star}_{q} {\pmb{\mathfrak{Z}} } ] =   \frac{(n+k)!}{n!k!}      
\end{array}
\right.
\]
as well as: 
\[
\left|
\begin{array}{l}
\displaystyle   \hbox{dim} (\dttP_{q} (z)) = \frac{(n+k)!}{n!k!} - nk    
\\
\displaystyle     \hbox{dim} (\dttP_{q}^{h} (z)) = \hbox{dim} (\dttP_{q} (z))  -1      
\end{array}
\right.
\]

{\em Grassmannian viewpoint for 2D-Maxwell theory}.  In this section we follow the method developed by F. H\'elein and J. Kouneiher  \cite{FH-01}      \cite{HKHK01} \cite{HK-02}  \cite{HK-03} for the general study of variational problem on maps.
In the case of Maxwell theory, 
we   understand the problem  via the study of the map $A: T^\star {\cal X} \rightarrow \Bbb{R} $. Here the multisymplectic manifold is  ${\pmb{\cal M}} =  \Lambda^{2} T^{\star} (T^{\star} {\cal X})  = \Lambda^{2} T^{\star}  {\pmb{\mathfrak{Z} }}  $.
We picture a map by its graph ${\bf G}$  a $2$-dimensional submanifold  of $ T^\star {\cal X}   $. Now we consider a point $(x,A) \in {\bf G} \subset T^{\star} {\cal X}$. At $(x,A)$, we consider the tangent plane to the graph ${\bf G}$, which is described by vectors $\displaystyle X_1 = \frac{\partial}{\partial x^{1}} + v_{1\mu} \frac{\partial}{\partial A_\mu} $ and $\displaystyle  X_2 = \frac{\partial}{\partial x^{2}} + v_{2\mu} \frac{\partial}{\partial A_\mu} $. 

  The set of local coordinates on ${\bf Gr}^{2} ( T^\star {\cal X} )$ is described by  $(x^{\mu} , A_{\nu} , v_{\mu\nu})$,  whereas the set of local  coordinates   on $\pmb{\mathfrak{Z}} = T^{\star} {\cal X}$ is $ (x^{\mu}$, $A_{\mu} ) $. A basis ${\pmb{\cal B}} = {\pmb{\cal B}}   [\Lambda^{2} T^{\star}_{(x,A)} \pmb{\mathfrak{Z}}]$ of the $6$-dimensional space $\Lambda^{2} T^{\star}_{(x,A)} \pmb{\mathfrak{Z}}$ is:
\bee
 \left.
\begin{array}{lcl}
\displaystyle  {\pmb{\cal B}}    & = &  \displaystyle   
{\Big{\{}}
\dd x^{1} \wedge \dd x^{2} , \dd A_{\nu} \wedge \dd x^{\mu} , \dd A_{1} \wedge \dd A_{2}
{\Big{\}}}_{1 \leq \mu,\nu \leq 2}
\\
 \displaystyle      & = &  \displaystyle   
{\Big{\{}}
\dd x^{1} \wedge \dd x^{2} , \dd A_{1} \wedge \dd x^{1} , \dd A_{1} \wedge \dd x^{2} , \dd A_{2} \wedge \dd x^{1} , \dd A_{2} \wedge \dd x^{2}  , \dd A_{1} \wedge \dd A_{2}
{\Big{\}}}.
\end{array}
\right.
\eee
 For the Maxwell {\sffamily 2D}-theory we have   the following dimensions for the key spaces involved in the Grassmannian construction: 
\bee
\left|
\begin{array}{l}
\displaystyle       \hbox{dim} [{\bf Gr}^{n}  {\pmb{\mathfrak{Z}} }  ] = 8  \\
 \displaystyle     \hbox{dim} [ \Lambda^{n} T^{\star} {\pmb{\mathfrak{Z}} } ] = 10   
\end{array}
\right.
\quad \quad \quad \quad
\left|
\begin{array}{l}
\displaystyle    \hbox{dim} [ {\bf D}^{\tiny{\hbox{\dd}} \mathfrak{y}}_{q} {\pmb{\mathfrak{Z}} }] = 6   \\ 
 \displaystyle      \hbox{dim} [ \Lambda^{n} T^{\star}_{q} {\pmb{\mathfrak{Z}} } ] =  4     
\end{array}
\right.
\quad \quad \quad \quad  
\left|
\begin{array}{l}
\displaystyle   \hbox{dim} (\dttP_{q} (z)) = 2   \\ 
 \displaystyle     \hbox{dim} (\dttP_{q}^{h} (z)) = 1      
\end{array}
\right.
\eee
    Any form $  \hbox{\sffamily\bfseries\slshape p}  \in \Lambda^{2} T^{\star}_{(x,A)} \pmb{\mathfrak{Z}}$ can be identified with the coordinates $(\mathfrak{e} , \pi^{A_{\mu} \nu} , { {\varsigma}})$ such that: 
\bee
 \hbox{\sffamily\bfseries\slshape p}  = \mathfrak{e} \vol +  \varepsilon_{\rho\nu} \pi^{A_{\mu} \rho} \dd A_{\mu} \wedge \dd x^{\nu}  + { {\varsigma}} \dd A_{1} \wedge \dd A_{2}.
\eee
Since,  $\varepsilon_{\rho\nu}\dd x^\nu =  \dd \mathfrak{y}_ {\rho}  $, we observe $  \varepsilon_{\rho\nu} \pi^{A_{\mu} \rho} \dd A_{\mu} \wedge \dd x^{\nu} 
 =  -  \varepsilon_{\rho\nu} \pi^{A_{\mu} \rho}  \dd x^{\nu} \wedge  \dd A_{\mu}
 =  \pi^{A_{\mu} \rho} \dd A_{\mu} \wedge \vol_{\rho} $. A tangent space is identified with coordinates on the Grassman bundle    $ \hbox{\sffamily\bfseries\slshape t} \cong (v_{\mu\nu})  \in {\bf Gr}^{2}_{(x,A)}{\pmb{\mathfrak{Z}}}$. We describe the pairing $\langle  \hbox{\sffamily\bfseries\slshape t}  , \hbox{\sffamily\bfseries\slshape p}  \rangle  =  \hbox{\sffamily\bfseries\slshape p}  (X_{1} , X_{2})$. Let notice that ${\big{\{}} X_{\mu} {\big{\}}}_{1 \leq \mu \leq 2}$ describes a basis of the tangent space $ \hbox{\sffamily\bfseries\slshape t} $. We have: 
 \[
 \left.
\begin{array}{lcl}
\displaystyle   
\langle  \hbox{\sffamily\bfseries\slshape t}  , \hbox{\sffamily\bfseries\slshape p}  \rangle  & = &  \displaystyle   
 \mathfrak{e}  \dd x^{1} \wedge \dd x^{2}  (X_{1} , X_{2}) +   \varepsilon_{\rho\nu} \pi^{A_{\mu} \rho} \dd A_{\mu} \wedge \dd x^{\nu}   (X_{1} , X_{2})  + { {\varsigma}} \dd A_{1} \wedge \dd A_{2}  (X_{1} , X_{2})
 \\
 \displaystyle & = &  \displaystyle  \mathfrak{e}  + \pi^{A_{\mu} \nu} v_{\mu\nu} + {{\varsigma}} (v_{11}v_{22} - v_{12} v_{21}).
 \end{array}
 \right.
\]
and we define the function:
\bee
{\cal W} (x,A,  \hbox{\sffamily\bfseries\slshape t}  , \hbox{\sffamily\bfseries\slshape p} )
=
\langle  \hbox{\sffamily\bfseries\slshape t}  , \hbox{\sffamily\bfseries\slshape p}  \rangle  
- L (x,A , \hbox{\sffamily\bfseries\slshape t} ).
\eee
Notice that the Lagrangian density $L (x,A , \hbox{\sffamily\bfseries\slshape t} )$ is identified with a function on $ {\bf Gr}^{2}_{(x,A)}{\pmb{\mathfrak{Z}}}$. The tangent space $\hbox{\sffamily\bfseries\slshape t}$ is in correspondence with $\hbox{\sffamily\bfseries\slshape p}$  -  denoted   $\hbox{\sffamily\bfseries\slshape t}  {\pmb{\leftrightarrow}}  \hbox{\sffamily\bfseries\slshape p} $ - if and only if 
\bee\label{gldo830}  \frac{\partial {\cal W}}{\partial \hbox{\sffamily\bfseries\slshape t}}  (x,A,  \hbox{\sffamily\bfseries\slshape t}  , \hbox{\sffamily\bfseries\slshape p} ) = 0 
\eee
Now we are looking for the enlarged pseudofiber $\dttP_{q} (z)$ and the pseudofiber $\dttP_{q}^{h} (z)$. A parametrization of $\displaystyle {\big{\{}}   \hbox{\sffamily\bfseries\slshape z}  \in {\bf D}^2_{(x,A)} (T^{\star} {\cal X}) \ / \   \vol ( \hbox{\sffamily\bfseries\slshape z} ) > 0  {\big{\}}}$ is described via coordinates $(t,v_{\mu\nu})$ with: 
\bee
 \hbox{\sffamily\bfseries\slshape z} = t^2 \partial_1 \wedge \partial_2 + t \varepsilon^{\mu\nu} v_{\mu\rho} \frac{\partial }{\partial A_\rho}
\wedge \frac{\partial}{\partial x^{\nu}} + (v_{11} v_{22} - v_{12} v_{21}) \frac{\partial}{\partial A_1 }
\wedge \frac{\partial}{\partial A_2 }.
\eee
Elements $\delta  \hbox{\sffamily\bfseries\slshape z}   \in T_z {\bf D}^{n}_{q} (T^{\star} {\cal X}) $ are described by coordinates $\delta t$ and  $ \delta v_{\mu\rho}$:
\bee
\delta  \hbox{\sffamily\bfseries\slshape z}  = \delta t {\big{(}}  2t \partial_1 \wedge \partial_2  + 
 \varepsilon^{\mu\nu} v_{\mu\rho} \frac{\partial }{\partial A_\rho}
\wedge \frac{\partial}{\partial x^{\nu}}
{\big{)}}
+ \delta v_{\mu\rho}  {\big{(}}   t \varepsilon^{\mu\nu}   \frac{\partial }{\partial A_\rho}
\wedge \frac{\partial}{\partial x^{\nu}} +   \varepsilon^{\mu\nu} \varepsilon^{\rho\sigma} v_{\nu\sigma} \frac{\partial}{\partial A_1 }
\wedge \frac{\partial}{\partial A_2 }   {\big{)}}
\eee
We also consider   the parametrization 
${\big{\{}}   \hbox{\sffamily\bfseries\slshape z}^{\tiny{\vol}}  \in {\bf D}^{\tiny{\vol}}_{(x,A)} (T^{\star} {\cal X}) \ / \   \vol (z) = 1  {\big{\}}}$ in such a context, $  \hbox{\sffamily\bfseries\slshape z} $ is written:
\bee
  \hbox{\sffamily\bfseries\slshape z}^{\tiny{\vol}} =   \partial_1 \wedge \partial_2 +  \varepsilon^{\mu\nu} v_{\mu\rho} \frac{\partial }{\partial A_\rho}
\wedge \frac{\partial}{\partial x^{\nu}} + (v_{11} v_{22} - v_{12} v_{21}) \frac{\partial}{\partial A_1 }
\wedge \frac{\partial}{\partial A_2 }.
\eee
Now we consider $\delta   \hbox{\sffamily\bfseries\slshape z}^{\tiny{\vol}}  \in T_z {\bf D}^{\tiny{\vol}}_{q} (T^{\star} {\cal X})$: 
\bee
\delta  \hbox{\sffamily\bfseries\slshape z}^{\tiny{\vol}}  =    \delta v_{\mu\rho}  {\big{(}}    \varepsilon^{\mu\nu}   \frac{\partial }{\partial A_\rho}
\wedge \frac{\partial}{\partial x^{\nu}} +   \varepsilon^{\mu\nu} \varepsilon^{\rho\sigma} v_{\nu\sigma} \frac{\partial}{\partial A_1 }
\wedge \frac{\partial}{\partial A_2 }   {\big{)}}
\eee
Then, we describe the space $  [ T_{z}{\bf D}^{n}_{q}  {\pmb{\mathfrak{Z}}} ]^{\perp} 
   =   
 {\Big{\{}} \hbox{\sffamily\bfseries\slshape p}  \in  \Lambda^{n} T^{\star}_{q}  {\pmb{\mathfrak{Z}}}  \ / \ \forall \delta  \hbox{\sffamily\bfseries\slshape z}   \in  {T_{z}{\bf D}^{n}_{q}  {\pmb{\mathfrak{Z}}} } \ , \ \hbox{\sffamily\bfseries\slshape p} (\delta  \hbox{\sffamily\bfseries\slshape z}  ) = 0 {\Big{\}}}
$, So that: $\hbox{\sffamily\bfseries\slshape p}  \in {\big{(}}  T_z {\bf D}^{\tiny{\vol}}_{q} ( {\pmb{\mathfrak{Z}}} ) {\big{)}}^{\perp} $ is equivalent to $\forall \delta  \hbox{\sffamily\bfseries\slshape z} \in  
 T_z {\bf D}^{n}_{q} (T^{\star} {\cal X})   \ , \  \hbox{\sffamily\bfseries\slshape p}(\delta  \hbox{\sffamily\bfseries\slshape z}) = \langle \delta   \hbox{\sffamily\bfseries\slshape z}  ,   \hbox{\sffamily\bfseries\slshape p}   \rangle =  0
$. Notice that, 
\bee
\left.
\begin{array}{rcl}
\displaystyle      \langle \delta   \hbox{\sffamily\bfseries\slshape z}  ,   \hbox{\sffamily\bfseries\slshape p}   \rangle   & = &  \displaystyle
\Big{\langle}   \delta t {\big{(}}  2t \partial_1 \wedge \partial_2  + 
 \varepsilon^{\mu\nu} v_{\mu\rho} \frac{\partial }{\partial A_\rho}
\wedge \frac{\partial}{\partial x^{\nu}}
{\big{)}}
 ,   \hbox{\sffamily\bfseries\slshape p} 
 \Big{\rangle}
   \\
\displaystyle     & &  \displaystyle 
 + 
\Big{\langle}
 \delta v_{\mu\rho}  {\big{(}}   t \varepsilon^{\mu\nu}   \frac{\partial }{\partial A_\rho}
\wedge \frac{\partial}{\partial x^{\nu}} +   \varepsilon^{\mu\nu} \varepsilon^{\rho\sigma} v_{\nu\sigma} \frac{\partial}{\partial A_1 }
\wedge \frac{\partial}{\partial A_2 }   {\big{)}} , 
 \hbox{\sffamily\bfseries\slshape p} 
  \Big{\rangle} 
 \\
 \displaystyle     & = &  \displaystyle 
  \delta t (2 \mathfrak{e} + v_{\mu\nu} \pi^{A_{\mu} \nu} )
    +
     \delta v_{\mu\rho} (\pi^{A_\mu \rho} + \varepsilon^{\mu\nu} \varepsilon^{\rho\sigma} v_{\nu\sigma} \varsigma ).
\end{array}
\right.
\eee
On the other side, 
\[
 {\big{[}}  T_z {\bf D}^{\tiny{\vol}}_{q} (\mathfrak{Z})  {\big{]}}^{\perp} =  {\big{[}}  T_z {\bf D}^{\tiny{\vol}}_{q} (T^{\star} {\cal X})  {\big{]}}^{\perp}
= {\Big{\{}}   \hbox{\sffamily\bfseries\slshape p} \in \Lambda^{n} T^{\star}_{q} \mathfrak{Z} \ / \ \forall \delta  \hbox{\sffamily\bfseries\slshape z}^{\tiny{\vol}}   \in  
 T_z {\bf D}^{\tiny{\vol}  }_{q} (T^{\star} {\cal X})   \ , \  \hbox{\sffamily\bfseries\slshape p}( \delta  \hbox{\sffamily\bfseries\slshape z}^{\tiny{\vol}}  ) = 0
{\Big{\}}};
\]
gives
\bee
\left.
\begin{array}{rcl}
\displaystyle      \langle \delta   \hbox{\sffamily\bfseries\slshape z}  ,   \hbox{\sffamily\bfseries\slshape p}   \rangle   & = &    \displaystyle 
\Big{\langle}
 \delta v_{\mu\rho}  {\big{(}}   t \varepsilon^{\mu\nu}   \frac{\partial }{\partial A_\rho}
\wedge \frac{\partial}{\partial x^{\nu}} +   \varepsilon^{\mu\nu} \varepsilon^{\rho\sigma} v_{\nu\sigma} \frac{\partial}{\partial A_1 }
\wedge \frac{\partial}{\partial A_2 }   {\big{)}} , 
 \hbox{\sffamily\bfseries\slshape p} 
  \Big{\rangle} 
 \\
 \displaystyle     & = &  \displaystyle 
    \delta v_{\mu\rho} (\pi^{A_\mu \rho} + \varepsilon^{\mu\nu} \varepsilon^{\rho\sigma} v_{\nu\sigma} \varsigma ).
     \\
\end{array}
\right.
\eee
   The  Legendre correspondence  related to the   system~\eqref{abibi07}($\mathfrak{i}$)  is characterized by the functional  determinant 
  $ \displaystyle
   \pmb{\Delta} =   \left|
 \begin{array}{c}
 \displaystyle  \pmb{\Delta}^{\mu\nu}_{\rho\sigma}   \\
  \end{array}
  \right|
   = \left|
 \begin{array}{c}
 \displaystyle \frac{\partial \pi^{A_\mu \nu} }{ \partial \partial_{\rho} A_{\sigma} }  \\
  \end{array}
  \right|
  $, see \cite{Katstrup} \cite{HK-02} \cite{HK-03}.  We have   the following terms: 
  \[
\left|
\begin{array}{rcl}
\displaystyle       { \partial \pi^{{A}_{1}1} }/{\partial (\partial_{1}A_{1} ) }     & = &  \displaystyle 0 \\
\displaystyle       { \partial \pi^{{A}_{1}1} }/{\partial (\partial_{1}A_{2} )}  & = &  \displaystyle 0 \\
\displaystyle     { \partial \pi^{{A}_{1}1} }/{\partial (\partial_{2}A_{1} )}   & = &  \displaystyle 0 \\
\displaystyle      { \partial \pi^{{A}_{1}1} }/{\partial ( \partial_{2}A_{2}) }   & = &  \displaystyle  - {{\varsigma}} \\
\end{array}
\right.
\quad \quad \quad \quad  
\left|
\begin{array}{rcl}
\displaystyle       { \partial \pi^{{A}_{2}1} }/{\partial (\partial_{1}A_{1} ) }     & = &  \displaystyle 0 \\
\displaystyle       { \partial \pi^{{A}_{2}1} }/{\partial (\partial_{1}A_{2} )}  & = &  \displaystyle 1 \\
\displaystyle     { \partial \pi^{{A}_{2}1} }/{\partial (\partial_{2}A_{1} )}   & = &  \displaystyle  - (1 - \varsigma)  \\
\displaystyle      { \partial \pi^{{A}_{2}1} }/{\partial ( \partial_{2}A_{2}) }   & = &  \displaystyle  0 \\
\end{array}
\right.
\]
\[ 
\left|
\begin{array}{rcl}
\displaystyle       { \partial \pi^{{A}_{2}2} }/{\partial (\partial_{1}A_{1} ) }     & = &  \displaystyle - \varsigma \\
\displaystyle       { \partial \pi^{{A}_{2}2} }/{\partial (\partial_{1}A_{2} )}  & = &  \displaystyle 0 \\
\displaystyle     { \partial \pi^{{A}_{2}2} }/{\partial (\partial_{2}A_{1} )}   & = &  \displaystyle 0 \\
\displaystyle      { \partial \pi^{{A}_{2}2} }/{\partial ( \partial_{2}A_{2}) }   & = &  \displaystyle  0 \\
\end{array}
\right.
\quad \quad \quad \quad 
\left|
\begin{array}{rcl}
\displaystyle       { \partial \pi^{{A}_{1}2} }/{\partial (\partial_{1}A_{1} ) }     & = &  \displaystyle 0 \\
\displaystyle       { \partial \pi^{{A}_{1}2} }/{\partial (\partial_{1}A_{2} )}  & = &  \displaystyle - (1 - \varsigma)  \\
\displaystyle     { \partial \pi^{{A}_{1}2} }/{\partial (\partial_{2}A_{1} )}   & = &  \displaystyle 1 \\
\displaystyle      { \partial \pi^{{A}_{1}2} }/{\partial ( \partial_{2}A_{2}) }   & = &  \displaystyle  0 \\
\end{array}
\right.
  \]
Then, we have: 
\[
\left.
\begin{array}{rcl}
\displaystyle    \pmb{\Delta}     & = &  \displaystyle   
\left|
 \begin{array}{c@{}c@{}c@{}c}
 \displaystyle
 0   \ & \   \displaystyle  0    \ & \   \displaystyle 0  \ & \   \displaystyle  - {{\varsigma}}
   \\
    \displaystyle
0   \ & \   \displaystyle   1  \ & \    \displaystyle  - (1 - \varsigma)   \ & \  \displaystyle  0
   \\
    \displaystyle
0  \ & \   \displaystyle  - (1 - \varsigma)     \ & \    \displaystyle 1 \ & \  \displaystyle 0   
   \\
    \displaystyle
- {\varsigma}  \ & \   \displaystyle 0   \ & \    \displaystyle 0 \ & \  \displaystyle 0   
\end{array} \right| 
\\
 \displaystyle     & = &  \displaystyle        \varsigma 
\left|
 \begin{array}{c@{}c@{}c}
     \displaystyle
0   \ & \   \displaystyle   1  \ & \    \displaystyle  - (1 - \varsigma)    
   \\
    \displaystyle
0  \ & \   \displaystyle  - (1 - \varsigma)     \ & \    \displaystyle 1   
   \\
    \displaystyle
- {\varsigma}  \ \ & \ \   \displaystyle 0   \ \ & \ \    \displaystyle 0   
\end{array} \right|  = -  \varsigma^{2} 
\left|
 \begin{array}{c@{}c}
   \displaystyle   1  \ & \    \displaystyle  - (1 - \varsigma)    
   \\
   \displaystyle  - (1 - \varsigma)     \ & \    \displaystyle 1    
\end{array} \right|   
\\
 \displaystyle     & = &  \displaystyle    -  \varsigma^2  (1 -  (1-\varsigma)^{2}  )
=     \varsigma^2 ( 1 -   (1 + \varsigma^{2} - 2 \varsigma ) ) =  - \varsigma^4  +  2   \varsigma^{3}  
= (2-\varsigma) \varsigma^3  \\
\end{array}
\right.
\]
 Finally,
$ \displaystyle
  \pmb{\Delta} =   \left|
 \begin{array}{c}
 \displaystyle  \pmb{\Delta}^{\mu\nu}_{\rho\sigma}   \\
  \end{array}
  \right|
     \neq 0  \quad \hbox{if and only if} \quad {\Big{\{}} \varsigma \neq 0,2  {\Big{\}}}
$.
If we generally denote $  \forall q \in {\pmb{\mathfrak{Z}}}   $ the object  $  {\cal P}_{q}   = \bigcup_{ z \in {\bf D}_{q}  {\pmb{\mathfrak{Z}}} }  \dttP_{q} (z) $, we now find for the Maxwell $\hbox{\sffamily 2D}$ theory.
\bee\label{grass01}
\left.
\begin{array}{rcl}
\displaystyle     {\cal P}_{q}    & = &  \displaystyle  {\Big{\{}}  (\mathfrak{e} , \pi^{A_{\mu} \nu} , {\varsigma} )  \in \Lambda^{2} T^{\star}_{(x,A)} \pmb{\mathfrak{Z}} \ / \  \varsigma \neq 0 , 2  {\Big{\}}}  
\\
\displaystyle     &  &  \displaystyle  \bigcup
 {\Big{\{}}  (\mathfrak{e} , \pi^{A_{\mu} \nu} , 0 )  \in \Lambda^{2} T^{\star}_{(x,A)} \pmb{\mathfrak{Z}} \ / \   \pi^{A_{1} 1}   = \pi^{A_{2} 2} = \pi^{A_1 2}  + \pi^{A_2 1} = 0    {\Big{\}}}
  \\
 \displaystyle     &  &  \displaystyle   \bigcup
  {\Big{\{}}  (\mathfrak{e} , \pi^{A_{\mu} \nu} ,  2 )  \in \Lambda^{2} T^{\star}_{(x,A)} \pmb{\mathfrak{Z}} \ / \  \pi^{A_1 2}  -  \pi^{A_2 1} = 0   {\Big{\}}} . \\
\end{array}
\right.
\eee

 \section{{\hbox{\sffamily\bfseries\slshape{Conclusion}}}}

  In this note we have described the DeDonder-Weyl $n$-plectic theory for the Maxwell variational problem via the general method developed by F. H\'elein and J. Kouneiher in the series  of papers \cite{FH-01} \cite{H-03} \cite{H-02} \cite{HK-01} \cite{HKHK01} \cite{HK-02} \cite{HK-03}. The main focus is on the determination of algebraic observables, dynamical observables and their related observable functionals. We have also detailed some calculations related to the copolarization of  the Maxwell theory. The main result is the Poisson bracket  \eqref{abffebe16}~between canonical forms  $(A,{\pmb{\pi}})$. A strong result that merges the physical needs and the mathematical construction of copolarization is that we can {\em not}  include  $ \displaystyle \textswab{C}^{1}T^\star{\pmb{\cal M}}^{\tiny{\hbox{\sffamily Maxwell}}}    $ in   observable $1$-forms.  There is obviously two directions for further studies. The first is the issue of quantization whereas the second is a more detailed treatment of higher Lepagean equivalent theories. The quantization theory for $n$-plectic geometry is still at its infancy. We refer to some recent works: F. H\'elein \cite{FH-01} \cite{H-02}, R.D. Harrivel \cite{FFFDCX} \cite{FFFDCX2} and also the ''precanonical quantization''  developed by I.V. Kanatchikov  \cite{Kanaquantiz}  \cite{Kanaquantiz1}  \cite{Kanaquantiz2}. Higher Lepagean equivalent theories for the Maxwell variational problem is currently in preparation \cite{Veypro}.  
 
 \

${\hbox{\sffamily\bfseries\slshape{Acknowledgments}}}$   {I am   grateful to Fr\'ed\'eric H\'elein and Joseph Kouneiher for discussions   about the topic of multisymplectic geometry. Also, for his help and support, I address many thanks to Volodya Rubtsov. Finally, I am   grateful to Igor V. Kanatchikov for his interest in the manuscript and for some comments.}

 \appendix 
 
 \section{\sffamily\bfseries\slshape Calculation of $\langle p , v  \rangle = {\pmb{\theta}}^{\tiny{\hbox{\sffamily DDW}}}_p ({\cal Z})$}\label{appendixA}

  {Here we give the detailed  calculation for $\langle p , v  \rangle = {\pmb{\theta}}^{\tiny{\hbox{\sffamily DDW}}}_p ({\cal Z})$.   We denote
  \bee
   {\cal Z}_{\nu} = \frac{\partial }{\partial x^{\nu}} + \sum_{1 \leq {{\mu}} \leq n}  {\cal Z}_{\nu\mu} \frac{\partial}{\partial A_{\mu} } =  \sum_{1 \leq {\pmb{\mu}} \leq 2n  } {\cal Z}_{\nu}^{ {\pmb{\mu}} } \frac{\partial }{\partial q^{\pmb{\mu}}}.
   \eee
 We have $
q^{\pmb{\mu}} = x^{{\pmb{\mu}}} = x^{\nu}  \ \hbox{if} \  1 \leq {\pmb{\mu}} = \nu  \leq n \quad \hbox{and} \quad  q^{\pmb{\mu}} = A_{{\pmb{\mu}} - n } = A_{\mu} \ \hbox{if} \  1 \leq {\pmb{\mu}} - n = \mu  \leq n
 $. The bold index $1 \leq {\pmb{\mu}} \leq 2n$ is a multi-index such that ${\cal Z}_{\nu}^{ {\pmb{\mu}} }  = \delta^{{\pmb{\mu}}}_{\nu} $ for $1 \leq {\pmb{\mu}} \leq n$ and 
${\cal Z}_{\nu}^{ {\pmb{\mu}} }  = {\cal Z}_{\nu\mu} $ for $n+1 \leq {\pmb{\mu}} \leq 2 n$.   
\[
\left.
\begin{array}{ccl}
\displaystyle  {\cal Z}   &  =  & \displaystyle
  {\cal Z}_1 \wedge  {\cal Z}_2 \wedge  {\cal Z}_3 \wedge  {\cal Z}_4 =  \sum_{ {{\pmb{\mu}}_1} < \cdots < {\pmb{\mu}}_4 } {\pmb{\cal Z}}^{{{\pmb{\mu}}_1} \cdots {{\pmb{\mu}}_4} }_{1 \cdots 4}
\frac{\partial}{\partial q^{{{\pmb{\mu}}_1}}} \wedge \cdots \wedge \frac{\partial}{\partial q^{{{\pmb{\mu}}_4}}}
\\
 \displaystyle    &   =  & \displaystyle
 \sum_{ {{\pmb{\mu}}_1} < \cdots < {\pmb{\mu}}_4 }
\left| \begin{array}{ccc}
{ {\cal Z}^{{\pmb{\mu}}_1}_{1} } & \cdots & { {\cal Z}^{{\pmb{\mu}}_1}_{4}   }\\
\vdots & & \vdots \\
{{\cal Z}^{{\pmb{\mu}}_4}_{1} } & \dots & {{\cal Z}^{{\pmb{\mu}}_4}_{4} }
\end{array}\right|  
\frac{\partial}{\partial q^{{{\pmb{\mu}}_1}}} \wedge \cdots \wedge \frac{\partial}{\partial q^{{{\pmb{\mu}}_4}}}
\end{array}
\right.
\]
We expand the expression: 
\[
\left.
\begin{array}{ccl}
\displaystyle  {\cal Z}   &  =  & \displaystyle  {\pmb{\cal Z}}^{{1} 2 34 }_{1234}   \partial_1 \wedge \partial_2 \wedge \partial_3 \wedge \partial_4
+  \underbrace{ \sum_{  n < {\pmb{\mu}}_4 }
{\pmb{\cal Z}}^{{1} 2 3 {{\pmb{\mu}}_4} }_{1234}
\partial_1 \wedge \partial_2 \wedge \partial_3 \wedge \frac{\partial}{\partial q^{{{\pmb{\mu}}_4}}}   
}_{(\mathfrak{i})}
\\
 \displaystyle    &     & \displaystyle 
+  \underbrace{  \sum_{  n < {\pmb{\mu}}_4 }
{\pmb{\cal Z}}^{{1} 2 4 {{\pmb{\mu}}_4} }_{1234}
\partial_1 \wedge \partial_2 \wedge \partial_4 \wedge \frac{\partial}{\partial q^{{{\pmb{\mu}}_4}}} 
}_{[{\bf II}]}
 +  \underbrace{   \sum_{  3 < {\pmb{\mu}}_4 }
{\pmb{\cal Z}}^{{1}  3 4 {{\pmb{\mu}}_4} }_{1234}
\partial_1 \wedge \partial_3 \wedge \partial_4 \wedge \frac{\partial}{\partial q^{{{\pmb{\mu}}_4}}}  
}_{[{\bf III}]} 
\\
 \displaystyle    &     & \displaystyle  +
   \underbrace{  \sum_{  3 < {\pmb{\mu}}_4 }
{\pmb{\cal Z}}^{{2}  3 4 {{\pmb{\mu}}_4} }_{1234}
\partial_2 \wedge \partial_3 \wedge \partial_4 \wedge \frac{\partial}{\partial q^{{{\pmb{\mu}}_4}}}  
}_{[{\bf IV}]}
\end{array}
\right.
\]
  Now we detail the different terms involved: ${\pmb{\cal Z}}^{{1} 2 34 }_{1234} =  1$,
\[
{\pmb{\cal Z}}^{ 123 {{\pmb{\mu}}_4} }_{1234} =
 \left| \begin{array}{cccc}
{ 1 } & { 0 }  & { 0 } & {0 }\\
{  0 } & { 1 }  & { 0 } & { 0   }\\
{ 0 } & { 0}  & { 1 } & { 0 }\\
{ {\cal Z}^{{\pmb{\mu}}_4}_{1} } & { {\cal Z}^{{\pmb{\mu}}_4}_{2} }  & { {\cal Z}^{{\pmb{\mu}}_4}_{3} } & { {\cal Z}^{{\pmb{\mu}}_4}_{4}   }\\
\end{array}\right|  = {\cal Z}^{{\pmb{\mu}}_4}_{4 }  ,
\quad \quad  
{\pmb{\cal Z}}^{ 124 {{\pmb{\mu}}_4} }_{1234} =
\left| \begin{array}{cccc}
{ 1 } & { 0 }  & { 0 } & {0 }\\
{  0 } & { 1 }  & { 0 } & { 0   }\\
{  0 } & { 0 }  & { 0 } & { 1   }\\
{ {\cal Z}^{{\pmb{\mu}}_4}_{1} } & { {\cal Z}^{{\pmb{\mu}}_4}_{2} }  & { {\cal Z}^{{\pmb{\mu}}_4}_{3} } & { {\cal Z}^{{\pmb{\mu}}_4}_{4}   }\\
\end{array}\right|  =  - {\cal Z}^{{\pmb{\mu}}_4}_{3}    
\]
\[
{\pmb{\cal Z}}^{ 13 4 {{\pmb{\mu}}_4} }_{1234} =
\left| \begin{array}{cccc}
{ 1 } & { 0 }  & { 0 } & {0 }\\
{  0 } & { 0 }  & { 1 } & { 0   }\\
{  0 } & { 0 }  & { 0 } & { 1   }\\
{ {\cal Z}^{{\pmb{\mu}}_4}_{1} } & { {\cal Z}^{{\pmb{\mu}}_4}_{2} }  & { {\cal Z}^{{\pmb{\mu}}_4}_{3} } & { {\cal Z}^{{\pmb{\mu}}_4}_{4}   }\\
\end{array}\right|  =  {\cal Z}^{{\pmb{\mu}}_4}_{2},
\quad \quad  
{\pmb{\cal Z}}^{ 234  {{\pmb{\mu}}_4} }_{1234} =
\left| \begin{array}{cccc}
{ 0 } & { 1 }  & { 0 } & {0 }\\
{  0 } & { 0 }  & { 1 } & { 0}\\
{  0 } & { 0 }  & { 0 } & { 1}\\
{ {\cal Z}^{{\pmb{\mu}}_4}_{1} } & { {\cal Z}^{{\pmb{\mu}}_4}_{2} }  & { {\cal Z}^{{\pmb{\mu}}_4}_{3} } & { {\cal Z}^{{\pmb{\mu}}_4}_{4}   }\\
\end{array}\right|  =  -   {\cal Z}^{{\pmb{\mu}}_4}_{1}.
\]
Therefore we obtain: 
\[
(\mathfrak{i}) =  {\cal Z}^{{\pmb{\mu}}_4}_{4 }  \partial_1 \wedge \partial_2 \wedge \partial_3 \wedge \frac{\partial}{\partial q^{{{\pmb{\mu}}_4}}} ,
 \quad   (\mathfrak{ii}) =  - {\cal Z}^{{\pmb{\mu}}_4}_{3}\partial_1 \wedge \partial_2 \wedge \partial_4 \wedge \frac{\partial}{\partial q^{{{\pmb{\mu}}_4}}} ,
 \quad   (\mathfrak{iii}) =   {\cal Z}^{{\pmb{\mu}}_4}_{2}\partial_1 \wedge \partial_3 \wedge \partial_4 \wedge \frac{\partial}{\partial q^{{{\pmb{\mu}}_4}}}  
 \]
 and
 \[
 (\mathfrak{iv}) =  -  {\cal Z}^{{\pmb{\mu}}_4}_{1} \partial_2 \wedge \partial_3 \wedge \partial_4 \wedge  \frac{\partial}{\partial q^{{{\pmb{\mu}}_4}}}.  \ \hbox{Then we obtain the expression for ${\cal Z}$}:
\]
\[
{\cal Z} = 
\partial_1 \wedge \partial_2 \wedge \partial_3 \wedge \partial_4
+
  {\cal Z}_{4\mu} \partial_1 \wedge \partial_2 \wedge \partial_3 \wedge \frac{\partial}{\partial   A_{\mu} }  
-
 {\cal Z}_{3\mu} \partial_1 \wedge \partial_2 \wedge   \partial_4 \wedge  \frac{\partial}{\partial   A_{\mu} }   
 +
 {\cal Z}_{2\mu} \partial_1 \wedge \partial_3 \wedge \partial_4 \wedge \frac{\partial}{\partial   A_{\mu} } 
 -
{\cal Z}_{1\mu} \partial_2 \wedge \partial_3 \wedge \partial_4 \wedge \frac{\partial}{\partial   A_{\mu} } 
\]
Since, 
 $
\langle p , v  \rangle = {\pmb{\theta}}^{\tiny{\hbox{\sffamily DDW}}}_p ({\cal Z}) = \mathfrak{e} \hbox{d} \mathfrak{y} ({\cal Z})   + {\pmb{\pi}}^{{A}_{\mu}\nu}
\hbox{d}{A}_{\mu} \wedge \hbox{d} \mathfrak{y}_{\nu}  ({\cal Z}) 
$, 
we expand it\footnote{Notice that $\vol_1 = \partial_1 \iN \vol = (-1)^{1-1} \dd x^{2} \wedge \dd x^{3} \wedge \dd x^{4}  = \dd x^{2} \wedge \dd x^{3} \wedge \dd x^{4}  $ as well as $\vol_2 = - \dd x^{1} \wedge \dd x^{3} \wedge \dd x^{4} $, also $\vol_3 =  \dd x^{1} \wedge \dd x^{2} \wedge \dd x^{4}$ and finally $\vol_4 = - \dd x^{1} \wedge \dd x^{2} \wedge \dd x^{3} $.} as
\[
\langle p , v  \rangle = \mathfrak{e} \hbox{d} \mathfrak{y} ({\cal Z})     + {\pmb{\pi}}^{{A}_{\mu}1}
\hbox{d}{A}_{\mu} \wedge \hbox{d} \mathfrak{y}_{1}  ({\cal Z}) 
 + {\pmb{\pi}}^{{A}_{\mu}2}
\hbox{d}{A}_{\mu} \wedge \hbox{d} \mathfrak{y}_{2}  ({\cal Z}) 
 + {\pmb{\pi}}^{{A}_{\mu}3}
\hbox{d}{A}_{\mu} \wedge \hbox{d} \mathfrak{y}_{3}  ({\cal Z}) 
 + {\pmb{\pi}}^{{A}_{\mu}4}
\hbox{d}{A}_{\mu} \wedge \hbox{d} \mathfrak{y}_{4}  ({\cal Z}) ,
\]
so that: 
\[
\begin{array}{lll}
 \displaystyle \langle p , v  \rangle  & = & \displaystyle \mathfrak{e} \hbox{d} \mathfrak{y} (\partial_1 \wedge \partial_2 \wedge \partial_3 \wedge \partial_4 )
 \\
    &   & \displaystyle     + {\pmb{\pi}}^{{A}_{\mu}1}
\hbox{d}{A}_{\mu} \wedge \hbox{d} \mathfrak{y}_{1}  ( -
{\cal Z}_{1\mu} \partial_2 \wedge \partial_3 \wedge \partial_4 \wedge \frac{\partial}{\partial   A_{\mu_4} }  ) 
  + {\pmb{\pi}}^{{A}_{\mu}2}
\hbox{d}{A}_{\mu} \wedge \hbox{d} \mathfrak{y}_{2}  (  {\cal Z}_{2\mu}\partial_1 \wedge \partial_3 \wedge \partial_4 \wedge \frac{\partial}{\partial  A_{\mu}}  ) 
 \\
  \displaystyle    &  & \displaystyle     + {\pmb{\pi}}^{{A}_{\mu}3}
\hbox{d}{A}_{\mu} \wedge \hbox{d} \mathfrak{y}_{3}  (- {\cal Z}_{3\mu}\partial_1 \wedge \partial_2 \wedge \partial_4 \wedge \frac{\partial}{\partial q^{{{\pmb{\mu}}_4}}}  ) 
   + {\pmb{\pi}}^{{A}_{\mu}4}
\hbox{d}{A}_{\mu} \wedge \hbox{d} \mathfrak{y}_{4}  (  {\cal Z}_{4\mu }  \partial_1 \wedge \partial_2 \wedge \partial_3 \wedge \frac{\partial}{\partial  A_{\mu} }  ) ,  
\end{array}
\]
 
 so that  $  \displaystyle   \langle p , v  \rangle   =   \mathfrak{e}     + {\pmb{\pi}}^{{A}_{\mu}1}
{\cal Z}_{1\mu}
 + {\pmb{\pi}}^{{A}_{\mu}2}
 {\cal Z}_{2\mu} 
 + {\pmb{\pi}}^{{A}_{\mu}3}
  {\cal Z}_{3\mu} 
 + {\pmb{\pi}}^{{A}_{\mu}4}
   {\cal Z}_{4\mu } . \quad      \rfloor
$
   
 \section{\hbox{\sffamily\bfseries\slshape{Explicit 2D coordinates calculation for $X \iN  {\pmb{\Omega}}^{{\tiny\hbox{\sffamily DDW}}}  $}}      }\label{appendixB}

 {\em First case $n$-vector fields on  $  {\bf D}^{n}_{m} {\pmb{\cal M}}^{\hbox{\tiny{\sffamily DDW}}} \subset  \Lambda^{n} T {\pmb{\cal M}}^{\hbox{\tiny{\sffamily DDW}}} $} \bee
\left.
\begin{array}{rcl}
\displaystyle    X    & = &  \displaystyle {\big{(}} \partial_{1}   + \Theta_{1 {\mu}}   \partial^{A_{\mu}}  + {\Upsilon}_{1} {\pmb{\partial}}_{\mathfrak{e} }
+ {\Upsilon}_{1}^{A_\mu \nu}  {\pmb{\partial}}_{A_{\mu} \nu}    {\big{)}} \wedge {\big{(}}  \partial_{2}   + \Theta_{2 {\mu}}   \partial^{A_{\mu}}  + {\Upsilon}_{2} {\pmb{\partial}}_{\mathfrak{e} }
+ {\Upsilon}_{2}^{A_\mu \nu}  {\pmb{\partial}}_{A_{\mu} \nu}  
 {\big{)}}
 \\
\displaystyle     & = &  \displaystyle  \partial_1  \wedge {\big{(}}  \partial_2   + \Theta_{2 {\mu}}   \partial^{A_{\mu}}  + {\Upsilon}_{2} {\pmb{\partial}}_{\mathfrak{e} }
+ {\Upsilon}_{2}^{A_\mu \nu}   {\pmb{\partial}}_{A_{\mu} \nu}   
 {\big{)}}
  \\
\displaystyle     &  &  \displaystyle
 +    \Theta_{1 {\rho}}   \partial^{A_{\rho}}  
 \wedge {\big{(}}  \partial_2   + \Theta_{2 {\mu}}   \partial^{A_{\mu}}  + {\Upsilon}_{2} {\pmb{\partial}}_{\mathfrak{e} }
+ {\Upsilon}_{2}^{A_\mu \nu}   {\pmb{\partial}}_{A_{\mu} \nu}   
 {\big{)}}
 \\
\displaystyle     &  &  \displaystyle  + {\Upsilon}_{1} {\pmb{\partial}}_{\mathfrak{e} }
 \wedge
 {\big{(}}  \partial_2   + \Theta_{2 {\mu}}   \partial^{A_{\mu}}  + {\Upsilon}_{2} {\pmb{\partial}}_{\mathfrak{e} }
+ {\Upsilon}_{2}^{A_\mu \nu}  {\pmb{\partial}}_{A_{\mu} \nu}   
 {\big{)}} 
 \\
\displaystyle     &  &  \displaystyle
  + {\Upsilon}_{1}^{A_\rho \sigma} {\pmb{\partial}}_{A_{\rho} \sigma}  
 \wedge
 {\big{(}}  \partial_2   + \Theta_{2 {\mu}}   \partial^{A_{\mu}}  + {\Upsilon}_{2} {\pmb{\partial}}_{\mathfrak{e} }
 {\big{)}}   + {\Upsilon}_{1}^{A_\rho \sigma}   {\pmb{\partial}}_{A_{\rho} \sigma}  
 \wedge
 {\Upsilon}_{2}^{A_\mu \nu} {\pmb{\partial}}_{A_{\mu} \nu}   .
 \end{array}
\right.
\eee
 Since we work on a subset of  the {\sffamily (DDW)} multisymplectic manifold, we only keep track of the concerned terms: 
 \[
\left.
\begin{array}{rcl}
\displaystyle  X      & = &  \displaystyle \partial_1  \wedge   \partial_2   + \Theta_{2 {\mu}}  \partial_1  \wedge   \partial^{A_{\mu}}  + {\Upsilon}_{2}   \partial_1  \wedge {\pmb{\partial}}_{\mathfrak{e} }
+ {\Upsilon}_{2}^{A_\mu \nu}   \partial_1  \wedge  {\pmb{\partial}}_{A_{\mu} \nu} 
 +    \Theta_{1 {\mu}}   \partial^{A_{\mu}}  
 \wedge   \partial_2      
+ \Theta_{1 {\rho}}   \partial^{A_{\rho}}  
 \wedge  {\Upsilon}_{2}^{A_\mu \nu}  {\pmb{\partial}}_{A_{\mu} \nu}   
  \\
\displaystyle     &  &  \displaystyle  + {\Upsilon}_{1} {\pmb{\partial}}_{\mathfrak{e} }
 \wedge
 \partial_2   
  + {\Upsilon}_{1}^{A_\mu \nu}  {\pmb{\partial}}_{A_{\mu} \nu}   
 \wedge
   \partial_2    + {\Upsilon}_{1}^{A_\mu \nu}   {\pmb{\partial}}_{A_{\mu} \nu}    \wedge  \Theta_{2 {\rho}}   \partial^{A_{\rho}} 
  +  \  \hbox{terms} \  {\pmb{\partial}}_{\mathfrak{e} }\wedge  {\pmb{\partial}}_{A_{\mu} \nu}  \ \hbox{and} \ {\pmb{\partial}}_{A_{\mu} \nu}  \wedge {\pmb{\partial}}_{A_{\rho} \sigma} .
    \\
\end{array}
\right.
\]
Then we expand the indices: 
 \[
\left.
\begin{array}{rcl}
\displaystyle  X      & = &  \displaystyle \partial_1  \wedge   \partial_2   + \Theta_{2 {1}}  \partial_1  \wedge   \partial^{A_{1}} 
 +  \Theta_{2 {2}}  \partial_1  \wedge   \partial^{A_{2}}  +    \Theta_{1 {1}}   \partial^{A_{1}}  
 \wedge   \partial_2  +
\Theta_{1 {2}}   \partial^{A_{2}}  
 \wedge   \partial_2       
  + {\Upsilon}_{2}   \partial_1  \wedge {\pmb{\partial}}_{\mathfrak{e} } 
  \\
  \displaystyle     &  &  \displaystyle 
   + {\Upsilon}_{1} {\pmb{\partial}}_{\mathfrak{e} }
 \wedge
 \partial_2   
+ {\Upsilon}_{2}^{A_1 1}   \partial_1  \wedge  {\pmb{\partial}}_{A_{1} 1}  
  + {\Upsilon}_{2}^{A_1 2}   \partial_1  \wedge   {\pmb{\partial}}_{A_{1} 2}    
 + {\Upsilon}_{2}^{A_2 1}   \partial_1  \wedge  {\pmb{\partial}}_{A_{2} 1}   
  + {\Upsilon}_{2}^{A_2 2}   \partial_1  \wedge  {\pmb{\partial}}_{A_{2} 2}    
   \\
\displaystyle     &  &  \displaystyle    
 + \Theta_{1 {1}}   \partial^{A_{1}}  
 \wedge  
 {\big{(}}    {\Upsilon}_{2}^{A_1 1} {\pmb{\partial}}_{A_{1} 1} + {\Upsilon}_{2}^{A_1 2} {\pmb{\partial}}_{A_{1} 2}  + {\Upsilon}_{2}^{A_2 1} {\pmb{\partial}}_{A_{2} 1} + {\Upsilon}_{2}^{A_2 2} {\pmb{\partial}}_{A_{2} 2} 
  {\big{)}}  
    \\
\displaystyle     &  &  \displaystyle    
 + \Theta_{1 {2}}   \partial^{A_{2}}   
 \wedge  
 {\big{(}}    {\Upsilon}_{2}^{A_1 1} {\pmb{\partial}}_{A_{1} 1} + {\Upsilon}_{2}^{A_1 2} {\pmb{\partial}}_{A_{1} 2}  + {\Upsilon}_{2}^{A_2 1} {\pmb{\partial}}_{A_{2} 1} + {\Upsilon}_{2}^{A_2 2} {\pmb{\partial}}_{A_{2} 2}  
  {\big{)}}  
 \\
\displaystyle     &  &  \displaystyle      + {\Upsilon}_{1}^{A_1 1}  {\pmb{\partial}}_{A_{1} 1}   
 \wedge
   \partial_2   + {\Upsilon}_{1}^{A_1 2}  {\pmb{\partial}}_{A_{1} 2}   
 \wedge
   \partial_2   + {\Upsilon}_{1}^{A_2 1}  {\pmb{\partial}}_{A_{2} 1}   
 \wedge
   \partial_2   + {\Upsilon}_{1}^{A_2 2}  {\pmb{\partial}}_{A_{2} 2}   
 \wedge
   \partial_2  \\
\displaystyle     &  &  \displaystyle    
    + {\big{(}}   {\Upsilon}_{1}^{A_1 1}   {\pmb{\partial}}_{A_{1} 1}  +  {\Upsilon}_{1}^{A_1 2}   {\pmb{\partial}}_{A_{1} 2}  +  {\Upsilon}_{1}^{A_2 1}   {\pmb{\partial}}_{A_{2} 1}  +  {\Upsilon}_{1}^{A_2 2}   {\pmb{\partial}}_{A_{2} 2}      {\big{)}}   \wedge    \Theta_{2 {1}}   \partial^{A_{1}} 
    \\
\displaystyle     &  &  \displaystyle    
    + {\big{(}}   {\Upsilon}_{1}^{A_1 1}   {\pmb{\partial}}_{A_{1} 1}  +  {\Upsilon}_{1}^{A_1 2}   {\pmb{\partial}}_{A_{1} 2}  +  {\Upsilon}_{1}^{A_2 1}   {\pmb{\partial}}_{A_{2} 1}  +  {\Upsilon}_{1}^{A_2 2}   {\pmb{\partial}}_{A_{2} 2}      {\big{)}}      \wedge  \Theta_{2 {2}}   \partial^{A_{2}} 
 \\
 \displaystyle     &  &  \displaystyle +  \  \hbox{terms involving} \  {\pmb{\partial}}_{\mathfrak{e} }\wedge  {\pmb{\partial}}_{A_{\mu} \nu}  \ \hbox{and} \ {\pmb{\partial}}_{A_{\mu} \nu}  \wedge {\pmb{\partial}}_{A_{\rho} \sigma} 
    \\
\end{array}
\right.
\]
Therefore we explicitly have the following calculation: 
 \[
\left.
\begin{array}{rcl}
\displaystyle      X \iN  {\pmb{\Omega}}^{{\tiny\hbox{\sffamily DDW}}}    & = &  \displaystyle 
X \iN {\big{(}}   \hbox{d} \mathfrak{e} \wedge   \dd x^{1} \wedge \dd x^{2} + \dd  {\pmb{\pi}}^{A_{1} \nu } 
 \wedge \dd  {A}_{1} \wedge   \hbox{d} \mathfrak{y}_{\nu}  + \dd  {\pmb{\pi}}^{A_{2} \nu } 
 \wedge \dd  {A}_{2} \wedge   \hbox{d} \mathfrak{y}_{\nu}  {\big{)}}
 \\
\displaystyle     & = &  \displaystyle  X \iN {\big{(}}   \hbox{d} \mathfrak{e} \wedge   \dd x^{1} \wedge \dd x^{2} {\big{)}}  +   X \iN {\big{(}} 
  \dd  {\pmb{\pi}}^{A_{1} 1 } 
 \wedge \dd  {A}_{1} \wedge   \hbox{d} \mathfrak{y}_{1}  + \dd  {\pmb{\pi}}^{A_{2} 1 } 
 \wedge \dd  {A}_{2} \wedge   \hbox{d} \mathfrak{y}_{1}  {\big{)}}
\\
\displaystyle     &  &  \displaystyle 
+   X \iN {\big{(}} 
 \dd  {\pmb{\pi}}^{A_{1} 2 } 
 \wedge \dd  {A}_{1} \wedge   \hbox{d} \mathfrak{y}_{2}  
 + \dd  {\pmb{\pi}}^{A_{2} 2 } 
 \wedge \dd  {A}_{2} \wedge   \hbox{d} \mathfrak{y}_{2}  {\big{)}}
 \\
\displaystyle     & = &  \displaystyle   X \iN ( \hbox{d} \mathfrak{e} \wedge  \hbox{d} \mathfrak{y} ) +        X \iN (  \dd  {\pmb{\pi}}^{A_{1} 1 } 
 \wedge \dd  {A}_{1} \wedge   \hbox{d}  x^{2}    -  \dd  {\pmb{\pi}}^{A_{1} 2 }  \wedge \dd  {A}_{1} \wedge  \dd x^{1}  
 )
  \\
\displaystyle     &  &  \displaystyle 
+ X \iN (
  \dd  {\pmb{\pi}}^{A_{2} 1 } 
 \wedge \dd  {A}_{2} \wedge \dd x^{2} -  \dd  {\pmb{\pi}}^{A_{2} 2 } 
 \wedge \dd  {A}_{2} \wedge   \hbox{d} x^{1}  ),
 \end{array}
\right.
\]
\[
\left.
\begin{array}{rcl}
\displaystyle    X \iN  {\pmb{\Omega}}^{{\tiny\hbox{\sffamily DDW}}}      & = &  \displaystyle  \hbox{d} \mathfrak{y} (X) \hbox{d}\mathfrak{e} - (\hbox{d} \mathfrak{e} \wedge  \hbox{d} \mathfrak{y}_1) (X) \dd x^1 - (\hbox{d} \mathfrak{e} \wedge  \hbox{d} \mathfrak{y}_2) (X) \dd x^2 \\ 
\displaystyle     &   &  \displaystyle 
+  (\hbox{d}{A}_{1}  \wedge \dd x^{2} ) (X)   \hbox{d} {\pmb{\pi}}^{{A}_{1}1} -  (  \hbox{d} {\pmb{\pi}}^{{A}_{1}1} \wedge
 \dd x^{2}   ) (X)\hbox{d}{A}_{1} + (   \hbox{d} {\pmb{\pi}}^{{A}_{1}1} \wedge
\dd{A}_{1}   ) (X) \dd x^2 
  \\
\displaystyle     &   &  \displaystyle 
-  (\hbox{d}{A}_{1}  \wedge \dd x^{1} ) (X)   \hbox{d} {\pmb{\pi}}^{{A}_{1}2} +  (  \hbox{d} {\pmb{\pi}}^{{A}_{1}2} \wedge
 \dd x^{1}   ) (X)\hbox{d}{A}_{1} - (   \hbox{d} {\pmb{\pi}}^{{A}_{1}2} \wedge
\dd{A}_{1}   ) (X) \dd x^1 
  \\
\displaystyle     &  &  \displaystyle +  (\hbox{d}{A}_{2}  \wedge \dd x^{2} ) (X)   \hbox{d} {\pmb{\pi}}^{{A}_{2}1} -  (  \hbox{d} {\pmb{\pi}}^{{A}_{2}1} \wedge
\dd x^{2}  ) (X)\hbox{d}{A}_{2}   + (   \hbox{d} {\pmb{\pi}}^{{A}_{2}1} \wedge
\dd{A}_{2}    ) (X) \dd x^2 
  \\
  \displaystyle     &   &  \displaystyle 
-  (\hbox{d}{A}_{2}  \wedge \dd x^{1} ) (X)   \hbox{d} {\pmb{\pi}}^{{A}_{2}2} +  (  \hbox{d} {\pmb{\pi}}^{{A}_{2}2} \wedge
 \dd x^{1}   ) (X)\hbox{d}{A}_{2} - (   \hbox{d} {\pmb{\pi}}^{{A}_{2}2} \wedge
\dd{A}_{2}   ) (X) \dd x^1 
  \\
\displaystyle     & =  &  \displaystyle   \hbox{d}\mathfrak{e}  - \Upsilon_{1} \dd  x^1 - \Upsilon_{2} \dd x^2       + \Theta_{1 {1}}    \hbox{d} {\pmb{\pi}}^{{A}_{1}1}  + \Theta_{2 {1}}    \hbox{d} {\pmb{\pi}}^{{A}_{1}2}  + \Theta_{1 {2}}    \hbox{d} {\pmb{\pi}}^{{A}_{2}1}  + \Theta_{2 {2}}    \hbox{d} {\pmb{\pi}}^{{A}_{2}2}  
  \\
\displaystyle     &    &  \displaystyle    -  (    {\Upsilon}_{1}^{A_1 1}  +  {\Upsilon}_{2}^{A_1 2}    ) \hbox{d}{A}_{1}    -  (    {\Upsilon}_{1}^{A_2 1}  +  {\Upsilon}_{2}^{A_2 2}    ) \hbox{d}{A}_{2}
  \\
\displaystyle     &    &  \displaystyle  
     -  {\big{(}}  { {{\Upsilon}}}_{1}^{A_1 2}       { {\Theta}}_{21}  + { {{\Upsilon}}}_{1}^{A_2 2}       { {\Theta}}_{22}    -  \Theta_{1 {1}}   {\Upsilon}_{2}^{A_1 2}  -  \Theta_{1 {2}}   {\Upsilon}_{2}^{A_2 2}      
      {\big{)}} \dd x^{1}
       \\
\displaystyle     &    &  \displaystyle
  -  {\big{(}}  { {{\Upsilon}}}_{2}^{A_1 1}       { {\Theta}}_{11} +  { {{\Upsilon}}}_{2}^{A_2 1}       { {\Theta}}_{12}   - { {{\Upsilon}}}_{1}^{A_1 1}  { {\Theta}}_{21}     - { {{\Upsilon}}}_{1}^{A_2 1}  { {\Theta}}_{22}  {\big{)}}  \dd  x^2.
\end{array}
\right.
\]

  {\em Second case: $n$-vector fields on  $ {\pmb{\cal M}}^{\hbox{\tiny{\sffamily Maxwell}}} \subset  \Lambda^{n} T {\pmb{\cal M}}^{\hbox{\tiny{\sffamily Maxwell}}}$}. Now, we consider $X   = X_1 \wedge X_2   \in  {\bf D}^{n}_{m} {\pmb{\cal M}}^{\hbox{\tiny{\sffamily Maxwell}}} \subset  \Lambda^{n} T {\pmb{\cal M}}^{\hbox{\tiny{\sffamily Maxwell}}} $ is written: 
\bee
\left.
\begin{array}{rcl}
\displaystyle    X    & = &  \displaystyle {\big{(}} \partial_{1}   + \Theta_{1 {\mu}}   \partial^{A_{\mu}}  + {\Upsilon}_{1} {\pmb{\partial}}_{\mathfrak{e} }
+ {\Upsilon}_{1}^{A_\mu \nu} {\big{(}} {\pmb{\partial}}_{A_{\mu} \nu}  - {\pmb{\partial}}_{A_{\nu} \mu}  {\big{)}} {\big{)}} 
  \\
\displaystyle     &  &  \displaystyle 
\wedge {\big{(}}  \partial_{2}   + \Theta_{2 {\mu}}   \partial^{A_{\mu}}  + {\Upsilon}_{2} {\pmb{\partial}}_{\mathfrak{e} }
+ {\Upsilon}_{2}^{A_\mu \nu} {\big{(}} {\pmb{\partial}}_{A_{\mu} \nu}  - {\pmb{\partial}}_{A_{\nu} \mu}  {\big{)}}
 {\big{)}}
 \\
\displaystyle     & = &  \displaystyle  \partial_1  \wedge {\big{(}}  \partial_2   + \Theta_{2 {\mu}}   \partial^{A_{\mu}}  + {\Upsilon}_{2} {\pmb{\partial}}_{\mathfrak{e} }
+ {\Upsilon}_{2}^{A_\mu \nu} {\big{(}} {\pmb{\partial}}_{A_{\mu} \nu}  - {\pmb{\partial}}_{A_{\nu} \mu}  {\big{)}}
 {\big{)}}
  \\
\displaystyle     &  &  \displaystyle 
 +    \Theta_{1 {\rho}}   \partial^{A_{\rho}}  
 \wedge {\big{(}}  \partial_2   + \Theta_{2 {\mu}}   \partial^{A_{\mu}}  + {\Upsilon}_{2} {\pmb{\partial}}_{\mathfrak{e} }
+ {\Upsilon}_{2}^{A_\mu \nu} {\big{(}} {\pmb{\partial}}_{A_{\mu} \nu}  - {\pmb{\partial}}_{A_{\nu} \mu}  {\big{)}}
 {\big{)}}
 \\
\displaystyle     &  &  \displaystyle  + {\Upsilon}_{1} {\pmb{\partial}}_{\mathfrak{e} }
 \wedge
 {\big{(}}  \partial_2   + \Theta_{2 {\mu}}   \partial^{A_{\mu}}  + {\Upsilon}_{2} {\pmb{\partial}}_{\mathfrak{e} }
+ {\Upsilon}_{2}^{A_\mu \nu} {\big{(}} {\pmb{\partial}}_{A_{\mu} \nu}  - {\pmb{\partial}}_{A_{\nu} \mu}  {\big{)}}
 {\big{)}} 
  \\
\displaystyle     &  &  \displaystyle 
  + {\Upsilon}_{1}^{A_\rho \sigma} {\big{(}} {\pmb{\partial}}_{A_{\rho} \sigma}  - {\pmb{\partial}}_{A_{\sigma} \rho}  {\big{)}}
 \wedge
 {\big{(}}  \partial_2   + \Theta_{2 {\mu}}   \partial^{A_{\mu}}  + {\Upsilon}_{2} {\pmb{\partial}}_{\mathfrak{e} }
 {\big{)}}   
   \\
\displaystyle     &  &  \displaystyle 
+ {\Upsilon}_{1}^{A_\rho \sigma} {\big{(}} {\pmb{\partial}}_{A_{\rho} \sigma}  - {\pmb{\partial}}_{A_{\sigma} \rho}  {\big{)}}
 \wedge
 {\Upsilon}_{2}^{A_\mu \nu} {\big{(}} {\pmb{\partial}}_{A_{\mu} \nu}  - {\pmb{\partial}}_{A_{\nu} \mu}  {\big{)}}.
\end{array}
\right.
\eee
  Since we work on a subset of the {\sffamily (DDW)} multisymplectic manifold, we only keep track of the concerned terms:  
 \[
\left.
\begin{array}{rcl}
\displaystyle  X      & = &  \displaystyle \partial_1  \wedge   \partial_2   + \Theta_{2 {\mu}}  \partial_1  \wedge   \partial^{A_{\mu}}  + {\Upsilon}_{2}   \partial_1  \wedge {\pmb{\partial}}_{\mathfrak{e} }
+ {\Upsilon}_{2}^{A_\mu \nu}   \partial_1  \wedge {\big{(}} {\pmb{\partial}}_{A_{\mu} \nu}  - {\pmb{\partial}}_{A_{\nu} \mu}  
 {\big{)}}
 \\
 \displaystyle     &  &  \displaystyle 
 +    \Theta_{1 {\mu}}   \partial^{A_{\mu}}  
 \wedge   \partial_2      
+ \Theta_{1 {\rho}}   \partial^{A_{\rho}}  
 \wedge  {\Upsilon}_{2}^{A_\mu \nu} {\big{(}} {\pmb{\partial}}_{A_{\mu} \nu}  - {\pmb{\partial}}_{A_{\nu} \mu}  {\big{)}}
  \\
\displaystyle     &  &  \displaystyle  + {\Upsilon}_{1} {\pmb{\partial}}_{\mathfrak{e} }
 \wedge
 \partial_2   
  + {\Upsilon}_{1}^{A_\mu \nu} {\big{(}} {\pmb{\partial}}_{A_{\mu} \nu}  - {\pmb{\partial}}_{A_{\nu} \mu}  {\big{)}}
 \wedge
   \partial_2    + {\Upsilon}_{1}^{A_\mu \nu} {\big{(}} {\pmb{\partial}}_{A_{\mu} \nu}  - {\pmb{\partial}}_{A_{\nu} \mu}  {\big{)}} \wedge  \Theta_{2 {\rho}}   \partial^{A_{\rho}} 
       \\
\displaystyle     &  &  \displaystyle +  \  \hbox{terms involving} \  {\pmb{\partial}}_{\mathfrak{e} }\wedge  {\pmb{\partial}}_{A_{\mu} \nu}  \ \hbox{and} \ {\pmb{\partial}}_{A_{\mu} \nu}  \wedge {\pmb{\partial}}_{A_{\rho} \sigma}  .
    \\
\end{array}
\right.
\]
Then we expand the indices: 
 \[
\left.
\begin{array}{rcl}
\displaystyle  X      & = &  \displaystyle \partial_1  \wedge   \partial_2   + \Theta_{2 {1}}  \partial_1  \wedge   \partial^{A_{1}} 
 +  \Theta_{2 {2}}  \partial_1  \wedge   \partial^{A_{2}} 
   \\
\displaystyle     &  &  \displaystyle 
  + {\Upsilon}_{2}   \partial_1  \wedge {\pmb{\partial}}_{\mathfrak{e} } + {\Upsilon}_{1} {\pmb{\partial}}_{\mathfrak{e} }  \wedge
 \partial_2   +    \Theta_{1 {1}}   \partial^{A_{1}}  
 \wedge   \partial_2  +
\Theta_{1 {2}}   \partial^{A_{2}}  
 \wedge   \partial_2            
  \\
  \displaystyle     &  &  \displaystyle 
+ {\Upsilon}_{2}^{A_1 2}   \partial_1  \wedge {\big{(}} {\pmb{\partial}}_{A_{1} 2}  - {\pmb{\partial}}_{A_{2} 1}   
 {\big{)}}
  + {\Upsilon}_{2}^{A_2 1}   \partial_1  \wedge {\big{(}} {\pmb{\partial}}_{A_{2} 1}  - {\pmb{\partial}}_{A_{1} 2}   
 {\big{)}} 
  \\
\displaystyle     &  &  \displaystyle
 + {\big{(}} \Theta_{1 {1}}   \partial^{A_{1}}   +  \Theta_{1 {2}}   \partial^{A_{2}}     {\big{)}}
 \wedge     {\Upsilon}_{2}^{A_1 2} {\big{(}} {\pmb{\partial}}_{A_{1} 2}  - {\pmb{\partial}}_{A_{2} 1}  {\big{)}}
+  {\big{(}}  \Theta_{1 {1}}   \partial^{A_{1}}  +  \Theta_{1 {2}}   \partial^{A_{2}}   {\big{)}}  
 \wedge  {\Upsilon}_{2}^{A_2 1} {\big{(}} {\pmb{\partial}}_{A_{2} 1}  - {\pmb{\partial}}_{A_{1} 2}  {\big{)}}
 \\
\displaystyle     &  &  \displaystyle  
  + {\Upsilon}_{1}^{A_1 2} {\big{(}} {\pmb{\partial}}_{A_{1} 2}  - {\pmb{\partial}}_{A_{2} 1}  {\big{)}}
 \wedge
   \partial_2  
     + {\Upsilon}_{1}^{A_2 1} {\big{(}} {\pmb{\partial}}_{A_{2} 1}  - {\pmb{\partial}}_{A_{1} 2}  {\big{)}}
 \wedge
   \partial_2
 \\
\displaystyle     &  &  \displaystyle   + {\Upsilon}_{1}^{A_1 2} {\big{(}} {\pmb{\partial}}_{A_{1} 2}  - {\pmb{\partial}}_{A_{2} 1}  {\big{)}} \wedge {\big{(}}   \Theta_{2 {1}}   \partial^{A_{1}} + \Theta_{2 {2}}   \partial^{A_{2}}   {\big{)}}
         + {\Upsilon}_{1}^{A_2 1} {\big{(}} {\pmb{\partial}}_{A_{2} 1}  - {\pmb{\partial}}_{A_{1} 2}  {\big{)}} \wedge {\big{(}} \Theta_{2 {1}}   \partial^{A_{1}} + \Theta_{2 {2}}   \partial^{A_{2}}   {\big{)}}
 \\
 \displaystyle     &  &  \displaystyle +  \  \hbox{terms involving} \  {\pmb{\partial}}_{\mathfrak{e} }\wedge  {\pmb{\partial}}_{A_{\mu} \nu}  \ \hbox{and} \ {\pmb{\partial}}_{A_{\mu} \nu}  \wedge {\pmb{\partial}}_{A_{\rho} \sigma} .
    \\
\end{array}
\right.
\]
Therefore we explicitly have the following calculation: 
 \[
\left.
\begin{array}{rcl}
\displaystyle      X \iN  {\pmb{\Omega}}^{{\tiny\hbox{\sffamily DDW}}}    & = &  \displaystyle 
X \iN {\big{(}}   \hbox{d} \mathfrak{e} \wedge   \dd x^{1} \wedge \dd x^{2} + \dd  {\pmb{\pi}}^{A_{1} \nu } 
 \wedge \dd  {A}_{1} \wedge   \hbox{d} \mathfrak{y}_{\nu}  + \dd  {\pmb{\pi}}^{A_{2} \nu } 
 \wedge \dd  {A}_{2} \wedge   \hbox{d} \mathfrak{y}_{\nu}  {\big{)}}
 \\
\displaystyle     & = &  \displaystyle  X \iN {\big{(}}   \hbox{d} \mathfrak{e} \wedge   \dd x^{1} \wedge \dd x^{2} {\big{)}}  +   X \iN {\big{(}} 
  \dd  {\pmb{\pi}}^{A_{1} 1 } 
 \wedge \dd  {A}_{1} \wedge   \hbox{d} \mathfrak{y}_{1}  + \dd  {\pmb{\pi}}^{A_{2} 1 } 
 \wedge \dd  {A}_{2} \wedge   \hbox{d} \mathfrak{y}_{1}  {\big{)}}
\\
\displaystyle     &  &  \displaystyle 
+   X \iN {\big{(}} 
 \dd  {\pmb{\pi}}^{A_{1} 2 } 
 \wedge \dd  {A}_{1} \wedge   \hbox{d} \mathfrak{y}_{2}  
 + \dd  {\pmb{\pi}}^{A_{2} 2 } 
 \wedge \dd  {A}_{2} \wedge   \hbox{d} \mathfrak{y}_{2}  {\big{)}}
 \\
\displaystyle     & = &  \displaystyle   X \iN ( \hbox{d} \mathfrak{e} \wedge  \hbox{d} \mathfrak{y} ) +        X \iN ( -  \dd  {\pmb{\pi}}^{A_{1} 2 } 
 \wedge \dd  {A}_{1} \wedge  \dd x^{1}  + \dd  {\pmb{\pi}}^{A_{2} 1 } 
 \wedge \dd  {A}_{2} \wedge \dd x^{2}  ).
 \end{array}
\right.
\]
Since, ${\pmb{\pi}}^{A_{1}1 }  = {\pmb{\pi}}^{A_{2} 2} = 0$,
\[
\left.
\begin{array}{rcl}
\displaystyle    X \iN  {\pmb{\Omega}}^{{\tiny\hbox{\sffamily DDW}}}      & = &  \displaystyle  \hbox{d} \mathfrak{y} (X) \hbox{d}\mathfrak{e} - (\hbox{d} \mathfrak{e} \wedge  \hbox{d} \mathfrak{y}_1) (X) \dd x^1 - (\hbox{d} \mathfrak{e} \wedge  \hbox{d} \mathfrak{y}_2) (X) \dd x^2 \\ 
\displaystyle     &   &  \displaystyle 
-  (\hbox{d}{A}_{1}  \wedge \dd x^{1} ) (X)   \hbox{d} {\pmb{\pi}}^{{A}_{1}2} +  (  \hbox{d} {\pmb{\pi}}^{{A}_{1}2} \wedge
 \dd x^{1}   ) (X)\hbox{d}{A}_{1} - (   \hbox{d} {\pmb{\pi}}^{{A}_{1}2} \wedge
\dd{A}_{1}   ) (X) \dd x^1 
  \\
\displaystyle     &  &  \displaystyle +  (\hbox{d}{A}_{2}  \wedge \dd x^{2} ) (X)   \hbox{d} {\pmb{\pi}}^{{A}_{2}1} -  (  \hbox{d} {\pmb{\pi}}^{{A}_{2}1} \wedge
\dd x^{2}  ) (X)\hbox{d}{A}_{2}   + (   \hbox{d} {\pmb{\pi}}^{{A}_{2}1} \wedge
\dd{A}_{2}    ) (X) \dd x^2 
  \\
\displaystyle     & =  &  \displaystyle   \hbox{d}\mathfrak{e}  - (\hbox{d} \mathfrak{e} \wedge  \dd x^{2}) (X) \dd  x^1 + (\hbox{d} \mathfrak{e} \wedge  \dd  x^{1}) (X) \dd x^2
  \\
\displaystyle     &    &  \displaystyle
+ \Theta_{2 {1}}    \hbox{d} {\pmb{\pi}}^{{A}_{1}2} +  (  {\Upsilon}_{2}^{A_2 1}   -  {\Upsilon}_{2}^{A_1 2}    ) \hbox{d}{A}_{1}   - (   \hbox{d} {\pmb{\pi}}^{{A}_{1}2} \wedge
\dd{A}_{1}   ) (X) \dd  x^1   
  \\
\displaystyle     &  &  \displaystyle +  \Theta_{1 {2}}   \hbox{d} {\pmb{\pi}}^{{A}_{2}1} +  (  {\Upsilon}_{1}^{A_1 2}  - {\Upsilon}_{1}^{A_2 1}  )\hbox{d}{A}_{2}   + (   \hbox{d} {\pmb{\pi}}^{{A}_{2}1} \wedge
\dd{A}_{2}    ) (X) \dd x^2,
\end{array}
\right.
\]
\[
\left.
\begin{array}{rcl}
\displaystyle      X \iN  {\pmb{\Omega}}^{{\tiny\hbox{\sffamily DDW}}}   & = &  \displaystyle   \hbox{d}\mathfrak{e}    - \Upsilon_{1} \dd  x^1 - \Upsilon_{2} \dd x^2     
+ \Theta_{2 {1}}    \hbox{d} {\pmb{\pi}}^{{A}_{1}2} +  (  {\Upsilon}_{2}^{A_2 1}   -  {\Upsilon}_{2}^{A_1 2}    ) \hbox{d}{A}_{1}   - (   \hbox{d} {\pmb{\pi}}^{{A}_{1}2} \wedge
\dd{A}_{1}   ) (X) \dd x^1   
\\
\displaystyle     &   &  \displaystyle +  \Theta_{1 {2}}   \hbox{d} {\pmb{\pi}}^{{A}_{2}1} +  (  {\Upsilon}_{1}^{A_1 2}  - {\Upsilon}_{1}^{A_2 1}  )\hbox{d}{A}_{2}   + (   \hbox{d} {\pmb{\pi}}^{{A}_{2}1} \wedge
\dd{A}_{2}    ) (X) \dd x^2
 \\
\displaystyle     & = &  \displaystyle  \hbox{d}\mathfrak{e}    - \Upsilon_{1} \dd x^1 - \Upsilon_{2} \dd  x^2     
+ \Theta_{2 {1}}    \hbox{d} {\pmb{\pi}}^{{A}_{1}2} + \Theta_{1 {2}}   \hbox{d} {\pmb{\pi}}^{{A}_{2}1} 
 \\
\displaystyle     &  &  \displaystyle
+  (  {\Upsilon}_{2}^{A_2 1}   -  {\Upsilon}_{2}^{A_1 2}    ) \hbox{d}{A}_{1}   +  (  {\Upsilon}_{1}^{A_1 2}  - {\Upsilon}_{1}^{A_2 1}  )\hbox{d}{A}_{2}  
 \\
\displaystyle     &  &  \displaystyle   +  {\big{(}} ( {\Upsilon}_{2}^{A_1 2}  \Theta_{1 {2}}  - {\Upsilon}_{2}^{A_2 1}  \Theta_{1 {2}}   ) - (  \Theta_{2 {2}}   {\Upsilon}_{1}^{A_1 2}  -  \Theta_{2 {2}}   {\Upsilon}_{1}^{A_2 1} ) {\big{)}}  \dd x^2 
 \\
\displaystyle     &  &  \displaystyle
 +  {\big{(}} ( {\Upsilon}_{1}^{A_1 2}  \Theta_{2 {1}}  - {\Upsilon}_{1}^{A_2 1}  \Theta_{2 {1}}   ) - (  \Theta_{1 {1}}   {\Upsilon}_{2}^{A_2 1}  -  \Theta_{1 {1}}   {\Upsilon}_{2}^{A_1 2} ) {\big{)}} \dd  x^1 .  \\
\end{array}
\right.
\]

\section{\hbox{\sffamily\bfseries\slshape{Calculation of the $\hbox{\sffamily 2D}$ Lepage-Dedecker Hamiltonian}}}\label{appendixC}

 $ ^{\lceil}     $  
Let us examine each of these terms. We denote by ${\pmb{\varsigma}} =  {1}/ { \varsigma  {\big{(}}      2   -  \varsigma    {\big{)}}   } $. The terms ${{\bf k}}_{1} $-${{\bf k}}_{4} $ correspond to the terms $\langle  p , v   \rangle = {\pmb{\theta}}^{\tiny{\hbox{(DDW)}}}_{(q,p)} ({\cal Z})$:
\bee\label{gopqsz11}
\left|
\begin{array}{rcl}
\displaystyle {{\bf k}}_{1} & = &  \displaystyle       -  ( {  \varsigma}^{-1} )   \pi^{{A}_{1}1}  \pi^{{A}_{2}2}
\\
\displaystyle  {{\bf k}}_{2}  & = &  \displaystyle    \pmb{\varsigma}  \pi^{{A}_{1}2}    \bl      \pi^{{A}_{1}2}  +   {\big{(}}   1     -  \varsigma  {\big{)}}   \pi^{{A}_{2}1}      \br
\\
\displaystyle {{\bf k}}_{3}  & = &  \displaystyle       \pmb{\varsigma}  \pi^{{A}_{2}1}    \bl      \pi^{{A}_{2}1}  +   {\big{(}}   1     -  \varsigma  {\big{)}}   \pi^{{A}_{1}2}      \br
\\
\displaystyle  {{\bf k}}_{4}   & = &  \displaystyle      -  ( {  \varsigma}^{-1} )  \pi^{{A}_{2}2}  \pi^{{A}_{1}1}
\end{array}
\right.
\eee
Also the two terms which are related to the Lepage-Dedecker part:
\bee\label{gopqsz1166}
\left|
\begin{array}{rcl}
\displaystyle  {{\bf k}}_{5} & = &  \displaystyle  
  ( {  \varsigma}^{-1} )   \pi^{{A}_{2}2}    \pi^{{A}_{1}1}
\\
\displaystyle  
{{\bf k}}_{6} & = &  \displaystyle    -     \varsigma {\pmb{\varsigma}}^{2}    \bl      \pi^{{A}_{2}1}  +   {\big{(}}   1     -  \varsigma  {\big{)}}   \pi^{{A}_{1}2}      \br     \bl      \pi^{{A}_{1}2}  +   {\big{(}}   1     -  \varsigma  {\big{)}}   \pi^{{A}_{2}1}      \br 
\end{array}
\right.
\eee
And finally, the three terms  which come from the Lagrangian density:
\bee\label{gopqsz1177}
\left|
\begin{array}{rcl}
\displaystyle  
{{\bf k}}_{7}  & = &  \displaystyle  - ( {1}/{2} )  {\pmb{\varsigma}}^{2}  
 \bl      \pi^{{A}_{2}1}  +   {\big{(}}   1     -  \varsigma  {\big{)}}   \pi^{{A}_{1}2}      \br      \bl      \pi^{{A}_{2}1}  +   {\big{(}}   1     -  \varsigma  {\big{)}}   \pi^{{A}_{1}2}      \br 
\\
\displaystyle  
{{\bf k}}_{8} & = &  \displaystyle   - ( {1}/{2} ) {\pmb{\varsigma}}^{2}  
 \bl      \pi^{{A}_{1}2}  +   {\big{(}}   1     -  \varsigma  {\big{)}}   \pi^{{A}_{2}1}      \br    \bl      \pi^{{A}_{1}2}  +   {\big{(}}   1     -  \varsigma  {\big{)}}   \pi^{{A}_{2}1}      \br   
\\
\displaystyle  
{{\bf k}}_{9} & = &  \displaystyle  {\pmb{\varsigma}}^{2} 
 \bl      \pi^{{A}_{2}1}  +   {\big{(}}   1     -  \varsigma  {\big{)}}   \pi^{{A}_{1}2}      \br    \bl      \pi^{{A}_{1}2}  +   {\big{(}}   1     -  \varsigma  {\big{)}}   \pi^{{A}_{2}1}      \br 
\end{array}
\right.
\eee

Let us consider  the equations~${{\bf k}}_{1}$,~${{\bf k}}_{4}$~and~${{\bf k}}_{5}$ in~\eqref{gopqsz11}. We denote by $(\mathfrak{i}) = {{\bf k}}_{1} + {{\bf k}}_{4} + {{\bf k}}_{5} $ so that 
\bee
(\mathfrak{i})  =     -  ( {  \varsigma}^{-1} )   \pi^{{A}_{2}2}    \pi^{{A}_{1}1}.
\eee
We denote   $(\mathfrak{ii}) = {{\bf k}}_{2} + {{\bf k}}_{3} $, so that:
\bee\label{LDLDM2D20}
(\mathfrak{ii}) 
 =  {\pmb{\varsigma}}  \bl      \pi^{{A}_{1}2}   \pi^{{A}_{1}2}  +   \pi^{{A}_{1}2}  {\big{(}}   1     -  \varsigma  {\big{)}}   \pi^{{A}_{2}1}      +
 \pi^{{A}_{2}1}  \pi^{{A}_{2}1}  +   \pi^{{A}_{2}1}  {\big{(}}   1     -  \varsigma  {\big{)}}   \pi^{{A}_{1}2}  \br .
\eee
It remains the following equations ${{\bf k}}_{6}$-${{\bf k}}_{9}$.
 We have respectively: 
 \bee\label{gopqsz1144}
\left.
\begin{array}{rcl}
\displaystyle {{\bf k}}_{6} & = &  \displaystyle  \bl   -     \varsigma {\pmb{\varsigma}}^{2}     \br 
\bl  \pi^{{A}_{2}1}     \pi^{{A}_{1}2}  +  \pi^{{A}_{2}1}     {\big{(}}   1     -  \varsigma  {\big{)}}   \pi^{{A}_{2}1}      + {\big{(}}   1     -  \varsigma  {\big{)}}   \pi^{{A}_{1}2}     \pi^{{A}_{1}2}  +   {\big{(}}   1     -  \varsigma  {\big{)}}   \pi^{{A}_{1}2}  {\big{(}}   1     -  \varsigma  {\big{)}}   \pi^{{A}_{2}1}    
\br
\\
\displaystyle  & = &  \displaystyle \bl   -     \varsigma {\pmb{\varsigma}}^{2}     \br 
\bl    {\big{(}} 1 + {\big{(}}   1     -  \varsigma  {\big{)}}^2  {\big{)}}  \pi^{{A}_{2}1}     \pi^{{A}_{1}2}  + \bl  \pi^{{A}_{2}1}  \br^2   {\big{(}}   1     -  \varsigma  {\big{)}}     + {\big{(}}   1     -  \varsigma  {\big{)}}  \bl \pi^{{A}_{1}2}    \br^2 
\br 
\\
\displaystyle & = &  \displaystyle  \bl   -     \varsigma {\pmb{\varsigma}}^{2}     \br 
\bl   {\big{(}} 2 ( 1     -  \varsigma) + \varsigma^2    {\big{)}}  \pi^{{A}_{2}1}     \pi^{{A}_{1}2}  + \bl  \pi^{{A}_{2}1}  \br^2   {\big{(}}   1     -  \varsigma  {\big{)}}     + {\big{(}}   1     -  \varsigma  {\big{)}}  \bl \pi^{{A}_{1}2}    \br^2   
\br
\end{array}
\right.
\eee
The second and the third give
 \bee\label{gopqsz1155}
\left.
\begin{array}{rcl}
\displaystyle {{\bf k}}_{7} & = &  \displaystyle  \bl  - ( {1}/{2} )  {\pmb{\varsigma}}^{2}   \br  \bl \bl      \pi^{{A}_{2}1}   \br^2 +   {\big{(}}   1     -  \varsigma  {\big{)}}^2  \bl   \pi^{{A}_{1}2}  \br^2    + 2     \pi^{{A}_{2}1}    {\big{(}}   1     -  \varsigma  {\big{)}}   \pi^{{A}_{1}2}    \br
\\
\displaystyle {{\bf k}}_{8} & = &  \displaystyle  \bl  - ( {1}/{2} )  {\pmb{\varsigma}}^{2}    \br   \bl  \bl  \pi^{{A}_{1}2}  \br +   {\big{(}}   1     -  \varsigma  {\big{)}}^2  \bl  \pi^{{A}_{2}1}  \br^2 + 2       \pi^{{A}_{1}2}    {\big{(}}   1     -  \varsigma  {\big{)}}   \pi^{{A}_{2}1}    \br 
\end{array}
\right.
\eee
Now, we denote $ ( \mathfrak{iii}) =  {{\bf k}}_{6}   +   {{\bf k}}_{9}  $, so that:
 \bee\label{LDLDM2D21}
( \mathfrak{iii}) 
   = (1- \varsigma)   {\pmb{\varsigma}}^{2}  \BL  {\big{(}} 2 ( 1     -  \varsigma) + \varsigma^2    {\big{)}}  \pi^{{A}_{2}1}     \pi^{{A}_{1}2}  + \bl  \pi^{{A}_{2}1}  \br^2   {\big{(}}   1     -  \varsigma  {\big{)}}     + {\big{(}}   1     -  \varsigma  {\big{)}}  \bl \pi^{{A}_{1}2}    \br^2          \BR .
  \eee
  and finally we denote $ ( \mathfrak{iv})  = {{\bf k}}_{7}    + {{\bf k}}_{8}  $, then  :
\bee\label{biocxp00}
( \mathfrak{iv})  =   - \frac{1}{2}  {\pmb{\varsigma}}^{2} \BL  \bl      \pi^{{A}_{2}1}   \br^2 +   {\big{(}}   1     -  \varsigma  {\big{)}}^2  \bl   \pi^{{A}_{1}2}  \br^2    + 2     \pi^{{A}_{2}1}    {\big{(}}   1     -  \varsigma  {\big{)}}   \pi^{{A}_{1}2}     
 +  \bl  \pi^{{A}_{1}2}  \br^{2} +   {\big{(}}   1     -  \varsigma  {\big{)}}^2  \bl  \pi^{{A}_{2}1}  \br^2 + 2       \pi^{{A}_{1}2}    {\big{(}}   1     -  \varsigma  {\big{)}}   \pi^{{A}_{2}1}  \BR.
\eee
Finally, we     compute 
$(\mathfrak{ii})  + (\mathfrak{iii}) + (\mathfrak{iv})$.
We   introduce the following notations: 
\bee
 \pi^{\circ\circ}  =  \pi^{{A}_{1}2}   \pi^{{A}_{1}2}  = \bl  \pi^{{A}_{1}2}  \br^2,
\quad \quad 
 \pi^{\circ\bullet}  =  \pi^{{A}_{1}2}   \pi^{{A}_{2}1} ,
\quad \quad 
 \overline{\pi}^{\circ\bullet}  =  \pi^{{A}_{1}1}   \pi^{{A}_{2}2} ,
\quad \quad 
 \pi^{\bullet\bullet}  =  \pi^{{A}_{2}1}   \pi^{{A}_{2}1}  = \bl  \pi^{{A}_{2}1}  \br^2.
\eee
So that the equations~\eqref{LDLDM2D20}~\eqref{LDLDM2D21} and \eqref{biocxp00} are written~\eqref{LDLDM2D30}~$(\mathfrak{ii})$-$(\mathfrak{iv})$:
\bee\label{LDLDM2D30}
\left.
\begin{array}{ccl}
\displaystyle
 ( \mathfrak{ii})    & = &  \displaystyle  
   {\pmb{\varsigma}}   \BL   \pi^{\circ\circ}   +   2  {\big{(}}   1     -  \varsigma  {\big{)}}   \pi^{\circ\bullet}     +
 \pi^{\bullet\bullet}     \BR,
\\
\displaystyle
 ( \mathfrak{iii})    & = &  \displaystyle  
 (1- \varsigma) [ {\pmb{\varsigma}} ]^{2}    \BL  {\big{(}} 2 ( 1     -  \varsigma) + \varsigma^2    {\big{)}}   \pi^{\circ\bullet}  
    +   \pi^{\bullet\bullet}    {\big{(}}   1     -  \varsigma  {\big{)}}     + {\big{(}}   1     -  \varsigma  {\big{)}}    \pi^{\circ\circ}          \BR,
\\
\displaystyle
 ( \mathfrak{iv})    & = &  \displaystyle  
  -  {1}/{2} [ {\pmb{\varsigma}} ]^{2}  
   \BL   \pi^{\bullet\bullet} +   {\big{(}}   1     -  \varsigma  {\big{)}}^2   \pi^{\circ\circ}    + 2        {\big{(}}   1     -  \varsigma  {\big{)}}   \pi^{\circ\bullet}    
  +   \pi^{\circ\circ} +   {\big{(}}   1     -  \varsigma  {\big{)}}^2      \pi^{\bullet\bullet} + 2          {\big{(}}   1     -  \varsigma  {\big{)}}   \pi^{\circ\bullet}  \BR.
  \end{array}
  \right.
\eee
where we have denoted $ {\pmb{\varsigma}}  =    \bl  \varsigma {\big{(}}      2   -  \varsigma    {\big{)}}    \br^{-1}  $
So that~\eqref{LDLDM2D30}-$(\mathfrak{ii})$  is written:  
\[
 ( \mathfrak{ii})  
 =   [{\pmb{\varsigma}}]^{2} \bl    \pi^{\circ\circ}   +     {\big{(}}   2    -  2\varsigma  {\big{)}}   \pi^{\circ\bullet}     +
 \pi^{\bullet\bullet}     \br  \varsigma  {\big{(}}      2   -  \varsigma    {\big{)}}  
 =   [{\pmb{\varsigma}}]^{2}
   \bl
       \pi^{\circ\circ}   +    2  \pi^{\circ\bullet}     -  2\varsigma   \pi^{\circ\bullet}     +
 \pi^{\bullet\bullet}     \br    {\big{(}}      2 \varsigma   -  \varsigma^2    {\big{)}}  .
\]
If we denote $
{\varphi}  =  {\big{(}} 2  
     \varsigma^2  {\big{(}}      2   -  \varsigma    {\big{)}}^2   {\big{)}}^{-1}
$, we obtain:
\bee 
(\mathfrak{ii})   = 2 {\varphi} 
{\Big{(}}
2   \pi^{\circ\circ}   \varsigma 
-  \pi^{\circ\circ}    \varsigma^2 
+ 
4 \pi^{\circ\bullet}  \varsigma
- 2  \pi^{\circ\bullet}  \varsigma^2
-
4 \pi^{\circ\bullet} \varsigma^2
+ 2 \pi^{\circ\bullet} \varsigma^3
+
2 \pi^{\bullet\bullet} \varsigma
-
 \pi^{\bullet\bullet} \varsigma^2
 {\Big{)}}   .
 \nonumber
\eee
The equation \eqref{LDLDM2D30}-$(\mathfrak{iii})$ is written: 
\[
\left.
\begin{array}{rcl}
\displaystyle  
(\mathfrak{iii})       & = &  \displaystyle 
    2   {\varphi}  (1- \varsigma)  \BL  {\big{(}} 2 ( 1     -  \varsigma) + \varsigma^2    {\big{)}}   \pi^{\circ\bullet}  
    +   \pi^{\bullet\bullet}    {\big{(}}   1     -  \varsigma  {\big{)}}     + {\big{(}}   1     -  \varsigma  {\big{)}}    \pi^{\circ\circ}          \BR
\\
\displaystyle  
   & = &  \displaystyle 
   2  {\varphi}    (1- \varsigma)  \BL  {\big{(}} 2   \pi^{\circ\bullet}     -  2 \varsigma  \pi^{\circ\bullet} + \varsigma^2   \pi^{\circ\bullet}
    +   \pi^{\bullet\bullet}      -  \varsigma  \pi^{\bullet\bullet} 
         +   \pi^{\circ\circ}    -  \varsigma    \pi^{\circ\circ}          \BR
\\
\displaystyle  
    & = &  \displaystyle 
    2 {\varphi}  \BL   2   \pi^{\circ\bullet}     -  2 \varsigma  \pi^{\circ\bullet} + \varsigma^2   \pi^{\circ\bullet}
    +   \pi^{\bullet\bullet}      -  \varsigma  \pi^{\bullet\bullet} 
         +   \pi^{\circ\circ}    -  \varsigma    \pi^{\circ\circ}      
    - 2   \pi^{\circ\bullet}   \varsigma 
    \\
\displaystyle  
    &  &  \displaystyle 
      +  2 \varsigma^2  \pi^{\circ\bullet} 
     - \varsigma^3   \pi^{\circ\bullet}
    -  \varsigma   \pi^{\bullet\bullet} 
         +  \varsigma^2  \pi^{\bullet\bullet} 
         -  \varsigma   \pi^{\circ\circ}    
         +   \varsigma^2    \pi^{\circ\circ}          \BR .
\end{array}
\right.
\]
and finally  \eqref{LDLDM2D30}-$(\mathfrak{iv})$ is written: 
\[
\left.
\begin{array}{rcl}
\displaystyle  
(\mathfrak{iv})       & = &  \displaystyle   -     {\varphi}  \BL   \pi^{\bullet\bullet} +
  {\big{(}}   1     -  \varsigma  {\big{)}}^2      \pi^{\bullet\bullet} 
  +   {\big{(}}   1     -  \varsigma  {\big{)}}^2   \pi^{\circ\circ}       
  +   \pi^{\circ\circ} 
  + 4          {\big{(}}   1     -  \varsigma  {\big{)}}   \pi^{\circ\bullet}  \BR 
  \\
  \displaystyle      & = &  \displaystyle     -      {\varphi}   \BL  2 \pi^{\bullet\bullet} 
 - 2  \varsigma  \pi^{\bullet\bullet}
 +
 { \varsigma^2 }      \pi^{\bullet\bullet} 
+   2 \pi^{\circ\circ} 
 - 2  \varsigma  \pi^{\circ\circ}
 +
 { \varsigma^2 }      \pi^{\circ\circ} 
  + 4        \pi^{\circ\bullet}       - 4 \varsigma   \pi^{\circ\bullet}  \BR .
  \end{array}
  \right.
\]
We now writes ${\cal H} =  (\mathfrak{i})  +  (\mathfrak{ii})  + (\mathfrak{iii}) + (\mathfrak{iv}) $.
\[
\left.
\begin{array}{rcl}
\displaystyle  
{\cal H}       & = &  \displaystyle  
     (\mathfrak{i})  + 
 {\varphi}   
\BL  4  \pi^{\circ\circ}   \varsigma 
-  2 \pi^{\circ\circ}    \varsigma^2 
+ 
8 \pi^{\circ\bullet}  \varsigma
- 4  \pi^{\circ\bullet}  \varsigma^2
-
8 \pi^{\circ\bullet} \varsigma^2
+ 4 \pi^{\circ\bullet} \varsigma^3
+
4 \pi^{\bullet\bullet} \varsigma
-
2
 \pi^{\bullet\bullet} \varsigma^2
 +
   4   \pi^{\circ\bullet}    
   \\
\displaystyle       &  &  \displaystyle 
    -  4 \varsigma  \pi^{\circ\bullet} + 2 \varsigma^2   \pi^{\circ\bullet}
    + 2  \pi^{\bullet\bullet}      - 2 \varsigma  \pi^{\bullet\bullet} 
         +  2 \pi^{\circ\circ}    - 2 \varsigma    \pi^{\circ\circ}         
    - 4   \pi^{\circ\bullet}   \varsigma 
      +  4 \varsigma^2  \pi^{\circ\bullet} 
     -  2 \varsigma^3   \pi^{\circ\bullet}
    -   2 \varsigma   \pi^{\bullet\bullet} 
    \\
\displaystyle       &  &  \displaystyle   
         +  2 \varsigma^2  \pi^{\bullet\bullet} 
         - 2 \varsigma   \pi^{\circ\circ}    
         + 2   \varsigma^2    \pi^{\circ\circ}            
  -    2  \pi^{\bullet\bullet} 
+ 2\varsigma  \pi^{\bullet\bullet}
- 
 { \varsigma^2 }      \pi^{\bullet\bullet} 
-2 \pi^{\circ\circ} 
+ 2 \varsigma  \pi^{\circ\circ}
-  
 { \varsigma^2 }      \pi^{\circ\circ} 
  -4       \pi^{\circ\bullet}       
  +4  \varsigma   \pi^{\circ\bullet}   
\BR
\\
\displaystyle  
     & = &  \displaystyle 
 {\varphi}  \BL
 2 \varsigma  \pi^{\circ\circ} 
  -  
 { \varsigma^2 }      \pi^{\circ\circ}    
 + 2 \pi^{\bullet\bullet} \varsigma  
 -  { \varsigma^2 }      \pi^{\bullet\bullet}  
 + 2 \pi^{\circ\bullet} \varsigma^3
 -
6 \pi^{\circ\bullet} \varsigma^2 
+  4 \pi^{\circ\bullet} \varsigma 
\BR +  (\mathfrak{i})
\\
\displaystyle  
    & = &  \displaystyle 
 {\varphi} 
\BL {\big{(}} 2 - \varsigma   {\big{)}} \pi^{\circ\circ} 
 { \varsigma }      
 +  {\big{(}} 2 - \varsigma   {\big{)}}      \pi^{\bullet\bullet}  
  { \varsigma }      
 + 2 \varsigma^2  {\big{(}} \varsigma - 3 {\big{)}} \pi^{\circ\bullet}  +  4 \pi^{\circ\bullet} \varsigma \BR
 +
  (\mathfrak{i})
\\
\displaystyle  
     & = &  \displaystyle 
 {\varphi}  \BL
     {\big{(}} 2 - \varsigma    {\big{)}}   \varsigma    
  \BL   \pi^{\circ\circ}      
 +         \pi^{\bullet\bullet}    
 \BR
  + 2 \varsigma^2  {\big{(}} \varsigma - 3 {\big{)}} \pi^{\circ\bullet}  +  4 \pi^{\circ\bullet} \varsigma \BR
  +  (\mathfrak{i}) .
  \quad      \rfloor
\end{array}
\right.
\]

\protect\label{conclusion}

\protect\label{mvf}

\pdfbookmark[1]{References}{ref}

\LastPageEnding

\end{document}